# INTELLIGENT LOAD BALANCING IN CLOUD COMPUTER SYSTEMS

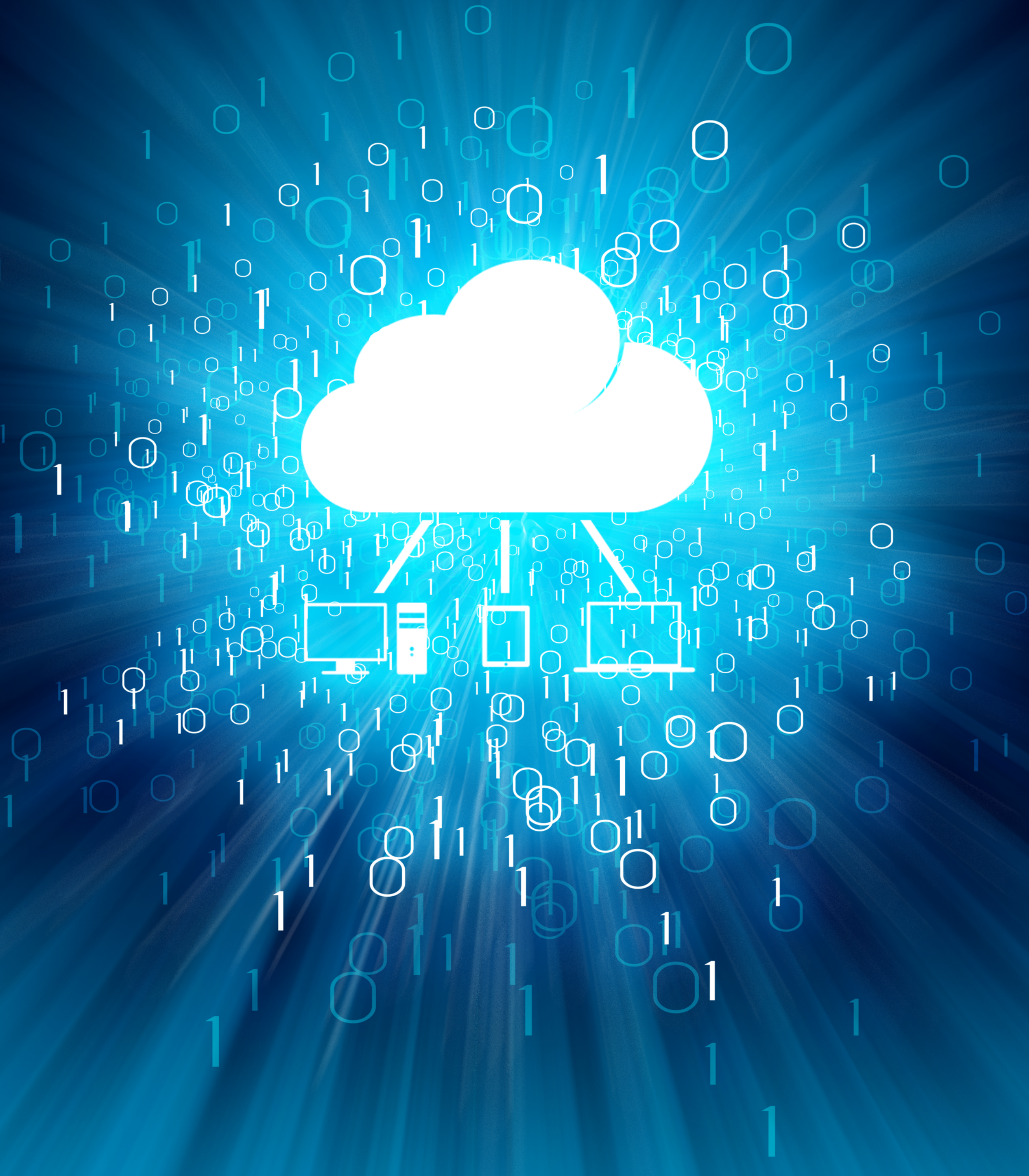

LESZEK SLIWKO

**A thesis submitted in partial fulfilment of the requirements of the University of Westminster for the degree of Doctor of Philosophy**

AUGUST 2018

*(CORRECTIONS: JANUARY 2019)*

## ABSTRACT

Cloud computing is an established technology allowing users to share resources on a large scale, never before seen in IT history. A cloud system connects multiple individual servers in order to process related tasks in several environments at the same time. Clouds are typically more cost-effective than single computers of comparable computing performance. The sheer physical size of the system itself means that thousands of machines may be involved. The focus of this research was to design a strategy to dynamically allocate tasks without overloading Cloud nodes which would result in system stability being maintained at minimum cost. This research has added the following new contributions to the state of knowledge: (i) a novel taxonomy and categorisation of three classes of schedulers, namely OS-level, Cluster and Big Data, which highlight their unique evolution and underline their different objectives; (ii) an abstract model of cloud resources utilisation is specified, including multiple types of resources and consideration of task migration costs; (iii) a virtual machine live migration was experimented with in order to create a formula which estimates the network traffic generated by this process; (iv) a high-fidelity Cloud workload simulator, based on a month-long workload traces from Google's computing cells, was created; (v) two possible approaches to resource management were proposed and examined in the practical part of the manuscript: the centralised metaheuristic load balancer and the decentralised agent-based system. The project involved extensive experiments run on the University of Westminster HPC cluster, and the promising results are presented together with detailed discussions and a conclusion.

# LIST OF CONTENTS

# LIST OF FIGURES

# LIST OF TABLES

# ACKNOWLEGEMENTS

I would like to express the deepest appreciation to Director of Studies, Professor Vladimir Getov, for his continuously upbeat attitude, his boundless support, patience and availability with regards to this research and scholarships, his excellence in teaching and all his other advice. I would especially like to thank him for introducing me to the amazing world of Cluster computing. Special thanks are also due to Dr Alexander Bolotov for his encouragement and invaluable help with this manuscript. Without his guidance and suggestions, this project would not have been possible.

I would also like to thank the University of Westminster for their permission to use their High-Performance Computing Centre, a massive help during this research. I would like to thank all the IT staff for all their support and incredible patience with setting up and executing research applications in this environment.

In addition, I would like to thank all researchers who have published and shared their work. In particular, I would like to thank Google's engineers and scientists: Joseph Hellerstein, Abhishek Verma, John Wilkes, Charles Reiss, Malte Schwarzkopf and many others, for freely releasing workload traces and explaining the workings of the Borg system in their publications.

This research was supported by a grant from the Student Development Fund 2016 at the University of Westminster, London.

# DECLARATION

I declare that all the material contained in this thesis is my own work. Parts of the work presented in this dissertation have been published in the following journals and conferences:

1. Sliwko, Leszek. "A Scalable Service Allocation Negotiation For Cloud Computing." Journal of Theoretical and Applied Information Technology, Vol.96. No 20, pp. 6751-6782, 2018.
2. Sliwko, Leszek, Vladimir Getov, and Alexander Bolotov. "Intelligent Load Balancing in Cloud Computer Systems." University of Westminster. Faculty of Science and Technology Doctoral Conference 2018, pp. 23.
3. Sliwko, Leszek, and Vladimir Getov. "Transfer Cost of Virtual Machine Live Migration in Cloud Systems." University of Westminster. Technical Report. November, 2017: 1-21.
4. Sliwko, Leszek, and Vladimir Getov. "AGOCS – Accurate Google Cloud Simulator Framework." In Scalable Computing and Communications Congress, 2016 Intl IEEE Conferences, pp. 550-558. IEEE, 2016.
5. Sliwko, Leszek, and Vladimir Getov. "A Meta-Heuristic Load Balancer for Cloud Computing Systems." In Computer Software and Applications Conference, 2015 IEEE 39th Annual, vol. 3, pp. 121-126. IEEE, 2015.
6. Sliwko, Leszek, and Vladimir Getov. "Workload Schedulers-Genesis, Algorithms and Comparisons." International Journal of Computer Science and Software Engineering 4, no. 6 (2015): 141-155.
7. Sliwko, Leszek, Vladimir Getov, and Alexander Bolotov. "Distributed Agent-Based Load Balancer for Cloud Computing." Automated Reasoning Workshop, 2015.
8. Sliwko, Leszek. "An Overview of Java Multi-Agent System Balancer." International Journal of Computational Intelligence and Information Security, Vol. 1, No. 2, pp. 4-11, 2010.

9. Sliwko, Leszek, and Aleksander Zgrzywa. "A Novel Strategy for Multi-Resource Load Balancing in Agent-Based Systems." International Journal of Intelligent Information and Database Systems 3, No. 2, pp. 180-202, 2009.
10. Sliwko, Leszek. "A Reinforced Evolution-Based Approach to Multi-Resource Load Balancing." Journal of Theoretical and Applied Information Technology, Vol. 4, No. 8, pp. 717-724, 2008.
11. Sliwko, Leszek, and Aleksander Zgrzywa. "Multi-Resource Load Optimization Strategy in Agent-Based Systems." Lecture Notes in Computer Science, 1(4496), pp. 348-357, 2007.
12. Sliwko, Leszek, and Ngoc Thanh Nguyen. "Using Multi-Agent Systems and Consensus Methods for Information Retrieval in Internet." International Journal of Intelligent Information and Database Systems 1, no. 2 (2007): 181-198.

The following relevant presentations have been delivered:

- Sliwko, Leszek. "Scala on 40-core HPC machine." Scala in the City #5. Signify Technology. June 27, 2018.
- Sliwko, Leszek. "Running Akka and Scala on a high-performance computing (HPC) system with 800 cores and 1.8 TB RAM." The Forge Talk. UK Home Office. May 31, 2018.

# 1. INTRODUCTION

The main focus of this project was to research and design a feasible strategy for managing and balancing a workload within a virtualised Cloud system – a system in which the computing cells are built from many thousands of networked nodes, and where the workload is significantly diversified and consists of short-lived batch jobs as well as long-lasting services. The shape of resources utilised by running tasks changes rapidly, thereby creating a very dynamic environment.

For a working solution to be designed, the first step required is to identify the challenges related to the allocation of tasks in different environments. Therefore, the initial part of this research focuses on analysing currently-utilised scheduling schemes and shortlisting areas which has potential for improvements.

Chapter 2 presents a novel taxonomy and categorisation of workload schedulers, focusing in particular on the key design factors that affect the scalability of a given solution, in addition to the features which improved the scheduler's architecture. This chapter describes their evolution, from early adoption to their modern implementations; in doing so, it sets out in detail their scheduling algorithms. This background review notes a trend towards the greater parallelisation of all three classes of examined schedulers, a factor which shaped the approaches adopted later in the research.

This introductory chapter explains the motivation behind this project, its research background, and how it has evolved over time.

## 1.1. PROJECT MOTIVATION

The biggest cloud systems offering elastic resource allocation are: Amazon EC2 (Jackson et al., 2010), Microsoft Azure (Li et al., 2010), Google Cloud Platform (Bedra, 2010), IBM Cloud (Kochut et al., 2011), Oracle Cloud (Jain and Mahajan, 2017), Alibaba Cloud (Zhou, 2017), Rackspace (Li et al., 2010) and GoGrid (ibid.). While the information on the size of the largest Cloud is not publicly available, Bloomberg Technology estimates that the Amazon EC2 consists of 1.5 million servers (Clark, 2014), while Gartner, Inc., an American technology research and advisory firm approximates its size to be more than two million servers. The total number of nodes in Google Cloud is estimated to be ca. 900k machines, making this market extremely competitive, with enormous forecasted market size growth. Gartner predicts a 17% annual growth in public spending for cloud services, reaching $411.4 billion by 2020 (Van der Meulen and Pettey, 2017). A few well-known examples of services backed up by cloud computing include Dropbox, Gmail, Twitter, Facebook and YouTube.

Clouds are typically more cost-effective than single computers of comparable speed, and usually enable applications to have higher availability than a single machine. This makes the software even more attractive as a service and is shaping the way applications are built today. Companies no longer need to be concerned with maintaining a huge infrastructure of thousands of servers in order to have enough computing power for those critical hours when their service is in highest demand. Instead, companies can simply rent a fleet of servers for a few hours (Wang et al., 2018).

Across the history of IT, such an elasticity of resources without paying a premium for a large-scale usage is exceptional. Recent developments in Big Data systems and Machine Learning technologies have fuelled growth in demand for cheap computing power; in response, several vendors have collaborated and the range of computing services offered to the market has significantly expanded. Prices have also been driven down and, as of 24 July 2018, the cost of renting a general-use instance of 16-core machine with 64GB memory was 80 cents per hour (data from aws.amazon.com/ec2/pricing website). These Cloud properties are particularly important for both small and medium-sized enterprises, who are able to minimise their initial outlay for building IT infrastructure. They can also focus on swiftly delivering the product to the market, a fact which is critical for any innovative proposal. The rapid development of Cloud technologies has introduced a new set of challenges and problems which require immediate solution. Cloud systems are usually made up of machines with different hardware configurations and capabilities (Mateescu et al., 2011), and these systems can be rapidly configured based on the user's requirements (Buyya et al., 2009). Therefore, dynamic resource sharing is a necessity. Resource management has been an active research area for a considerable period of time and the systems often feature a highly specialised load balancing strategies such as Google’s Borg (Burns et al., 2016), Microsoft’s Apollo (Boutin et al., 2014) or Alibaba’s Fuxi (Zhang et al., 2014b). Since larger computing cells are likely to be required in the near future (Wilkes, 2016), Cloud load balancing is a topic worthy of dedicated research.

The focus of this project was to examine possible solutions to allocating and managing many concurrently running tasks in a Cloud system. The initial assumption was that existing Cloud management software could be improved by deploying intelligent load balancing routines and therefore, achieving a better allocation quality and higher system scalability. The main novel aspects of this approach were to schedule the incoming tasks, which allows running programs to be offloaded to alternative system nodes on the fly (hence the name ‘load balancing’), in addition to designing algorithms capable of proactively managing a workload in such a dynamic environment (hence the name ‘intelligent’). This research breaks with the concept that the execution of a task in a cluster is immovable or unstoppable, and instead examines the available technology to implement such a strategy. Since none of the commercially available cluster schedulers realise such a feature, the objective of this research is to implement a working prototype for the Cloud load balancer, and to evaluate their performance advances emerging out of the designed solution.

## 1.2. RESEARCH PROBLEM

The background review and crafting of the schedulers' taxonomy (Chapter 2) helped to formally define the D-Resource System Optimisation Problem (D-RSOP) which is later discussed in Chapter 3. The presented model consists of nodes and tasks with the main function of the load balancer being to keep a good load balance through resource vectors comparisons. D-RSOP belongs to the NP-Hard problems class which are believed to be unsolvable in polynomial time, i.e. the 'P versus NP' problem (Frieze, 1986). Cloud systems are focused on maintaining the continuity of third party operations with minimum disturbance; therefore, this model also considers the cost of change when deploying new tasks or when re-allocating existing tasks.

Following the study presented in Chapter 3, the two first goals of the load balancing solution were formed, namely:

- Goal (I) – maintaining a global balance across the Cloud system so that an individual node is not overloaded. In virtualised Cloud environments, this goal is achieved through the Virtual Machine Live Migration (VM-LM) allowing a running program to be migrated to alternative nodes without stopping their execution. The formal definition is presented as (2) in section 3.3.
- Goal (II) – minimising the System Transformation Cost (STC) which is the global cost of task re-allocations on Cloud infrastructure, i.e. minimising the total size of data transferred across the Cloud's network during VM-LM process. For detailed explanation see (5) in section 3.3.

These goals will now shape the initial concepts of the balancing strategy presented in the subsequent sections. However, the D-RSOP model relies mainly on the VM-LM feature to re-allocate tasks between nodes. Further research was required to establish a reliable technique for measuring the VM-LM cost within the Cloud infrastructure.

## 1.3. RESEARCH PLAN

With the expected outcome established, the project was then able to move to the planning phase in which further challenges were identified. This included the unknown impact of offloading a running task and the lack of sufficiently detailed Cloud simulation tools. Considering the diversity of the research areas involved, the decision was made to execute the project in consecutive stages as presented in Figure 1:

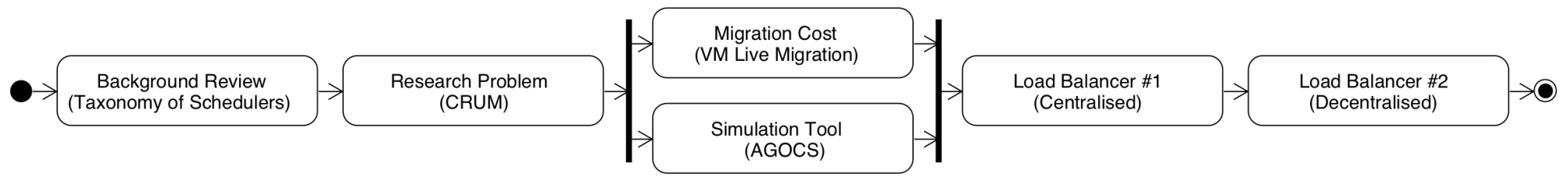

Figure 1: Research project stages

The above plan of action allowed the gradual refinement of the project goals as more knowledge was acquired. As the steps were completed, the foundations of the load balancer prototype were incrementally built. The flow was followed with the exception of a few selected routines in the centralised metaheuristic load balancer code implemented early in the project as a proof of concept.

In order to improve the readability of the manuscript, each stage of the project has a dedicated chapter (Chapters 2 to 7) which contains a literature review, core content and a detailed summary. The below sections summarise the main outcomes realised in those stages and list the main achievements, while the overall conclusions are covered in Chapter 8.

## 1.4. MIGRATION COST

Chapter 4 details the VM-LM feature which forms the backbone of the proposed solution. It allows a working application, within a VM instance, to be migrated to an alternative node without stopping its execution. This technology allows the dynamic balancing of the workload between suitable nodes within the Cloud system.

The existing research focused on examining the impact of VM-LM on the VM instance, such as: (i) the impact of allocated VM memory size on migration time (Zhao and Figueiredo, 2007; Salfner et al., 2011; Dargie, 2014); (ii) the impact of memory page dirtying rate on migration time (Verma et al., 2011, Rybina et al., 2015) and the downtime length (Salfner et al., 2011; Liu et al., 2013); (iii) the effect of available network bandwidth on migration time (Akoush et al., 2010; Zhang et al., 2016; Deshpande and Keahey, 2017); (iv) the energy overhead required to perform VM-LM (Huang et al., 2011; Liu et al., 2013; Callau-Zori et al., 2017), (v) determining the Quality of Service specifications for migrated VMs and applying resource control mechanisms during VM-LM (Abali et al., 2017), (vi) a strategy for parallel migrations of multiple VMs (Sun et al., 2016), (vii) various memory transfer optimisations as presented in Noel and Tsirkin (2016), Tsirkin and Noel (2016), Ramasubramanian and Ahmed (2017).

However, the migration of VM instances causes disruptions at the infrastructure level when non-trivial volumes of data need to be transferred and network bandwidth which could be allocated to alternative processes is consumed. The research work presented in Chapter 4 evaluates the overall cost of this process on the network, rather than only on individual nodes. Figure 2 visualises the process of VM-LM:

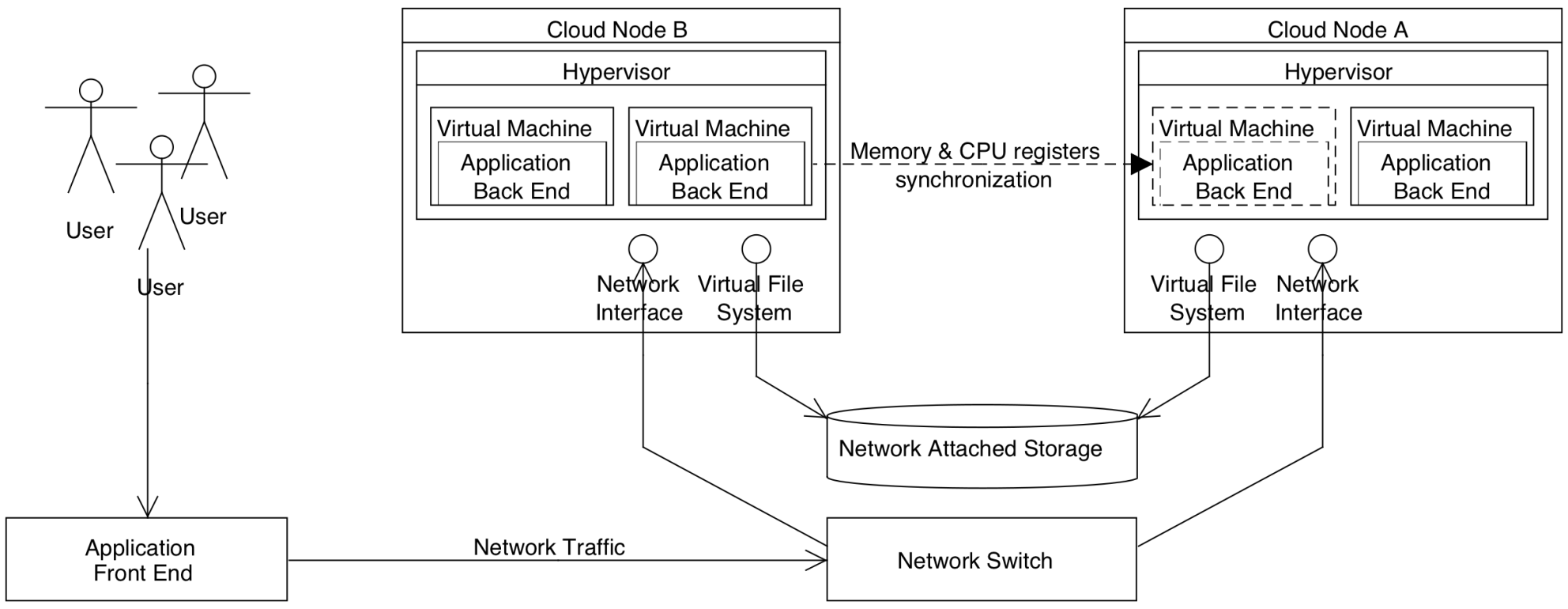

Figure 2: Virtual Machine Live Migration process

Chapter 4 presents an analysis of the five major areas of the VM-LM process, namely: CPU registers, memory, permanent storage, network switching and code structure and dynamics – and analyses their impact on the size of the migrated data. However, to provide a reliable VM-LM cost estimation technique, actual practical experiments were required.

The next phase of the project involved setting up an isolated network of several machines with VirtualBox installed, in addition to measuring the size of the transferred data during VM-LM between them. VirtualBox was chosen due to its universal compatibility with hardware, popularity and easy-to-use GUI management console. Additionally, VirtualBox is an Open Source project and its code could be analysed with a focus on VM-LM. During experiments, a Live Migration Data Transfer (LMDT) formula was devised which could be successfully used to estimate data transferred during VM-LM.

## 1.5. SIMULATION TOOL

The D-RSOP model was based on a conceptual analysis and it was clear that a more practical approach was necessary since the project could not progress further without workload data from a real-world Cloud environment. As discussed in Chapter 5, realistic workload data input could be obtained via two main approaches:

- Using an artificial Cloud workload generator (Beitch et al., 2010; Ganapathi et al., 2010; Wang et al., 2011; Malhotra and Jain, 2013).
- Acquiring and parsing real-world workload traces (Iosup et al., 2008; Hellerstein et al., 2010; Kavulya at al., 2010; Klusáček, 2014; Feitelson et al., 2014) to a format which could be used in further research.

Upon detailed examination, existing artificial workload generators such as CloudSim, GreenCloud and EMUSIM did not provide the necessary resource utilisation statistics that could be used in this project. Such accurate and realistic parameters could be obtained only from actual workload traces.

Given this scenario, the best option was to acquire and parse real-world workload traces and base the simulation on these. Additionally, one of the project's practical activities was to examine actual workload traces to better understand challenges in workload planning. For this, it was possible to retrieve and analyse traces from the Google Cluster Data (GCD) project (Hellerstein et al., 2010). GCD workload traces are month-long, and contain processing data from a computing cell of ca. 12.5k nodes. Google services are constantly utilised, 24-hours a day, from any location around the globe. As such, they provide a good variety of tasks found within the production environment. Additionally, GCD are generally of a high quality and only a small number of anomalies are present.

Cloud environments can have a very complex structure. This is the result of not only the sheer size of workload, but also the relationships between the nodes and tasks executed on them. One should also consider the overall high dynamicity of a typical Cloud environment where running programs dynamically allocate and release resources such as memory and CPU cores. During examination of GCD workload traces structure (Reiss et al., 2013), additional major complications were noted. Based on this, two further load balancing strategy goals were added:

- Goal (III) – aside from being able to allocate enough resources, nodes should also match the constraints of tasks. The four tasks' constraints types were defined as in the GCD structure: equal, not equal, greater than and less than. For example, a task might require a node with an external IP address. In such cases it will define a constraint which requires the IP address flag to be equal to true. Subsection 5.5.2 introduces the concept of Task Constraints.
- Goal (IV) – the solution should handle the occurrences of Resource Usage Spikes (RUS), where a running program significantly increases its resource consumption in a short period of time. User-defined required resources (i.e. resources not currently being used but which are defined in task specification) from all production tasks allocated to a given node should never exceed this node capacity. The node, therefore, should always be able to execute all its production tasks at full capacity. See subsection 7.5.4 for detailed explanation of RUS.

The project's focus has shifted into creating Accurate Google Cloud Simulator (AGOCS) framework, a high-fidelity Cloud workload simulator which could reliably replay month-long GCD workload traces and simulate a Cloud environment. Given the sheer size of GCD data, the main requirement for AGOCS was a highly parallel design. Therefore, AGOCS was built upon functional programming concepts with a Scala and Akka Actors/Streams framework (Roestenburg et al., 2015). AGOCS inherited many beneficial features from this technology stack, such as native support for objects immutability, lock-free collections and components, native agents' supervision strategies for recovery from data corruption errors, thread-safe TrieMap (Prokopec et al., 2012), and a mature test-kit. In order to guarantee a reasonably bug-free code, an extensive suite of test units was created.

During the research, AGOCS was deployed at the University of Westminster's HPC cluster (see Appendix C), where most of further experiments took place. AGOCS allowed the running of simulations where a given solution could be tested if and how well, it satisfies D-RSOP goals.

## 1.6. LOAD BALANCER DESIGNS

Finally, with a solid simulation environment up and running, the project reached a state where it could progress further with the design of load balancing solutions. The next steps, as presented in Chapter 6 and 7, involved designing and implementing two main load balancer prototypes:

- Centralised load balancing strategy with the use of metaheuristic algorithms – this approach has been already examined by previous researchers (Józefowska et al., 2002; Leung, 2004), yielding a satisfactory quality of results. However, it was found out the given algorithms could be slightly improved. Chapter 6 covers details of this solution.
- Decentralised load balancing with the use of an agent-based network – this approach is based on utilising the technology of software agents, cooperating to find allocations for a number of tasks on a set of machines (Kim et al., 2004; Leung et al., 2010). This work is presented in Chapter 7.

The preliminary analysis focused on the pros and cons of the above solutions. The findings are summarised in Tables 1 and 2:

| Advantages | Disadvantages |
|---|---|
| • Well-studied approach;<br>• Better control over job execution and centralised management of failover and restarting controls;<br>• Predicable behaviour;<br>• Supports complex scheduling policies and fairness. | • Single point of failure – prone to 'head-of-line' blocking job;<br>• Complex strategies imply scheduler's high overheads;<br>• Metaheuristic algorithms might not scale well enough to support huge systems. |

Table 1: Advantages of Metaheuristic Load Balancer

| Advantages | Disadvantages |
|---|---|
| • Very scalable – scheduling decision computations are distributed over several independent nodes;<br>• Possibility of deploying advanced scheduling strategies (for example artificial intelligence and autonomy of an agent);<br>• No single point of failure. | • Unpredictable and difficult to control – difficult to enforce scheduling policies and fairness;<br>• Communication overhead of an agent-based system;<br>• Overall performance might be lower than using a centralised approach. |

Table 2: Advantages of Decentralised Agent-based Load Balancer

The design of the centralised load balancing strategy assumed that metaheuristic algorithms would be able to dynamically balance the Cloud workload – an approach known as 'monolithic' scheduling (Schwarzkopf et al., 2013). A variety of metaheuristic algorithms were tested, such as Greedy, Genetic Algorithms (GA), Tabu Search (TS) and Simulated Annealing (SA). A novel variant of the Seeded Genetic Algorithms (SGA) which seeded the initial GA population with results from Greedy, TS and SA performed substantially better than their counterparts.

However, after extensive experiments, it was noted that this approach did not scale well because of the high computation overhead of metaheuristic algorithms. The centralised metaheuristic load balancer could efficiently support around sixty tasks executed on twelve nodes; however, as more tasks and nodes were added, and the solution search space grew, the quality of returned allocations rapidly decreased. None of the tested designs could scale to reliably schedule ca. 140k tasks on 12.5k nodes, as required in the data from GCD workload traces.

Therefore, further research was focused on designing a decentralised load balancing strategy, where nodes represented by software agents could negotiate task allocations between themselves and Service Allocation Negotiation (SAN) protocol was created. In the prototype implementation of the Multi-Agent System Balancer (MASB) system, each node is represented by Node Agent (NA) which monitors its node's resources allocations levels and makes sure that the node is not overloaded. When the allocated tasks exceed the node's resources, NA will communicate with other NAs and attempt to offload overloading tasks.

The MASB could support scheduling 140k tasks on 12.5k nodes. However, the decentralisation of scheduling logic also removed the centrally available store object with the state of the system. Each scheduling decision had to be made only on the partial information of the computing cell state. Therefore, another software agent component was introduced – Broker Agents (BA). BA's task was to gather information about the state of the nodes and to provide a quoting mechanism to initially retrieve the best available candidates for a given task. The SAN protocol was extended accordingly, and the capability of forced-migrations was added to better support restrictive constraints on some of the tasks.

In brief, the SAN protocol could be seen as the process of narrowing down the selection of candidate nodes. At first, randomly selected BA provides a quote with a number of candidate node recommendations, and since BA uses its own cache of node states, the recommendations most likely do not represent the current state of the node. After this, the source NA messages all the NAs of those candidate nodes, receiving information as to whether NA would accept task migration. Having collected all the replies, the source NA decides which of the candidate nodes is the best fit for a given task and attempts to migrate task there.

This step might be repeated if a selected candidate node is no longer accepting a task – in such a case, the second-best candidate node is selected.

Moving away from the concept of the centralised load balancing and offloading the actual scheduling logic to the nodes themselves resulted in more time available for the execution of allocation routines. As such, more sophisticated algorithms could be deployed, such as metaheuristic methods. Both BA and NA use Service Allocation Score (SAS) functions to calculate their Allocation Score (AS) value. This determines how well the nodes' resources are utilised, with more proportional allocations given a higher value. Both BAs and NAs, when making task allocation recommendations and decisions, tend to gravitate towards more desirable allocations. It was found that using different functions for Service Initial Allocation Score (SIAS) and Service Re-allocation Score (SRAS) was beneficial. This pattern improved the tightness of task allocations, which resulted in lower resource waste. NAs were given more autonomy in deciding which tasks to accept and which to offload in order to preserve their node's stable state.

Ultimately, a working solution was found, and the remaining part of this project was focused on testing the suitability and scalability of the MASB prototype and on introducing enhancements to improve the performance of the proposed solution. At peak times, almost all nodes of HPC Cluster at the University of Westminster was running experimental simulations which allowed the MASB to be rapidly reiterated and improved.

## 1.7. CONTRIBUTIONS TO KNOWLEDGE

This research added the following new contributions to the knowledge:

i. A novel taxonomy and categorisation of three classes of schedulers, namely OS-level, Cluster and Big Data, which highlight their unique evolution and underline their different objectives (Chapter 2);
ii. An abstract model of cloud resources utilisation is specified, including multiple types of resources and consideration of task migration costs (Chapter 3);
iii. A virtual machine live migration was experimented with in order to create a formula which estimates the network traffic generated by this process (Chapter 4);
iv. A high-fidelity Cloud workload simulator, based on a month-long workload traces from Google's computing cells, was created (Chapter 5);
v. Two possible approaches to resource management were proposed and examined in the practical part of the manuscript: the centralised metaheuristic load balancer (Chapter 6) and the decentralised agent-based system (Chapter 7).

In addition, the practices of running a Scala-based computation-intensive application on HPC machines are summarised and presented in Sliwko (2018a) and Sliwko (2018b).

# 2. TAXONOMY OF SCHEDULERS

Although managing workload in a Cloud system is a modern challenge, scheduling strategies are a well-researched field as well as being an area where there has been considerable practical implementation. This background review started by analysing deployed and actively used solutions, and presents a taxonomy in which schedulers are divided into several hierarchical groups based on their architecture and design. While other taxonomies do exist (e.g. Krauter et al., 2002; Yu and Buyya, 2005; Pop et al., 2006; Smanchat and Viriyapant, 2015; Rodriguez and Buyya, 2017; Zakarya and Gillam, 2017; Tyagi and Gupta, 2018), this review has focused on the key design factors that affect the throughput and scalability of a given solution, as well as the incremental improvements which bettered such an architecture.

Figure 3 visualises how the schedulers' groups are split. Each of these groups is separately discussed in the sections which follow.

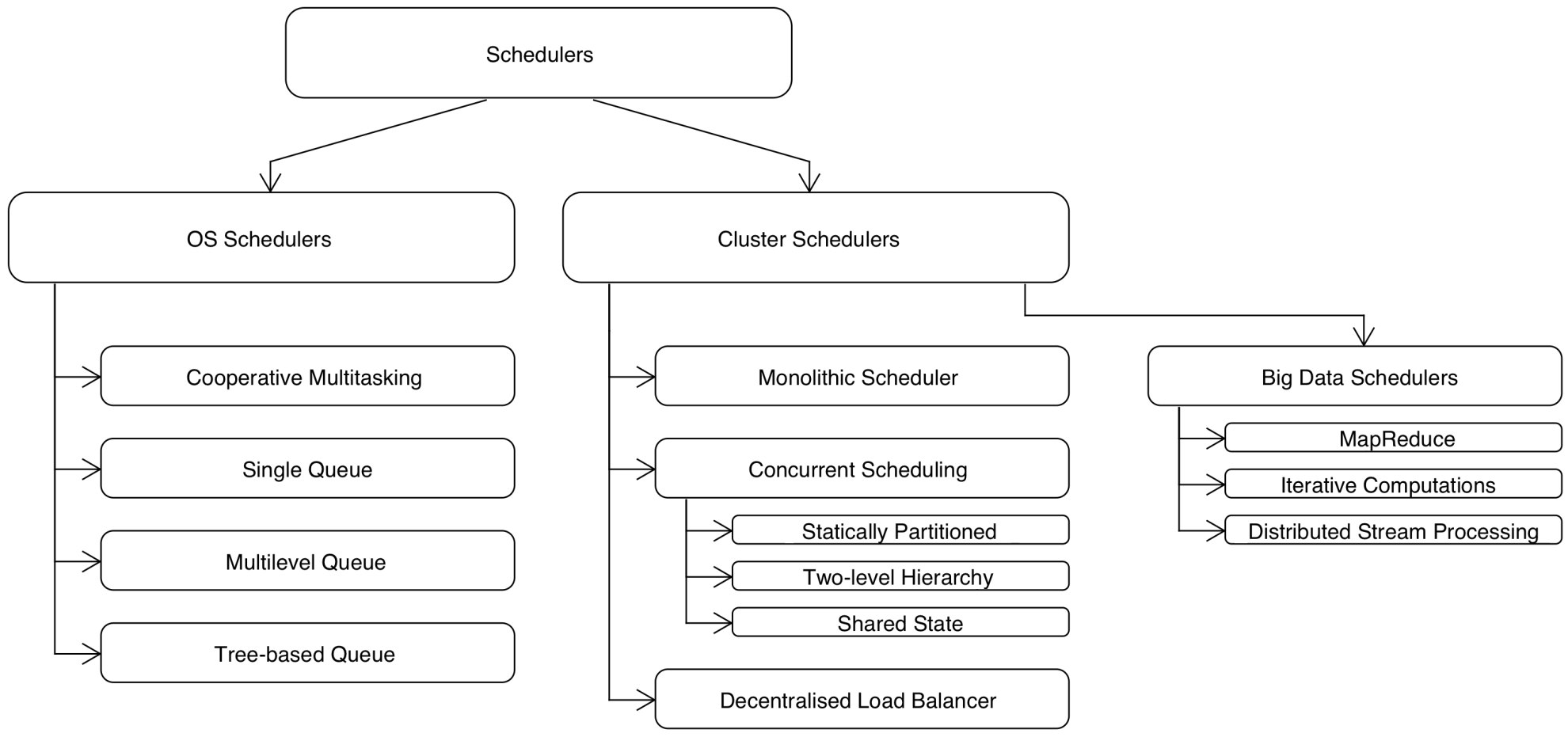

Figure 3: Schedulers taxonomy

It should be noted that this chapter is based partially on work already published in Sliwko and Getov (2015b).

## 2.1. METACOMPUTING

The concept of connecting computing resources has been an active area of research for a considerable period of time. The term ‘metacomputing’ was established as early as 1987 (Smarr and Catlett, 2003) and since then the topic of scheduling has been one of the key subjects in many research projects, such as (i) service localising idle workstations and utilising their spare CPU cycles – HTCondor (Litzkow et al., 1988); (ii) the Mentat – a parallel run-time system developed at the University of Virginia (Grimshaw, 1990); (iii) blueprints for a national supercomputer (Grimshaw et al., 1994), and (iv) the Globus metacomputing infrastructure toolkit (Foster and Kesselman, 1997).

Prior to the work of Foster et al. (2001), there was no clear definition of what ‘grid’ systems referred to. Following this publication, the principle that grid systems should allow a set of participants to share a number of connected computer machines and their resources became established. These shared system policies are defined by a list of rules, for example the resources which are shared, who (and the extent to which) they can use those resources, and the kind of quality of service that might be expected.

As shown in the following sections, the requirements of a load balancer in a decentralised system varies significantly compared to scheduling jobs on a single machine (Hamscher et al., 2000). One important difference are network resources, in that the machines are usually geographically distributed and transferring data from one machine to another is costly. In addition to the effective spreading of tasks across networked machines, the load balancer in Clusters generally provides a mechanism for fault-tolerance and user session management. The sections below also explain the workings of several selected current and past schedulers and distributed frameworks. Understanding these will help to develop the knowledge about how scheduling algorithms were developed over time, and how they have been conceptualised in different ways. This is by no means a complete taxonomy of all available designs, but rather an analysis of some of the landmark features and ideas in the history of schedulers.

## 2.2. OS SCHEDULERS

The Operating System (OS) Scheduler, also known as a 'short-term scheduler' or 'CPU scheduler', works within very short time frames, i.e. time-slices. During scheduling events, an algorithm must examine planned tasks and assign them appropriate CPU times (Bulpin, 2005; Arpaci-Dusseau and Arpaci-Dusseau, 2015). This requires schedulers to use highly optimised algorithms with very small overheads. Process schedulers have the difficult task of maintaining a delicate balance between responsiveness (minimum latency) and throughput. This is generally achieved by prioritising the execution of processes with a higher sleep/processing ratio (Pabla, 2009).

At the time of writing, the most advanced strategies also take into consideration the latest CPU core where the process ran the previous time. This is known as 'Non-Uniform Memory Access (NUMA) awareness', where the aim is to reuse the same CPU cache memory wherever possible (Blagodurov et al., 2010). The memory access latency differences can be very substantial, for example ca. 3-4 cycles for L1 cache, ca. 6-10 cycles for L2 cache and ca. 40-100 cycles for L3 cache (Drepper, 2007). NUMA awareness also involves prioritising the act of choosing a real idle core which must occur prior to its logical SMT sibling, also known as 'Hyper-Threading (HT) awareness'. Given this, NUMA awareness is a crucial element in the design of modern OS schedulers. With a relatively high data load to examine in a short period of time, implementation needs to be strongly optimised to ensure faster execution.

OS Schedulers tend to provide only a very limited set of tuneable parameters, wherein the access to modify them is not straightforward. Some of the parameters can change only during the kernel compilation process and require rebooting, such as compile-time options CONFIG_FAIR_USER_SCHED and CONFIG_FAIR_CGROUP_SCHED, or on the fly using the low-level Linux kernel's tool 'sysctl'.

### 2.2.1. COOPERATIVE MULTITASKING

Early multitasking Operating Systems, such as Windows 3.1x, Windows 95, 96 and Me, Mac OS prior to X, adopted a concept known as Cooperative Multitasking or Cooperative Scheduling (CS). In early implementations of CS, applications voluntarily ceded CPU time to one another. This was later supported natively by the OS, although Windows 3.1x used a non-pre-emptive scheduler which did not interrupt the program, wherein the program needed to explicitly tell the system that it no longer required the processor time. Windows 95 introduced a rudimentary pre-emptive scheduler, although this was for 32-bit applications only (Hart, 1997). The main issue in CS is the hazard caused by the poorly designed program. CS relies on processes regularly giving up control to other processes in the system, meaning that if one process consumes all the available CPU power, it causes all the systems to hang.

### 2.2.2. SINGLE QUEUE

Prior to Linux kernel version 2.4, the simple Circular Queue (CQ) algorithm was used to support the execution of multiple processes on the available CPUs. The selection of the next process to run was based on a Round Robin policy (Shreedhar, 1995). In kernel version 2.2, processes were further split into non-real/real-time categories, and scheduling classes were introduced. This algorithm was replaced by O(n) scheduler in Linux kernel versions 2.4-2.6. In O(n), processor time is divided into epochs, and within each epoch every task can execute up to its allocated time slice before being pre-emptied. The time slice is given to each task at the start of each epoch, and is based on the task's static priority added to half of any remaining time-slices from the previous epoch (Bulpin, 2005). Thus, if a task does not use its entire time slice in the current epoch, it can execute for longer in the next epoch. O(n) scheduler requires iteration through all currently planned processes during a scheduling event (Jones, 2009) – this can be seen as a weakness, especially for multi-core processors.

Between Linux kernel versions 2.6-2.6.23 came the implementation of the O(1) scheduler. O(1) further splits the processes list into active and expired arrays. Here, the arrays are switched once all the processes from the active array have exhausted their allocated time and have been moved to the expired array. The O(1) algorithm analyses the average sleep time of the process, with more interactive tasks being given higher priority in order to boost system responsiveness. The calculations required are complex and subject to potential errors, where O(1) may cause non-interactive behaviour from an interactive process (Wong et al., 2008; Pabla, 2009).

### 2.2.3. MULTILEVEL QUEUE

With Q(n) and O(1) algorithms failing to efficiently support the interactivity of applications, the design of OS Scheduler evolved into a multilevel queue in which repeatedly sleeping (interactive) processes are pushed to the top of queue and executed more frequently. At the same time, background processes are pushed down and run less frequently, although for longer periods.

Perhaps the most widespread scheduler algorithm is Multilevel Feedback Queue (MLFQ), which is implemented in all modern versions of Windows NT (2000, XP, Vista, 7 and Server), Mac OS X, NetBSD and Solaris kernels (up to version 2.6, when it was replaced with O(n) scheduler). MLFQ was first described in 1962 in a system known as the Compatible Time-Sharing System (Corbató et al., 1962). Fernando Corbató was awarded the Turing Award by the ACM in 1990 'for his pioneering work organizing the concepts and leading the development of the general-purpose, large-scale, time-sharing and resource-sharing computer systems, CTSS and Multics'. In MLFQ, jobs are organised into a set of queues $Q_0$, $Q_1$, …, $Q_i$ wherein a job is promoted to a higher queue if it does not finish within $2^i$ time units. The algorithm processes the job from the front of the lowest queue at all times, meaning that short processes are given preference. While having a very poor worst-case scenario, MLFQ turns out to be very efficient in practice (Becchetti et al., 2006).

Staircase Scheduler (Corbet, 2004), Staircase Deadline Scheduler (Corbet, 2007), Brain F. Scheduler (Groves et al., 2009) and Multiple Queue Skiplist Scheduler (Kolivas, 2016) constitute a line of successive schedulers developed by Con Kolivas since 2004 which are based on a design of Fair Share Scheduler from Kay and Lauder (1988). None of these schedulers have been merged into the source code of mainstream kernels and they are available only as experimental '-ck' patches. Although the concept behind those schedulers is similar to MLFQ, the implementation details differ significantly. The central element is a single, ranked array of processes for each CPU ('staircase'). Initially, each process (P1, P2, …) is inserted at the rank determined by its base priority; the scheduler then picks up the highest ranked process (P) and runs it. When P has used up its time slice, it is reinserted into the array but at a lower rank, where it will continue to run but at a lower priority. When P exhausts its next time-slice, its rank is lowered again. P then continues until it reaches the bottom of the staircase, at which point it is moved up to one rank below its previous maximum, and is assigned two time-slices. When P exhausts these two time-slices, it is reinserted once again in the staircase at a lower rank. When P once again reaches the bottom of the staircase, it is assigned another time-slice and the cycle repeats. P is also pushed back up the staircase if it sleeps for a predefined period. This means that interactive tasks which tend to sleep more often should remain at the top of the staircase, while CPU-intensive processes should continuously expend more time-slices but at a lower frequency. Additionally, each rank level in the staircase has its own quota, and once the quota is expired all processes on that rank are pushed down.

Most importantly, Kolivas' work introduced the concept of 'fairness', in which each process gets a comparable share of CPU time to run, proportional to the priority. If the process spends much of its time waiting for I/O events, then its spent CPU time value is low, meaning that it is automatically prioritised for execution. This means that interactive tasks which spend most of their time waiting for user input get execution time when they need it. This represents the notion of 'sleeper fairness'. This design also prevents a situation in which the process is 'starved', i.e. never executed.

### 2.2.4. TREE-BASED QUEUE

While the work of Con Kolivas has never been merged into the mainstream Linux kernel, it has introduced the key concept of 'fairness', which is the crucial feature of the design of most current OS schedulers. At the time of writing, Linux kernel implements Completely Fair Scheduler (CFS), which was developed by Ingo Molnár and introduced in kernel version 2.6.23. A central element in this algorithm is a self-balancing red-black tree structure in which processes are indexed by spent processor time. CFS implements the Weighted Fair Queueing (WFQ) algorithm, in which the available CPU time-slices are split between processes in proportion to their priority weights ('niceness'). WFQ is based on the idea of the 'ideal processor', meaning that each process should have an equal share of CPU time adjusted for their priority and total CPU load (Jones, 2009; Pabla, 2009).

Lozi et al. (2016) offers an in-depth explanation of the algorithm's workings, noting potential issues regarding the CFS approach. The main criticism revolves around the four problematic areas:

- Group Imbalance – the authors' experiments have shown that not every core of their 64-core machine is equally loaded: some cores run either only one process or no processes at all while the rest of the cores were overloaded. It was found that the scheduler was not balancing the load because of the hierarchical design and complexity of the load tracking metric. To limit the complexity of the scheduling algorithm, the CPU cores are grouped into scheduling groups, i.e. nodes. When an idle core attempts to steal work from another node, it compares only the average load of its node with that of its victim's node. It will steal work only if the average load of its victim's group is higher than its own. This creates inefficiency since idle cores will be concealed by their nodes' average load.
- Scheduling Group Construction – this concern relates to the way scheduling groups are constructed which is not adapted to modern NUMA machines. Applications in Linux can be pinned to a subset of available cores. CFS might assign the same cores to multiple scheduling groups with those groups then being ranked by distance, for example nodes one hop apart, nodes two hops apart and so on. This feature was designed to increase the probability that processes would remain close to their original NUMA node. However, this could result in the application being pinned to particular cores which are separated by more than one hop, with work never being migrated outside the initial core. This might mean that an application uses only one core.
- Overload-on-Wakeup – this problem occurs when a process goes to sleep on a particular node and is then awoken by a process on the same node. In such a scenario, only cores in this scheduling group will be considered to run this process. The aim of this optimisation is to improve cache utilisation by running a process close to the waker process, meaning that there is the possibility of them sharing the last-level memory cache. However, the might be the scheduling of a process on a busy core when there are idle cores in alternative nodes, resulting in the severe under-utilisation of the machine.

- Missing Scheduling Domains – this is the result of a line of code omission while refactoring the Linux kernel source code. The number of scheduling domains is incorrectly updated when a particular code is disabled and then enabled, and a loop exits early. As a result, processes can be run only on the same scheduling group as their parent process.

Lozi et al. (2016) have provided a set of patches for the above issues, and have presented experimental results after applying fixes. They have also provided a set of tools on their site which could be used to detect those glitches early in the Linux kernel lifecycle. Moreover, it has been argued (Lozi et al., 2016) that the sheer number of optimisations and modifications implemented into CFS scheduler changed the initially simple scheduling policy into one which was very complex and bug-prone – as of 26$^{th}$ June 2018, there were 742 commits to CFS source code (‘fair.c’ file in github.com/torvalds/linux repository) since November 2011. As such, an alternative approach is perhaps required, such as a scheduler architecture based on pluggable components. This work clearly demonstrates the immerse complexity of scheduling solutions catering to the complexities of modern hardware.

## 2.3. CLUSTER SCHEDULERS

Distributed computing differs from traditional computing in many ways. The sheer physical size of the system itself means that thousands of machines may be involved, with thousands of users being served and millions of API calls or other requests needing processed. While responsiveness and low overheads tend to be the focus of process schedulers, the focus of cluster schedulers is to focus upon high throughput, fault-tolerance and scalability. Cluster schedulers usually work with queues of jobs spanning to hundreds of thousands, and indeed sometimes even millions of jobs. They also seem to be more customised and tailored to the needs of organisation which is using them.

Cluster schedulers usually provide complex administration tools with a wide spectrum of tuneable parameters and flexible workload policies. All configurable parameters can usually be accessed through configuration files or via the GUI interface. However, it has been documented that site administrators only rarely stray from a default configuration (Etsion and Tsafrir, 2005). The most used scheduling algorithm is simply a First-Come-First-Serve (FCFS) strategy with backfilling optimisation.

The most common issues which cluster schedulers must deal with are:

- Unpredictable and varying load (Moreno et al., 2013);
- Mixed batch jobs and services (ibid.);
- Complex policies and constraints (Adaptive Computing, 2002);
- Fairness (ibid.);
- A rapidly increasing workload and cluster size (Isard et al., 2007);
- Legacy software (ibid.);
- Heterogeneous nodes with a varying level of resources and availability (Thain et al., 2005);
- The detection of underperforming nodes (Zhang et al., 2014b);
- Issues related to fault-tolerance (ibid.) and hardware malfunctions (Gabriel et al., 2004).

Another interesting challenge, although one which is rarely tackled by commercial schedulers, is minimising total power consumption. Typically, idle machines consume around half of their peak power (McCullough et al., 2011). Therefore, the total power consumed by a Data Centre can be lowered by concentrating tasks on a reduced number of machines and powering down the remaining nodes (Pinheiro et al., 2001; Lang and Patel, 2010).

The proposed grouping of Cluster schedulers loosely follows the taxonomy presented in Schwarzkopf et al. (2013).

### 2.3.1. MONOLITHIC SCHEDULER

The earliest Cluster schedulers were built with a centralised architecture in which a single scheduling policy allocated all incoming jobs. The tasks would be picked from the head of the queue and scheduled on system nodes in a serial manner by an allocation loop. Examples of centralised schedulers include Maui (Jackson et al., 2001) and its successor Moab (Adaptive Computing, 2015), Univa Grid Engine (Gentzsch, 2001), Load Leveler (Kannan et al., 2001), Load Sharing Facility (Etsion and Tsafrir, 2005), Portable Batch System (Bode et al., 2000) and its successor TORQUE (Klusáček et al., 2013), Alibaba's Fuxi (Zhang et al., 2014b), Docker Swarm (Naik, 2016), Kubernetes (Vohra, 2017) and several others.

Monolithic schedulers implement a wide array of policies and algorithms, such as FCFS, FCFS with backfilling and gang scheduling, Shortest Job First (SJF), and several others. The Kubernetes (Greek: 'κυβερνήτης') scheduler implements a range of scoring functions such as node or pod affinity/anti-affinity, resources best-fit and worst-fit, required images locality, etc. which can be additionally weighted and combined into node's score values (Lewis and Oppenheimer, 2017). As an interesting note – one of the functions (BalancedResourceAllocation routine) implemented in Kubernetes evaluates the balance of utilised resources (CPU and memory) on a scored node.

Monolithic schedulers are often plagued with a ‘head-of-queue’ blocking problem in which as a long job is awaiting a free node, the shorter jobs which follow are held. To partially counter this problem, the schedulers often implement ‘backfilling’ optimisation, where shorter jobs are allowed to execute while the long job is waiting. Perhaps the most widespread scheduler is Simple Linux Utility for Resource Management (SLURM) (Yoo et al., 2003). SLURM uses a best-fit algorithm which is based on either Hilbert curve scheduling or fat tree network topology; it can scale to thousands of CPU cores (Pascual, 2009). At the time of writing, the fastest supercomputer in the world is Sunway TaihuLight (Chinese: ‘神威·太湖之光’), which uses over 40k CPU processors, each of which contains 256 cores. Sunway TaihuLight’s workload in managed by SLURM (TOP500 Project, 2017).

The Fuxi (Chinese: ‘伏羲’) scheduler presents a unique strategy in that it matches newly-available resources against the backlog of tasks rather than matching tasks to available resources on nodes. This technique allowed Fuxi to achieve a very high utilisation of Cluster resources, namely 95% utilisation of memory and 91% utilisation of CPU. Fuxi has been supporting Alibaba’s workload since 2009, and it scales to ca. 5k nodes (Zhang et al., 2014b).

While Cluster scheduler designs have generally moved towards more parallelised solutions, as demonstrated in the next subsection, centralised architecture is still the most common approach in High-Performance Computing. Approximately half the world’s supercomputers use SLURM as their workload manager, while Moab is currently deployed on about 40% of the top 10, top 25 and top 100 on the TOP500 list (TOP500 Project, 2017).

The research presented in Chapter 6 attempted to improve a centralised scheduler’s design by introducing metaheuristic algorithms as a fundamental component of scheduling logic. The resulting metaheuristic load balancer prototype is presented together with the experimental results and discussion.

### 2.3.2. CONCURRENT SCHEDULING

Historically, monolithic schedulers were frequently built on the premise of supporting a single 'killer-application' (Barroso et al., 2003). However, the workload of the data centre has become more heterogeneous as systems and a modern Cluster system runs hundreds of unique programs with distinctive resource requirements and constraints. A single code base of centralised workload manager means that it is not easy to add a variety of specialised scheduling policies. Furthermore, as workload size is increased, the time to reach a scheduling decision is progressively limited. The result of this is a restriction in the selection of scheduling algorithms to less sophisticated ones, which affects the quality of allocations. To tackle those challenges, the Cluster schedulers evolved into more parallelised designs.

#### 2.3.2.1. STATICALLY PARTITIONED

The solution to the numerous policies and the lack of parallelism in central schedulers was to split Cluster into specialised partitions and manage them separately. Quincy (Isard et al., 2009), a scheduler managing workload of Microsoft's Dryad, follows this approach.

The development of an application for Dryad is modelled as a Directed Acyclic Graph (DAG) model in which the developer defines an application dataflow model and supplies subroutines to be executed at specified graph vertices. The scheduling policies and tuning parameters are specified by adjusting weights and capacities on a graph data structure. The Quincy implements a Greedy strategy. In this approach, the scheduler assumes that the currently scheduled job is the only job running on a cluster and so always selects the best node available. Tasks are run by remote daemon services which periodically update the job manager about the vertex's execution status. A vertex might be re-executed in case of failure. If any task has failed more than a configured number of times, the entire job is marked as failed (Isard et al., 2007).

Microsoft has built several frameworks on top of Dryad, such as COSMOS (Helland and Harris, 2011) which provided SQL-like language optimised for parallel execution. COSMOS was designed to support data-driven search and advertising within the Windows Live services owned by Microsoft, such as Bing, MSN and Hotmail. It analysed user behaviours in multiple contexts, such as what people searched for, what links they clicked, what sites they visited, the browsing order, and the ads they clicked on (ibid.). Although the Dryad project had several preview releases, it was ultimately dropped when Microsoft shifted its focus to the development of Hadoop.

The main criticism of the static partitioning is inflexibility – the exclusive sets of machines in a Cluster are dedicated to certain types of workload. That might result in a part of scheduler being relatively idle, while other nodes are very active. This leads to the Cluster's fragmentation and the suboptimal utilisation of available nodes since no machine sharing is allowed.

#### 2.3.2.2. TWO-LEVEL HIERARCHY

The solution to the inflexibility of static partitioning was to introduce two-level architecture in which a Cluster is partitioned dynamically by a central coordinator. The actual task allocations take place at the second level of architecture in one of the specialised schedulers. The first two-level scheduler was Mesos (Hindman et al., 2011), developed at the University of California (Berkeley), and is now hosted in the Apache Software Foundation. Mesos was a foundation base for other

Cluster systems such as Twitter's Aurora (Aurora, 2018) and Marathon (Mesosphere, 2018).

Mesos introduces a two-level scheduling mechanism in which a centralised Mesos Master acts as a resource manager that dynamically allocates resources to different scheduler frameworks, for example Hadoop, Spark and Kafka, via Mesos Agents. Mesos Agents are deployed on cluster nodes and use Linux's cgroups or Docker container (depending upon the environment) for resource isolation. Resources are distributed to the frameworks in the form of 'offers' which contain currently unused resources. Scheduling frameworks have autonomy in deciding which resources to accept and which tasks to run on them.

Mesos works most effectively when tasks are relatively small, short-lived and have a high resource 'churn rate', i.e. they relinquish resources more frequently. In the current version 1.4.1, only one scheduling framework can examine a resource offer at any given time. This resource is effectively locked for the duration of a scheduling decision, meaning that concurrency control is pessimistic. Several practical considerations for using Mesos in the production environment as well as best practices advice are presented in Campbell (2017).

Two-level schedulers offered a working solution to the lack of parallelisation found in central schedulers and the low efficiency of statically partitioned Clusters. Nevertheless, the mechanism used causes resources to remain locked while the resources offer is being examined by a specialised scheduler. This means the benefits from parallelisation are limited due to pessimistic locking. In addition, the schedulers do not coordinate between each other and must rely on a centralised coordinator to make them offers, which further restricts their visibility of the resources in a Cluster.

#### 2.3.2.3. SHARED STATE

To address the limited parallelism of the two-level scheduling design, the alternative approach taken by some organisations was to redesign schedulers' architecture into several schedulers, all working concurrently. The schedulers work on a shared Cluster's state information and manage their resources' reservations using an optimistic concurrency control method. A sample of such systems includes: Microsoft's Apollo (Boutin et al., 2014), Omega – the Google Borg's spinoff (Schwarzkopf et al., 2013), HashiCorp's Nomad (HashiCorp, 2018), and also Borg (Burns et al., 2016) itself which has been refactored from monolithic into parallel architecture after the experimentations with Omega.

By default, Nomad runs one scheduling worker per CPU core. Scheduling workers pick job submissions from the broker queue and then submit it to one of the three schedulers: a long-lived services scheduler, a short-lived batch jobs scheduler and a system scheduler, which is used to run internal maintenance routines. Additionally, Nomad can be extended to support custom schedulers. Schedulers process and generate an action plan, which constitutes a set of operations to create new allocations, or to evict and update existing ones (HashiCorp, 2018).

Microsoft's Apollo design seems to be primarily tuned for high tasks churn, and at peak times is capable of handling more than 100k of scheduling requests per second on a ca. 20k nodes cluster. Apollo uses a set of per-job schedulers called Job Managers (JM) wherein a single job entity contains a multiplicity of tasks which are then scheduled and executed on computing nodes. Tasks are generally short-lived batch jobs (Boutin et al., 2014). Apollo has a centralised Resource Monitor (RM), while each node runs its own Process Node (PN) with its own queue of tasks. Each PN is responsible for local scheduling decisions and can independently reorder its job queue to allow smaller tasks to be executed immediately, while larger tasks wait for resources to become available.

Additionally, PN computes a wait-time matrix based on its queue which publicises the future availability of the node's resources. Scheduling decisions are made optimistically by JMs based on the shared cluster's resource state, which is continuously retrieved and aggregated by RM. This design helps to avoid decisions which are suboptimal and conflicting were the architecture to be completely decentralised (ibid.).

Furthermore, Apollo splits tasks into those which are regular and those which are opportunistic. Opportunistic tasks are used to fill resource gaps left by regular tasks. The system also prevents overloading the cluster by limiting the total number of regular tasks that can be run on a cluster. Apollo implements locality optimisation by taking into consideration the location of data for a given task. For example, the system will score nodes higher if the required files are already on the local drive as opposed to machines needing to download data (ibid.).

Historically, Omega was a spinoff from Google's Borg scheduler. Despite the various optimisations acquired by Borg over the years, including internal parallelism and multi-threading, in order to address the issues of head-of-line blocking and scalability problems, Google decided to create an Omega scheduler from the ground up (Schwarzkopf et al., 2013). Omega introduced several innovations, such as storing the state of the cluster in a centralised Paxos-based store that was accessed by multiple components simultaneously. The eventual conflicts were resolved by optimistic locking concurrency control. This feature allowed Omega to run several schedulers at the same time and improve the scheduling throughput. Many of Omega's innovations have since been folded into Borg (Burns et al., 2016).

Omega's authors highlight the disadvantages of the shared state and parallel reservation of resources, namely: (i) the state of a node could have changed considerably when the allocation decision was being made, and it is no longer possible for this node to accept a job; (ii) two or more allocations to the same node could have conflicted and both scheduling decisions are nullified; and (iii) this strategy introduces significant difficulties when gang-scheduling a batch of jobs as (i) or (ii) are happening (Schwarzkopf et al., 2013).

In this research, special attention was given to Google's Borg, one of the most advanced and published schedulers. Moreover, while other schedulers are designed to support either a high churn of short-term jobs, for example Microsoft's Apollo (Boutin et al., 2014), Alibaba's Fuxi (Zhang et al., 2014b), or else a limited number of long-term services, such as Twitter's Aurora (Aurora, 2018), Google's engineers have created a system which supports a mixed workload. Borg has replaced two previous systems, Babysitter and the Global Work Queue, which were used to manage long-running services and batch jobs separately (Burns et al., 2016). Given the significance of Borg's design for this research, it is discussed separately in section 2.4.

### 2.3.3. DECENTRALISED LOAD BALANCER

This research proposes a new type of Cluster's workload orchestration model in which the actual scheduling logic is processed on nodes themselves, which is a significant step towards completely decentralised Cluster orchestration. The cluster state is retrieved from a subnetwork of BAs, although this system does not rely on the accuracy of this information and uses it exclusively to retrieve an initial set of candidate nodes where a task could potentially run. The actual task to machine matching is performed between the nodes themselves. As such, this design avoids the pitfalls of the concurrent resource locking, which includes conflicting scheduling decisions and the non-current state of nodes' information. Moreover, the decentralisation of the scheduling logic also lifts complexity restrictions on scheduling logic, meaning that a wider range of scheduling algorithms can be used, such as metaheuristic methods.

Chapter 7 presents MASB – a decentralised agent-based load balancer prototype – in which the TS algorithm supports making scheduling decisions separately on each node. Furthermore, MASB breaks with the concept that the execution of a task is immovable or unstoppable. As a result of the advances of the virtualisation technology and the introduction of the VM-LM feature, a running program can now be offloaded to an alternative node without stopping its execution. Therefore, MASB is not only scheduling the coming tasks, it is also actively moving the currently existing tasks so that they can fit better on the available resources of the Cluster, hence the name 'load balancer'.

### 2.3.4. BIG DATA SCHEDULERS

In taxonomy presented in this chapter, Big Data schedulers are visualised as a separate branch from Cluster Schedulers. Although it could be argued that Big Data Schedulers belong to one of the Cluster schedulers designs discussed previously, this separation signifies their over-specialisation, and that only a very restricted set of operations is supported (Isard et al., 2007; Zaharia et al., 2010). The scheduling mechanisms are often intertwined with the programming language features, with Big Data frameworks often providing their own API (Zaharia et al., 2009; White, 2012) and indeed sometimes even their own custom programming language, as seen with Skywriting in CIEL (Murray et al., 2011).

Generally speaking, Big Data frameworks provide very fine-grained control over how data is accessed and processed over the cluster, such as Spark RDD objects persist operations or partitioners (Zaharia et al., 2012). Such a deep integration of scheduling logic with applications is a distinctive feature of Big Data technology. At the time of writing, Big Data is also the most active distributed computing research area, with new technologies, frameworks and algorithms being released on a regular basis.

Big Data is the term given to the storage and processing of any data sets so large and complex that they become unrealistic to process using traditional data processing applications based on relational database management systems. It depends on the individual organisation as to how much data is described as Big Data, but the following examples may be considered to get an idea of scale:

- The NYSE (The New York Stock Exchange) produces about 15 TB of new trade data per day (Singh, 2017);
- Facebook warehouse stores upwards of 300 PB of data, with an incoming daily rate of about 600 TB (Vagata and Wilfong, 2014);
- The Large Hadron Collider (Geneva, Switzerland) produces about fifteen petabytes of data per year (White, 2012).

As a result of a massive size of the stored and processed data, the central element of a Big Data framework is its distributed file system, such as Hadoop Distributed File System (Gog, 2012), Google File System (Ghemawat et al., 2003) and its successor Colossus (Corbett et al., 2013). The nodes in a Big Data cluster fulfil the dual purposes of storing the distributed file system parts, usually in a few replicas for fault-tolerance means, and also providing a parallel execution environment for system tasks. The speed difference between locally-accessed and remotely stored input data is very substantial, meaning that Big Data schedulers are very focused on providing 'data locality' which means running a given task on a node where input data are stored or are in the closest proximity to it.

The origins of the Big Data technology are in the 'MapReduce' programming model, which implements the concept of Google's inverted search index. Developed in 2003 (Dean and Ghemawat, 2010) and later patented in 2010 (U.S. Patent 7,650,331), the Big Data design has evolved significantly since, and is presented in the subsections below.

#### 2.3.4.1. MAPREDUCE

MapReduce is the most widespread principle which has been adopted for processing large sets of data in parallel. The name MapReduce originally referred only to the Google's proprietary technology, but the term is now broadly used to describe a wide range of software, such as Hadoop, CouchDB, Infinispan and MongoDB. The key features of MapReduce are its scalability and fine-grained fault-tolerance. The original thinking behind MapReduce was inspired by the 'map' and 'reduce' operations present in Lisp and other functional programming languages (Dean and Ghemawat, 2010):

- 'Map' is an operation used in the first step of computation and is applied to all available data that performs the filtering and transforming of all key-value pairs from the input data set. The 'map' operation is executed in parallel on multiple machines on a distributed file system. Each 'map' task can be restarted individually and a failure in the middle of a multi-hour execution does not require restarting the whole job from scratch.
- The 'Reduce' operation is executed after 'map' operations complete. It performs finalising operations, such as counting the number of rows matching specified conditions and yielding fields frequencies. The 'Reduce' operation is fed using a stream iterator, thereby allowing the framework to process list of items one at the time, thus ensuring that the machine memory is not overloaded (Dean and Ghemawat, 2010; Gog, 2012).

Following the development of the MapReduce concept, Yahoo! engineers began the Open Source project Hadoop. In February 2008, Yahoo! announced that its production search index was being generated by a 10k-core Hadoop cluster (White, 2012). Subsequently, many other major Internet companies, including Facebook, LinkedIn, Amazon and Last.fm, joined the project and deployed it within their architectures. Hadoop is currently hosted in the Apache Software Foundation as an Open Source project.

As in Google's original MapReduce, Hadoop's users submit jobs which consist of 'map' and 'reduce' operation implementations. Hadoop splits each job into multiple 'map' and 'reduce' tasks, which subsequently process each block of input data, typically 64MB or 128MB (Gog, 2012). Hadoop's scheduler allocates a 'map' task to the closest possible node to the input data required – so-called 'data locality' optimisation. In so doing, the following allocation order is used: the same node, the same rack and finally a remote rack (Zaharia et al., 2009). To further improve performance, the Hadoop framework uses 'backup tasks' in which a speculative copy of a task is run on a separate machine in order to finish the computation faster. If the first node is available but behaving poorly, it is known as a 'straggler', with the result that the job is as slow as the misbehaving task. This behaviour can occur for many reasons, such as faulty hardware or misconfiguration. Google estimated that using 'backup tasks' could improve job response times by 44% (Dean and Ghemawat, 2010).

At the time of writing, Hadoop comes with a selection of schedulers, as outlined below:

- 'FIFO Scheduler' is a default scheduling system in which the user jobs are scheduled using a queue with five priority levels. Typically, jobs use the whole cluster, so they must wait their turn. When another job scheduler chooses the next job to run, it selects jobs with the highest priority, resulting in low-priority jobs being endlessly delayed (Zaharia et al., 2009; White, 2012).
- 'Fair Scheduler' is part of the cluster management technology Yet Another Resource Negotiator (YARN) (Vavilapalli et al., 2013), which replaced the original Hadoop engine in 2012. In Fair Scheduler, each user has their own pool of jobs and the system focuses on giving each user a proportional share of cluster resources over time. The scheduler uses a version of 'max-min fairness' (Bonald et al., 2006) with minimum capacity guarantees that are specified as the number of 'map' and 'reduce' task slots to allocate tasks across users' job pools. When one pool is idle, and the minimum share of the tasks slots is not being used, other pools can use its available task slots.

- 'Capacity Scheduler' is the second scheduler introduced within the YARN framework. In essence, this can be seen as a number of separate MapReduce engines with FCFS scheduling for each user or organisation. Those queues can be hierarchical, with a queue having children queues, and with each queue being allocated task slots capacity which can be divided into 'map' and 'reduce' tasks. Task slots allocation between queues is similar to the sharing mechanism between pools found in Fair Scheduler (White, 2012).

The main criticism of MapReduce is the acyclic dataflow programming model. The stateless 'map' task must be followed by a stateless 'reduce' task, which is then executed by the MapReduce engine. This model makes it challenging to repeatedly access the same dataset, a common action during the execution of iterative algorithms (Zaharia et al., 2009).

#### 2.3.4.2. ITERATIVE COMPUTATIONS

Following the success of Apache Hadoop, a number of alternative designs were created to address Hadoop's suboptimal performance when running iterative MapReduce jobs. Examples of such systems include HaLoop (Bu et al., 2010) and Spark (Zaharia et al., 2010).

HaLoop has been developed on top of Hadoop, with various caching mechanisms and optimisations added, and making the framework loop-aware, for example by adding programming support for iterative application and storing the output data on the local disk. Additionally, HaLoop's scheduler keeps a record of every data block processed by each task on physical machines, and tries to schedule subsequent tasks taking inter-iteration locality into account. This feature helps to minimise costly remote data retrieval, meaning that tasks can use data cached on a local machine (Bu et al., 2010; Gog, 2012).

Similar to HaLoop, Spark's authors noted a suboptimal performance of iterative MapReduce jobs in the Hadoop framework. In certain kinds of application, such as iterative Machine Learning algorithms and interactive data analysis tools, the same data are repeatedly accessed in multiple steps and then discarded; therefore, it does not make sense to send it back and forward to central node. In such scenarios Spark will outperform Hadoop (Zaharia et al., 2012).

Spark is built on top of HDSF, but it does not follow the two-stage model of Hadoop. Instead, it introduces resilient distributed datasets (RDD) and parallel operations on these datasets (Gog, 2012):

- 'reduce' - combines dataset elements using a provided function;
- 'collect' - sends all the elements of the dataset to the user program;
- 'foreach' - applies a provided function onto every element of a dataset.

Spark provides two types of shared variables:

- 'accumulators' - variables onto each worker can apply associative operations, meaning that they are efficiently supported in parallel;
- 'broadcast variables' - sent once to every node, with nodes then keeping a read-only copy of those variables (Zecevic, 2016).

The Spark job scheduler implementation is conceptually similar to that of Dryad's Quincy. However, it considers which partitions of RDD are available in the memory. The framework then re-computes missing partitions, and tasks are sent to the closest possible node to the input data required (Zaharia et al., 2012).

Another interesting feature implemented in Spark is the concept of 'delayed scheduling'. In situations when a head-of-line job that should be scheduled next cannot launch a local task, Spark's scheduler delays the task execution, and lets other jobs start their tasks instead. However, if the job has been skipped long enough, typically up to ten seconds, it launches a non-local task. Since a typical Spark workload consists of short tasks, meaning that it has a high task slots churn, tasks have a higher chance of being executed locally. This feature helps to achieve almost optimal 'data locality' with a minimal impact on fairness, and the cluster throughput can be almost doubled, as shown in an analysis performed on Facebook's workload traces (Zaharia et al., 2010).

#### 2.3.4.3. DISTRIBUTED STREAM PROCESSING

The core concept behind distributed stream processing engines is the processing of incoming data items in real time by modelling a data flow in which there are several stages which can be processed in parallel. Other techniques include splitting the data stream into multiple sub-streams, and redirecting them into a set of networked nodes (Liu and Buyya, 2017).

Inspired by Microsoft's research into DAG models (Isard et al., 2009), Apache Storm (Storm) is a distributed stream processing engine used by Twitter following extensive development (Toshniwal et al., 2014). Its initial release was 17 September 2011, and by September 2014 it had become open-source and an Apache Top-Level Project.

The defined topology acts as a distributed data transformation pipeline. The programs in Storm are designed as a topology in the shape of DAG, consisting of ‘spouts’ and ‘bolts’:

- ‘Spouts’ read the data from external sources and emit them into the topology as a stream of ‘tuples’. This structure is accompanied by a schema which defines the names of the tuples’ fields. Tuples can contain primitive values such as integers, longs, shorts, bytes, strings, doubles, floats, booleans, and byte arrays. Additionally, custom serialisers can be defined to interpret this data.
- The processing stages of a stream are defined in ‘bolts’ which can perform data manipulation, filtering, aggregations, joins, and so on. Bolts can also constitute more complex transforming structures that require multiple steps (thus, multiple bolts). The bolts can communicate with external applications such as databases and Kafka queues (Toshniwal et al., 2014).

In comparison to MapReduce and iterative algorithms introduced in the subsections above, Storm topologies, once created, run indefinitely until killed. Given this, the inefficient scattering of application’s tasks among Cluster nodes has a lasting impact on performance. Storm’s default scheduler implements a Round Robin strategy and for resource allocation purposes, Storm assumes that every worker is homogenous. This design results in frequent resource over-allocation and inefficient use of inter-system communications (Kulkarni et al, 2018). To remedy this phenomenon, more complex solutions are proposed such as D-Storm (Liu and Buyya, 2017). D-Storm’s scheduling strategy is based on a metaheuristic algorithm Greedy, which also monitors the volume of the incoming workload and is resource-aware.

Typical examples of Storm’s usage include:

- processing a stream of new data and updating databases in real time, e.g. in trading systems wherein data accuracy is crucial;
- continuously querying and forwarding the results to clients in real time, e.g. streaming trending topics on Twitter into browsers, and
- a parallelisation of computing-intensive query on the fly, i.e. a distributed Remote Procedure Call (RPC) wherein a large number of sets are being probed (Marz, 2011).

Storm has gained widespread popularity and is being used by companies such as Groupon, Yahoo!, Spotify, Verisign, Alibaba, Baidu, Yelp, and many more. A comprehensive list of users is available at the storm.apache.org website.

At the time of writing, Storm is being replaced at Twitter by newer distributed stream processing engine – Heron (Kulkarni et al, 2018) which continues the DAG model approach, but focuses on various architectural improvements such as reduced overhead, testability and easier access to debug data.

## 2.4. GOOGLE'S BORG

To support its operations, Google utilises a high number of data centres around the world, which at the time of writing number sixteen. Borg admits, schedules, starts, restarts and monitors the full range of applications run by Google. Borg users are Google developers and system administrators, and users submit their workload in the form of jobs. A job may consist of one of more tasks that all run the same program (Burns et al., 2016).

### 2.4.1. DESIGN CONCEPTS

The central module of the Borg architecture is BorgMaster, which maintains an in-memory copy of most of the state of the cell. This state is also saved in a distributed Paxos-based store (Lamport, 1998). While BorgMaster is logically a single process, it is replicated five times in order to improve fault-tolerance. The main design priority of Borg was resilience rather than performance. Google services are seen as very durable and reliable, the result of multi-tier architecture, where no component is a single point of failure exists. Current allocations of tasks are saved to Paxos-based storage, and the system can recover even if all five BorgMaster instances fail. Each cell in the Google Cluster in managed by a single BorgMaster controller. Each machine in a cell runs BorgLet, an agent process responsible for starting and stopping tasks and also restarting them should they fail. BorgLet manages local resources by adjusting local OS kernel settings and reporting the state of its node to the BorgMaster and other monitoring systems.

The Borg system offers extensive options to control and shape its workload, including priority bands for tasks (i.e. monitoring, production, batch, and best effort), resources quota and admission control. Higher priority tasks can pre-empt locally-running tasks in order to obtain required resources. The exception is made for production tasks which cannot be pre-empted. Resource quotas are part of admission control and are expressed as a resource vector at a given priority, for a period of time (usually months). Jobs with insufficient quotas are rejected immediately upon submission. Production jobs are limited to actual resources available to BorgMaster in a given cell. The Borg system also exposes a web-based interface called Sigma, which displays the state of all users' jobs, shows details of their execution history and, if the job has not been scheduled, also provides a 'why pending?' annotation where there is guidance about how to modify the job's resource requests to better fit the cell (Verma et al., 2015).

The dynamic nature of the Borg system means that tasks might be started, stopped and then rescheduled on an alternative node. Google engineers have created the concept of a static Borg Name Service (BNS) which is used to identify a task run within a cell and to retrieve its endpoint address. The BNS address is predominantly used by load balancers to transparently redirect RPC calls to a given task's endpoints. Meanwhile, the Borg's resource reclamation mechanisms help to reclaim under-utilised resources from cell nodes for non-production tasks. Whilst in theory users may request high resource quotas for their tasks, in practice they are rarely fully utilised in a continuous manner; rather, they have peak times of day or are used in this way when coping with a denial-of-service attack. BorgMaster has routines that estimate resource usage levels for a task and reclaim the rest for low-priority jobs from the batch or the best effort bands (Verma et al., 2015).

### 2.4.2. JOBS SCHEDULERS

Early versions of Borg had a simple, synchronous loop that accepted jobs requests and evaluated on which node to execute them. The current design of Borg deploys several schedulers working in parallel – the scheduler instances use a share state of the available resources, but the resource offers are not locked during scheduling decisions (optimistic concurrency control). In case of conflict, when two or more schedulers allocate jobs to the same resources, all the jobs involved are returned to the jobs queue (Schwarzkopf et al., 2013).

When allocating a task, Borg's scheduler scores a set of available nodes and selects the most feasible machine for this task. Initially, Borg implemented a variation of Enhanced Parallel Virtual Machine algorithm (E-PVM) (Amir et al., 2000) for calculating the task allocation score. Although this resulted in the fair distribution of tasks across nodes, it also resulted in increased fragmentation and later difficulties when fitting large jobs which required the most of the node's resources or even the whole node itself. An opposite to the E-PVM approach is a best-fit strategy, which, in turn, packs tasks very tightly. The best-fit approach may result in the excessive pre-empting of other tasks running on the same node, especially when the resources required are miscalculated by the user, or when the application has frequent load spikes. The current model used by Borg's scheduler is a hybrid approach that tries to reduce resource usage gaps (Verma et al., 2015).

Borg also takes advantage of resources pre-allocation using 'allocs' (short for allocation). Allocs can be used to pre-allocate resources for future tasks in order to retain resources between restarting a task or to gather class-equivalent or related tasks, such as web applications and associated log-saver tasks, onto the same machine. If an alloc is moved to another machine, its tasks are also rescheduled.

One point to note is that, similar to MetaCentrum users (Klusáček and Rudová, 2010), Google's users tend to overestimate the memory resources needed to complete their jobs, in order to prevent jobs being killed due to exceeding the allocated memory. In over 90% of cases, users tend to overestimate the amount of resources required, wasting in some cases close to 98% of the requested resource (Moreno et al., 2013; Ray et al., 2017).

### 2.4.3. OPTIMISATIONS

Over the years, Borg design has acquired a number of optimisations, namely:

- Score caching – checking the node's feasibility and scoring it is a computation-expensive process. Therefore, scores for nodes are cached and small differences in the required resources are ignored;
- Equivalence classes – submitted jobs often consist of a number of tasks which use the same binary and which have identical requirements. Borg's scheduler considers such a group of tasks to be in the same equivalence class. It evaluates only one task per equivalence class against a set of nodes, and later reuses this score for each task from this group;

- Relaxed randomisation – instead of evaluating a task against all available nodes, Borg examines machines in random order until it finds enough feasible nodes. It then selects the highest scoring node in this set.

While the Borg architecture remains heavily centralised, this approach does seem to be successful. Whilst this eliminates head-of-line job blocking problems and offers better scalability, it also generates additional overheads for solving resource collisions. Nevertheless, the benefits from better scalability often outweigh the incurred additional computation costs which arise when scalability targets are achieved (Schwarzkopf et al., 2013).

## 2.5. SUMMARY AND CONCLUSIONS

This chapter has presented a taxonomy of available schedulers, ranging from early implementations to modern versions. Aside from optimising throughput, different class schedulers have evolved to solve different problems. For example, while OS schedulers maximise responsiveness, Cluster schedulers focus on scalability, provide support a wide range of unique (often legacy) applications, and maintain fairness. Big Data schedulers are specialised to solve issues accompanying operations on large datasets and their scheduling mechanisms are often extensively intertwined with programming language features.

Table 3 presents a comparison of the presented schedulers with their main features and deployed scheduling algorithms:

| Scheduler class | Requirements known pre-execution | Fault-tolerance mechanisms | Configuration | Common algorithms | Scheduling decision overhead | Design focus (aside throughput) |
|---|---|---|---|---|---|---|
| OS Schedulers | No | No | Simple (compile-time and runtime parameters) | CS, CQ, MLFQ, O(n), O(1), Staircase, WFQ | very low - low | • single machine<br>• NUMA awareness<br>• responsiveness<br>• simple configuration |
| Cluster Schedulers | Yes[1] | Yes | Complex (configuration files and GUI) | FCFS (backfilling and gang-scheduling), SJF, Best-Fit, Scoring Functions | low - high | • distributed nodes<br>• fairness<br>• complex sharing policy<br>• power consumption<br>• fault-tolerance |
| Big Data Schedulers | Yes[2] | Yes | Complex (configuration files and GUI) | Best-Fit, FCFS (locality and gang-scheduling), Greedy, Fair Scheduler, Round Robin | low - medium | • specialised frameworks<br>• parallelism<br>• distributed data storage<br>• massive data |

1. Cluster users are notorious in overestimating resources needed for the completion of their tasks, which results in cluster system job schedulers often over-allocating resources (Klusáček and Rudová, 2010; Moreno et al., 2013).
2. MapReduce jobs tend to have consistent resource requirements, i.e. in majority of cases, every 'map' task processes roughly the same amount of data (input data block size is constant), while 'reduce' task requirements shall be directly correlated to the size of returned data.

Table 3: Schedulers comparison

OS schedulers have evolved in such a way that their focus is on maximising responsiveness while still providing good performance. Interactive processes which sleep more often should be allocated time-slices more frequently, while background processes should be allocated longer, but less frequent execution times. CPU switches between processes extremely rapidly which is why modern OS scheduling algorithms were designed with a very low overhead (Wong et al., 2008; Pinel et al., 2011). The majority of end-users for this class of schedulers are non-technical. As such, those schedulers usually have a minimum set of configuration parameters (Groves et al., 2009).

OS scheduling was previously deemed to be a solved problem (Torvalds, 2001), but the introduction and popularisation of multi-core processors by Intel (Intel Core™2 Duo) and AMD (AMD Phenom™ II) in the early 2000s enabled applications to execute in parallel. This mean that scheduling algorithms needed to be re-implemented in order to once again be efficient. Modern OS schedulers also consider NUMA properties when deciding which CPU core the task will be allocated to. Furthermore, the most recent research explores the potential application of dynamic voltage and frequency scaling technology in scheduling to minimise power consumption by CPU cores (Sarood et al., 2012; Padoin et al., 2014). It is not a trivial matter to build a good universal solution which caters to the complexities of modern hardware; therefore, it would be reasonable to develop the modular scheduler architecture suggested in (Lozi et al., 2016).

Cluster schedulers have a difficult mission in ensuring ‘fairness’, that is, sharing cluster resources proportionally to every user while maintaining stable throughput in a very dynamic environment consisting of variety of applications. Cluster systems tend to allow administrators to implement complex resource sharing policies with multiple input parameters (Adaptive Computing, 2002). Cluster systems implement extensive fault-tolerance strategies and sometimes also focus on minimising power consumption (Lang and Patel, 2010). Surprisingly, the most popular approach to scheduling is a simple FCFS strategy with variants of backfilling. However, due to the rapidly increasing cluster size, the current research focuses on parallelisation, as seen with systems such as Google’s Borg and Microsoft‘s Apollo.

Big Data systems are still rapidly developing. Nodes in Big Data systems fulfil the dual purposes of storing distributed file system parts and providing a parallel execution environment for system tasks. Big Data schedulers inherit their general design from cluster system’s jobs schedulers. However, they are usually much more specialised for the purpose of the framework and are also intertwined with the programming language features. Big Data schedulers are often focused on ‘locality optimisation’ or running a given task on a node where input data is stored or in the closest proximity to it.

The design of modern scheduling strategies and algorithms is a challenging and evolving field of study. While early implementations were often based on simplistic approaches, such as a CS, it is the case that modern solutions use complex scheduling schemas. Moreover, the literature frequently mentions the need for a modular scheduler architecture (Vavilapalli et al., 2013; Lozi et al., 2016) which could customise scheduling strategies to hardware configuration or applications.

This project's key research question was to investigate possible advances to the designs of Cloud load balancers. Two discrete research tracks emerged during this background review, namely: (i) further improvements to the existing monolithic scheduler design and (ii) a novel decentralised architecture based on agent system. These approaches are subsequently detailed in Chapters 6 and 7 respectively. However, before those load balancer designs could be experimented with, the research had to complete other crucial steps, such as formally defining the research problem as presented in the following chapter.

# 3. CLOUD RESOURCES UTILISATION MODEL

The examination of existing schemas in Chapter 2 provided a clear outlook of how a scheduler should work and what its main design goals should be, insofar as the workload model could be defined. Of all the solutions researched, the most similar was the orchestration software used in Clusters. However, while Cluster schedulers are predominantly focused on the fair use of available resources, the main purpose of commercial Cloud systems is to keep third party operations working continuously and with minimal disturbance. Grid or Cluster systems have the capacity to queue jobs when requested resources are not immediately available and to process them when they become available, while Cloud systems must provide or deny resources with the minimum possible delay to compute the decision (Hacker, 2010).

Therefore, in the majority of problem instances, it may be assumed that the system already has the capacity to process all current jobs, although the system should be able to detect and handle situations where existing resources are insufficient. The main challenge is to allocate those jobs properly so that no single node is overloaded and the system is stable, as understood in (2) in section 3.3.

In recent years, services provided by Cloud data centres have become gradually more diversified (Kanev et al., 2015) as well as bigger (Verma et al., 2015, Burns et al., 2016). Given this, the scheduler system should be able to cope with such an effect. This research will focus mainly on providing system stability combined with optimal minimal cost. Other features, such as fairness and data locality, will be considered only as secondary objectives.

This chapter introduces the Cloud Resource Utilisation Model (CRUM), and is based on work published in Sliwko (2008), Sliwko and Getov (2015a) and Sliwko et al. (2015).

## 3.1. NODES AND TASKS

The CRUM consists of nodes and tasks where the purpose of the load balancer is to keep a good load balance through resource vector comparisons. The Cloud Computing definition recognises four distinctive service models (Mell and Grance, 2011; Burnett et al., 2011; Limoncelli et al., 2014):

- Software as a Service (SAAS), where the consumer uses a provider's applications running on a Cloud infrastructure. The applications might be accessed directly by consumers, such as through web-based email and web services, or via a specialised program, as with most mobile applications. The cloud infrastructure is completely transparent for the end-user.
- Platform as a Service (PAAS), where the consumer is provided with the ability to deploy and run its applications within a Cloud system. However, the consumer does not control the underlying cloud infrastructure, but has control over the deployed applications and limited control over the hosting environment's configuration and settings.
- Infrastructure as a Service (IAAS), where the consumer is provided with a variety of fundamental computing resources – usually network-based – as with DNS, routing, storage, databases and firewalls.
- Unified Communication as a Service (UCAAS), where the service provider packages multi-platform communications channels. These services might include physical devices including mobile devices, IP telephony or video conferencing modules.

This research will focus on the PAAS model. In considering what is actually constituted as a 'task' in a Cloud environment, an example may be seen in a popular Cloud environment such as Amazon's EC2, where applications are deployed within the Virtual Machine (VM). Those VM instances often carry much more than they need in order to support different hardware configurations, execution environments, and varying user tasks (Younge et al., 2010). Depending on the design, some long-lived tasks might come also with preinstalled local database such as PostgreSQL. This schema has many benefits, such as the almost complete separation of the execution contexts and OS environment parameters, although the tasks might still share the same hardware if deployed on the same node. Depending on the type of contract signed, tasks might have guaranteed execution environment parameters, which would generally be for a fixed price, or share resources with other VM, which would generally be pay-as-you-go (e.g. Amazon EC2 Spot Instances).

Tasks require resources which are provided by the nodes. Every node has a certain quantity of variable resources available, referred to in this manuscript as the available resources. All resources on nodes are considered renewable and continuous, meaning that these resources do not expire and cannot be depleted: assigning a task to a node only lowers the available resource levels temporarily. To simplify the definition, both the resources needed by the task and the resources available on the node are described by the vector of non-negative real numbers. Several types of resources exist which can be utilised by the task, such as memory, CPU cycles and disk I/O operations. The model also supports artificial resources, called 'virtual resources'. Given this, the number of defined resources is potentially unlimited.

Tasks may have their resource needs shaped differently. There will be tasks experiencing hourly, daily or weekly variability in usage (Mao and Humphrey, 2011). Some Cloud systems introduce features that ensure an application will be able to cope with increasing traffic to maintain performance (Namjoshi and Gupte, 2009). One such example of this is the Amazon EC2 AutoScale, which can automatically start and stop additional application instances during demand spikes and lulls in order to minimise costs. AutoScale is part of Amazon's

CloudWatch package, which can monitor a range of basic system values such as CPU utilisation, data transfer and disk usage activity. Additionally, in can handle several complex metrics including DynamoDB tables, EBS volumes, Elastic Load Balancers and Amazon SQS queues.

## 3.2. SYSTEM TRANSFORMATION COST

A Cloud system environment is characterised by very dynamic changes in resource availability. To name just a few possible scenarios during its operation, some nodes might become idle or overloaded, additional resources might become available, new nodes might be added to the network, the demand for particular service may decrease, or part of a cloud network could go offline. Therefore, it is critical to provide a mechanism to proactively migrate tasks to alternative nodes.

Distributed systems often store or process large amounts of 'state'. State consists of data such as databases, files, relations, session data and identifiers which are frequently updated (Qiao et al., 2013). On the other hand, 'corpus' is a body of data that is rarely updated and relatively static, as with a library index which may be updated once per month, as opposed to an email system, which is constantly updated as new messages arrive continuously. The system might store all state on one machine, although this strategy quickly reaches its limit as one machine might be able to store only a limited amount of state, and may be unable to serve data requests fast enough. Distributed computing system designers have created several strategies to deal with this issue. Most use replication, sharing and sharding, which brings problems of consistency, availability and partitioning of data (Limoncelli et al., 2014).

In modern virtualised Cloud environments, programs are usually deployed in VMs (Limoncelli et al., 2014). VM state transfer is performed via VM-LM where VM instance is transferred on the fly to another machine. This process, known as ‘teleporting’ in VirtualBox (Oracle, 2018) and ‘vMotion’ in VMware (Marshall, 2015), transfers CPU, registry, memory, network connections and mappings to the persistent storage of another machine without stopping the execution of the original instance. In addition to the saved task’s state, the cloud system might also need to copy huge VM system image files, such as custom Linux VM images in Amazon EC2 cloud, wherein the size varies according to which system image is used. The size of VM image might significantly impact on the migration cost, meaning that image trimming is always advised. Younge et al. (2010) identified one case in which the system image was significantly decreased from 4GB to 636MB without any loss in the functionality provided.

In this research, it is assumed that every virtualised application deployed on the Cloud is available to be migrated live. In this model, every task has a cost value assigned, which can be seen as an abstract representation of the impact the task migration will have. The model considers the task migration cost to be the total data transferred over a network such that it can move the already running application to an alternative node. Chapter 4 discusses this approach and details the LMDT formula which can be used to estimate the total size of the transferred data during VM-LM.

The task migration cost value is considered to be constant in a given time window since the migration of a certain task to any node will cause the same impact throughout the whole system. Furthermore, the deployment process is standardised and automated regardless of the vendor – in most cases it is just enough to start VM instance in listening mode on the target node, and then to point the currently running VM instance to this location. The VM manager will take care of the proper allocation itself. In other words, the amount of work required to initiate the same program in different environments is either the same, or there is very little variation.

Figure 4 visualises the system transformation process, and highlights the migration costs incurred by re-allocating tasks.

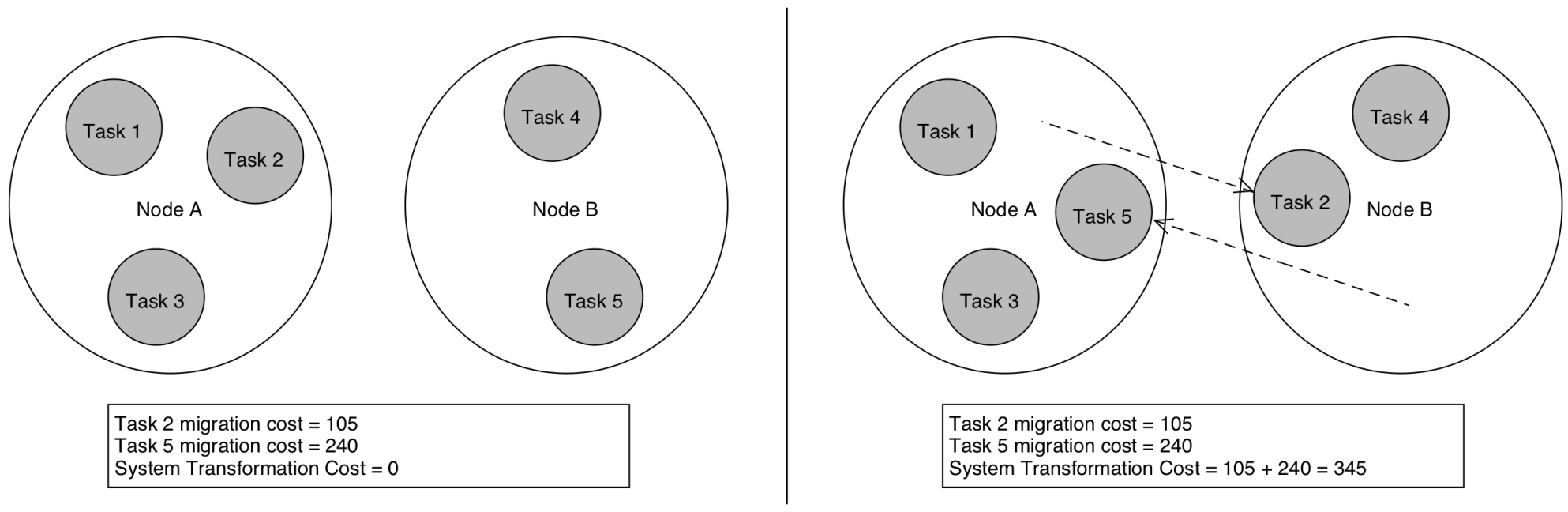

Figure 4: System Transformation Cost

Here, the left side of the figure presents the initial state of the system in which tasks 1, 2 and 3 are being executed on Node A, and tasks 4 and 5 are run on Node B. The system undergoes transformation, with the nodes exchanging tasks 2 and 5 (on the right side of the figure). Tasks are being migrated via VM-LM and this process incurs the following migration costs: 105MB for the migration of Task 2 and 240MB for the migration of Task 5. The total size of data transferred while re-allocating tasks to alternative nodes is called System Transformation Cost (STC). In this sample, STC is 345MB. STC is formally defined as (4) in the following section.

## 3.3. PROBLEM FORMULATION

To better introduce the D-Resource System Optimisation Problem (D-RSOP), Figure 5 visualises how the node's resources are utilised by tasks and shows how the node's state is evaluated as being stable or overloaded. Both Figures 4 and 5 present the same scenario; however, the former highlights the resources utilisation changes as the system is transformed (i.e. tasks are re-allocated):

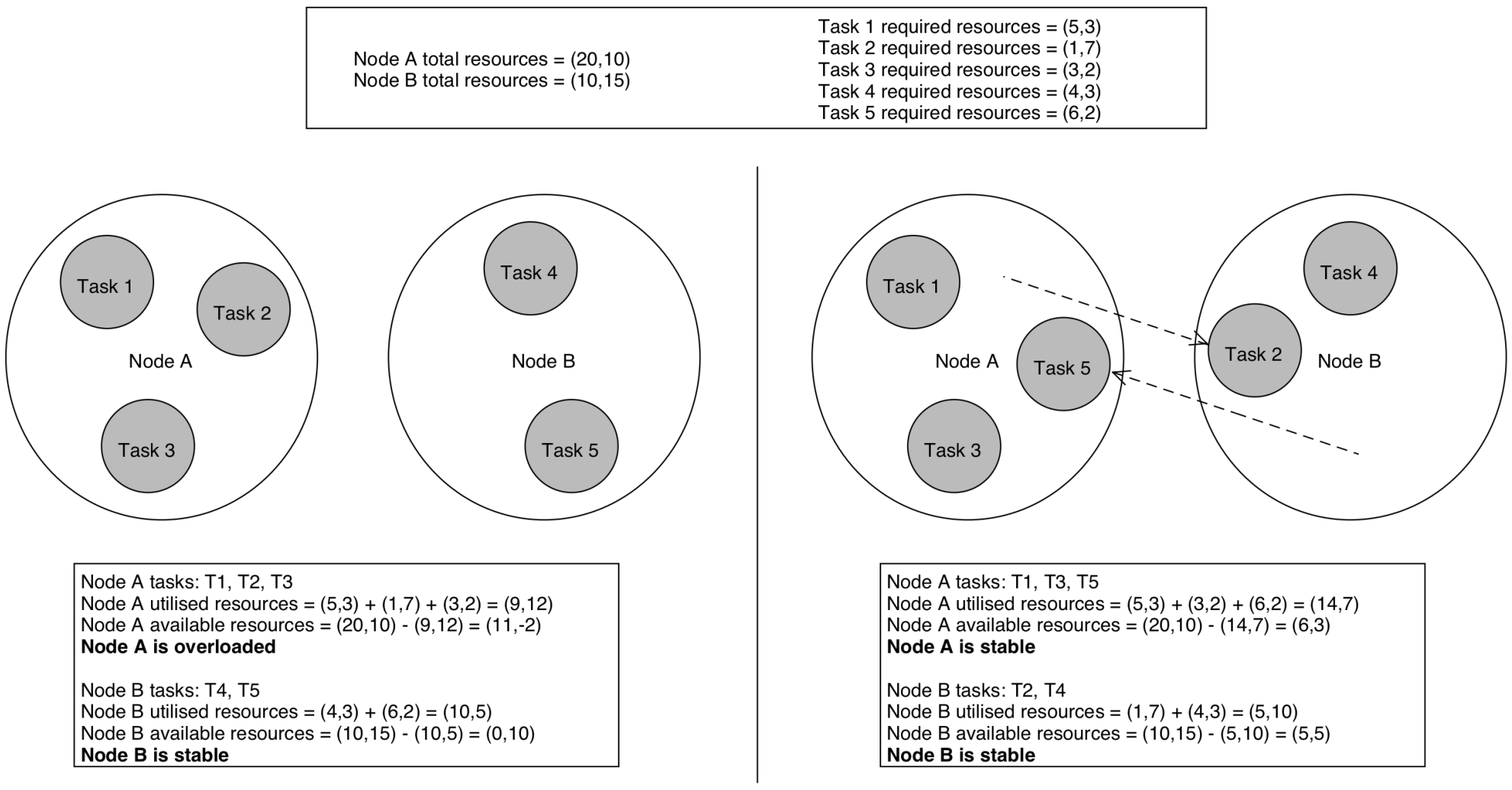

Figure 5: Sample system transformation

Here, two resource types are defined in the system. For example, Task 1 requires (5,3) of resources, which could be CPU and memory. Again, the left side of the figure presents the initial state of the system in which Node A is overloaded due to the available resource levels being negative, i.e. Node A available resources equal (11,-2) with the second value in pair being negative. In such a setup, Node B is stable, but since Node A is overloaded, the system state is overloaded.

After the system transformation (the right side of the figure), which consists of migrations Task 2 to Node B and Task 5 to Node A, both system nodes are stable. As such, the system state is stable. It should be also noted that this system transformation incurs STC – a total of all migrations costs of re-allocated tasks (see Figure 4 for a relevant sample).

In the D-RSOP, let us define:

- $\eta = \{n_1, n_2, \ldots, n_m\}$ as a set of all nodes in the system;
- $\tau = \{t_1, t_2, \ldots, t_p\}$ as a set of all tasks in the system;

- $\psi = \{i_1, i_2, \dots, i_d\}$ as a set of all kinds of resource types defined in the system such as CPU, memory, network bandwidth, and so on. Please note that the below definitions are subscripted with '$i$' as a function for a resource type $i$. The model also supports 'virtual resources' (see discussion in section 3.1).
  E.g. for a system with three resource types we could define $i \in \{CPU, memory, network\}$;

- $a: \psi \times \eta \to \mathbb{R}_0^+$ as a node's total resources function and $a_i(n)$ as a total level (non-negative real number) of a resource $i$ on the node $n$.
  E.g. $a_{CPU}(n_1) = 2$ specifies node $n_1$ as having two CPU cores installed and dedicated to use by tasks;

- $r: \psi \times \tau \to \mathbb{R}_0^+$ as a task's required resources function and $r_i(t)$ as a required level (non-negative real number) of a resource $i$ of task $t$.
  E.g. $r_{CPU}(t_1) = 0.5$ specfies task $t_1$ as requiring half of CPU core's time to run;

- $c: \tau \to \mathbb{R}^+$ as a task migration cost function and $c(t)$ as a task migration cost (the size of data in MB) for a task $t$, namely the cost incurred migrating task's executable and its state.
  E.g. $c(t_1) = 210$ specifies task $t_1$ as needing to transfer 210MB of data during task migration process (see Chapter 4 for details of LMDT formula);

- $\mu: \tau \to \eta$ as a task assignment function where a task has to be assigned to a node, i.e. $\mu$ is defined for each $t \in \tau$. Each task $t$ is initially assigned by task assignment function $\mu_0$ to some node $n \in \eta$. During the system transformation, any number of tasks can be re-allocated to different nodes and a new task assignment function $\mu_1$ is created. Such a system transformation is referred to as $(\mu_0 \to \mu_1)$.
  E.g. $\mu(t_1) = n_1$ specifies that task $t_1$ is assigned to node $n_1$. It is assumed that task $t_1$ consumes the resources available on node $n_1$;

- $\Lambda = (\tau, \eta, \psi, a, r, c)$ is considered as a problem space and pair $(\Lambda, \mu)$ as a system. Please note that when computing the system transformation, $\Lambda$ remains unchanged while $\mu$ is modified;
- For every node $n \in \eta$ we define a set $\tau_n = \{t \in \tau: \mu(t) = n\}$ of all tasks assigned to the node $n$.
  E.g. if the system consists of node $n_1$ and tasks $\tau = \{t_1, t_2\}$, and $\mu(t_1) = n_1$ and $\mu(t_2) = n_1$ (meaning that tasks $t_1$ and $t_2$ are assigned to node $n_1$), then $\tau_{n_1} = \{t_1, t_2\}$;
- $f: \psi \times \eta \to \mathbb{R}$ as the available resources levels function on the nodes

$$f_i(n) = a_i(n) - \sum_{t \in \tau_n} r_i(t) \tag{1}$$

E.g. if the system consists of node $n_1$ and tasks $\tau = \{t_1, t_2\}$, and $\tau_{n_1} = \{t_1, t_2\}$, $a_{CPU}(n_1) = 2$, $r_{CPU}(t_1) = 0.5$ and $r_{CPU}(t_2) = 0.2$, then $f_{CPU}(n_1) = a_{CPU}(n_1) - \big(r_{CPU}(t_1) + r_{CPU}(t_2)\big) = 2 - (0.5 + 0.2) = 1.3$, meaning that node $n_1$ has 1.3 CPU core available (as a CPU idle time).

We consider system $(\Lambda, \mu)$ as stable, if:

$$f_i(n) \geq 0, \text{ i.e. } \textstyle\sum_{t \in \tau_n} r_i(t) \leq a_i(n) \text{ for every } n \in \eta,\ i \in \psi \tag{2}$$

Meaning, that every node in the system is stable (has no overloaded resources). Otherwise, the system $(\Lambda, \mu)$ is overloaded. This consideration is referred to as Goal (I).

During the system transformation $(\mu_0 \to \mu_1)$, a task may be re-allocated to a different node. This process is referred to as task migration. Definition (3) specifies the cost of migration for task $t$ within the system transformation $(\mu_0 \to \mu_1)$:

$$c_{(\mu_0 \to \mu_1)}(t) = \begin{cases} 0, & \mu_0(t) = \mu_1(t) \\ c(t), & \mu_0(t) \neq \mu_1(t) \end{cases} \tag{3}$$

This denotes that task migration cost is incurred only if the task changes the node it is assigned to. E.g. if within $(\mu_0 \to \mu_1)$ we re-allocate task $t_1$ and $t_2$, but don't re-allocate task $t_3$, then $c_{(\mu_0 \to \mu_1)}(t_3) = 0$, meaning only migrated tasks incur migration costs.

Every system transformation process $(\mu_0 \to \mu_1)$ has STC defined as a sum of all incurred migration costs (unmigrated tasks have zero migration cost):

$$s_{(\mu_0 \to \mu_1)} = \sum_{t \in \tau} c_{(\mu_0 \to \mu_1)}(t) \tag{4}$$

Considering the initial task assignment $\mu_0$, the task assignment $\mu^*$ is optimal for $\mu_0$, if $\mu^*$ renders system $(\Lambda, \mu^*)$ stable and:

$$s_{(\mu_0 \to \mu^*)} \leq s_{(\mu_0 \to \mu)}, \text{ for every stable system } (\Lambda, \mu) \tag{5}$$

N.b. when $(\Lambda, \mu_0)$ is stable for initial task assignment $\mu_0$, the STC equals zero as it is considered optimal. Minimising the STC is referred to as Goal (II).

We also consider two task assignment functions $\mu_0$ and $\mu_1$ to be neighbours if:

$$|\{t \in \tau: \mu_0(t) \neq \mu_1(t)\}| = 1 \tag{6}$$

This means that, only a single task has changed node within the system transformation $(\mu_0 \to \mu_1)$. Definition (6) has been introduced in order to better support a design based on selected metaheuristic algorithms in which a single step evaluates a set of neighbour solutions (algorithms are listed in subsections 6.1.1 to 6.1.5).

## 3.4. PROBLEM ANALYSIS

The class of problems for which an algorithm can provide an answer in polynomial time is called ‘class P’. For some problems, there is no known way to find an answer ‘fast’ in polynomial time; nevertheless, the answer can be verified in polynomial time, for example the subset sum problem ‘given a set of integers, does some nonempty subset of them sum to zero?’ (Frieze, 1986). A class of problems for which an answer cannot be verified in polynomial time is called NP.

NP-Hard class problems are those which are ‘at least as difficult as problems in NP’ (Schirmer, 1995); all NP problems can be reduced in polynomial time to NP-Hard class. NP-Hard problems need not be in NP since they need not have solutions verifiable in polynomial time.

NP-Complete problems (Karp, 1972) can be solved through an exhaustive search, although the time to wait for the solution grows unacceptably with the problem size as the number of iterations needed to solve the problem becomes enormous (Schirmer, 1995). In such cases, the best scenario is to use super-polynomial time algorithms. The ‘P versus NP’ problem (Levin, 1973; Cook, 1975) is one of the seven open Millennium Prize Problems of the Clay Mathematics Institute, and is considered by many to be the most important open problem in the field (Fortnow, 2009). It is now commonly believed that P ≠ NP, and that it is rather unlikely that any efficient (Polynomial Time) exact algorithms will be able to solve NP-hard problems. NP-hard problems may be of any type, ranging from search, decision, or optimisation problems to feasibility problems (Schirmer, 1995), although discrete optimisation problems are generally NP-hard problems.

The D-RSOP is a variant of a classical Resource-Constrained Project Scheduling Problem (RCPSP), meaning that D-RSOP also belongs to the NP-hard (Nondeterministic Polynomial-time hard) problems class. Since its advent, RCPSP has been examined numerous times by researchers, with numerous solutions having been proposed, implemented and tested (Boctor, 1990; Demeulemeester and Herroelen, 1992; Józefowska et al., 2002; Bouleimen and Lecocq, 2003; Brucker et al., 2003; Lim et al., 2013).

RCPSP can be solved using simple heuristics, such as the algorithms H1m and HCRA (Józefowska et al., 2002), but the result quality is low. Although exact methods have been explored, either they have a limitation of problem size such as Branch and Bound (Bouleimen and Lecocq, 2003) and Constraint-Propagation-Based Cutting Planes (Demassey et al., 2005), or focus only on deriving new lower bounds. The reason for this is that the optimal solution can be found and verified only in small problem instances (Józefowska et al., 1998; Lim et al., 2013). Interesting examples include X-Pass methods (Davis and Patterson, 1975; Cooper, 1976), Scatter Search (Debels et al., 2006; Mobini et al., 2009) and Filter-and-Fan (Ranjbar, 2008). An exhaustive survey on the various methods employed to solve RCPSP problems can be found in Boctor (1990) and Kolisch and Hartmann (2006), where the standard benchmark data (Kolisch and Sprecher, 1997) is used for performance evaluation.

## 3.5. NP-HARDNESS PROOF

The defined D-RSOP can be compared to the so-called bin-packing problem in Computational Complexity Theory. The bin-packing relates to the questions of packing a number of objects of different volumes into a finite number of bins of a known capacity in a way that minimises the number of bins used. Finding a solution for $k$ bins is known as an NP-complete problem (Coffman et al., 1996). Let us define it as a $P_{k_bin}$ problem.

Let us define the $P_{opt_bin}$ problem as the problem of minimising the number of bins which can contain a specified set of objects. We can find a solution of $P_{k_bin}$ by solving the $P_{opt_bin}$ problem: it is enough to compute an optimal number of bins $k_{opt}$ and compare it with $k$. Thus, $P_{opt_bin}$ is NP-hard as $P_{k_bin}$ can be reduced to $P_{opt_bin}$ by a polynomial-time many-one reduction.

Let us define the problem $P_{d_opt_bin}$ as an instance of a D-RSOP with the following re-definitions:

- $a: \psi \times \eta \rightarrow \{v\}$, where $v \in \mathbb{R}_0^+$ is the known bin capacity (7)
- $c: \tau \rightarrow \{0\}$, i.e. there is no task migration cost (8)

The above assumptions imply the STC is zero (as every task $t \in \tau$ can be freely re-allocated). In such a definition, (5) is always satisfied, and thus the task assignment optimality is subject only to (2).

If we add the additional consideration:

- $|\psi| = 1$, i.e. we consider only one kind of a resource (9)

we can see that $P_{opt_bin} \leq P_{d_opt_bin}$, i.e. $P_{d_opt_bin}$ is at least as hard as $P_{opt_bin}$. Then $P_{d_opt_bin}$ is NP-hard; consequently, D-RSOP is NP-Hard ■

## 3.6. SUMMARY AND CONCLUSIONS

Modelling the workings of a Cloud system is a non-trivial task. The CRUM presented here is the outcome of a review of the scheduling mechanisms which currently exist, and the related literature as presented in Chapter 2.

Based on this analysis, the D-RSOP was defined together with Goals (I) and (II), which are then used to evaluate the designed load balancer solution. Two crucial challenges were identified, which subsequently became the focal areas of the following research:

- Firstly, this chapter highlighted the key problem of calculating the STC, i.e. the impact of reassigning consecutive tasks to alternate nodes. Chapter 4 will focus on examining the dimensions of this problem, directly estimating the task migration cost via experimental work.
- Secondly, this analysis also highlighted the vast complexity of the load balancing in distributed systems, especially when considering the overall dynamicity of Cloud environments. It became apparent that, by itself, static modelling does not yield satisfactory results, and therefore that a more practical approach is required for the research. The resulting high-fidelity Cloud workload simulator is presented in Chapter 5.

Although very stimulating, the formal analysis presented in this chapter did not provide a definitive answer to the researched problem of large-scale load balancing. This said, it did yield a substantial stepping-stone, which was useful in research.

# 4. VIRTUAL MACHINE LIVE MIGRATION

Having identified the main challenges and requirements of a comprehensive load balancing strategy for a Cloud, the next area of study was to determine how to practically re-allocate running programs between nodes. CRUM, introduced in Chapter 3, requires that tasks can move across Cloud nodes without losing their execution state. Therefore, an additional study was needed in order to become more familiar with Cloud virtualisation layers.

Cloud systems are unique in their elasticity, that is, their ability to dynamically allocate available resources to applications is unprecedented in computer science history. This elasticity is backed up by large-scale virtualisation, where every aspect of an application runs in a dedicated VM environment. Applications are executed in VM instances and are no longer bounded to a physical node. This means it is possible to move the VM instances around and to place them easily within another node if the target node meets the requisite task constraints. VM instances can be migrated 'cold', whereby an instance is suspended, transferred to an alternative node and resumed, although it should be noted that during the migration progress services rendered by tasks are unavailable (Sapuntzakis et al., 2002). Modern VMs such as XenServer (Barham et al., 2003), VirtualBox, and VMware also support VM-LM, where VM instances are migrated on the fly. In this case there is no offline period, except for a short final synchronisation pause of around 60 to 300ms (Clark et al., 2005).

Historically, VM-LM technology was debuted by VMware with the introduction of vMotion and GSX Server in 2003. Soon, other vendors attempted to develop VM-LM features of their own, for example Microsoft added Quick Migration in its Windows Server 2008 Hyper-V (later renamed to Live Migration) (Savill, 2016) and Citrix released XenMotion for XenServer in the same year. There have been several studies modelling various aspects of the transfer cost for VM-LM since then. The most notable examples of related work include:

- The impact of allocated VM memory size on migration time (Zhao and Figueiredo, 2007; Salfner et al., 2011; Dargie, 2014);
- The impact of memory page dirtying rate on migration time (Verma et al., 2011, Rybina et al., 2015) and the downtime length (Salfner et al., 2011; Liu et al., 2013);
- The effect of available network bandwidth on migration time (Akoush et al., 2010; Zhang et al., 2016; Deshpande and Keahey, 2017);
- The energy overhead required to perform VM-LM (Huang et al., 2011; Liu et al., 2013; Callau-Zori et al., 2017);
- Determining the Quality of Service specifications for migrated VMs and applying resource control mechanisms during VM-LM (Abali et al., 2017);
- A strategy for parallel migrations of multiple VMs (Sun et al., 2016);
- Various memory transfer optimisations as presented in Noel and Tsirkin (2016), Tsirkin and Noel (2016), Ramasubramanian and Ahmed (2017).

While these approaches are valid in general, they focus solely on the impact for a particular VM instance and consider only factors such as loss of performance or network packages, length of downtime or impact on users. However, the migration of VM instances also causes disruptions on the infrastructure level, especially when non-trivial volumes of data need to be transferred and clutter network bandwidth, which could be allocated to alternative processes. Therefore, the research work presented in this article focuses on VM-LM and evaluates the total volume of information to be migrated. The main contributions of this chapter are the experiments and the Live Migration Data Transfer (LMDT) formula which helps to estimate the total size of data transferred over a network during the VM-LM process.

## 4.1. VIRTUAL MACHINES IN CLOUD COMPUTING

While a range of tools and virtualisation technologies for building Clouds exist (Jin et al., 2010; Luo et al., 2011), the virtualisation solutions currently deployed across service providers are unfortunately not standardised, and differ significantly in many aspects. In particular, larger and highly specialised solutions such as Google Cluster (Hellerstein, 2010) tend to be vastly customised in order to support their core operations. Based on the initial classification (Shirinbab et al., 2014), the existing virtualisation approaches can be divided into the following five categories:

- Full virtualisation relies upon an on the fly in-kernel translation of privileged instructions to user-level instructions. This results in significant performance drop since binaries of applications and their libraries must be analysed and transformed during the execution.
- Paravirtualisation requires modification to the source code of the Guest OS. All privileged instructions are replaced with function calls to the hypervisor services, i.e. 'hypercalls'. The biggest drawback of this method is the necessity to have access to the source code of the Guest OS, which is not always possible and may interfere with the intellectual property rights of commercial OS-es.
- Hybrid virtualisation generally offers superior performance in comparison to the types above. In this model the Guest OS uses paravirtualisation for certain hardware drivers and full virtualisation for other features. For example, the Guest OS can take advantage of hardware support for nested page tables, thereby reducing the number of hypercalls required for virtual memory operations. At the same time the Guest OS can benefit from fast I/O operations via lightweight access to paravirtualised devices as there is no need to rely on emulated hardware (Chisnall, 2008).

- Hardware-assisted virtualisation has the advantage of hardware-level support. Recent additions to hardware have introduced several processor-level and memory-level mechanisms which directly support virtualisation as part of the microarchitecture. Typical examples include Intel's VT-x and the AMD-V architectures at processor-level, while memory-level support is usually achieved within a memory management unit. This approach eliminates the need to hook and emulate privileged instructions by hypervisors, meaning the guest OS can run at its native privilege levels.
- Virtual containers (VC) is an OS level virtualisation methodology in which a specially patched kernel allows multiple isolated user-space instances. This solution is not a true hypervisor, but rather should be considered as an advanced implementation of the chroot operation. Nevertheless, from the users' point of view it is perceived as a real server (Dua et al., 2014). VCs impose almost none of virtualisation overhead costs since they operate inside a single kernel and require no hardware support to run efficiently. VCs are generally locked to a single kernel version such as Docker, LXC or OpenVZ, which makes this technology more suitable for running multiple instances of a single application (Tang et al., 2014; Smith, 2017). User-space instances are separated only by a container abstraction layer, and the VC security is considerably lower than in other virtualisation techniques.

For the purposes of this research work, six widespread VMs that support VM-LM have been shortlisted. The main selection criterion was to include only mature and optimised implementations of the VM-LM technology. While VM-LM was first introduced as far back as 2009, this feature is still being added and is available only as an experimental feature in many VMs. Therefore, the preference has been given to VMs supported by established corporations or a vast open-source community. Additionally, all selected VMs support a variety of platforms and generally have good compatibility with commonly available hardware, with the exception of XenServer, which requires certain hardware features to be available. The shortlisted VMs are as follows:

- XenServer (Barham et al., 2003) has become a very popular choice for Cloud systems, and is currently being used as a primary VM hypervisor in several Cloud providers including Amazon EC2, IBM SoftLayer, Liquid Web, GoGrid, Fujitsu Global Cloud Platform, Rackspace Cloud, and CloudEx (Jin et al., 2010).
- VirtualBox supports a wide set of Host OS-es, namely Linux, Mac OS X, Windows XP and in its later versions, Solaris, and OpenSolaris. In addition, there are ports to FreeBSD and Genode. Natively supported Guest OS-es are almost all versions of Windows, Linux, BSD, OS/2, Solaris, and so on. To achieve the best possible integration, VirtualBox comes with a set of native drivers and system applications called 'Guest Additions' that optimise the Guest OS for better performance and usability.
- The WMware product is a line of hypervisors. Type 1 runs directly on hardware while Type 2 runs on OS such as Windows, Linux and Mac OS X. VMware supports VM-LM, but it does not emulate instruction sets for hardware components that are not physically present. Instead, it focuses on running CPU instructions directly on the machine. However, this feature might cause problems when a VM instance is migrated to a machine which has a different hardware setup, such as using different instruction sets or having a different number of CPU cores (Mashtizadeh et al., 2011; Marshall, 2015).
- A KVM (Kernel-based VM) component was merged into Linux mainline kernel version 2.6.20. For KVM to work, the CPU must offer hardware-assisted virtualisation support: Intel's VT-x for the x86 architecture and VT-i for Itanium architecture or AMD-V for AMD processors. KVM currently supports saving/restoring the VM state and offline/online migrations. In addition, the VM-LM can be performed between AMD and Intel hosts.

- Hyper-V is also known as Windows Server Virtualisation services. It provides virtualisation services in Windows 8 or newer versions. Hyper-V is capable of running several unique instances, called 'partitions', each with its own kernel. Although VM-LM is supported, it is quite restricted and has several limitations, chief of which is that a VM instance can be migrated only between identical versions of Windows Server 2008. Furthermore, only x64 or Itanium architectures are supported, and all cluster nodes must be on the same TCP/IP sub-net.
- Docker works on the principles of VCs and relies on the resource isolation features of the Linux kernel, such as cgroups and kernel namespaces. Docker enables the creation of multiple independent VCs to run within a single Linux instance (Merkel, 2014), in which each is seen as a full OS capable of running services, handling logging, and so on. At the time of writing, Docker did not support Live Migration; the integration with the Checkpoint/Restore In Userspace (CRIU) tool does not allow the migration of a running application to the alternative container on the fly. However, recent publications (Yu and Huan, 2015) describe early experiments with Live Migration feature, while a working prototype was also demonstrated in 2016 (Estes and Murakami, 2016).

Table 4 presents a comparison of the selected VMs based on their core characteristics. It should be noted that the Host OS and the Guest OS lists are not exhaustive; other OS-es may work without modifications. XenServer and WMware are implemented as type-1 (bare-metal) hypervisors which can work directly on hardware without the need for a Host OS. Unfortunately, further information, including more details about design decisions and operational principles, tend to be proprietary, and as such not publicly available.

| Virtual Machine | Virtualisation approach | Guest OS performance | Live Migration technology | Host OS | Guest OS | License |
|---|---|---|---|---|---|---|
| XenServer | Paravirtualisation | Native | XenMotion, Storage XenMotion | Windows, OS X x86, Linux | Windows 2008/7/8/10, CentOS, Red Hat/SUSE/ Oracle/Scientific/Debian Linux, Ubuntu, CoreOS | Open Source, Commercial |
| VirtualBox | Full virtualisation, Paravirtualisation | Close to native, Native | Teleporting | Windows, OS X x86, Linux, Solaris, FreeBSD | Windows 98/NT/XP/2000/ Vista/2008/7/8/10, DOS, OS/2, FreeBSD, Mac OS, Solaris, Linux, Syllable, Ubuntu | Open Source |
| WMware | Full virtualisation | Close to native | vMotion | Windows, OS X x86, Linux | Windows, Linux, Solaris, FreeBSD, Netware, Mac OS, OS/2, DOS, CoreOS, BeOS, Darwin, Ubuntu, SUSE/Oracle Linux | Commercial |
| KVM | Full virtualisation *(requires AMD-V or Intel-VT-x)* | Close to native | Live Block Migration | Linux, FreeBSD, illumos | Windows Vista/2008/7/8/ 2012, CentOS, Red Hat, SUSE/Oracle Linux, Solaris | Open Source |
| Hyper-V | Full virtualisation | Close to native | Live Migration *(formerly: Quick Migration)* | Windows 8/Server 2012 (R2), Microsoft Hyper-V Server | Windows Vista/7/8/10/2008/ 2012/2016, Red Hat, Oracle/ SUSE/Debian Linux, Cent/OS, FreeBSD | Commercial |
| Docker | Virtual Containers | Native | *working prototype (but not Open Source yet)* | Windows Server 2016 *(only worker node)*, CentOS, Red Hat/SUSE/ Oracle Linux | *(same as Host OS)* | Open Source, Commercial |

Table 4: Virtual Machines comparison

In this research VirtualBox is considered as a representative VM in VM-LM research for the following reasons:

- VirtualBox does not require any dedicated hardware while XenServer Type 1, for example, runs only on a limited set of hardware.
- VM-LM has been supported since version 3.1, meaning it should be considered as a mature solution. Upon detailed inspection of VirtualBox's source code (see detailed discussion in subsection 4.2.2), it was established that this feature's design does not differ substantially from other VM implementations.
- During VM-LM, VirtualBox transfers only what is currently allocated and being used by the VM memory, i.e. not the total memory configured for the VM. The difference in transferred data might be substantial if applications within the VM instance do not fully utilise the available VM memory (Hu et al., 2013).

- VirtualBox is widely used. There are a number of freely available ports for all major operating systems such as Windows, Linux, Mac OS X, Solaris and FreeBSD in existence. There are many publicly available guides and resolutions to issues available on the Internet, as well as many free out-of-the-box and preinstalled VirtualBox images, such as VirtualBoxImages.com and VirtualBoxes.org websites.
- One of the biggest strengths of VirtualBox is its ability to virtualise nearly any type of platform, including Linux, Windows, OS X, Solaris, Android and Chromium OS. The VirtualBox community is very active and continuously improves compatibility with currently supported platforms as well as adding new ones. Furthermore, VirtualBox has universally good compatibility with hardware due to Guest Additions package.
- VirtualBox is used as a foundation for VC systems such as Docker. For example, on Mac OS X, docker-machine tool provisions a specialised VirtualBox VM instance to run its kernel.
- In comparison to other listed VMs, VirtualBox instances generally have fewer problems being migrated to nodes with a different Host OS and different processor architecture. As an example of the opposite, Hyper-V and WMware require the same type of CPU architecture between the source and target hosts.

In addition, VirtualBox is also easy to set up since binaries are provided directly from the Oracle site. VirtualBox provides a GUI management console and is generally easy to use even for inexperienced users, which makes experiments much easier. Since version 4, VirtualBox has been released as an Open Source project, leading to many fixes, improved stability and optimised performance patches being added to its source code.

## 4.2. LIVE MIGRATION

Both cold migration and VM-LM are techniques for moving one VM instance from one node to another, usually of the same type. Depending on the vendor, various restrictions might apply, although in general the same CPU architecture is required and the VM will also perform a check of available features and extensions. A number of VMs also support open formats, such as the Open Virtualisation Format, which allows the distribution of virtual appliances in a manner not tied to any particular hypervisor or processor architecture (Bernstein et al., 2009).

Cold migration requires stopping a VM and saving its state to snapshot files, moving these files to the destination and then restoring the VM instance from a previously saved state. An obvious disadvantage of this approach is the unavoidable downtime required to stop the VM, transfer the state files and start the VM on the target node during which the service is not accepting requests. Furthermore, Shirinbab et al. (2014) distinguish between cold migration, where the VM instance is actually shutdown, and hot migration, where the VM instance is only suspended, whereby the application preserves most of its state.

However, modern VMs support a more advanced on the fly migration. The VM-LM feature – called ‘teleporting’ in VirtualBox (Oracle, 2016) or ‘vMotion’ in VMware (Marshall, 2015) – is a powerful functionality of modern VMs. The VM-LM dramatically reduces the downtime needed for virtualised applications and is the most suitable approach to achieving high-availability services. In essence, a continuously running VM instance is transferred to another VM running on a different physical machine without stopping its execution. This is done by transferring all VM memory and CPU registers on the fly, switching network devices to a new network address, and then either transferring the whole local disk storage content or reopening interfaces used in the Network Attached

Storage (NAS). Therefore, the transfer cost of VM-LM depends on the specific application workload for the major areas of the migration process – CPU registers, memory, permanent storage, and network switching. In the following subsections, key VM-LM challenges, as well as a performance analysis of their impact on the total migration cost for those four major categories of application workload, are discussed.

### 4.2.1. COMPUTATION-INTENSIVE APPLICATIONS

Wu and Zhao (2011) have shown that the amount of available CPU cycles on a machine has a substantial impact on the migration time. One of their tests demonstrates that assigning more CPU cycles to the VM-LM process often results in an exponential reduction in the total migration time, but only to a point of around 50% CPU utilisation. In this research, assigning more than 50% of CPU utilisation did not shorten the migration time any further. Furthermore, the experiments have also shown that changes between 40% and 80% in the CPU utilisation for different applications did not noticeably affect the migration time. This can be explained by the relatively small size of the CPU registers and the L1/L2/L3/L4 caches that need to be copied over.

### 4.2.2. MEMORY-INTENSIVE APPLICATIONS

Memory usage intensity has a huge impact on migration time. Memory migration can be achieved by directly copying memory to a target node. This process consists of copying all memory pages one by one onto a target node. If the content of a memory page is altered during migration, this memory page is marked as dirty, which will result in another attempt to copy it again in the future.

Generally speaking, during every migration the majority of pages which are not modified frequently will be copied either once, or a small number of times.

However, a subset of memory pages, commonly referred to as a Writable Working Set (WWS), will be altered in a very rapid manner – much faster than the speed at which network adapters can exchange information. These pages are often used for the stack and the local variables of the running processes as well as for network and disk buffers.

The usual solution in this scenario is for the VM-LM mechanisms to detect and mark hot pages by counting how many times a memory page has been dirtied during migration (Jin et al., 2009), and then to synchronise those remaining memory pages at the final stage while both VMs are paused. In the first migration round all memory pages are copied to the target VM while the source VM continues to run. During that round, some of the pages will have their content modified and will be marked as dirty. In the next round there will be a further attempt to copy them. The next few rounds, i.e. round 2 and round 3 as shown in Figure 6, will attempt to copy previously dirtied pages, thereby hopefully decreasing the number of dirtied memory pages.

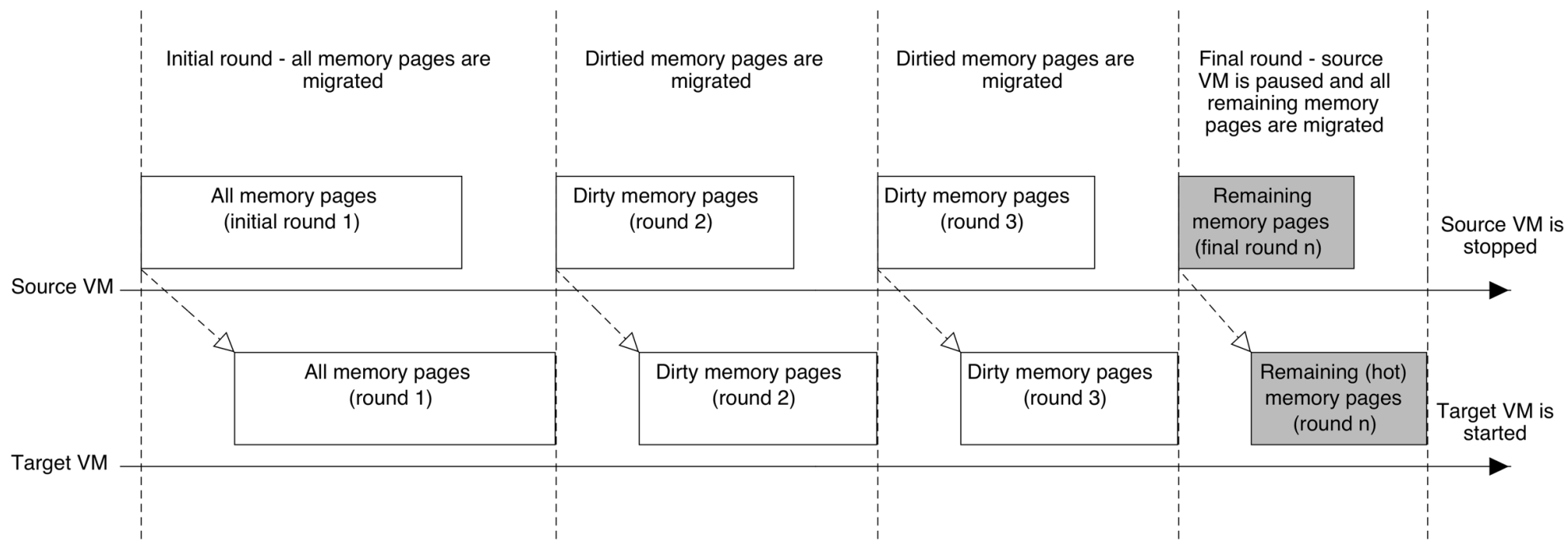

Figure 6: Memory migration rounds

However, some memory pages are dirtied so rapidly that network adapters cannot transfer them over the network fast enough. Therefore, the final round pauses the source VM and all remaining pages are copied onto the target VM, which subsequently starts while the source VM stops and shuts down. Clarke et al. (2005) provide an analysis and estimation of the size of WWS. Furthermore, specific research in Wu and Zhao (2011) has substantively examined what kinds of memory operations have an impact on WWS size. One of their testing scenarios has been split into two variants, that is, memory-read-intensive applications and memory-write-intensive applications.

The memory size initially allocated to an application had a linear impact on the migration time which is expected since more data had to be copied and there was no additional impact on the migration time when the memory-read-intensity increased. However, increases in memory-write-intensity did significantly slow down the migration process, albeit not linearly. When enough CPU cycles had been assigned to benchmark and the memory page dirtying ratio was high enough, the XenServer momentarily entered its final synchronisation, and the migration time was not increased any further. It is difficult to provide actual numbers as to where the memory writing rate reaches a critical point. The existing research results in Liu et al. (2013) show that the memory writing rate started to significantly impact upon the migration time at around 800Mbit/s, although these results are isolated to the specific research testing machine.

In practice, however, WWS is usually relatively small and VM downtime is barely noticeable for most applications. VM-LM usually requires a minimum stoppage before the next VM continues its execution at its final destination, providing a practically seamless migration of a running VM. For example, the XenServer requires only 60-300ms to perform final memory synchronisation when migrating the Quake 3 server (Clark et al., 2005).

The same research shows, that during the VM-LM of the VM running SPECweb99 benchmark, only 18.2MB out of a total 676.8MB allocated memory was transferred in the final synchronisation phase. The Quake 3 server needed to suspend the VM to transfer only 148KB of a total of 56.3MB allocated memory.

Nevertheless, it is possible to design extreme scenarios where the VM is writing to a certain region of the memory faster than the network can transfer it, for example the custom program 'MMuncher' resulted in the transfer of a huge 222.5MB out of a total of 255.4MB allocated memory in the final synchronisation phase. It should be noted that the VM migration process itself generally does not consume much memory. In the worst-case scenario, recent research has shown that one of the tested VMs required just a little over 200MB to migrate 2GB of virtual memory (Hu et al., 2013).

There is a further significant difference depending on the implemented memory transfer method. More specifically, VMs such as KVM, VMware and VirtualBox transfer only the currently allocated and used by the VM memory, while other VMs such as Hyper-V and XenServer transfer the total configured memory. This difference in transferred data might be substantial, potentially even one order of magnitude (247-354MB vs. 2255-2266MB) of the total transferred migration data (Hu et al., 2013).

In VirtualBox, the number of VM-LM rounds is explicitly controlled by the VM downtime. VirtualBox implements a voting mechanism, where all defined modules – 'units' in VirtualBox's source code – vote in each VM-LM round for the completion of the live data transferring stage and the suspension of VM. From VM-LM's point of view, the most interesting modules in VirtualBox are 'Saved State Manager' and 'Page Manager and Monitor' in files 'SSM.cpp' and 'PGMSavedState.cpp' respectively, which contain decision logic for triggering the VM suspension and for entering the final round of VM-LM. The decision is based on the estimated remaining dirty pages migration time, separately for the short-term average over the last four rounds and the long-term average based on all previous rounds. This algorithm then computes if the migration of all currently dirty pages exceeds the hardcoded 250ms maximum downtime. It should be noted that the mentioned source code has not been updated since its initial implementation in 2010, and could be improved further, potentially reducing the VM-LM overhead.

Researchers have also proposed certain optimisation techniques which could reduce memory migration time. Two such techniques are identified below:

- Jin et al. (2009) propose the adaptive compression of migrated memory pages based on word similarity. Experiments show that most memory pages fall into either low or high word similarity, with the former usually a good candidate for fast compression algorithm.
- Du et al. (2010) suggest an ordered transfer of memory pages based on their dirtying rates as well as factoring in the available network bandwidth. The next migration iteration starts as soon as the accumulated dirtying rate exceeds the available bandwidth.

### 4.2.3. DISK I/O-INTENSIVE APPLICATIONS

Storage migration either involve transferring the whole storage over the network or only reopening the connection to a configured NAS device. Modern Cloud systems typically implement NAS as a preferable storage option since it has many advantages such as centralised administration, a variety of standardised solutions across many different vendors and reduced failure rate features.

If any virtual devices are mapped on local storage media, they also need to be migrated together with the VM instance. The VM is generally not aware of higher level data structures from the Guest OS, such as files or memory segments. The VM only reads and/or writes its serialised form as a stream of bytes. Therefore, the VM does not recognise which data is left over from previous operations but which is now marked as clean, meaning that the process saves everything as shown in Figure 7:

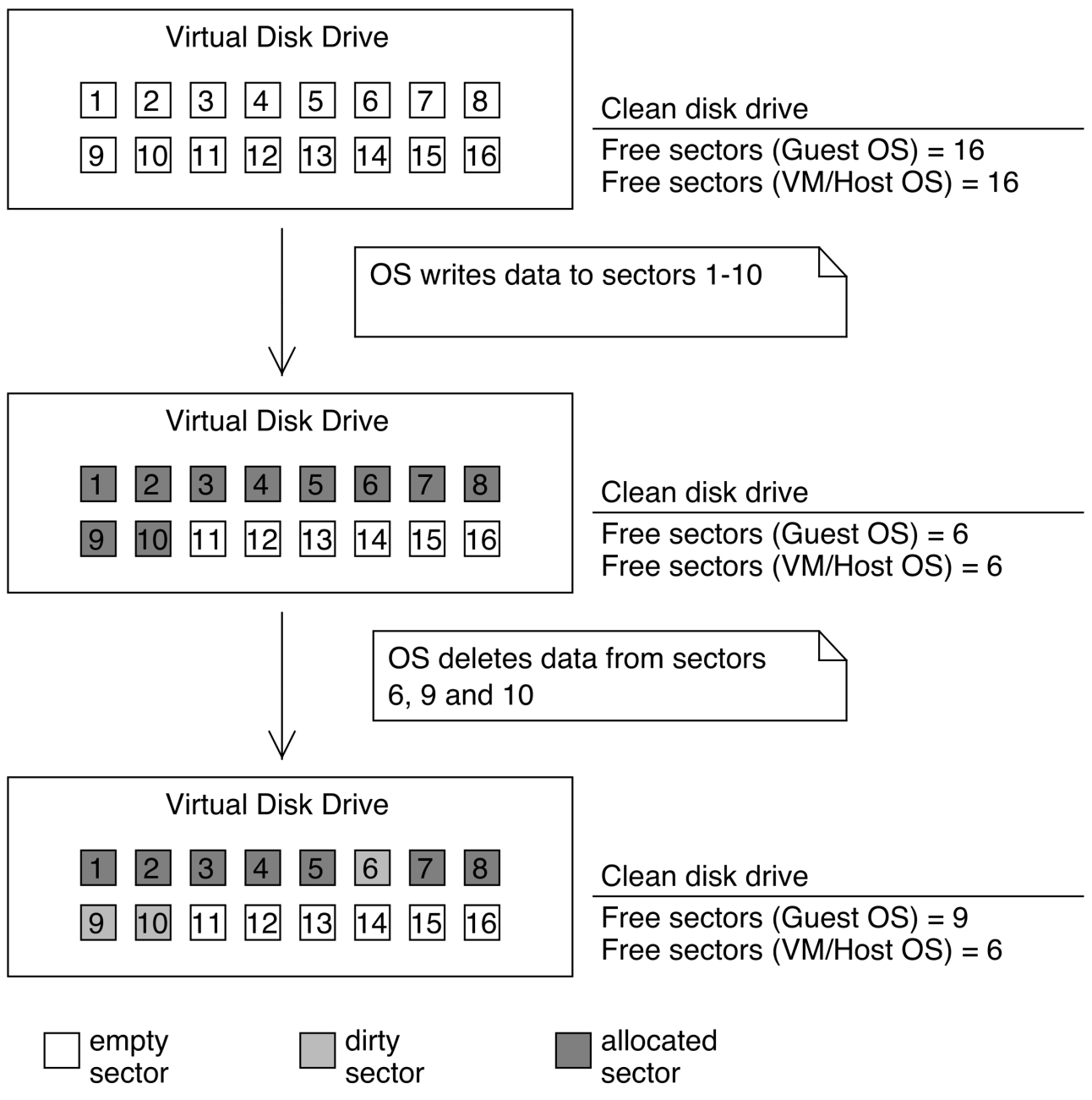

Figure 7: Virtual disk read/write operations

Following the write/delete operation, the Guest OS keeps only sectors 1-5 and 7-8 as allocated, even though the VM/Host OS does not know that dirty sectors 6, 9 and 10 are unused and saves them as valid data in its state file. Therefore, upon migration, more data than indicated by the Guest OS might need to be migrated since it involves copying the unused/dirty sectors.

Disk I/O operations are not easy to measure correctly since any modern OS successfully caches files in its memory. In addition, modern persistent storage drives may have substantial caches ('disk buffers') built-in. At the time of writing, hard drives come with 128MB of such memory such as Seagate STBD6000100 6TB SATA, while solid-state drives come with up to 1GB cache memory such as Samsung SSD 850 PRO.

Previous research has shown that it is possible to significantly exceed the available disk cache memory in order to force the VM to save data to actual persistent storage (Wu and Zhao, 2011). In this test, a sequential read pattern was used as it can generate more consistent I/O Operations Per Second (IOPS) compared to either sequential write or random read or write patterns. As with memory, an increase in IOPS caused an exponential increase in migration time, but the authors did not notice the existence of a 'break point', after which time further increases do not occur. This could be explained by a lack of monitoring of the disk sectors dirtying ratio implemented in VM. Generally speaking, memory operations have a bandwidth a several orders of magnitude wider than respective I/O operations, especially for random access transfers.

When migrating local storage data, it should be noted that generally storage data is less dynamically modified than memory. As such, the time needed to migrate storage media is more linear. Hu et al. (2013) advise Cloud administrators to ensure their virtualisation system supports delta migration, where the target node has a snapshot or a recent state image, and there is a need to migrate and then merge only delta of the new data.

### 4.2.4. NETWORK-INTENSIVE APPLICATIONS

For network resources, VM-LM relies on maintaining all open connections without the involvement of any kind of network traffic forwarding mechanism on the original node, since this may go offline and slow down responses from the host. VM-LM also copies the TCP stack state and, in addition, will carry IP addresses.

One suggested method to resolve these challenges is to keep all network interfaces on a single switched LAN (Clark et al. 2005). This would allow the target node to generate an unsolicited ARP reply, broadcasting that the IP has moved to a new location and has a new Ethernet MAC Address. This would cause all routers to update their ARP cache and direct all network packages to the new host. An alternative solution is to migrate the Ethernet MAC Address together with other resources, relying on the network switch to detect the port change, although this may result in lost packages.

The impact of network utilisation intensity on migration time has also been studied (Liu et al., 2013). Migrating the VM's state between different physical nodes requires transferring a significant amount of data, and therefore the available network bandwidth is a valid concern. It can be noted that increases in network bandwidth utilisation exponentially decreases migration time. Additional complexity which is not tackled in this research, comes from the physical placement of Cloud servers. The transfer rate between two servers from the same rack which would tend to be connected by high-speed fibre optic cables isolated from Cloud network noise, will generally be faster than servers from two different racks. This would also be much faster than the connection between server racks in two geographically different data centres. Faster connections could significantly reduce the time as well as reduce the number of rounds needed to perform VM-LM.

### 4.2.5. CODE STRUCTURE AND DYNAMICS

The discussion in subsection 4.2.2 reveals that the number of VM-LM rounds and the amount of data transferred in each of those rounds is directly related to the size of WWS. Previous research (Clark et al., 2005; Wu and Zhao, 2011; Hu et al., 2013) suggests that the memory write operations are the core cause of the repeated migrations of memory pages forming WWS. However, those investigations do not specify which applications are more prone to migrate harder and how the VM-LM is actually impacted.

The WWS is heavily impacted by the exact instruction composition of the application's compiled native code, or, as in the case of Java Virtual Machine (JVM), the interpreted byte code which is then compiled to the executable code.

Frameworks equipped with Automatic Memory Management (AMM) such as Java’s Garbage Collector (GC) (Urma et al., 2014) frequently move memory pages as new objects are created and are then released, which is a very common pattern within object-oriented programming. Furthermore, the frequency and the distribution of modified memory addresses strongly depend on the code's design and its style of programming. For example, the frequent use of immutable objects and singletons results in somewhat more static memory, while the common use of mutable objects and dynamically growing collections, such as linked lists, lead to bigger WWS, which means that the application will be migrated harder. The reuse of existing objects from object pools (‘flyweight’ pattern) has the potential to lower the WWS size. Nevertheless, it is extremely challenging to predict what kind of behaviour VM-LM will demonstrate without knowledge of the technology stack and a detailed analysis of the application code.

While experimenting with the VM-LM process of various applications, a non-linear relation of the size of WWS to LMDT was noted. However, as shown below in the details of the experiments, this phenomenon is quite specific to an application. For some programs it is barely noticeable, while for others is clearly visible. As demonstrated in sections below, rapid exponential increases in LMDT are especially visible for the busy backend application running in VMs which use AMM, such as JVM’s GC. This is the result of massive memory movements during the memory reclamation phases of GC and, therefore, higher amount of dirtied memory pages which need to be copied in the next VM-LM round.

## 4.3. EXPERIMENTS

This research focuses on finding which parameters would significantly impact the size of the LMDT. Therefore, based on the above analysis, several experiment scenarios were designed. These are presented below.

### 4.3.1. CONFIGURATION

The scope of this experiment has been necessarily limited to hardware which is accessible in the laboratory and the available network infrastructure as described below. The tests were designed with a focus on Open Source solutions, which gave the option to examine the source code and better understand the inner workings of migrated applications. Experiments were performed on a 100BASE-T network consisting of two servers (Intel Celeron 1.8 GHz with 2GB memory), a router and a NAS device exposing a shared folder via Samba server (CIFS). As such, the testing network was fully isolated, and the network noise was minimised. The testing machines had Kubuntu 15.04 installed as the Host OS, using the default configuration. The experiments used VectorLinux 6.0 Standard 'Voyager' as the Guest OS, this being a lightweight and slimmed down Linux distribution designed to avoid resources overhead on existing system services and daemons. VectorLinux 6.0 uses a Linux 2.6.27 kernel which proved to be most stable during teleportations.

In order to perform the experiments, it was necessary to define VM instances in VirtualBox. Each instance consists of two objects, the VM instance configuration and the virtual disk image (VDI) file. The VDI is usually stored on NAS device and is available to every node in this system as a remote share. Here, the VDI file is mapped via the Common Internet File System (CIFS), shared via the Samba server. VirtualBox requires the identical definition of a particular VM so as to exist on both source and target hosts.

The example below shows the creation of 'VM_VectorLinux_512', prepared for a Linux 32-bit host system with 512MB memory and one virtual CPU core:

```
VBoxManage createvm --name VM_VectorLinux_512 --ostype Linux_32 --
register
VBoxManage modifyvm VM_VectorLinux_512 --memory 512
VBoxManage modifyvm VM_VectorLinux_512 --cpus 1
VBoxManage storageattach VM_VectorLinux_512 --storagectl
"IDE Controller" --port 0 --device 0 --type hdd --medium
/mnt/VM_Shares/VM_VectorLinux_512.vdi
```

The VM-LM process itself is seemingly effortless from a user's perspective – it is sufficient to start the target VM instance in listening mode:

```
VBoxManage modifyvm VM_VectorLinux_512 --teleporter on --
teleporterport 6000
```

The next step is to execute the teleportation command from the source VM hypervisor's command prompt, providing the target host's IP address:

```
VBoxManage controlvm VM_VectorLinux_512 teleport --host 192.168.1.210
--port 6000
```

VirtualBox will take care of the migration by itself and report any errors. VirtualBox performs a strict comparison between CPU models and types and it was necessary to disable strict CPU id checks so as to complete VM-LM. Nevertheless, regardless of whether a strict CPU identification check was enabled or disabled, the CPU on the target machine still had to support the same set of architectural features and extensions such as SSE, SSE2 and MMX, as the CPU on the source machine.

Data transfer measurements were taken with the help of the iptraf tool. Used bandwidth was measured separately for sent and received data and then totalled. The measurement error of iptraf was only 1.96%; when exactly 100MB of random traffic was sent, the iptraf recorded transferred data averaged 84790KB (n=20, s=1654KB). Figure 8 demonstrates a sample measurement in which the VM-LM was performed on port 6000, showing that the source host sent 85216593 bytes and the target host sent 1455690 bytes (the last pair of TCP Connections).

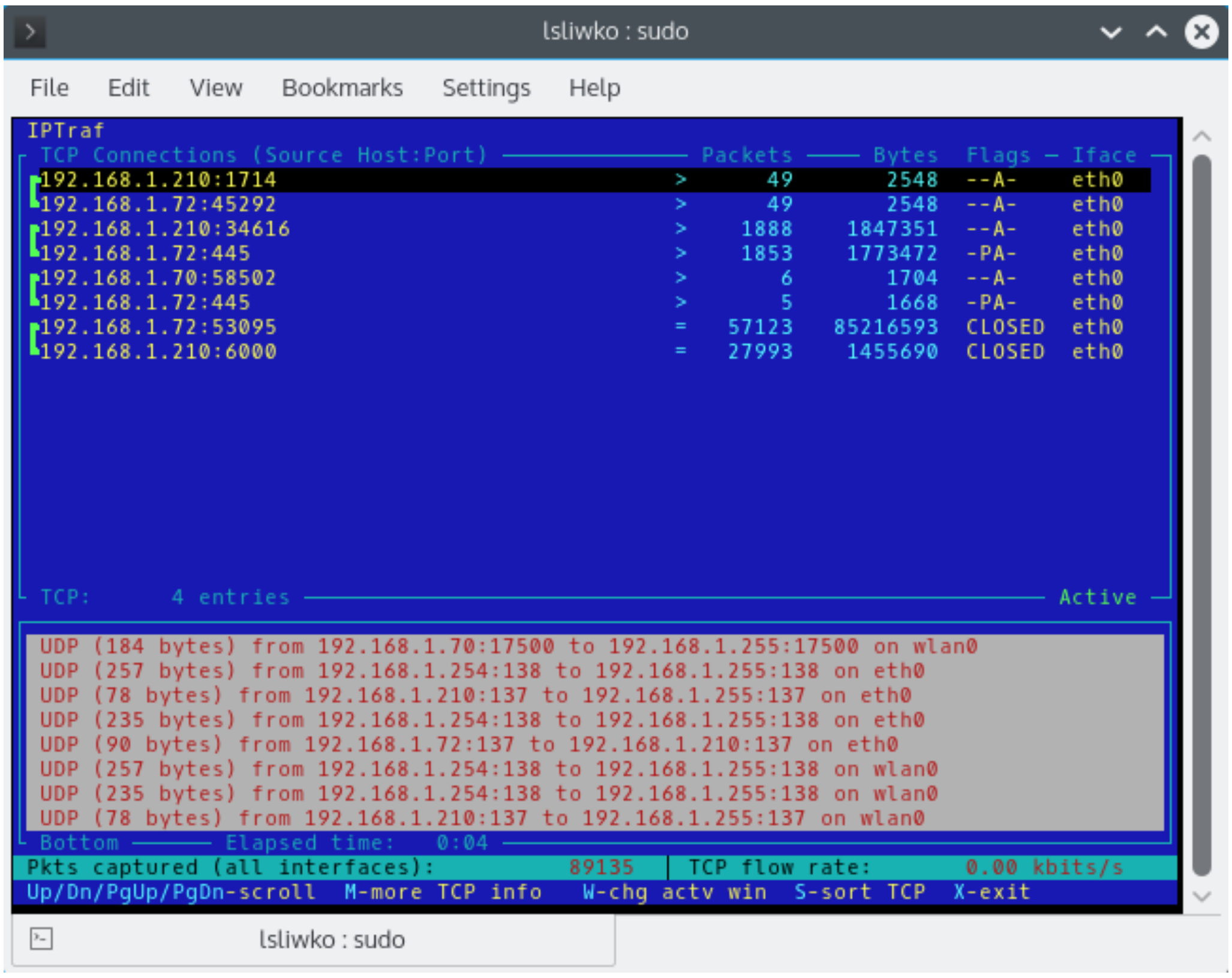

Figure 8: Measuring transferred data with the iptraf tool

### 4.3.2. EXPERIMENTAL SCENARIOS

Cloud systems allow users to deploy a very wide range of applications and services. In order to have a wide variety of applications this experiment was performed with the following configurations:

- An idle VM with only basic kernel canonical services running and a simple Xfce-4.4.3 Window Manager. During VM-LM, a whole environment is migrated – the OS and the running service itself. Therefore, this configuration was used as a reference point and it was possible to measure the impact of only Guest OS on migration.
- SPECjvm2008 is a benchmark program suite, released by the Standard Performance Evaluation Corporation in 2008, for measuring the Java runtime environments. It consists of 38 benchmark routines focusing on core java functionality, and is grouped into eleven categories. It has a wide variety of workloads, from computation-intensive calculations to XML file processors (Oi, 2009). The SPECjvm2008 workload mimics a variety of common general-purpose application computations. For a more detailed study of SPECjvm2008 performance, see Shiv et al. (2009). In the experiment, SPECjvm2008 benchmark ran on Java 1.6.0_06-b20. SPECjvm2008 is free to use, while newer suites such as SPECjbb2015, require a license.

- It is estimated that the Apache HTTP Server (Apache) serves about half of all active websites and is still the most widely deployed Internet web server. As of November 2017, Apache is running 44.55% of all active websites (Netcraft, 2017). Apache is often used with a range of technologies such as PHP, Perl, Python and frameworks such as WordPress. Apache is available as open-source software released under the Apache License for a wide number of Operating Systems such as UNIX-based, Microsoft Windows, NetWare and OS/2. In this experiment, static content was deployed and an external machine with a JMeter (v2.13) used to simulate user traffic.
- In a typical online system, the data are stored in a persistent storage component, usually a database. This experiment examined the impact of the VM-LM process performed while PostgreSQL version 9.2.24 database (Obe and Hsu, 2017) was running ‘select’ and ‘update’ queries on a large database table. PostgreSQL is a popular database, with a market share of 26.5% of all actively used databases worldwide (Stack Overflow, 2017).
- VM Allocator is a custom application used to simulate a running application with a static read only memory area and sizeable WWS memory. Such an approach enabled the configuration of an exact set of dirtying pages and their ratio to total memory; therefore, experiments could be conducted with higher confidence. To make the simulation more realistic, the VM Allocator ran several threads in parallel.

### 4.3.3. IDLE VIRTUAL MACHINE

To analyse the impact of allocated/used memory in VM-LM, the first experiment was the migration of the same Guest OS and running applications in three different VM configurations: 256MB, 512MB and 1024MB. No other parameters, such as the number of CPU cores, the enabled CPU features (PAE/NX, VT-s/AMD-V), and the configured peripherals, were altered.

Figure 9 presents the VM-LM of an idle VM. In this test, it was possible to observe a slight increase in the allocated kernel memory as well as an increase in the memory allocated to the VM. This is the effect of the Linux kernel requiring about 2MB of memory for managing every additional 256MB of memory. In this setup the memory stack size was set to 8192 bytes.

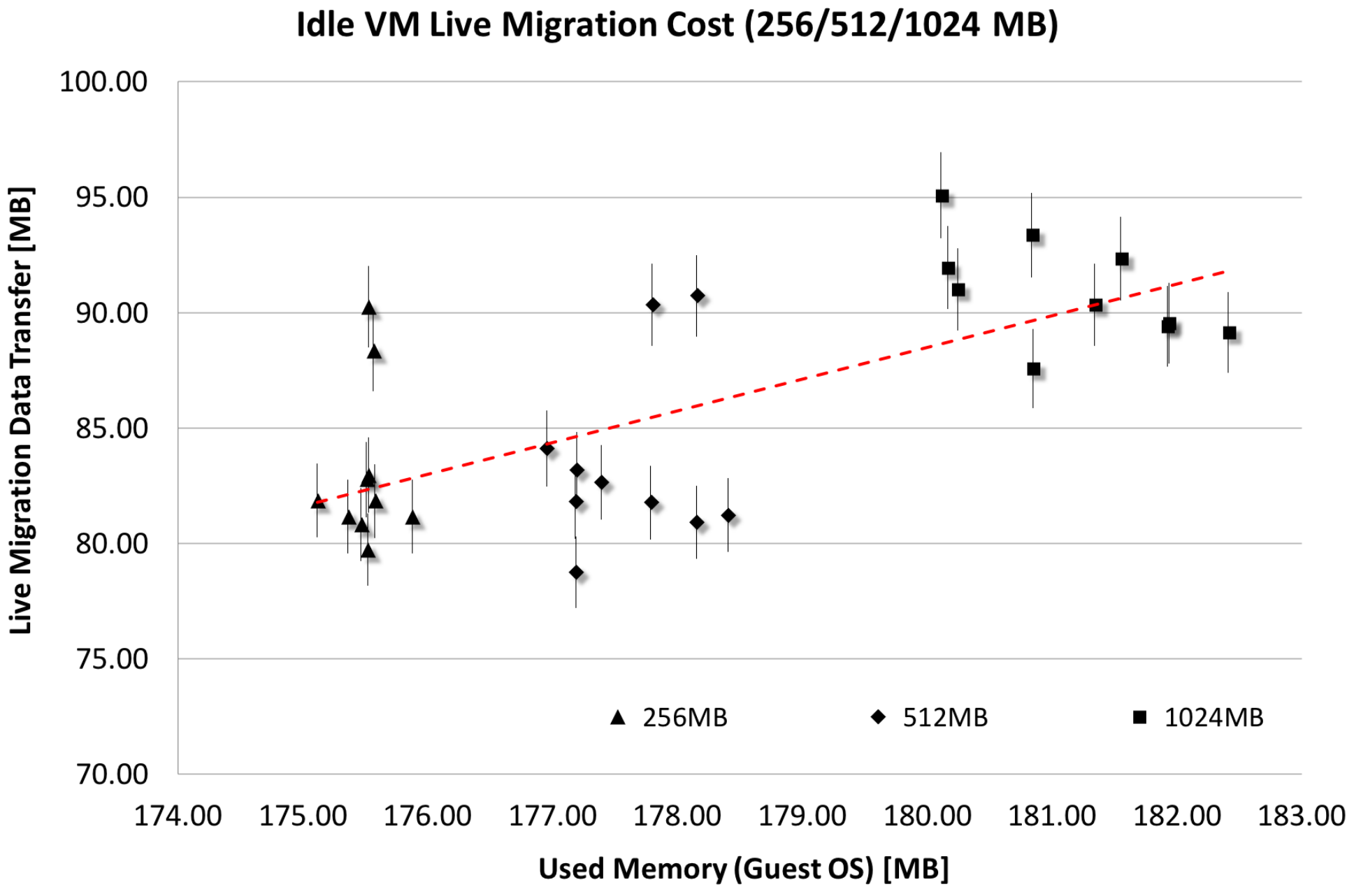

Figure 9: Idle VM Live Migration (256/512/1024MB)

It can also be noted that there was a minor increase in the total data transferred during the VM-LM. However, adding additional memory to idle the VM instance has only a minimal impact on the total transferred data: migrating an idle VM with 256MB memory transferred about 80MB and increasing the size of the configured VM memory from 256MB to 1024MB resulted in only around 10MB more data being transferred.

### 4.3.4. APACHE HTTP SERVER

In this experiment the Apache HTTP Server (v2.4.18 compiled with APR 1.5.2) was deployed within the Guest OS. The Apache server was configured to serve static content (10MB of images) over the HTTP channel. To have a reference point, an idle Apache HTTP Server instance was measured initially.

Transferring a VM instance using an idle Apache HTTP Server instance required ca. 170MB of network traffic. To simulate increasing user traffic, multiple continuous requests were generated with JMeter (v2.13) deployed on the external machine. JMeter is software designed to load test functional behaviour and measure performance of Web Applications such as web servers or services. In this research, JMeter simulated from 50 to 250 concurrent users, ca. 65 to 220 requests per second. It should be noted that the requested content was static, meaning that the additional allocated memory was mainly to support the increasing number of simultaneously open connections.

Figure 10 demonstrates the almost linear correlation between the total transferred data and memory utilisation. It should be noted that opening and handling additional TCP connections is processed in the system's TCP/IP stack, which could impact the size of the canonical memory allocated by the Linux kernel.

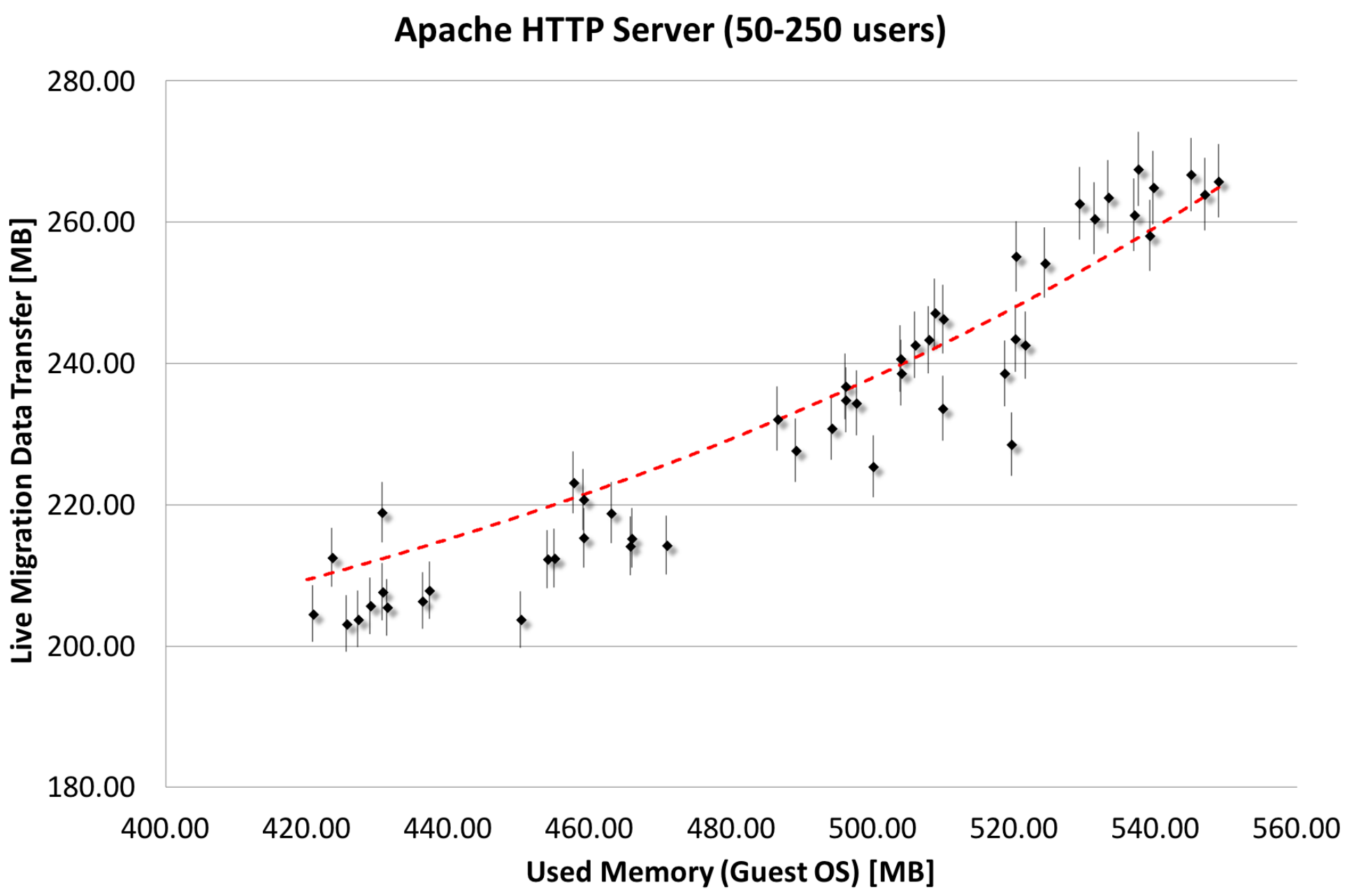

Figure 10: 50-250 users Apache HTTP Server Live Migration

This test scenario produces a near-linear correlation, as the migrated webserver is light on computations and the writable memory set size is rather constant. Therefore, an additional SPECjvm2008 experiment was used to examine how CPU-intensive applications behave during VM-LM.

### 4.3.5. SPECJVM2008 SUITE

The experiments with CPU-intensive applications involved the SPECjvm2008 benchmark suite executed on Java 1.6.0_06-b20 OpenJDK. SPECjvm2008 evaluates the performance of encryption and decryption implementations, various compression methods, floating point operations, objects serialisation/deserialisation, graphics visualisation, XML transformations and others. Therefore, this suite performs a set of backend-heavy operations.

Similar to the previous test with the Apache HTTP Server, it was necessary to firstly examine the impact of the VM memory size on data transfer. In order to force the loading and caching system libraries, a single SPECjvm2008 process was run initially. It should be noted that SPECjvm2008 batch files were configured with a 96MB maximum heap space. Java, by default, allocates a quarter of the available physical memory upon starting, which might interfere with the running of as many as eight benchmark processes in parallel.

The test deployed an increasing number of SPECjvm2008 instances which were executed on a single VM machine. The main reason for this test was that those processes are separated; therefore, each of them will increase the WWS by a comparable size. Figure 11 demonstrates the test results, which are visibly clustered in groups denouncing from 1 to 8 instances executed in parallel. Aside from the first SPECjvm2008 process which loads up about 32MB of libraries, each additional SPECjvm2008 process allocates additional ca. 15.5MB of memory. The increase in transferred data is visibly exponential.

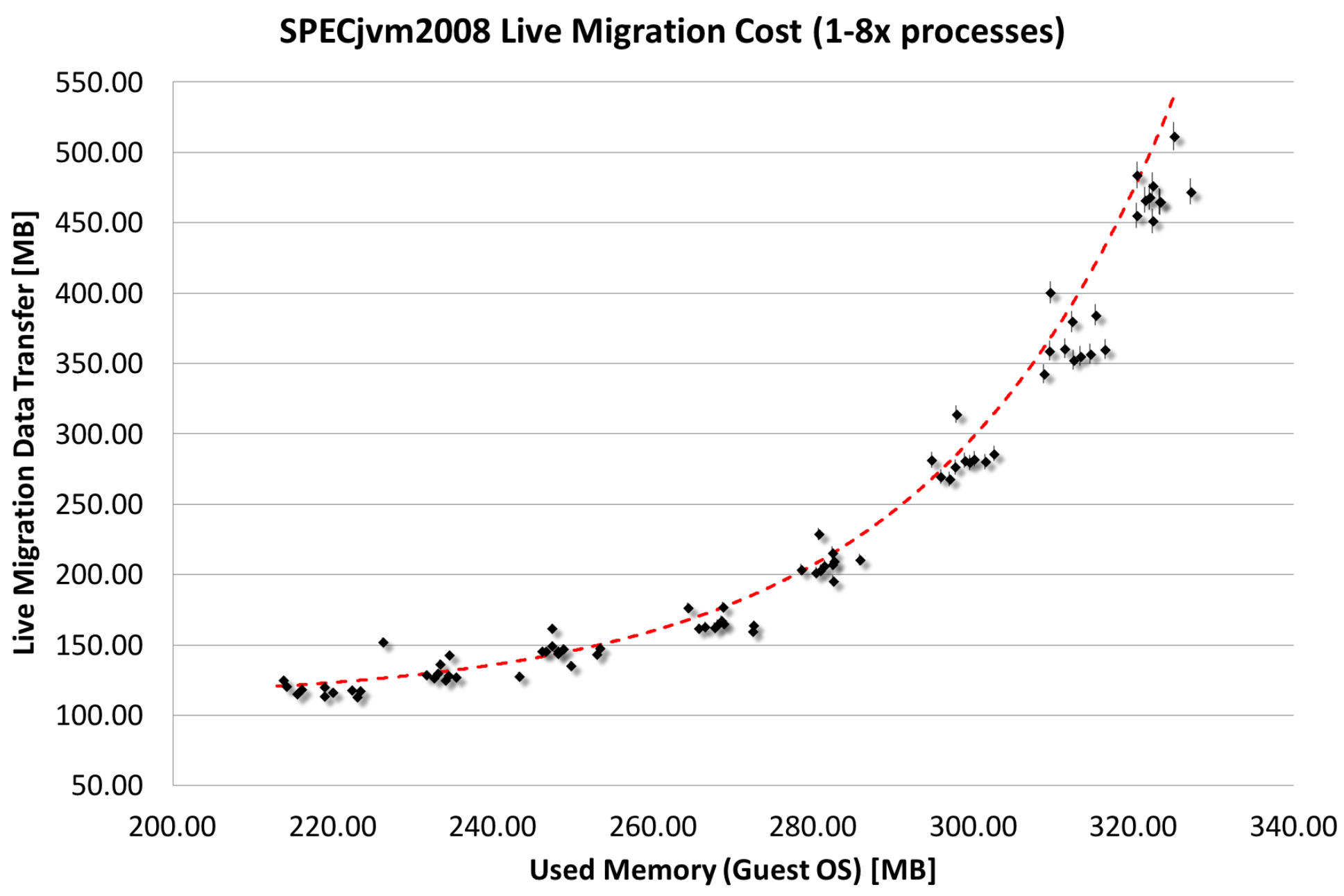

Figure 11: SPECjvm2008 Live Migration (1-8x processes)

This test also highlights a relative inefficiency when the active Java applications within a VM instance are being migrated. Good results are hard to achieve due to the increased memory movements caused by Java's GC. A solution to remedy such scenarios has been proposed by Hou et al. (2015), namely a custom kernel module which pauses JVM and invokes the GC just before each VM-LM round, then only objects from the tenured generation are being copied. In the final VM-LM round with the VM entirely paused, all objects from both the young and the tenured generation of the heap are being copied. Presented results show significant reduction of the total VM-LM time by over 90%, compared to the vanilla VM migration.

### 4.3.6. POSTGRESQL DATABASE

A typical online system consists of a frontend application, its backend services provider and a persistent storage component, usually in the form of a database. Having examined a popular frontend application (Apache HTTP Server) and a simulated backend application (SPECjvm2008), the persistent storage component now needs to be examined. While the current IT universe offers a wide range of specialised database solutions, such as graph-databases, NoSQL-databases, object and document-oriented databases which are suited to different data models, the most commonly used are still relational databases, such as Oracle, MySQL, Microsoft SQL Server and PostgreSQL, among others.

In this research, a PostgreSQL version 9.2.24 database (Obe and Hsu, 2017) was selected due to its popularity, ease of use, reliability, stability, wide range of supported platforms, and design fit for a high-volume environment, while also being an Open-source project. At the time of writing, the PostgreSQL is often rated as third or fourth in the popularity index, with a market share of 26.5% of all actively used databases worldwide (Stack Overflow, 2017). Given this, it is a representative choice for experimentation.

The following SQL commands were used to generate a test data table with one million rows of randomly generated strings, and then to apply an index to one of the columns:

```
SELECT
  generate_series(1,1000000) AS id,
  md5(random()::text) AS dataRaw,
  md5(random()::text) AS dataIndexed INTO TestData;
CREATE INDEX TestData_idx ON TestData(dataIndexed);
```

The test data came in two types: unindexed and indexed using default PostgreSQL B-tree. The SQL query optimiser can use database indexes when filtering and retrieving data from database tables, meaning a reduction in response time. B-tree indexes support general comparison operations such as >, <, =, etc., and also partially 'like' clauses. B-tree indexes can also be used to retrieve data in a sorted order. The test database which was generated allocated 2.11GB of disk space. It was stored remotely to the NAS device and mapped locally.

The next trials were designed to measure changes in the data transferred during VM-LM as PostgreSQL was running an SQL query. Those experiments were the most challenging to register consecutive results since the PostgreSQL database relies on multi-level caching to speed up its operations. It should be also noted that those tests did access files outside their VM as PostgreSQL was configured to store the bulk of its data on a remotely accessed NAS device and its cache buffers were cleaned between tests by restarting the server daemon service. OS was also forced to first synchronise and then drop disk caches within the same commands flow:

```
sync; /etc/init.d/postgresql-9.0 stop; echo 1 >
/proc/sys/vm/drop_caches; /etc/init.d/postgresql-9.0 start
```

The subsequent experiments were split into two groups, presenting different scenarios:

- Where 'select' queries were run first on an indexed data and then on unindexed data, with the 'like' clause appended in both scenarios. In the first scenario, the query engine used a previously created index and loaded only data for matching rows. In the second scenario, the query engine was forced to iterate through each of the table's rows to collect results. The main assumption for this group was that the additional memory activity from loading all the data would significantly increase the size of WWS and, as such, the first scenario would result in a smaller LMDT.
- Where an 'update' query modified parts of a test table. Five separate scenarios were designed, updating 20%, 40%, 60%, 80% and 100% of consecutive table rows respectively. Updating larger data sets involves building larger database transactions logs and requires more intensive memory operations which results in the expansion of the WWS. Furthermore, the 'update' operations require the modification of remote database files which are accessed over the network, and changes must be additionally committed via the CIFS protocol, the mechanism which is the additional source of memory activity.

During experiments using the 'select' query, the PostgreSQL processes allocated ca. 229MB in addition to the memory allocated by Guest OS. Predictably, queries involving the indexed data were executed much faster than queries executed on unindexed data, taking three and eight minutes respectively. Interestingly, there was no noticeable LMDT difference when executing 'select' queries on indexed and unindexed data columns, meaning that the size of the WWS remained roughly the same. The explanation for this behaviour is the extensive reuse of the memory cache buffers by PostgreSQL. Until buffers are not dirtied by data modifications, they can be rapidly released and reused. The PostgreSQL exposes

'pg_buffercache' view, which is useful for examining the inner workings of the buffering. The first noticeable aspect during the scenarios with 'update' operations was the considerable slowdown during the VM-LM process. Normally, the update of one million rows would take two minutes outside VM-LM, and five minutes while VM was being migrated.

Figure 12 presents the results of the experiments. Processing 'update' operations, which altered 20% of rows, resulted in ca. 310MB being allocated by PostgreSQL. Increasing the range of updated rows resulted in ca. 40-55MB memory being further allocated by database processes for each additional 20% of all the data rows processed. The allocated memory size changed very rapidly, and so only the range of memory changes is given. There was no noticeable difference between when an indexed or unindexed data was updated.

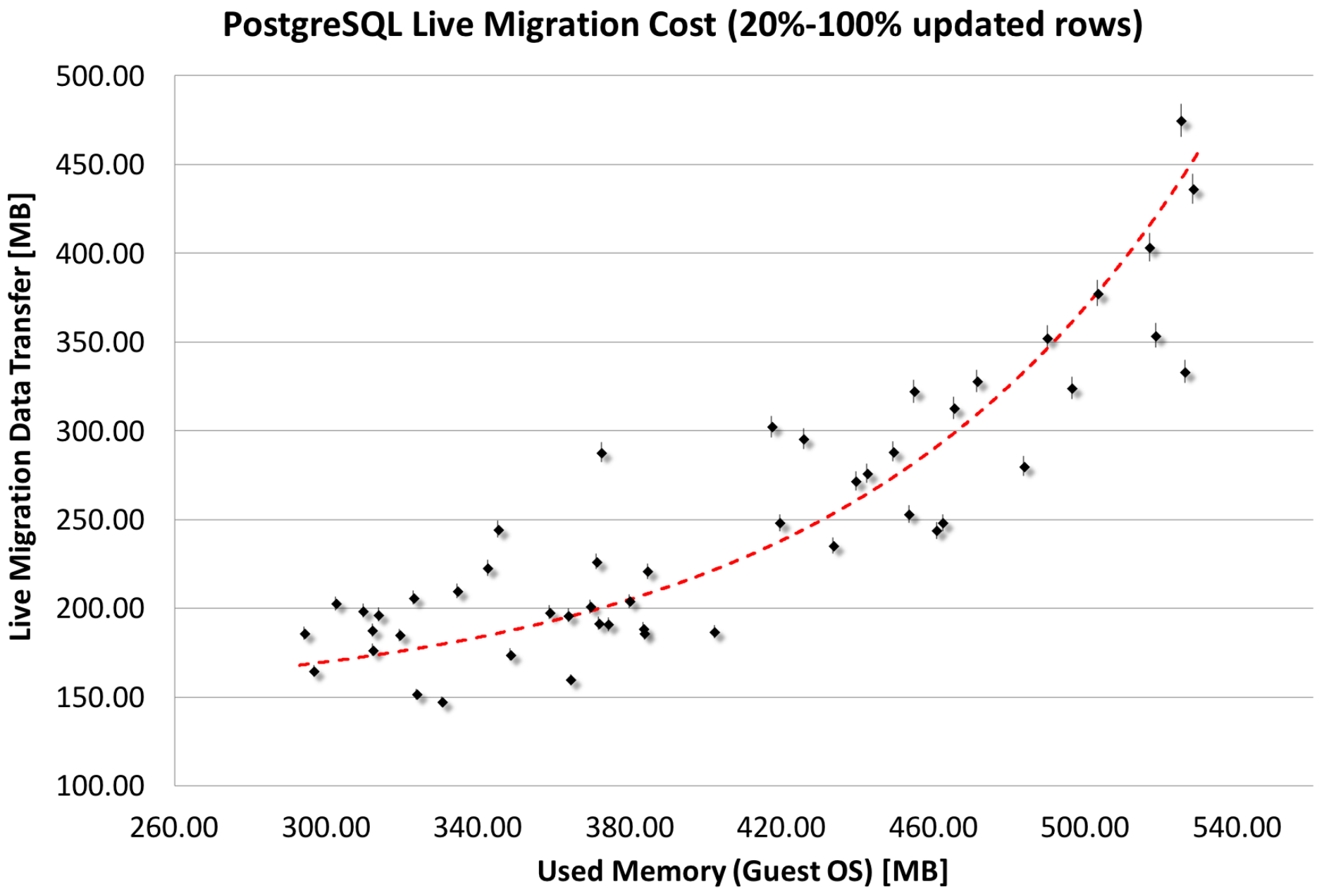

Figure 12: PostgreSQL Live Migration (20%-100% updated rows)

This test emphasised the exponential nature of LMDT while migrating rapidly changing data. However, it also highlighted the difficulties of measuring the exact size of allocated memory and isolating WWS. Considering those difficulties, the next series of tests has focused on creating a custom program generating an artificial WWS.

### 4.3.7. CUSTOM VM-ALLOCATOR

The above scenarios test real-world scenarios, but it is difficult to measure the exact parameters of the running applications, such as the WWS size. It was decided to implement a simple custom program (see Appendix D), VM-Allocator, to help with the experiment. VM-Allocator used the following parameters:

- Total Allocated Memory size – this memory is allocated and randomised upon the program's start. It should be noted that the setting of the VM-Allocator's memory pages remain fixed for the duration of test;
- WWS Size – this memory is part of the Total Allocated Memory and is continuously overwritten with random data. Several continuously working threads are used to write to WWS memory area;
- WWS Overwrite Interval – this parameter controls the speed of writing to the WWS memory area. In experiments the interval was set at one second;
- WWS Overwrite Threads Count – this parameter sets the number of concurrently memory overwriting threads. In test implementation, a single thread writes to the memory sequentially, thus in the experiments four threads were used to keep dirtying memory pages more randomly.

There was a preference to avoid overwriting memory faster than the network could transfer it. This would result in very linear VM-LM data transferred increase. Since the WWS memory area would be transferred every VM-LM round, the memory could only be finally synchronised in the final round.

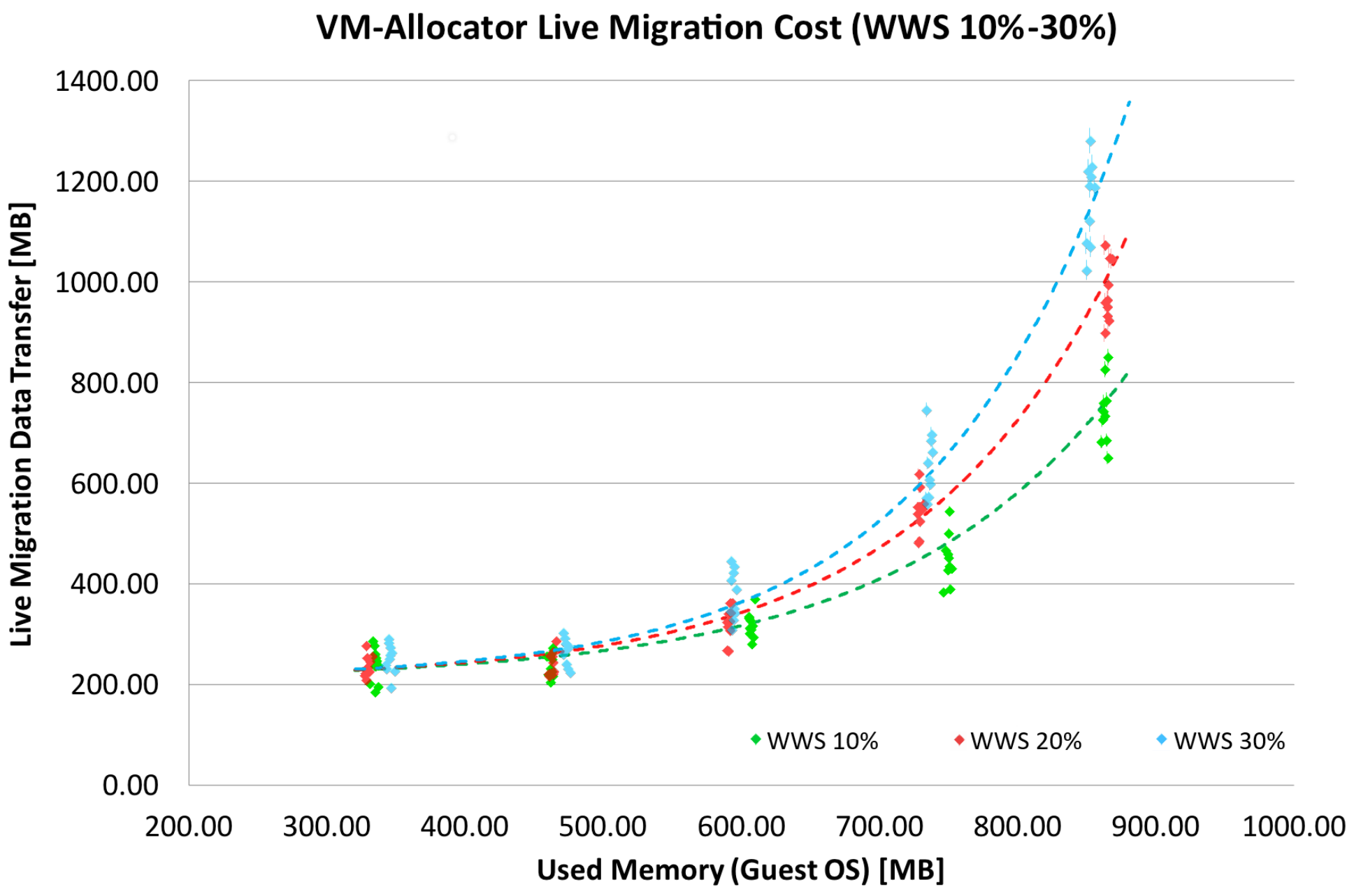

Figure 13: VM-Allocator Live Migration Cost (WWS 10%-30%)

Figure 13 presents the WWS test results. In this assessment, the aim was to measure the impact of the WWS size on the transferred data. Therefore, it was necessary to test several different ratios of WWS vs. Total Allocated Memory – 10%, 20% and 30%. The increase in WWS size (without an increase of the memory overwriting speed) exponentially increases the LMDT size. The VM Allocator initialises all memory only once upon starting. Therefore, the measured used memory varies only marginally. As in previous examinations, the increase in transferred data is exponential.

## 4.4. LIVE MIGRATION DATA TRANSFER FORMULA

Designing a method to accurately estimate live program migration time is not a trivial task. Nevertheless, a considerable amount of research has been done on the issue and several approximation models have been proposed (Clark et al., 2005; Jin et al., 2009; Akoush et al., 2010; Liu et al., 2013) with very good results. In Shirinbab et al. (2014), a large real-time telecommunication application was migrated in several scenarios and the results compared. In Zhang et al. (2016) the authors determined the network bandwidth required to guarantee a particular VM migration time.

It has been noted that the most feasible approach is to rely on historical data of memory dirtying rate for that particular program (Liu et al., 2013). In larger data centres, most of the workload is heterogeneous and service-oriented. The memory access pattern of each application may vary depending on the actual utilisation of the provided functionality. In such cases actual cost estimation may cause deviations if a model uses only previously sampled data. Experimental results have proven that when an adaptive model relies on historical data yields, the results have a higher than 90% accuracy.

However, historical migration data is not always available due to new applications, or programs not yet migrated, or where the utilisation of service has increased significantly, or where it has not been traced. In addition, changes in the environment may have a high impact on the migration time, for example the migration process itself consumes CPU cycles and network bandwidth, and a lack of these will slow down the migration process. Therefore, deriving a reliable VM-LM cost estimation formula is of utmost importance.

Designing a general use formula for migration time is not feasible in practice. While total migration time is consistent for the same VM software, experiments show huge variances when migrating the same applications between different vendors. For example, the migration of the virtualised Ubuntu instance took between 7 and 23 seconds depending on the VM used (Hu et al., 2013). This is also confirmed by further research (Che et al., 2010; Chierici and Veraldi, 2010; Feng et al., 2011; Huang et al., 2011; Tafa et al., 2011).

One of the less researched factors in VM-LM is the actual size of data transferred. This has a direct impact on the Cloud infrastructure because every additional transferred byte limits the available bandwidth, and introduces noise to Cloud operations. Other factors such as decreased service performance, downtime, and increased CPU usage are local and isolated to a single VM instance or a particular physical machine at most. This experiment resulted in the following observations:

- The total configured VM memory has a small effect on the total transferred data. The reason for this is that the Guest OS kernel allocates a small fraction of its memory to manage the total memory. In the test experiment, the kernel required about 2MB of memory to manage 256MB memory. Extending the VM memory from 256MB to 1024MB resulted only in the transference of only 10MB of additional data.
- The number of open network connections has minimal effect on the total data transferred. This is explained by the fact that current TCP/IP implementations are very mature, optimised and do not require many resources to perform network operations.
- Serving static content that does not require processing is economical in terms of VM-LM. Since data is mostly static, the majority of memory operations are read only. Therefore, transferring the memory page can be done only once, and thus the increase in transferred data during VM-LM is nearly linear.
- The high degree of computation activity by itself has no noticeable impact on the size of transferred data. However, the computation-intensive processes and programs that significantly alter the memory state have the most substantial impact on the VM-LM data transfer size. When a memory page is repeatedly changed by write operations, the VM would repeatedly transfer it over the network. When the speed of dirtying those memory pages exceeds the available network bandwidth, those pages will be marked and must be transferred in the final round of migration.

- In experiments, every node has pre-mapped remote storage upon start. Therefore, there is no additional cost for accessing shared drives. Such setup is widely used in clusters and Clouds.

From the above observations, the most significant factor in estimating the size of transferred data during the VM-LM process is the migrated application's allocated memory itself. Very rapid (i.e. faster than the network can transfer it) and fully overwriting WWS will result in this area of memory being fully migrated over and over again throughout all VM-LM rounds. Therefore, the increase of transferred data in this rare case is linear since WWS will always be transferred n-times. However, it is highly unlikely that the application will consistently overwrite WWS fully. Based on the experiments, the following formula for the size of the LMDT has been devised:

$$LMDT = CMDT + MF * e^{(AF*AM)} \tag{10}$$

- **AM** (Application Memory):

  $$AM = Total\ Used\ Memory - Canonical\ Memory \tag{11}$$

  Note that in the experiments, the Canonical (Kernel) Memory was measured together with any libraries loaded. System libraries and modules are loaded upon request and then are cached in the kernel memory and are shared by other applications.
- **AF** (Application Factor) – this parameter varies per application, and experiments show that the best approach is to estimate it by running several simulations. This parameter determines how steep the rise of LMDT is. In general, applications with complex logic modify memory pages more often and a significantly higher number of memory pages are being marked as dirty. This is especially true for AMM, which tends to re-allocate substantial amounts of data during the memory reclaim phase. The values presented in Table 5 were estimated from test experiments.

- **MF** (Migration Factor) – this parameter depends on infrastructure and it has been constant through all test experiments. In this experiment MF = 9.6MB.
- **CMDT** (Canonical/Kernel Memory Data Transferred) – each VM and the contained applications transfer a certain chunk of data every time. This is measured when both the VM and applications are idle, i.e. no user requests are served and no data is processed. It should also be noted that the first instance of the application increases the canonical memory since the required libraries are loaded and cached in the memory. The estimated values in the test experiments are summarised in Table 5.

When applied to the experiment data, the above formula closely estimates the total size of the transferred data. For less computation-intensive applications, such as the Apache HTTP Server, the average approximation error is about ±4%; the value is adjusted for iptraf measurement error margin. For computation-intensive codes like SPECjvm2008, PostgreSQL and VM-Allocator, the average approximation error is between ±8% and ±11%.

| Application (for 1024MB VM configuration) | CMDT [MB] | AF |
|---|---|---|
| Idle VM | 90 | - |
| Apache HTTP Server (v2.4.18 compiled with APR 1.5.2) | 175 | 0.00682 |
| SPECjvm2008 | 115 | 0.03305 |
| PostgreSQL (v9.2.24) | 145 | 0.01072 |
| VM-Allocator Test I (WWS is 10% of total memory) | 213 | 0.00620 |
| VM-Allocator Test II (WWS is 20% of total memory) | 213 | 0.00676 |
| VM-Allocator Test III (WWS is 30% of total memory) | 213 | 0.00714 |

Table 5: Application estimated LMDT values

The presented LMDT formula does not consider the cost of switching existing network connections to a new node. This is usually negligible since it is done in the hardware layer but depending on the implemented solution this additional cost might vary. In addition, network data compression and optimizations like adaptive memory pages' compression (Liu et al., 2013) might significantly reduce the size of the transferred data.

The input parameters in LMDT formula, such as AM, AF, MF and CMDT, were computed for a particular test configuration. These values will differ depending on environment, for example VM vendor and version, hardware specifications, network structure and available bandwidth, type and configuration of Guest OS, a particular migrated application, and so on. However, an analysis of the code reveals likeness of algorithms used in other VMs and the implemented network transfer method (i.e. TCP/IP packages streaming). It has been noted that, given similar environments, there is little variety in the VM-LM impact, and historical data can be used to accurately estimate these metrics and parameters.

The most practical way to obtain those values is through experimentation. Therefore, the working model should be capable of monitoring ongoing VM-LM and adjusting itself accordingly. Such a solution has been presented in the literature (Akoush et al., 2010; Verma et al., 2011) with good results. In practice, the best approach may be to initially benchmark all applications that are deployed on the Cloud system and then use these data to project the migration cost for this particular application based on recorded parameters.

## 4.5. SUMMARY AND CONCLUSIONS

This research demonstrates that while the methodology for estimating the VM-LM cost should be tailored to a particular environment such as the deployed VM, network configuration, used storage types, available CPU cycles for migration, it is possible to find a reliable approach for the purposes of estimation. From the previous research and the experimental results, several factors have been identified as having a significant impact on the task migration cost:

- The size of the allocated application memory and the size of WWS (or memory dirtying rate) – memory-write-intensive applications are difficult to migrate since frequently modified memory pages must be migrated repeatedly over and over again. As presented in experiments in this chapter, the size of WWS is often related to a specific application design and utilisation level.
- Frameworks equipped with AMM – this type of solutions, such as Java's GC, are considerably harder to migrate. This is due to the significant amounts of data being re-allocated during the memory reclaim phase which results in larger numbers of memory pages being marked as dirty in each VM-LM round. As mentioned in section 4.5 there exist efficient strategies trying to address this weakness, however they require much tighter coupling with the application deployed within the VM which might not be always possible.
- The available spare CPU cycles on both the target and the source machines – migration is an OS process executed on the same physical machine as a hypervisor. Low CPU cycles can create a bottleneck in the migration process itself (Wu and Zhao, 2011). It was also noted that assigning additional CPU cycles to the migration process reduces the total migration time but only to a certain point.
- The size of the virtual drivers mapped to the local storage media and the IOPS rate – if data is stored locally and not on a NAS device, it needs to be transferred together with the VM state. Persistent data size can be substantial such as several TBs of data and can be a dominant factor in the migration time (Hu et al., 2013).
- Migrating several VM instances in parallel – multiple concurrent migrations might interfere with each other. They are also likely to be slower than the same VM instances migrated sequentially one after another.

- The network bandwidth available for migration – the migration process consists of transferring amounts of serialised data over the network to another physical node. The more network throughput can be allocated to migration, the faster it will complete. Additionally, a slow transfer rate might result in larger WWS, i.e. the memory dirtying rate exceeds network capabilities (Clark et al., 2005; Liu et al., 2013). Earlier experiments were performed on a Wi-Fi network, where VM teleportation was very unstable.

Analysing the VM-LM cost is not simplistic and can have many dimensions and return various results based on the Cloud system. Therefore, it is necessary to apply appropriate limitations to the cost model and focus exclusively on the most important factors. As has been demonstrated in the experiments, estimating the LMDT size is not a trivial task, and many parameters need to be considered. Nevertheless, the results computed with LMDT formula have an acceptable approximations level of up to ±11% in the worst-case scenario.

The scope of this research has been necessarily limited to the available software and hardware. All tested applications were executed on a single VM instance, while present systems tend to have more complex dependencies, such as distributed databases and file systems and 3rd party systems. Especially, the design of modern systems seems to follow the micro-services architecture principles. In such environments, loss of performance in one system component could easily propagate to other parts of the system.

The experiments presented and the resulting LMDT formula completed the CRUM introduced in Chapter 3, meaning that the project could move to the practical part of research. The following chapters describe, firstly, (i) a simulation framework based on a real-world workload traces (Chapter 5), and then (ii) the actual implementation of two Cloud load balancer prototypes (Chapters 6 and 7).

# 5. ACCURATE GOOGLE CLOUD SIMULATOR

Theoretical investigations into the Cloud's resources model, in addition to practical experiments on the virtualisation layer (as detailed in Chapter 3 and 4), have highlighted the immense complexity of the workings of the Cloud environment. It became obvious that statistical data and a dry analysis by itself were insufficient for the research to move further, and that a more detailed and concrete approach was required, such as an analysis of the actually recorded Cloud workload.

The direct predecessors of Cloud systems were cluster and grid systems. The difference between a cloud and a cluster is that a cluster is a group of computers which are interconnected between themselves using high-speed networks, such as gigabit Ethernet, SCI and Myrinet, whereas Cloud servers can be geographically distributed. The main difference between the cloud and the grid is the resource distribution strategy applied. The grid model is decentralised and computation may occur over many administrative areas, whereas the cloud features a centralised resource distribution and resources are dynamically provisioned (Mateescu et al., 2011). In addition, clouds are usually a collection of computers owned by one party which are open to the public, the available computing power of which can be rented by anyone. In contrast to this open access, grid computers are owned by multiple parties and are usually closed to the public (Armbrust et al., 2009).

Correctly characterising user behaviour is of utmost importance when modelling Cloud workloads (Sharma et al., 2011; Malhotra and Jain, 2013; Shen et al., 2015). Cloud workloads have been researched in detail and are reasonably well understood (Mishra et al., 2010; Wang et al., 2011; Sharma et al., 2011; Zhang et al., 2011; Moreno et al., 2013; Reiss et al., 2012; Abdul-Rahman, et al., 2014). There have been limited attempts to accurately simulate Cloud workloads with consideration of detailed task parameters and constraints (Beitch et al., 2010; Ganapathi et al., 2010; Kliazovich et al., 2012; Calheiros et al., 2013), especially with consideration of workload scheduling.

Evaluating the performance of distributed applications and services without unrestricted access to existing Cloud environments is a very difficult task. The characteristics of a cloud workload in a data centre differ significantly from traditional grid computing (Di et al., 2012). There is only a limited number of publicly available cloud system workload traces in existence, and those are stripped of useful details (Mishra et al., 2010). The research community relies predominantly on simulations and models to conduct their experiments. The quality of input data and its realistic nature is a very important factor since it has a direct impact on the accuracy of results. Cloud systems are very dynamic, complex entities, and even the best simulation models must employ simplifications and are unable to provide realistic user configurations. This problem is even more visible when the studied area touches deep system-critical mechanisms such as task scheduling or fault handling and prevention schemes.

Normally, Cloud providers would not allow developers to alter core system components such as the scheduler or the provisioning services in a working system. In an ideal scenario, every Cloud system designer would have unconstrained access to a Cloud system of a considerable size which could be used as a test bed for developing models and strategies. However, in reality, a developer has to compete for access to a computing centre with many other business units.

Therefore, this part of the research has focused on building a flexible Cloud workload simulation framework which could be deployed in a local environment, i.e. the researcher's desktop machine or laptop, while providing at the same time high-quality, detailed and accurate workload parameters of the simulated Cloud system. Previous research and analysis of available workload traces show that Cloud workloads are usually highly variable and non-cyclical. Spread around the globe, Clouds' users are not constrained by predefined schedules, meaning that the workload is not correlated to season or time of day, in contrast to Grid and Cluster environments. Therefore, to research the proposed problem, a realistic Cloud workload simulation model is required. In order to setup a realistic scenario, two approaches can be used:

- Use an artificial Cloud workload generator (Beitch et al., 2010; Ganapathi et al., 2010; Wang et al., 2011; Malhotra and Jain, 2013).
- Acquire and parse real-world workload traces (Iosup et al., 2008; Kavulya et al., 2010; Hellerstein et al., 2010; Klusáček, 2014; Feitelson et al., 2014; El-Sayed et al., 2017) to a format which can be used in further research.

A number of simulators already exist addressing various aspects of the current Cloud systems such as computations, for example CloudAnalyst (Wickremasinghe, 2009), CloudSim (Garg et al., 2011) or networking/energy use, for example GreenCloud/NS2 platform (Kliazovich et al., 2012). However, those simulators, while very flexible, do not provide details about the fine-grained parameters that might be required in some types of simulation, such as memory page size, cache size, disk I/O time, cycles and memory access per instruction. Such low-level parameters can be obtained only from detailed real-world workload logs.

This chapter presents the AGOCS – a novel high-fidelity Cloud workload simulator which is based on parsing real-world workload traces. According to previous research and personal computing experience, building a high-fidelity workload generator is an extremely difficult task. The number of dependencies, constraints and other details required to capture the overall dynamicity of Cloud systems (Zhang et al., 2011) forces researchers to simplify models and make assumptions.

Therefore, a decision was made to base the test approach and simulation on real-world workload giving greater detail.

This chapter is based on previous work published in Sliwko and Getov (2016).

## 5.1. WORKLOAD TRACES ARCHIVES

The following workload traces are publicly available:

- Google Cluster Data (GCD) project (Hellerstein et al., 2010) – this repository (available from github.com/google/cluster-data) includes detailed traces over a month-long period (May 2011) from a 12.5K-node network. The statistics include CPU usage, memory usage, disk I/O operations (only for the first two weeks, after that the logs' configuration changed), network speed etc.;
- Grid Workload Archive (Iosup et al., 2008) – hosted in Delft University of Technology in the Netherlands. This repository contains workload traces from almost a dozen grid systems. The majority include CPU usage, memory usage and disk I/O operations;
- Parallel Workloads Archive (Feitelson et al., 2014) – this repository contains over 30 workload logs from around the world. The earliest traces are from 1993 and the latest from 2012. Those workload traces were thoughtfully cleaned of anomalies and data errors;
- MetaCentrum Workload Log (Klusáček, 2014) and CERIT-SC Grid Workload (Klusáček and Parák, 2017) – archive contains data sets generated from TORQUE workload traces, deployed in the Czech National Grid Infrastructure (22 clusters having 219 nodes with 1494 CPUs);
- Yahoo! M45 Supercomputing Project (Kavulya et al., 2010) – Yahoo! made its 4k-node Hadoop cluster's workload traces freely available to selected universities for academic research.

All the above repositories, except for the Yahoo! workload logs, can be used freely for research work since there are no legal restrictions and/or requirements for presenting these data or derived data in any kind of research work.

For the purposes of this research, real-world workload traces from the GCD project are used. The main reason for the selection of this repository is the high quality of workload traces. Traces are complete and contain a low number of anomalies which are thoughtfully explained including their schema and format (Reiss et al., 2013). They have been gathered from a large system over a significant period of time.

Google offers a variety of services, and therefore their backend systems are diversified and represent a complete spectrum of computation requirements. Few computing services require as much computation per request as search engines. On average, a single query requires the examination and processing of hundreds of megabytes of data, even when elegant optimisations like reverse-index (Byrne et al., 2001) applied. At the heart of every Google search query processing lays the PageRank algorithm (Richardson and Domingos, 2001) which tracks and evaluates the importance of every result, meaning that users can receive the most significant results at the top of results page. Additionally, every query is checked by the Spell Checker service and is also processed by the Ads server, where most of Google revenue comes from. For a detailed description of how the search query is processed at Google datacentres, please see additional research findings (Barroso et al., 2003).

Google engineers focused on designing a throughput-oriented framework in which ca. 80% of the workload consists of a high number of batch jobs, which have a runtime of 12 to 20 minutes (Schwarzkopf et al., 2013) and a smaller number of long-lived service jobs. This mixed longevity of submitted jobs creates very good testing field for the purpose of this research, where the aim is to design a flexible and scalable scheduler capable of handling very high workload on any number of nodes. Other designs seem to be focused on processing either the high churn of short-lived batch tasks (e.g. Microsoft's Apollo and Alibaba's Fuxi) or a smaller number of long-term services (e.g. Twitter's Aurora). Finally, Google Inc. is a global company and its data centres are working continuously 24-hours a day. Its clusters' workloads are not cyclical, which could be a problem if using traces from a more local data centre.

Almost all Google Cloud performance profiling is done with help of the Google Wide Profiling (GWP) framework (Ren et al., 2010). GWP is inspired by systems such as DCPI (Anderson et al., 1998) and is based on the premise of low overhead sampling both of the machines within the datacentre and of the execution time within a machine. Every day, GWP collectors randomly select a limited number of Cloud nodes to profile and start profiling routines via RPC calls. The profile data are collected via the 'perf' tool and are tagged with corresponding code locations. They are then aggregated with samples from other machines in the Dermel database (Melnik et al., 2010) for convenient analysis (Kanev et al., 2015).

The GCD workload traces are stored in the Cloud Storage in the bucket 'clusterdata-2011-2', and can be downloaded using the 'gsutil' tool. The compressed archives are approximately 41GB, while the uncompressed archives are about 191GB. Unfortunately, no logging system is perfect, and every workload trace examined has a proportion of anomalies. The GCD logs are of high quality, although there are a few known inconsistences:

- Disk time data is not logged after the first 14 days due to changes in the monitoring system;
- Approximately 0.003% of jobs are not listed as they run on nodes not included in the workload traces;

- Approximately 70 jobs have no task information. The explanation given is that those jobs run but the tasks were disabled;
- Approximately 0.013% of the task events and 0.0008% of the job events have missing fields;
- Fewer than 0.05% of job and task scheduling event records are missing and less than 1% of resource usage measurements are missing.

Some resource statistics data are inaccurate, for example the cycles per instruction and the memory accesses per instruction parameters have values out of range for the underlying micro-architecture. The cause of this might be bugs in the statistic measurement system.

Additionally, GCD traces were obfuscated from user data, operating system and platform details, job purpose information and special constraints' names and values. These characteristics would be a very interesting point of research. It is also important to point out that GCD workloads are delayed by ten minutes. This shift has been applied in order to split pre-existing cluster conditions such as already existing nodes from new incoming requests, for example scheduled tasks and resource utilisations.

## 5.2. GOOGLE CLOUD WORKLOAD

There are many ways of splitting the distributed computing system's workload into unique categories of tasks. Based on existing Google Cluster workloads analysis, the 80$^{th}$ percentile of batch jobs finish within 12 to 20 minutes, while the 80$^{th}$ percentile of tasks finish within 29 days. Most jobs (ca. 80%) are batch jobs. Batch jobs tend to have a fast turnaround with a short execution time; in Google Cluster the 80$^{th}$ percentile inter-arrival time is between 4 to 7 seconds. Therefore, a low-overhead and low-quality allocation algorithm is suitable for this type of job. Services execute for much longer, and usually involve some type of interactivity, meaning that high-quality allocation guaranteeing good performance is critical. In Google Cluster fewer than 20% of all jobs show the 80$^{th}$ percentile inter-arrival time as between 2 to 15 minutes. However, services tend to consume a majority of system resources – approximately 55-80% of all resources in Google Cluster (Schwarzkopf et al., 2013). Those values seem to be comparable with results from a similar analysis of available cluster workload traces like Yahoo! (Kavulya et al., 2010), Facebook (Chen et al., 2012) and Google (Mishra et al., 2010; Sharma et al., 2011; Zhang et al., 2011; Reiss et al., 2012).

The examined workload traces do not specify details about underlying architecture, although some details about Google Cloud's architecture and jobs specifications can be found in Kanev et al. (2015). From examining the workload logs, it was found that Google Cloud jobs show significant diversity in workload behaviour, with no single hotspot application. From a software point of view, the jobs' executables look sound and Google engineers seem to put a considerable amount of effort into profiling and optimising them. While a significant number of Google services are written in a variety of programming languages, such as C++, Java, Python and Go, it is the C++ code that consumes most of CPU cycles. Code sharing is frequent, but binaries are generally statistically linked in order to avoid dynamic dependency issues, as well as to gain a small performance boost at the expense of executable size, which often reaches 100MB. Almost all Google's datacentre software is stored in a single shared repository and is built using a single build system – Bazel (ibid.).

Google services are deployed on RedHat. Instances are heavily customised, and many OS-level libraries were modified to boost performance or to provide better security, for example in the replacement of malloc by tcmalloc (Ghemawat and Menage, 2005).

Google Cloud has an architecture formed of distributed, multi-tiered services, where services expose only limited sets of APIs. Such a pattern helps to reduce necessary testing. Communication between services is performed only via RPC calls, where requests and responses are serialised in a Protocol Buffers format optimised for reducing the size of data (Varda, 2008). Protocol Buffers do not explicitly implement any type of compression, although it supports 'varint encoding' – a variable-length encoding for integer data that means small values use less space, i.e. values 0-127 take one byte plus the header, even if the field type is bit-wise wider, for example 64-bit integer. As with many other Cloud systems, Google Cloud is gradually becoming more and more diversified. Data centres were initially built with a single 'killer-application' in mind (Barroso et al., 2003), but nowadays the utilisation model of typical Cloud system is to accept a continuously increasing pool of diverse applications and services.

Kanev et al. (2015) analysed workload traces over a three-year period and noted that during the earliest period examined (August 2011), the top 50 applications accounted for 80% of CPU cycles. Three years later (August 2014), the top 50 applications (not necessary the same binaries) consumed only 60% of CPU cycles. The authors argue that Google Cloud data centres are still more specialised in their operations than publicly available clouds, which are exposed to much more varied technology stacks. Jobs executed on Google Cloud nodes include (Reiss et al., 2013): (i) the processes responsible for content ad targeting which matches ads with web pages based on page content; (ii) scalable distributed storage ('bigtable') (Chang et al., 2008); (iii) flight search and pricing engine; (iv) gmail back-end server and front-end server ('gmail' and 'gmail-fe'); (v) the components of search indexing pipeline; (vi) search engine services ('search1', 'search2', etc.) (Meisner et al., 2011), (vi) video processing tasks ('video') such as transcoding and feature extraction, and so on.

The GCD workload traces are stored in Google Storage for Developers in the bucket 'clusterdata-2011-2'. Traces can be downloaded by the 'gsutil' tool, and the compressed archives are approximately 41GB. The uncompressed archives are about 191GB. The schema and format of available assets of GCD is detailed in Reiss et al. (2013); a description of the structure is presented in Table 6:

| Files | Description | Size (uncompressed) |
|---|---|---|
| machine_attributes/ part-00000-of-00001.csv | Aside from computational capacities (i.e. memory and CPU), each machine might have a set of machine attributes as key-value pairs. Those represent machine properties, such as kernel version, CPU clock speed or presence of an external IP address. Unfortunately, those values have been obfuscated, and only hashes are available. | 1.21GB |
| machine_events/ part-00000-of-00001.csv | The majority of nodes existed in the system before the logging process. However, during the cluster operation a number of nodes were shut down, upgraded or added. This part of the workload traces contains a list of events with those actions. | 2.9MB |
| job_events/<br>part-00000-of-00500.csv<br>part-00001-of-00500.csv<br>…<br>part-00499-of-00500.csv | The jobs queue is the base of all processing. Those files contain a sequential list of all submitted jobs to the cell. Entries in those files state which user submitted the task, the priority of the task and the local scheduling class (i.e. priority of access to local machine's resources). | 332MB |
| task_constraints/<br>part-00000-of-00500.csv<br>part-00001-of-00500.csv<br>…<br>part-00499-of-00500.csv | Tasks can specify constraints on machine attributes (detailed in subsection 5.5.2). There are four types of constraints:<br>• EQUAL – checks if attribute exists and has required value<br>• NOT EQUAL – checks if attribute is not defined or has different value than specified<br>• LESS THAN – checks if attribute (specified as integer number) is strictly less than passed value<br>• GREATER THAN – checks if attribute (specified as integer number) is strictly greater than passed value | 3.04GB |
| task_events/<br>part-00000-of-00500.csv<br>part-00001-of-00500.csv<br>…<br>part-00499-of-00500.csv | Every job contains a number of tasks to be scheduled and executed on networked computers. Tasks specify requested resources, such as CPU cores, memory and local disk space, priority and local scheduling class. | 16.55GB |
| task_usage/<br>part-00000-of-00500.csv<br>part-00001-of-00500.csv<br>…<br>part-00499-of-00500.csv | Task usage is the biggest (ca. 89% of data size-wise) and the most interesting piece of data, containing real metrics of resources used by tasks. This includes mean and maximum memory usage, mapped/unmapped page cache memory usage, mean and maximum disk I/O time and cycles per instruction. The usage values were gathered from each measurement window (usually 300 seconds), and in some cases were aggregated from several sub-containers. | 170.54GB |

Table 6: Google Cluster Data archive structure

It is also important to note that GCD workloads are delayed by ten minutes. This shift has been applied in order to split pre-existing cluster conditions such as already existing nodes from new incoming requests, e.g. scheduled tasks.

## 5.3. AGOCS ARCHITECTURE

The presented framework generates and accurately times workload events and feeds them to the designated scheduler instance (push model). The typical architecture for testing consists of a single stand-alone AGOCS server and a number of simultaneously running instances of scheduling algorithms.

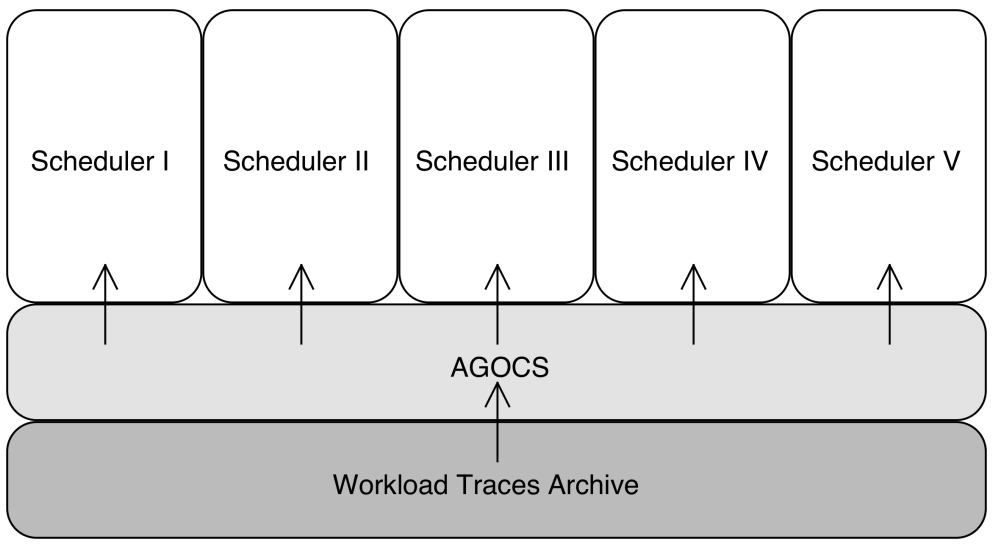

Figure 14: AGOCS use case

AGOCS framework was implemented in Scala functional programming language. Since Scala is based on JVM, it allows access to a wide range of mature Java libraries such as Google Guava and Apache Commons (see Appendix B for the specifications of runtime libraries). The detailed simulations of distributed computing models require a high degree of parallelisation in order to run effectively. To develop such a highly concurrent program, an Akka Actors/Streams framework is used because it has a very low overhead per instance, approximately 300 bytes. Additionally, given that Akka framework can be deployed in a distributed environment, AGOCS can be deployed on multiple machines through simple configuration changes. Additionally, Akka library uses Google's efficient Protocol Buffers (Varda, 2008) as its default serialisation mechanism for internal communications, i.e. between Akka cluster nodes.

The development of AGOCS (and also the load balancer prototypes detailed in Chapters 6 and 7) was done with the help of IntelliJ IDEA using native Scala plugin (see Appendix A). This is presented in Figure 15, where a part of NA source code can be seen:

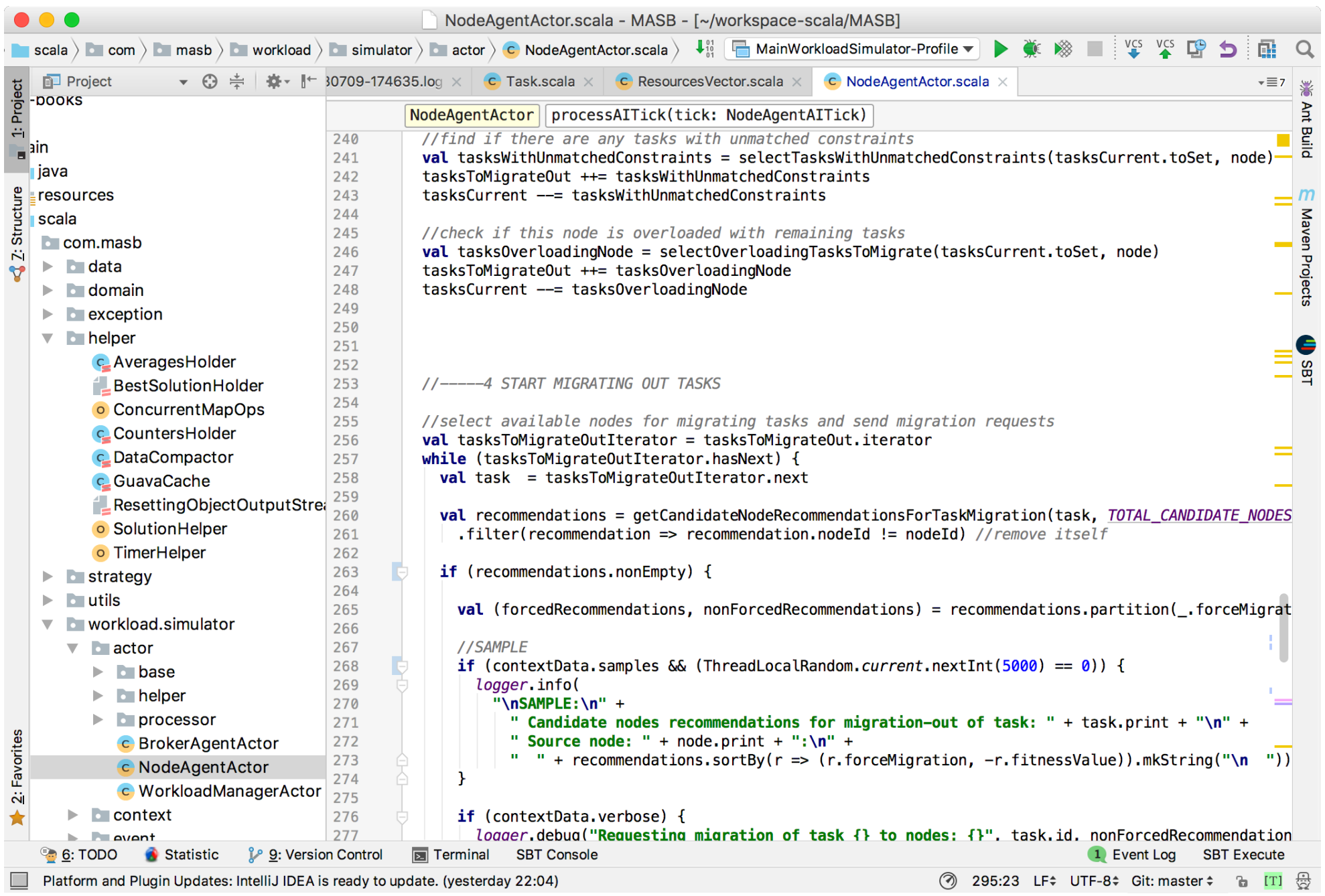

Figure 15: Scala IntelliJ IDEA

An interesting feature of AGOCS is that it can be paused at any time, allowing users to take a snapshot of current tasks distributions and the state of scheduled jobs. The snapshot files contain all simulation data in a serialised form, meaning that they can be stored and examined later. This approach enables researchers to conveniently and directly compare various scheduling algorithms at any time while they are running. The native Java's serialisation mechanism (which can be used in Scala) was initially used to store the simulation's state; however, it was found that it does not support very large context objects generated by experiments simulating 100k or more nodes, and was replaced by Kryo framework. While most simulation framework functionalities are enabled in configuration files or triggered by a command line, AGOCS also offers a graphical simulation monitor module implemented in JavaFX (the visual layout of nodes is generated procedurally from Halton sequence):

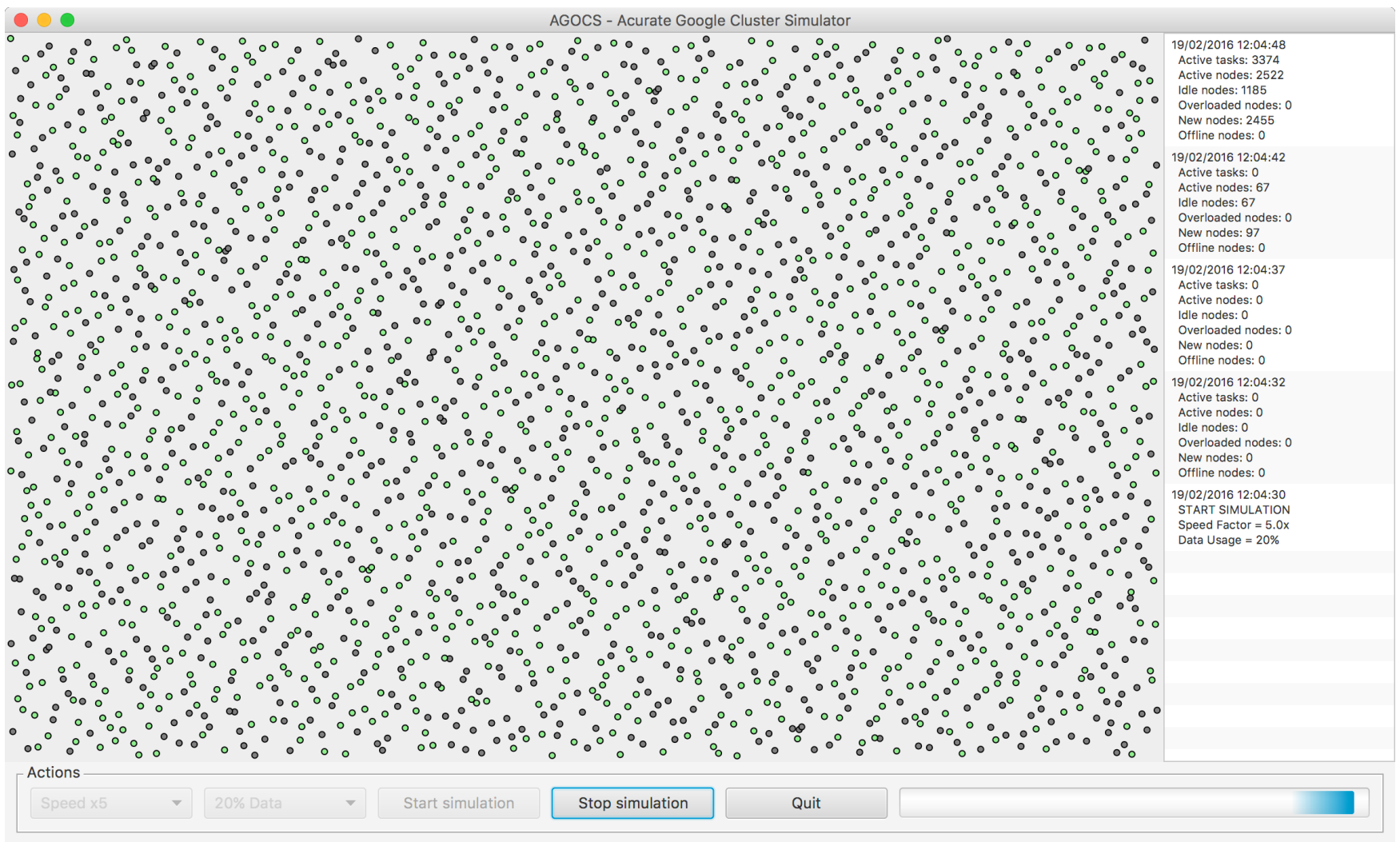

Figure 16: AGOCS simulation monitor

AGOCS was designed for common desktop machines even though it requires loading and processing a huge amount of workload traces data (191GB of uncompressed data). Therefore, running a simulation framework server is disk I/O-intensive and may interfere with OS's swap memory operations. Despite this, it was still possible to work around this issue by attaching an external disk drive to the test machine.

## 5.4. RELATED WORK

There currently exist a number of Cloud simulation frameworks such as CloudAnalyst (Wickremasinghe, 2009), GreenCloud (Kliazovich et al., 2012), Network CloudSim (Garg et al., 2011; Malhotra and Jain, 2013) and EMUSIM (Calheiros et al., 2013). Those frameworks were designed to cover a wide range of Cloud systems simulations, while AGOCS was designed with a focused goal of simulating a Google Computing Cloud cell environment with consideration of a very fine-grained and detailed aspect simulation such as tasks resource utilisation, task constraints, jobs queue simulation, jobs and tasks priority class, node's local scheduler simulation, detailed statistic as memory cache hit ratio and so on. A brief comparison between these frameworks is shown in Table 7:

| **Framework** | **AGOCS** | **CloudAnalyst** (Wickremasinghe, 2009) | **GreenCloud** (Kliazovich et al., 2012) | **Network CloudSim** (Garg et al., 2011) | **EMUSIM** (Calheiros et al., 2013) |
|---|---|---|---|---|---|
| **Platform** | Scala/Akka | CloudSim | NS2 | CloudSim | AEF |
| **Language** | Scala | Java | C++/OTCL | Java | Java |
| **Simulator Type** | Event Based | Event Based | Packet Level | Packet Level | Event Based |
| **Supported workload traces** | Google Cluster Data (CSV files) | Custom (ASCII/XML) | Loadable configuration settings (TCL) | Custom (ASCII/XML) | Custom (ASCII/XML) |
| **Networking** | Limited | Limited | Full | Full | Limited |
| **Resource constraints** | Yes | Yes | Yes | Yes | Yes |
| **Supported and reported resource types** | - CPU Cores (Used and requested)<br>- Canonical Memory (Used)<br>- Assigned Memory (Used and requested)<br>- Page Cache Memory (Used)<br>- Disk I/O Time (Used)<br>- Local and Remote Disk Space (Used)<br>- Cycles Per Instruction (Used)<br>- Memory Access Per Instruction (Used)<br>- Local Scheduler (Priority Class)<br>- Jobs Priority<br>- Tasks Priority | - CPU Cores (Requested)<br>- Bandwidth (Requested)<br>- Memory (Requested)<br>- Millions of Instructions Per Second | - Server Load factor (Used and requested)<br>- Bandwidth (Used and requested)<br>- Memory (Used and requested)<br>- Energy Used (split by servers, switches, etc.)<br>- Service Timeout | - CPU Cores (Requested)<br>- Bandwidth (Requested)<br>- RAM size (Requested)<br>- Millions of Instructions Per Second | - CPU Cores (Requested)<br>- Bandwidth (Requested)<br>- RAM size (Requested)<br>- Millions of Instructions Per Second |
| **Attribute constraints** | Yes | Limited | No | Limited | Limited |
| **Build-in scenarios** | Google Cluster (cell A), 12.5K nodes | Generator and examples | Examples | Generator and examples | Several predefined scenarios |

Table 7: Cloud simulators comparison

## 5.5. SIMULATION FRAMEWORK DESIGN

All workload traces come in similar formats, where a change in the environment state is reported as an event. In GCD the jobs queue is the base of all processing. All entries in the jobs queue traces relate to jobs submissions, jobs cancellations, changes in jobs' priorities and so on. Listed jobs contain a series of tasks, which are reported in separate log files, that are subsequently executed on available nodes. The configuration of available nodes is reported in yet another log file. Given this, the simulation framework must cope with several independent sources of system configuration and must process them in a synchronised manner. All entries in the traces files come with a timestamp or period range. There are four main sources of configuration state changes in the workload simulation model, namely:

- Dynamic resource usage of processes – the resources utilisation levels are not constant through the life on an application, and indeed sometimes vary greatly from their specified requirements;
- New jobs are scheduled and current jobs complete their processing or are cancelled – this is the core operation in any scheduled system. When a job is scheduled, the user specifies the required resources and constraints;
- Changes in jobs resource requirements and/or constraints – during execution, tasks might have their resource requirements and constraints altered. This may result in a node no longer being suitable for certain types of task;
- Changes in nodes' configurations – during a cluster system lifecycle, nodes might be taken offline for maintenance or upgraded, with new nodes being added or old nodes removed. This scenario is rarely visible in smaller data centres, for example MetaCentrum infrastructure is relatively infrequently updated (Klusáček, 2014). It is more common to find frequent alterations to configuration within larger systems such as Google Cluster.

### 5.5.1. WORKLOAD EVENTS

To be able to handle this highly concurrent environment, all workload state updates, such as new tasks, updated constraints, new nodes and removed nodes, are done via immutable events, as shown in Figure 17.

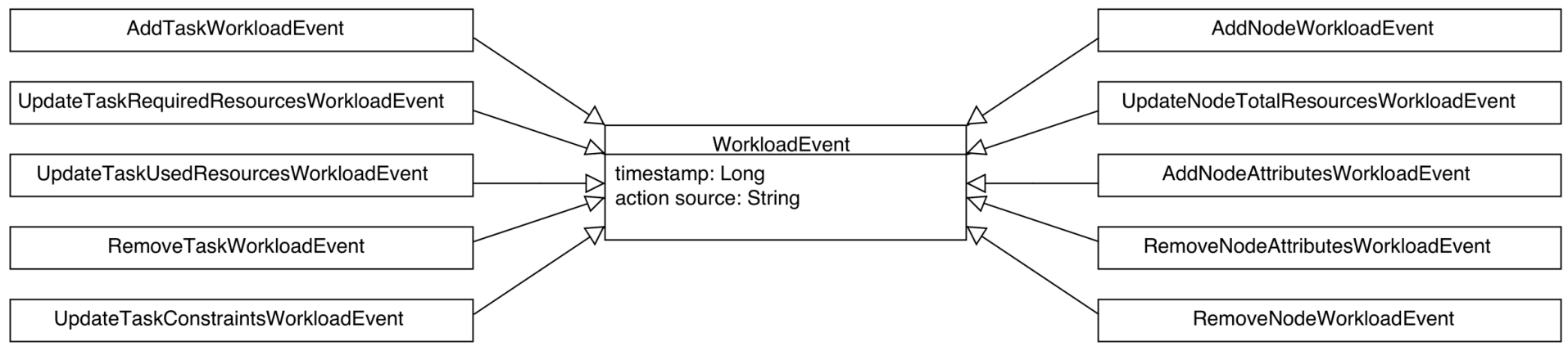

Figure 17: Workload events class diagram

Every event is marked with a timestamp to sort event batches from several parsers in the correct execution order. Such a setup enables the simulation system to maintain consistency of state even under a heavy load. Detailed descriptions of all workload events are presented below:

- AddTaskWorkloadEvent – generated for each new task. Tasks are always generated with initial resource requirements and constraints;
- UpdateTaskRequiredResourcesWorkloadEvent – in the majority of cases, requested resources values do not change after initial value. However, in several instances tasks get their required resources updated and this event will be generated;
- UpdateTaskUsedResourcesWorkloadEvent – upon execution, tasks dynamically allocate various amounts of memory, utilise storage space in different levels and so on. This event is generated to keep track of currently allocated resources;

- UpdateTaskConstraintsWorkloadEvent – task constraints are a set of logical operators set on node attributes and their values enable or disable execution of that task on certain node;
- RemoveTaskWorkloadEvent – this event is generated when the task finishes its execution or is killed by the system or user. When examining Google workload traces it was observed that significant parts of the tasks were killed by the native system;
- AddNodeWorkloadEvent – this event is generated when a new node is added to the cluster. The majority of these events are generated upon the start of the simulation;
- UpdateNodeTotalResourcesWorkloadEvent – during simulation lifecycle, certain nodes are taken offline and their resources updated (i.e. new memory banks are added). As with AddNodeWorkloadEvent, the majority of such events occur at the start of the simulation;
- AddNodeAttributesWorkloadEvent – this event is generated if node attributes are updated or new attributes are added. The GCD project does not specify the meaning of attributes as values and names are obfuscated, but it does suggest features like the existence of external IP address and the specific version of Linux kernel;
- RemoveNodeAttributesWorkloadEvent – as in the event above, node attributes can be removed, for example a given node might have lost its external IP address;
- RemoveNodeWorkloadEvent – during the recorded period certain nodes were taken down for maintenance, or were completely removed from the cluster. This event removes the node from the available nodes pool.

GCD traces keep record of all tasks updates and action. A task might have only two states with a number of transformations between those states: (i) pending (task is awaiting allocation to a node), or (ii) running (task is running on a node).

Once allocated, the running task cannot go back to pending state. If a task is evicted or lost a clone task is created added to the queue. Figure 18 presents the lifecycle of a task and Table 8 demonstrates how those updates are directly mapped to workload events in the simulator:

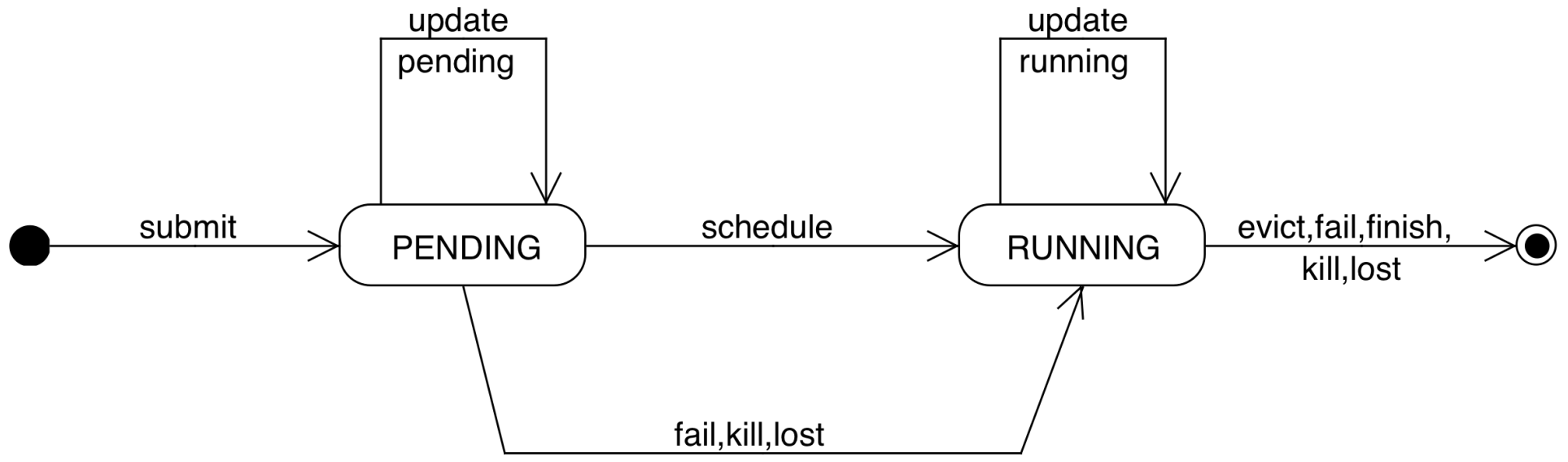

Figure 18: Workload events lifecycle diagram

| Action | Description | Workload Event Type |
|---|---|---|
| SUBMIT | Task has been submitted to cluster scheduler | AddTaskWorkloadEvent<br>(action creates a task in the queue and sends it to Workload Manager – this is where the initial task resource requirements are registered in) |
| SCHEDULE | Task has been scheduled by cluster scheduler | (actions are the results of the actions of the internal Google scheduler, therefore the simulator is ignoring them and no event is generated) |
| EVICT | Task has been evicted (and killed) from node. Reasons include: (i) higher priority task was scheduled on this node, (ii) node was taken offline, (iii) hardware malfunctions and (iv) task used resources exceeded node capacity. | RemoveTaskWorkloadEvent<br>(actions mark the end of a task and the simulator will delete task's definition and remove all references to it) |
| FAIL | Task failed or became unresponsive (i.e. execution crashed) | |
| FINISH | Task finished normally | |
| KILL | Task has been killed (by user or system) | |
| LOST | Task was terminated, but there is no record indicating that | |
| UPDATE PENDING | Task priority, resource levels or constraints were updated | UpdateTaskRequiredResourcesWorkloadEvent<br>(actions mark changes in task priority, required resources and constraints; these are often the result of users changing the requirements of already submitted tasks) |
| UPDATE RUNNING | Task priority, resource levels or constraints were updated during execution | |

Table 8: Tasks to workload events mapping

### 5.5.2. TASK CONSTRAINTS

Changes in task constraints are independently managed via UpdateTaskConstraintsWorkloadEvent. Aside from the required resources, incoming tasks provide a set of constraints for the node they can run on. These data can be found in the task_constraints files and is in the format of triple: attribute name, attribute value and logical constraint operator. The logical constraint operator can be:

- Equal (both numeric and text values are allowed) – the attribute has to be present on the node and have a value equal to the specified constraint value, or empty if no value has been specified;
- Not Equal (both numeric and text values are allowed) – the attribute has to be either missing from node attributes list or have a different value than the specified constraint value;
- Less Than (only numeric values are allowed) – the node's attribute value must be strictly less than the specified constraint value;
- Greater Than (only numeric values are allowed) – the node's attribute value must be strictly greater than the specified constraint value.

Figure 19 shows the implemented design of Task Constraints:

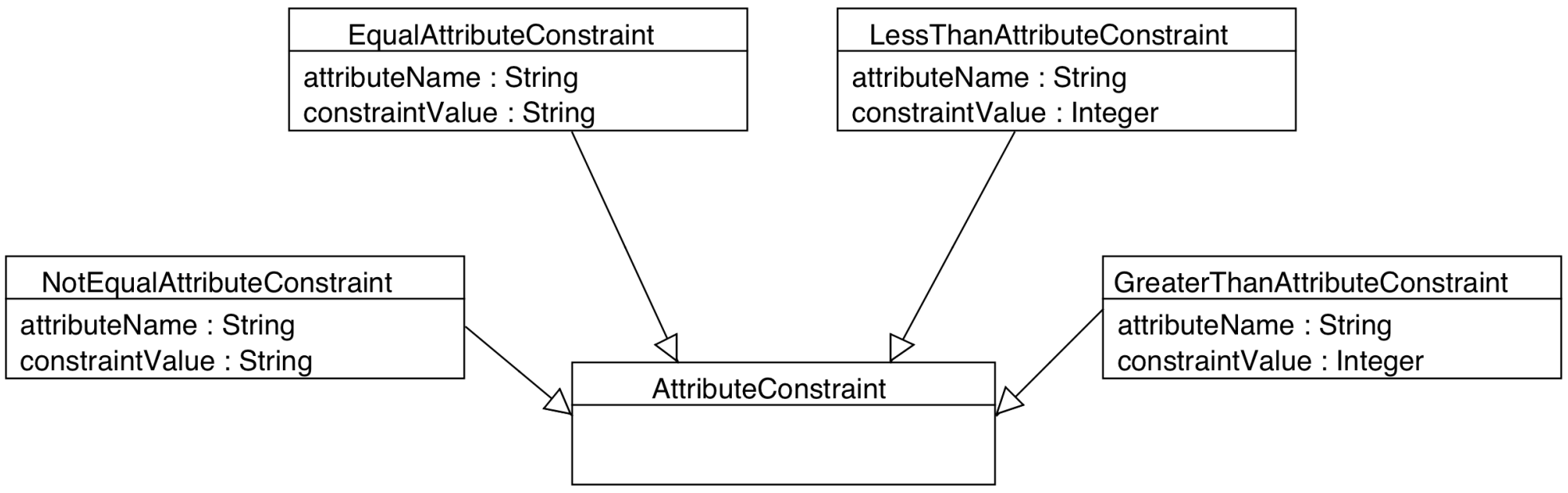

Figure 19: Task Constraints

It should further be pointed out that task constraints, as well as resource requirements, can be dynamically updated. Ensuring that tasks are only executed on nodes with attributes which match Task Constraints is referred to as Goal (III).

### 5.5.3. EVENT PARSERS

The simulator uses five independent Events Parser, which are implemented as the Actors from Akka framework, which read and parse workload traces data files and generate Workload Events. Each Events Parser holds a buffer of events, thirty minutes ahead of simulation time to avoid synchronous methods' calls. When a worker remains idle and the system usage is low, it will fill the events buffer. Table 9 lists Event Parsers:

| **MachineEventsGoogleClusterDataFileEventParser** |
|---|
| Reads /machine_events files and generates nodes configuration events:<br>• AddNodeWorkloadEvent<br>• RemoveNodeWorkloadEvent<br>• UpdateNodeTotalResourcesWorkloadEvent |

| **TaskEventsGoogleClusterDataFileEventParser** |
|---|
| Reads /task_events files and generates tasks events:<br>• AddTaskWorkloadEvent<br>• RemoveTaskWorkloadEvent<br>• UpdateTaskRequiredResourcesWorkloadEvent |

| **TaskUsageGoogleClusterDataFileEventParser** |
|---|
| Reads /task_usage files and generates task migration cost events:<br>• UpdateTaskUsedResourcesWorkloadEvent |

| **TaskConstraintsGoogleClusterDataFileEventParser** |
|---|
| Reads /task_constraints files and generates task constraints events:<br>• UpdateTaskConstraintsWorkloadEvent |

| **MachineAttributesGoogleClusterDataFileEventParser** |
|---|
| Reads /machine_attributes files and generates machine attributes events:<br>• AddNodeAttributesWorkloadEvent<br>• RemoveNodeAttributesWorkloadEvent |

Table 9: Workload events parsers

Every five seconds WorkloadGenerator collects events from events parsers and updates the system state in shared system object ContextData. The ContextData object is repeatedly read by various system elements, meaning that it has been designed so that it can support highly concurrent scenarios. All workload state is stored in TrieMap, which is a set of thread-safe lock-free implementations of a hash array mapped trie (Odersky et al., 2016). The TrieMap structure is more detailed as researched (Prokopec et al., 2012). The workload simulator is highly concurrent, and updating the shared system state is performed with custom non-blocking operations where the main part of the code is executed in parallel. Such a design allows the full utilisation of all available testing machine CPU cores.

Due to the size of the archive (191GB) it was not possible to fit it into the memory, and so it was decided that the best way forward was to continuously keep reading and parsing trace files on runtime while keeping a set of events in fast-access memory. The main purpose of these buffers is to minimise blocking operations while reading and preparing the next set of events. Each events parser keeps a buffer of events in memory (thirty minutes of events ahead and no more than one million events) and releases them to the Workload Manager on request every five seconds. If events parser does not have the requested set events, it will block the request (synchronous call) until enough events are loaded from the data files. While idle, each events parser will passively keep reading the data files, parsing them into buffered events.

Such a lightweight design allows the model to comfortably run a month-long simulation on the testing machine (see Appendix A for specifications) in ca. nine hours with 100x speed factor, which is equal to processing ca. 21.22GB of data per hour. The majority (ca. 89%) of data (170.54GB) comes from resource usage log files. After the initial loading and buffering of data (ca. 20 seconds), the system runs with consistent ca. 10-15% CPU usage.

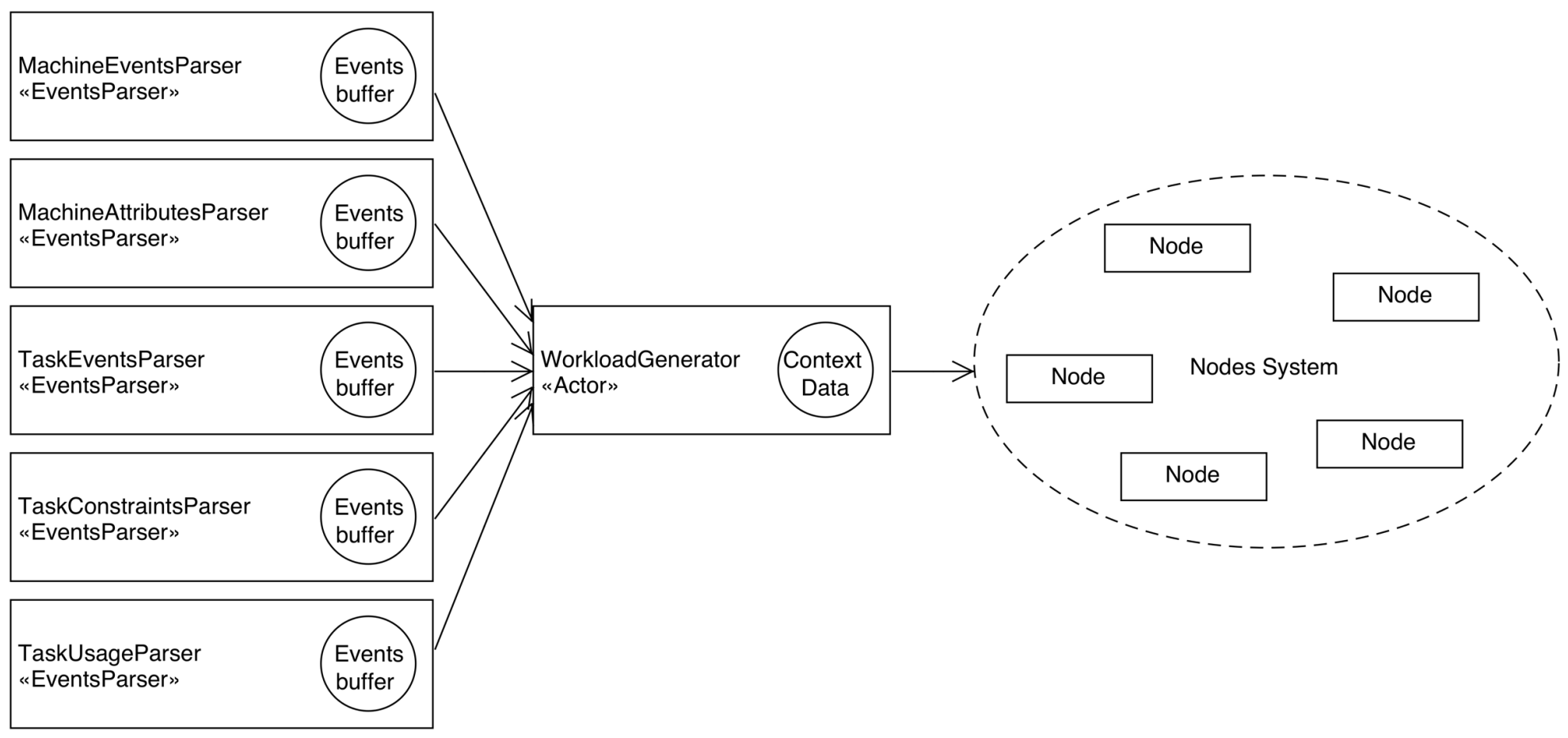

Figure 20: Workload events generation

## 5.6. ALTERNATIVE DESIGNS

During this experiment, the design of the simulation framework significantly evolved based on project requirements and experiences. While Scala and other high-level languages offer a variety of out-of-the-box routines and functions, especially related to concurrency, it is sometimes better to implement certain functionalities to maintain better control of data flow. In the case of this workload simulator the design featured fine-details process for generating and maintaining workload state, while leaving the task of running processes in parallel to the Akka framework.

Nevertheless, there are several optimisations and strategies that could be applied to build an even more efficient workload simulator. The subsections below discuss alternative designs for a simulator.

### 5.6.1. PRE-PROCESSING OF DATA FILES

The current implementation of simulator reads directly from GCD work traces files and parses data on the fly. However, only a fraction of all available data was actually employed. These data were especially visible when reading and parsing task resource usage (task_usage) files, where a majority of fields were disregarded (i.e. mapped/unmapped page cache memory usage, mean and maximum disk I/O time, cycles per instruction and so on). The simulator flow could be refactored into two stages:

- First, where all available workload traces data are read, processed and then stored into a list of events. The result list of events could be persisted to file or other persistence media in a serialised form which could be directly read into objects;
- Second, where workload generator reads, and replays stored previously stored events. This step could be additionally improved by streaming techniques as explained in the following subsection.

Such an approach would help avoid processing massive amounts of original data and significantly reduce the overhead from parsing logic. However, the trade-off would be higher complexity of the code and less flexibility during the experiment.

### 5.6.2. STREAMING EVENTS GENERATION

Java 8 (and also Scala as it is based on JVM) introduced a number of features and optimisations for streaming operations. Most implemented transformations, such as parsing files, filtering bogus events or sorting by timestamp, could be natively converted into parallel operations. Such a stream would be split into separate pipelines, with each event created and examined in a separate process and the OS would execute each pipeline in parallel on multiple CPU cores.

Research literature includes information about Java 8 streams with an explanation about their processing (Urma et al., 2014). It is difficult to estimate performance gains from this approach, although it could significantly reduce the complexity of the existing code base. The trade-off would be less control over a generation of events – i.e. proper timing and staging would be difficult to achieve when framework controls the utilisation of pipelines.

The flow of proposed stream operations is presented below. The Workload Events stream would start with reading and streaming lines from all archive files in parallel (flatMap operation). Generating a single Workload Event might (i) require combining several lines from several files, the logic of which is encapsulated in Event Parsers (map operation). Each Events Parser would (ii) generate a uniformed Workload Event object which is then (iii) accepted/denied by Events Filter (filter operation). Filtering events is critical in order to avoid bogus data. Finally, a collection of those objects (iv) is ordered by timestamp (sortBy operation) and (v) feed to Workload Generator (collect operation), which then distributes them to destination services. This kind of flow-based processing is popular among functional programmers.

Pre-processed data could be stored either in a set of files or a database. Storing data in a database would provide the additional benefit of a mature query interface (i.e. SQL or NoSQL APIs) which could be used to directly examine workload data by external applications.

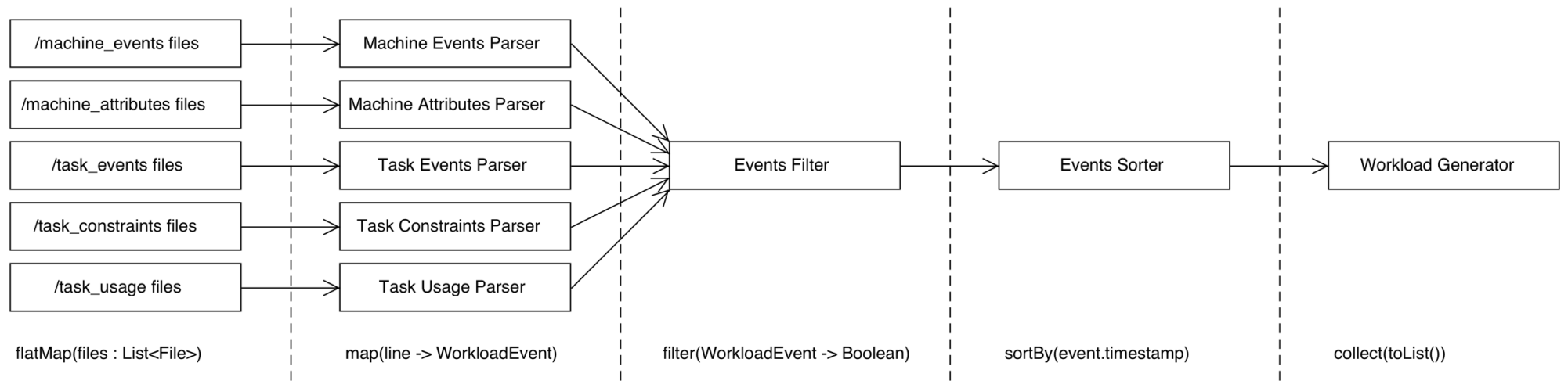

Figure 21: Stream-based simulator

## 5.7. DATA CORRUPTION

Reiss et al. (2013) states a number of reported anomalies in GCD traces. However, during experiments several further irregularities were found. The subsections below present these findings, detailing the adjustment made to the simulation framework to mitigate those glitches.

### 5.7.1. REPORTED RESOURCE USAGE IRREGULARITIES

The AGOCS framework tracked global resource usage ratios, i.e. how much of the available resource is allocated globally across all nodes. Figure 22 visually presents global usage ratios in GCD, separately for CPU and memory, over a full-month simulation period. A single bar corresponds to one minute.

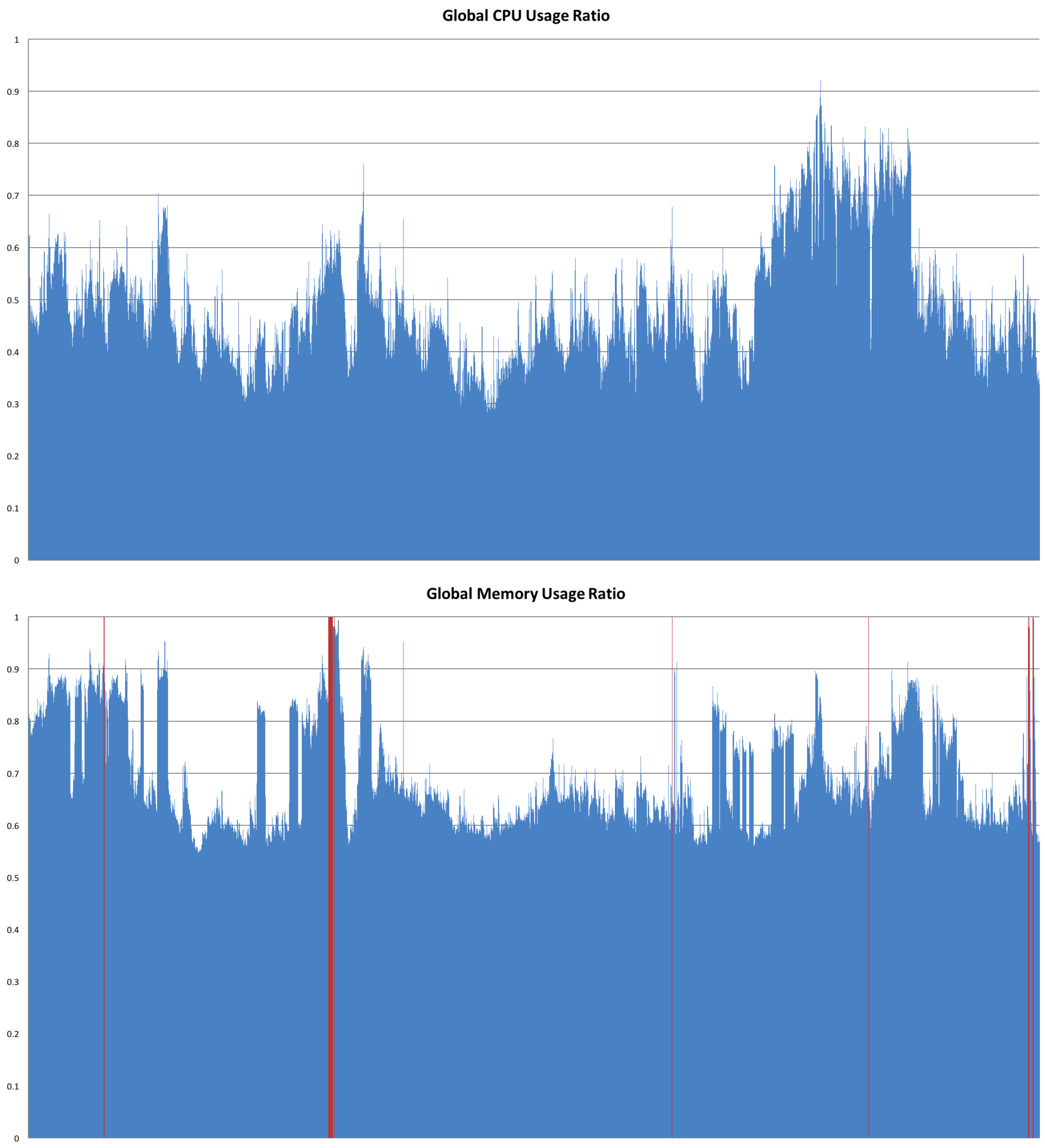

Figure 22: Global CPU and memory usage ratios (per minute)

Occasionally, the total memory allocated by all tasks is higher than the total available memory on the system nodes, as reported in GCD. During simulations in this research, five occurrences of this phenomenon were identified:

- The first spike at minute 3118 of the simulation, reporting around 130% memory usage;
- A seven-hour period between minute 12387 and minute 12799 of the simulation, reporting ca. 104% of peak memory usage;
- The second spike at minute 26589 of the simulation, reporting ca. 120% memory usage;
- The third spike at minute 34694 of the simulation, reporting ca. 145% memory usage;
- A half-hour period between minute 41479 and minute 41516 of the simulation, reporting ca. 106% peak memory usage.

Following analysis, it was discovered that a large number of non-production tasks were killed and then immediately restarted. Due to a ten-minute reporting window in GCD traces, those spikes resulted in abnormal usage reports. Therefore, the highlighted periods should be treated as examples of data corruption and are excluded from measurements.

### 5.7.2. USER-DEFINED RESOURCE REQUIRED IRREGULARITIES

User-defined requirements ratios for production tasks for CPU and memory create a more flattened pattern compared to the task resource usage in Figure 22, oscillating around 80-90% and 60-70% respectively. The memory spike at minute 15489 was caused by a glitch in the monitoring system when a wide batch of production tasks were cancelled and the reported values from two ten-minute time windows overlapped.

Figure 23 presents CPU and memory as user-defined requirements global ratios for production tasks:

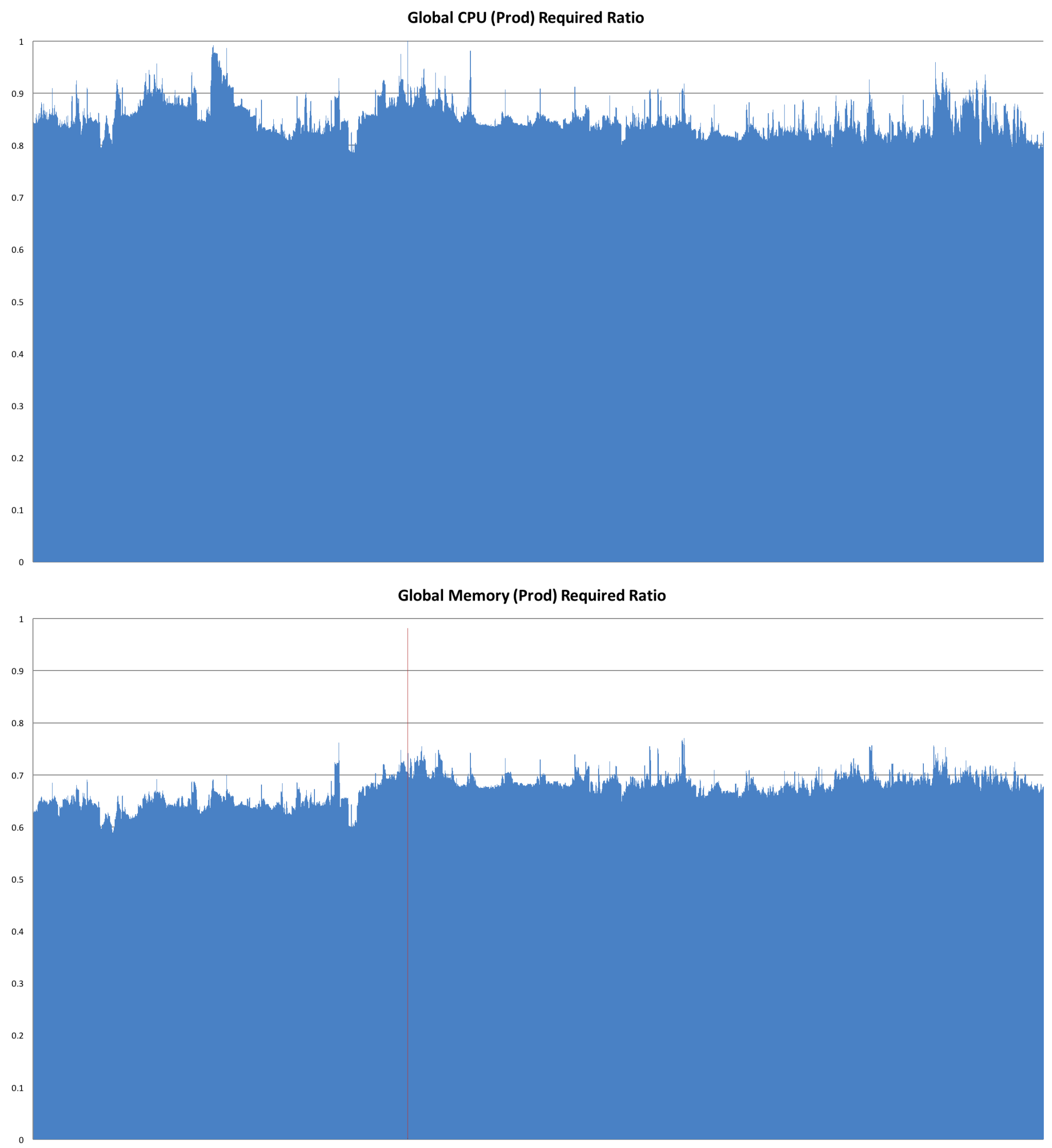

Figure 23: Global CPU and memory required ratios (per minute)

### 5.7.3. TASKS' CONSTRAINTS IRREGULARITIES

GCD workload traces contain a number of internal irregularities in task constraints events. This point was discussed briefly with Google engineers over email exchanges, with one possible cause identified as a bug which collapsed 'greater or equal' and 'less or equal' constraints into 'greater than' and 'less than' constraints while the data were being obfuscated. As such, it was not possible for some tasks to be matched to any node at any time. During experiments, a total of ca. 22.4k unique tasks with unmatchable constraints were found, grouped into 1903 execution batches. Those irregularities represent less than 0.01% of all tasks and are easy to filter out. In this and in following simulations, those tasks are reported as an error and ignored.

## 5.8. SIMULATION ACCURACY

It is virtually impossible to estimate error margins of simulation without knowing the exact tools used to monitor workload in GCD traces. One major bottleneck in the examined workload is memory, while CPU cores are relatively unallocated in comparison to user-defined requirements. The high memory footprint comes from Borg programs that are statically linked to reduce dependencies on their runtime environment (Verma et al., 2015). The result of this would be the allocation of more memory than if they were using shared libraries.

Additionally, GCD traces provide a very comprehensive array of memory usage parameters, such as canonical (kernel) memory used, page cache memory used, memory access per instruction and so on. Barring several unusual occurrences in workload traces (see section 5.7), the traces seem to accurately report true memory usage. This assumes that memory readings were very accurate and were drawn directly from the kernel. This is also supported by data, where memory is often allocated at exactly the maximum level for a given node. In such a scenario, the error margin for the presented simulations must have been negligible.

The GCD data has been obfuscated, and the size of the requested and used memory and the CPU are available only in normalised form, with a value of 0.5 being the most frequent for a node. As such, determining the amount of memory allocated for a program is not trivial. While Google does not disclose the hardware parameters of its node servers, in 2009 CNET's reporter captured a rare photo of Google server mainboard Gigabyte GA-9IVDP (Shankland, 2009), equipped with eight filled DIMM slots of memory and with the custom-made 12V battery attached (used to reduce the impacts of power outages):

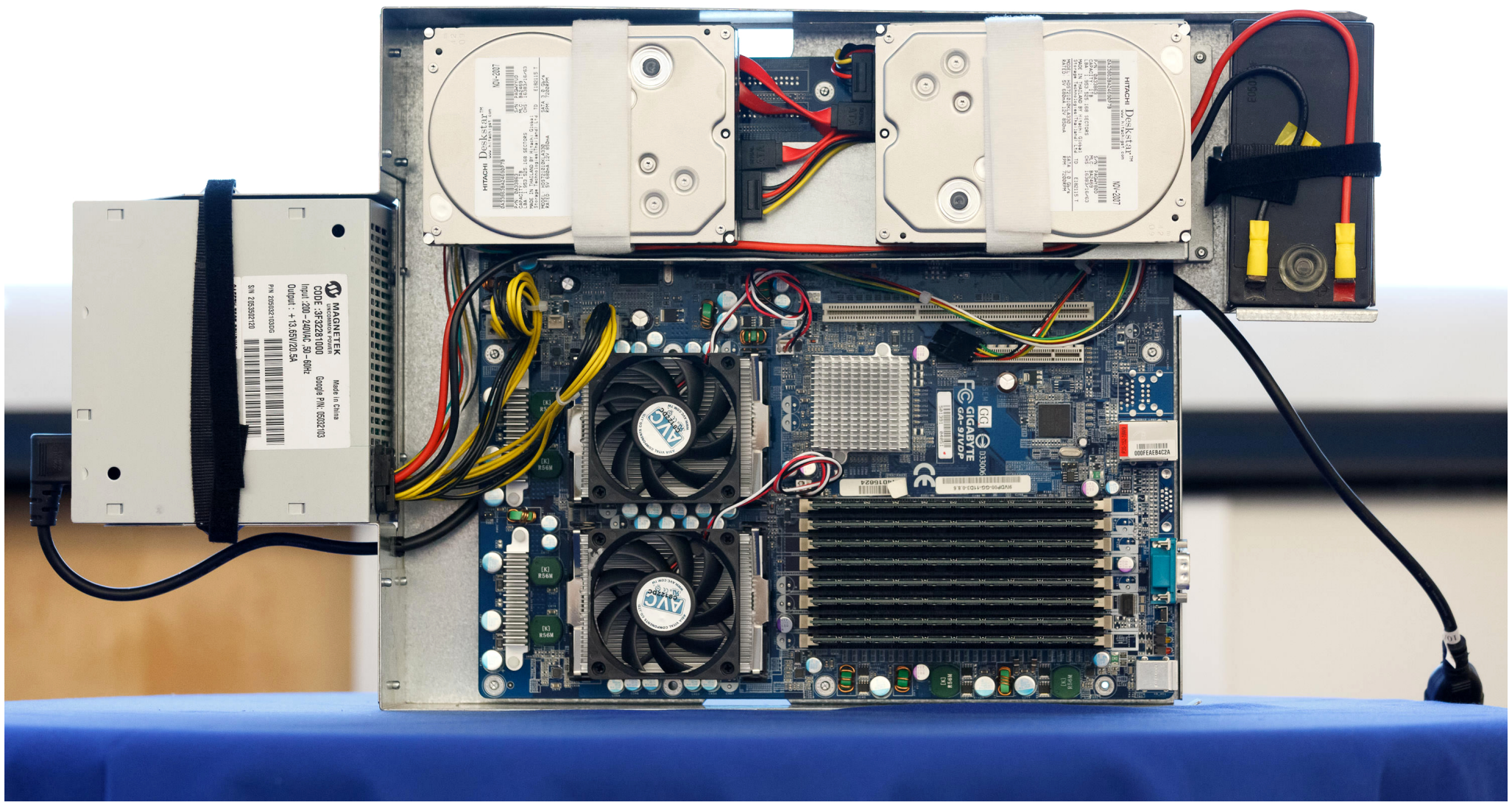

*(Image reproduced with permission of the rights holder, Stephen Shankland/CNET)*

Figure 24: Google node server photography (2009)

GCD traces were recorded in May 2011, meaning that it is safe to assume that the presented server was commonly used as a node in Google data centres over this period. Using this hardware, a typical fully-equipped GCD node would be able to support up to 64GB of memory. The assumption about this value is shared in Zhang et al. (2014a) and is used as an input for calculating the task migration cost.

As an interesting aside, it should be noted that Google's hardware has advanced significantly since GCD traces were first recorded. Since 2015, Google has deployed Tensor Processing Units (TPU) in its data centres. TPUs are used to support deep learning algorithms which often have very high computing power requirements, especially for speech recognition services (Jouppi et al., 2017). In November 2015, Google released TensorFlow, an open source library for supporting Machine Learning routines such as defining models and training them (Abadi et al., 2016).

The AGOCS framework itself records minuscule variances in nodes and tasks counts that are the result of concurrent update operations, while modifying shared context data objects (see the note in Appendix F). In comparing results from distinctive simulations, no instance has been identified where node counts differ by more than one node in one-minute intervals. Regarding task counts, the highest difference found was nineteen tasks. Considering the insignificance of this effect, it was therefore accepted as a trade-off for better performance.

## 5.9. PERFORMANCE EVALUATION

The closest alternative Cloud simulator to AGOCS is CloudSim package (Calheiros et al., 2011), created by Melbourne 'Clouds' Lab. CloudSim. For the purpose of this experiment the 3.1 version available was used. Both tools are created using JVM-based technologies and therefore the testing environment is identical – see Appendix A for detailed hardware and software specification.

CloudSim offers greater flexibility when setting up an environment. Nodes and tasks, referred to as 'cloudlets' in the CloudSim package, are set up in Java classes which are then compiled to separate jar package and run together with the main jar file. This approach is advantageous as compiled classes can be further automatically optimised by JVM even during execution (HotSpot technology).

On the other hand, AGOCS is not configured statistically, but continuously reads workload traces files and updates its state. That ensures simulation is very scalable; however, those parsing operations are quite expensive and might create bottlenecks on some machines.

In order to realise comparable input data sizes, both frameworks in presented tests were configured to initialise the same number of tasks and nodes in a single step. During simulation, on average, GCD schedules ca. 140k tasks on ca. 12.5k nodes. This means that the experiments tended to preserve this ratio of eleven tasks per one node in the below performance evaluation, for example 5500 tasks were submitted to 500 nodes. CloudSim was configured to assign a single VM to a single host machine. AGOCS was run with the highest possible speed factor that the testing machine (see Appendix A) was capable of running, i.e. 12x. Figure 25 presents the simulation time results.

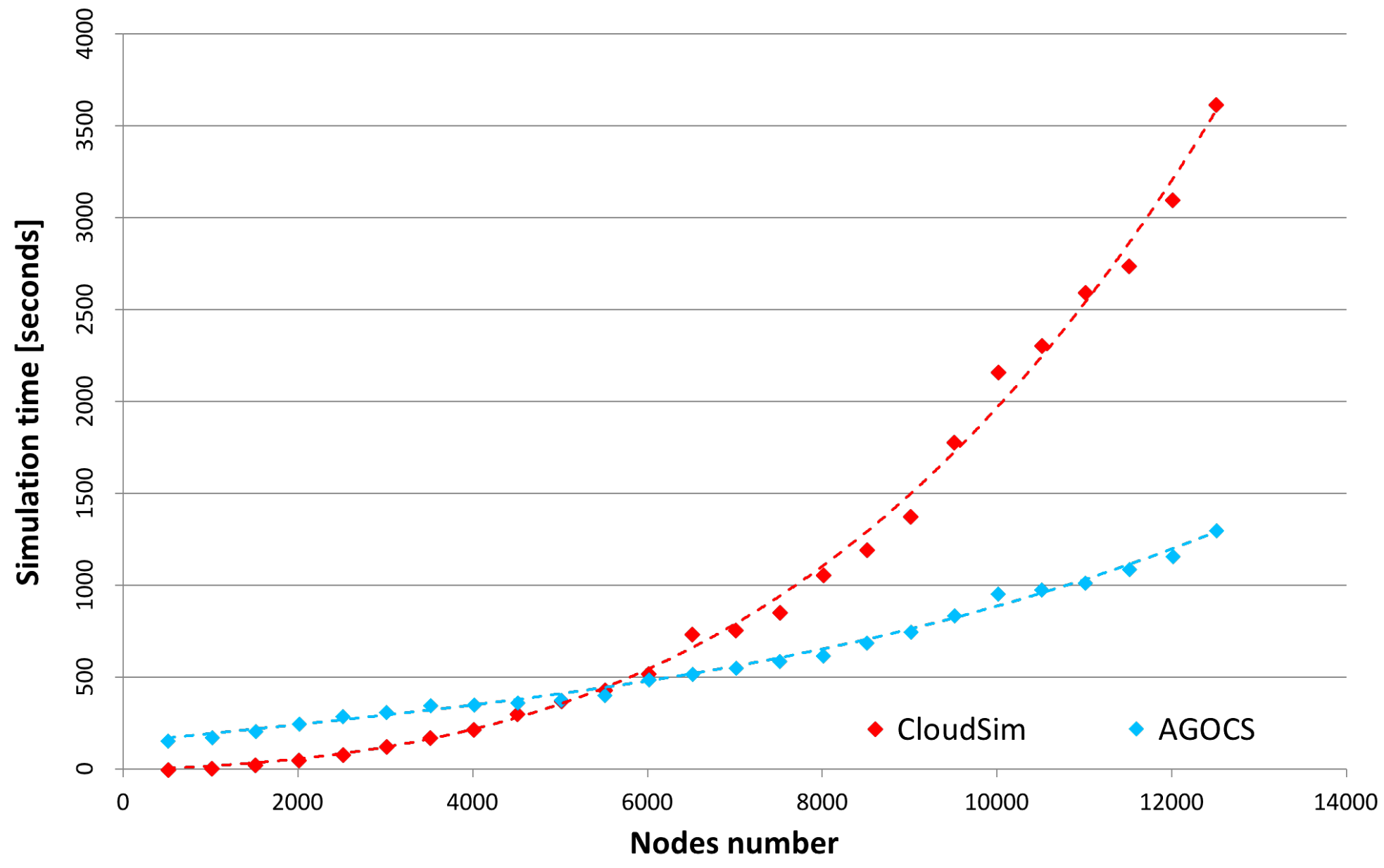

Figure 25: Simulator performance comparison

CloudSim performs better in smaller sets of data, although the computation time increases significantly for more complex sets. AGOCS's computation time increase is less rapid, although it requires initial time to preload data to its buffers. AGOCS was designed with multi-threading in mind and can take advantage of all available CPU cores. Its main bottleneck is workload traces' reading speed. CloudSim is completely memory-driven but implemented as a single-threaded application and utilises only one CPU core. CloudSim's code is prone to over-use of Java’s ArrayList class, while HashSet would work much faster – a significant amount of CloudSim’s simulation time is spent in Java's ArrayList.removeAll method. Overall, AGOCS simulates more layers of complexity and it is more accurate when describing secondary machine parameters:

- AGOCS supports adding and removing nodes during simulation. In smaller Cloud systems such as MetaCentrum (Klusáček, 2014), this may not be an issue as machine configuration is very static. However, in a larger environment, nodes are frequently modified and/or exchanged;
- Tasks being executed have actual values for both requested and actually used resources. This is very important factor as the task resources utilisation level is not constant and usually fluctuates depending on activity tasks currently performed. Therefore, to obtain a realistic and current view on node machine utilisation a simulation framework needs to consider actually used resources rather than requested ones. It should also be noted that the CloudSim framework also supports several resource utilisation models defined on task: full, stochastic and predefined (based on PlanetLab datacentre’s traces);
- Tasks and nodes are a representation of real machines and tasks run on Google Cluster. While CloudSim can generate random parameters based on statistical analysis, this approach will widen error margins and uncommon machine configurations might be missed;
- AGOCS simulation provides not only values for defined resources but also a number of secondary parameters such as disk I/O time, cycles per instruction and memory access per instruction. This makes the simulation more realistic and might serve as an input for processes;
- AGOCS's tasks have sets of constraints while nodes have sets of attributes. While a node may have enough resources to run certain tasks, it might be missing some features required to successfully complete the task fully, for example the availability of external IP address.

Therefore, AGOCS provides more complex and accurate simulations than CloudSim package, albeit at the expense of performance and flexibility. However, AGOCS has limitations and constraints that the researcher should be aware of, namely:

- While providing a relatively detailed description of the physical layer and requested resources as well as rich set of secondary parameters, AGOCS does not provide values for bandwidth utilisation. Unfortunately, GCD workload traces do not provide values for network transfer, which could be a critical missing feature in certain research projects;
- CloudSim package is easily extendable by a third party, and several other tools have been already built upon this framework (Wickremasinghe, 2009; Garg et al., 2011; Malhotra and Jain, 2013), often adding new features and new resource types. AGOCS is based on already existing workload traces and until Google decides to repeat this experiment and potentially extend set of monitored parameters, new modules are highly unlikely;
- The main risk of using AGOCS for research is that the generated workload does not change between interactions. Simulation is replayed the same way every time, with features such as timings of tasks and changes of nodes always identical. This may lead to developing over-specialised or over-trained algorithms that work on a single particular set of data only. However, the length of provided traces (one-month) is more than enough to evaluate the researched product in wide variety of situations. This said, the researcher must be aware of the above limitations in an attempt to achieve an accurate simulation.

AGOCS and CloudSim share many similarities and features even though they represent quite different approaches to the same research problem. CloudSim provides an informative top view of a Cloud system and is strong in testing high-level algorithms and strategies, while AGOCS is very suitable to fine-tune those algorithms and running simulations that are as close as possible to real Cloud systems. Load balancing strategies need to consider very fine-grained details and effects, often originating from the physical layer of a tested system. Due to implemented complexity, AGOCS is very appropriate for this class of research.

## 5.10. SUMMARY AND CONCLUSIONS

There are several notable aspects of the design of the workload simulator which are identified below:

- Simulating Cloud workloads on a complex network is not simplistic. The considerable number of parameters and dependencies require a well-designed domain model. Goals (III) and (IV) were added to the load balancing solution's feasibility criteria. This acquired knowledge was crucial when designing and implementing the load balancer prototypes, which are presented in the following chapters;
- The system should be able to cope with data anomalies and data corruption. The available traces are of high quality, but anomalies exist in the data provided. The system should be able to continue upon receiving bogus data. Common data errors include: corrupted state of task (i.e. task is marked as running when job has already finished), corrupted usage logs (i.e. reporting task resource usage, when task has not been created yet) and the global usage of any reported resource exceeding cluster capabilities. . Therefore, a working simulation framework must gracefully handle those errors without crashing, as highlighted in section 5.7;
- Several sources of workload state updates exist – for example task required resources and task constraints are provided by two different sets of files which are not synced time-wise. In experimental simulations, the GCD workload traces were split into one-minute intervals as detailed in section 7.6;

- Due to the complexity of data, it is difficult to properly test the created simulator. The design should allow for simulator 'testability' in mind. Every object and state should have appropriate unit tests during implementation. A sample test units suite is presented in Appendix E.

AGOCS was designed with usability and performance in mind. As a very lightweight framework it is capable of being run on typical desktop machine available in any laboratory. During the experiment it was found that it was comfortable to operate a month-long simulation test in approximately nine hours with 100x speedup factor, which is equal to processing ca. 21GB of workload traces data per hour. As a result of these characteristics, the AGOCS framework successfully serves as a foundation for both metaheuristic load balancer and distributed agent-based load balancer prototypes, as detailed in Chapters 6 and 7 respectively.

# 6. METAHEURISTIC LOAD BALANCER

Having examined the existing scheduling strategies (Chapter 2), defined the theoretical and practical restrictions concerning load balancing strategies (Chapter 3 and 4) and, more importantly, developed a robust simulation framework (Chapter 5), this study can now explore the applicable approaches to task allocations. Examining real-world workload traces from a working Cloud provides a valuable outlook into the mechanics of a massively distributed system. It also highlights additional challenges that were not identified during the modelling phase, namely tasks' execution constraints and RUS. It should further be noted that this chapter is based on work published in Sliwko (2008) and Sliwko and Getov (2015a).

Initial experiments with metaheuristic algorithms were performed at a very early stage of this research, helping to shape its general direction, prior to formally defining CRUM and then D-RSOP and Goals (I) and (II). The original design idea of this project was to extend existing load balancing strategies, such as FCFS, SJF, Round Robin and 'best-fit', towards more complex algorithms designed to solve NP-Hard problems. It was assumed that by introducing a holistic approach, where the scheduler tries to improve tasks' allocations globally by reducing resource usage gaps, the overall throughput of the Cluster would increase. At the same time, the load balancer should ensure that the STC stays within designed parameters. Given those design goals, the research focused on implementing those procedures, and the sections below describe this approach in detail.

Although a simple heuristic can be used to solve D-RSOP, their results are of poor quality (Sliwko, 2008; Sliwko and Zgrzywa, 2009). As such, the general approach to solve job-scheduling problems is to employ metaheuristic strategies. One can argue that metaheuristics might not be acceptable as a solution for load balancing problems given that each scheduling event can be very time consuming and have high overheads. However, the resources management in a distributed system has nowhere near the dynamic and robustness level required on scheduling processes on CPU cores. As long as the load balancer provides viable configuration changes or changes in tasks assignment within ten minutes, this strategy may be considered successful.

In the field of approximation algorithms, various strategies have been designed to find near optimal solutions to NP-Hard problems (Buyya et al., 2009; Ausiello, 2012; Pooranian et al., 2015) including metaheuristic algorithm. The term 'metaheuristic' originally derives from the Greek 'μετά' (a higher level) and 'ευρισκειν' (to discover) and is a scientific method that solves a problem with the help of iterative stochastic processes. A heuristic algorithm usually sacrifices the optimality of the solution in order to finish within a satisfactory timeframe. Generally speaking, it is possible to find a reasonably satisfactory solution, but there is no proof that the result could not be better or that the solution found by the heuristic algorithm would be feasible in the first place.

Many different metaheuristic algorithms are present in the literature, and new variants are continually being proposed. Some of the most significant contributions to the field are Evolutionary and GA, TS, Ant Colony, SA and Quantum Annealing (QA), Particle and Swarm Intelligence and Immune Systems.

Metaheuristic strategies tend to avoid iterating through the whole solution space by testing candidate solutions only in close proximity to a current state (for example the 'crossover' step in GA) with occasional attempts to escape local optima via a 'mutation' step in GA or 'tunnelling' in QA, for example. Such approaches result in reasonably good solutions reasonably quickly.

Based on previous research, including Józefowska et al. (1998), Józefowska et al. (2001), Sliwko (2008), Sliwko and Zgrzywa (2009), Kalra and Singh (2015),

Pooranian et al. (2015), and Fanjul-Peyro et al. (2017) not every algorithm will perform well in the context of the D-RSOP problem. The main issue in this model is the fact that not every solution is feasible, and that in fact the majority of candidates are not feasible at all. Such a setup proves to be difficult for the majority of existing strategies since usually only a small percentage of neighbour solutions are acceptable. Additionally, there is usually no starting state, with the strategy having to find this point by itself.

## 6.1. LOAD BALANCER DESIGN

The load balancer prototype was implemented in the functional programming language Scala. The core of the load balancer is a decision-making module based on metaheuristic algorithms which assigns tasks to nodes. The load balancer sequence was designed as shown in Figure 26 below:

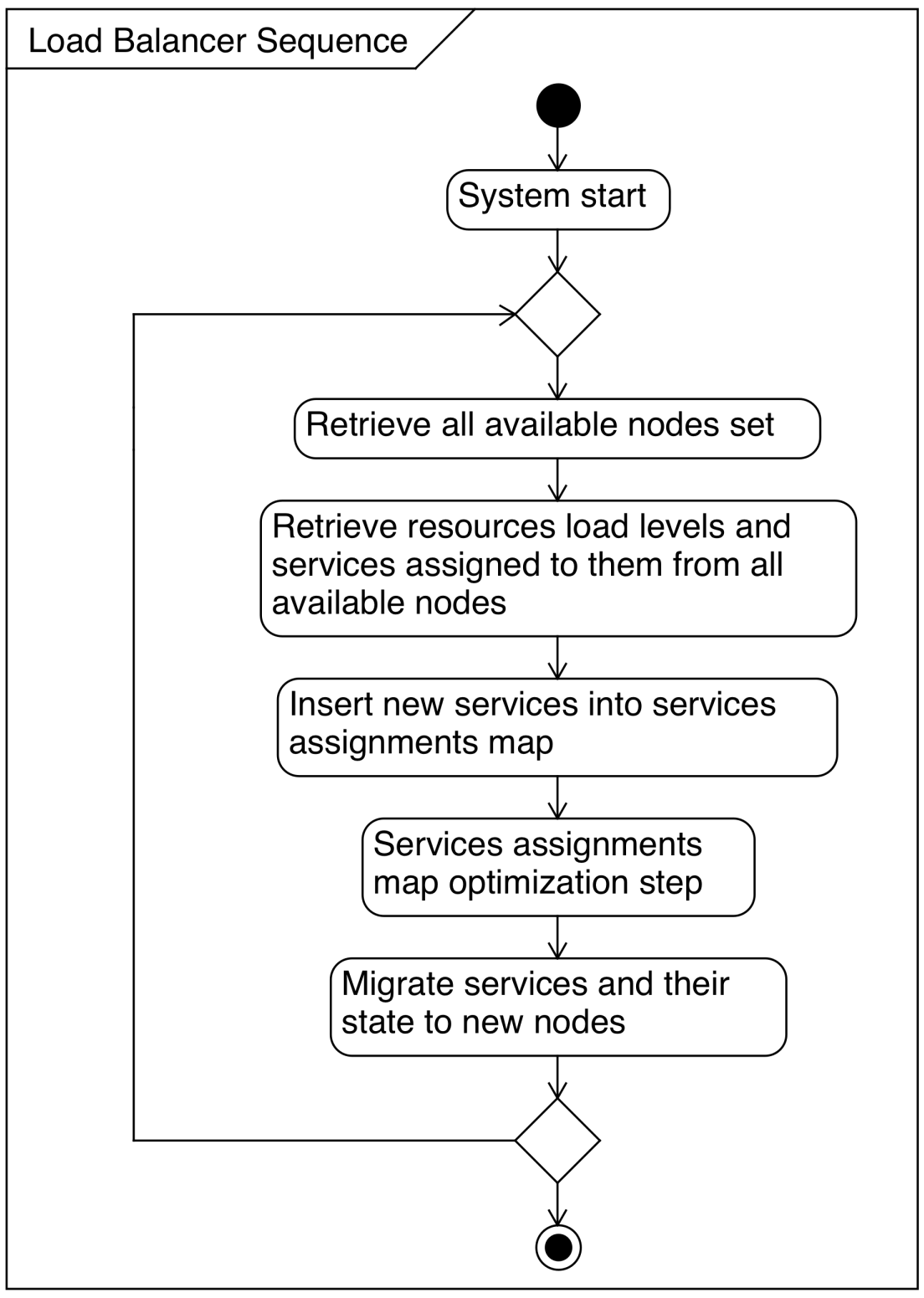

Figure 26: Load balancer sequence

The load balancer must maintain a difficult balance between the speed and quality of its decisions since badly assigned tasks can cause global system instability. The selection of the most efficient algorithm is crucial. For the purpose of the experiment several of the most promising strategies were studied, as outlined in the following subsections.

### 6.1.1. GREEDY

Greedy is an algorithm that follows the problem-solving heuristic of making the locally optimal choice at each stage with the hope of finding a global optimum (Chvatal, 1979). In many problems, a greedy strategy is effective, although it does not usually produce a globally-good solution in this research. Nevertheless, a greedy heuristic will yield locally optimal solutions in a very quick time. Greedy relies on examining immediate neighbourhood for better solutions (as per definition (6) in section 3.3).

### 6.1.2. TABU SEARCH

TS was introduced by Fred W. Glover in 1986 (Glover, 1986) and further formalised in 1989 (Glover, 1989). This algorithm has been suggested by previous research pertaining to a similar problem (Józefowska et al., 2002). Like Greedy, TS searches for an improved solution in its neighbours. TS enhances its performance by maintaining a list of visited solutions so that the algorithm does not consider that possibility repeatedly.

### 6.1.3. SIMULATED ANNEALING

SA is a general method for finding the global optimum via a process inspired from annealing, a metallurgical process where a material is heated and cooled in a controlled way so as to increase the size of its crystals and reduce their defects (Weinberger, 1990). This effect is implemented in the SA algorithm by a slow decrease in the probability of accepting worse solutions as it explores the solution space. Previous research concerning the use of this strategy in load balancing can be found in Józefowska et al. (2001).

### 6.1.4. GENETIC ALGORITHM

GA is a search heuristic that mimics the process of natural selection. GA belongs to the larger class of evolutionary algorithms which generate solutions to optimisation problems using techniques inspired by natural evolution, such as inheritance, mutation, selection, and crossover. Unmodified GA has been previously examined with good results (Józefowska et al., 1998). In this research, a variant of Genetic Drift step was developed (Sliwko, 2008).

### 6.1.5. SEEDED GENETIC ALGORITHM

SGA is the generation of random solutions that represents the costliest step in GA strategy, sometimes taking up to 60-70% of a total computation time. Therefore, a novel approach was implemented, where Genetic Drift step (Sliwko, 2008) is replaced with locally optimal solutions (i.e. solutions seeding) found by Greedy, TS and SA algorithms. This approach was calculated to allow for a dramatic lowering of the total size of population since individual genotypes are of higher quality.

To test this approach, respective strategy variations were created, namely SGA-Greedy, SGA-TS and SGA-SA.

### 6.1.6. FULL SCAN

Full Scan (FS) is the strategy which performs a full search over all available configurations. FS strategy is convergent, meaning that it is able to find the globally optimal solution in finite time, under appropriate modelling assumptions.

Multiple optimisation techniques have been implemented in this algorithm, such as shaving and path-cut (Demassey et al., 2005), and task-with-largest-migration-cost-first, i.e. the algorithm sorts tasks by their migration costs and the tasks with the highest migration cost are selected to be re-allocated first; the algorithm returns as soon as current STC is greater or equal to any previously found STC.

## 6.2. EXPERIMENTS SETUP

This experiment generated test configuration based on previous research (Mishra et al., 2010; Iosup et al., 2011; Di et al., 2012; Moreno et al., 2013) and also on personal professional experience while working with Amazon EC2 cloud instances. Tables 10 and 11 present test configurations.

Three strategies (Greedy, TS and SA) were designed with the end state, i.e. that no more steps were possible. If a strategy finished before a given time it was continuously re-run and the best result was selected. The number of runs significantly varied per strategy, especially in the lower sizes of the solutions space. Each algorithm creates a number of candidate solutions during their run. Deciding whether a candidate solution is stable, meaning that no nodes are overloaded, tends to be the most expensive step in computations: around 50-70% of CPU time depending on the strategy tested, is spent on the validation of solution feasibility routines.

As an optimisation, implementations were caching newly created solutions (see subsection 6.4.2), meaning that the same tasks assignment setup is never tested twice for being stable since the result is retrieved from memory.

| Task | Initial node | Migration cost | Resource I (CPU) | Resource II (Memory) | Resource III (Network) | Resource IV (I/O speed) |
|---|---|---|---|---|---|---|
| 01 | A | 4 | 1 | 10 | 4 | 2 |
| 02 | C | 5 | 1 | 6 | 5 | 2 |
| 03 | G | 4 | 5 | 2 | 5 | 6 |
| 04 | A | 17 | 10 | 17 | 1 | 2 |
| 05 | D | 10 | 14 | 10 | 1 | 1 |
| 06 | C | 3 | 3 | 12 | 3 | 8 |
| 07 | C | 6 | 15 | 2 | 18 | 3 |
| 08 | F | 6 | 1 | 4 | 8 | 4 |
| 09 | D | 4 | 4 | 3 | 17 | 10 |
| 10 | D | 4 | 8 | 19 | 19 | 8 |
| 11 | B | 8 | 5 | 9 | 18 | 4 |
| 12 | I | 6 | 16 | 14 | 3 | 2 |
| 13 | G | 4 | 6 | 5 | 17 | 11 |
| 14 | E | 5 | 18 | 11 | 13 | 4 |
| 15 | F | 1 | 10 | 9 | 12 | 8 |
| 16 | A | 9 | 12 | 17 | 14 | 1 |
| 17 | D | 5 | 3 | 6 | 8 | 6 |
| 18 | B | 5 | 8 | 12 | 3 | 11 |
| 19 | C | 7 | 15 | 12 | 8 | 9 |
| 20 | G | 1 | 4 | 8 | 6 | 12 |
| 21 | F | 7 | 12 | 10 | 5 | 1 |
| 22 | G | 2 | 3 | 16 | 16 | 2 |
| 23 | H | 5 | 6 | 19 | 1 | 4 |
| 24 | D | 3 | 16 | 11 | 2 | 3 |
| 25 | F | 4 | 14 | 8 | 15 | 9 |
| 26 | G | 10 | 4 | 15 | 7 | 8 |
| 27 | B | 2 | 20 | 19 | 5 | 2 |
| 28 | B | 8 | 16 | 2 | 3 | 5 |
| 29 | G | 6 | 16 | 10 | 3 | 1 |
| 30 | F | 5 | 1 | 1 | 3 | 10 |
| 31 | J | 4 | 19 | 18 | 1 | 8 |
| 32 | H | 10 | 6 | 14 | 3 | 7 |
| 33 | I | 2 | 3 | 10 | 3 | 2 |
| 34 | E | 4 | 2 | 8 | 1 | 8 |
| 35 | E | 9 | 8 | 9 | 9 | 5 |
| 36 | F | 4 | 8 | 15 | 13 | 1 |
| 37 | G | 2 | 12 | 8 | 5 | 3 |
| 38 | E | 2 | 16 | 11 | 1 | 2 |
| 39 | D | 8 | 13 | 8 | 6 | 4 |
| 40 | G | 3 | 6 | 9 | 10 | 1 |
| 41 | I | 6 | 14 | 1 | 11 | 8 |
| 42 | H | 7 | 8 | 3 | 10 | 3 |
| 43 | F | 9 | 9 | 9 | 10 | 9 |
| 44 | A | 8 | 11 | 8 | 12 | 11 |
| 45 | C | 2 | 5 | 5 | 7 | 18 |
| 46 | G | 6 | 2 | 7 | 3 | 2 |
| 47 | J | 5 | 4 | 3 | 10 | 16 |
| 48 | H | 8 | 5 | 2 | 14 | 8 |
| 49 | B | 2 | 6 | 7 | 1 | 1 |
| 50 | I | 1 | 1 | 9 | 6 | 13 |
| 51 | G | 4 | 4 | 11 | 9 | 6 |
| 52 | L | 3 | 7 | 2 | 7 | 5 |
| 53 | E | 12 | 6 | 6 | 10 | 12 |
| 54 | J | 10 | 3 | 9 | 8 | 10 |
| 55 | K | 8 | 5 | 5 | 4 | 8 |
| 56 | H | 7 | 6 | 3 | 5 | 7 |
| 57 | A | 3 | 8 | 12 | 2 | 6 |
| 58 | F | 1 | 12 | 17 | 1 | 9 |
| 59 | F | 6 | 10 | 8 | 6 | 14 |
| 60 | C | 5 | 9 | 2 | 3 | 8 |

Table 10: Experiment data – Tasks configuration

| Node | Resource I (CPU) | Resource II (Memory) | Resource III (Network) | Resource IV (I/O speed) |
|---|---|---|---|---|
| A | 100 | 50 | 100 | 70 |
| B | 70 | 40 | 70 | 50 |
| C | 50 | 80 | 70 | 50 |
| D | 60 | 60 | 50 | 80 |
| E | 50 | 90 | 80 | 40 |
| F | 60 | 100 | 50 | 60 |
| G | 80 | 50 | 50 | 40 |
| H | 80 | 80 | 80 | 90 |
| I | 60 | 60 | 50 | 80 |
| J | 40 | 50 | 80 | 100 |
| K | 50 | 80 | 80 | 40 |
| L | 50 | 50 | 60 | 80 |

Table 11: Experiment data – Nodes configuration

The following chart plots the average number of unique candidate solutions created in each test scenario:

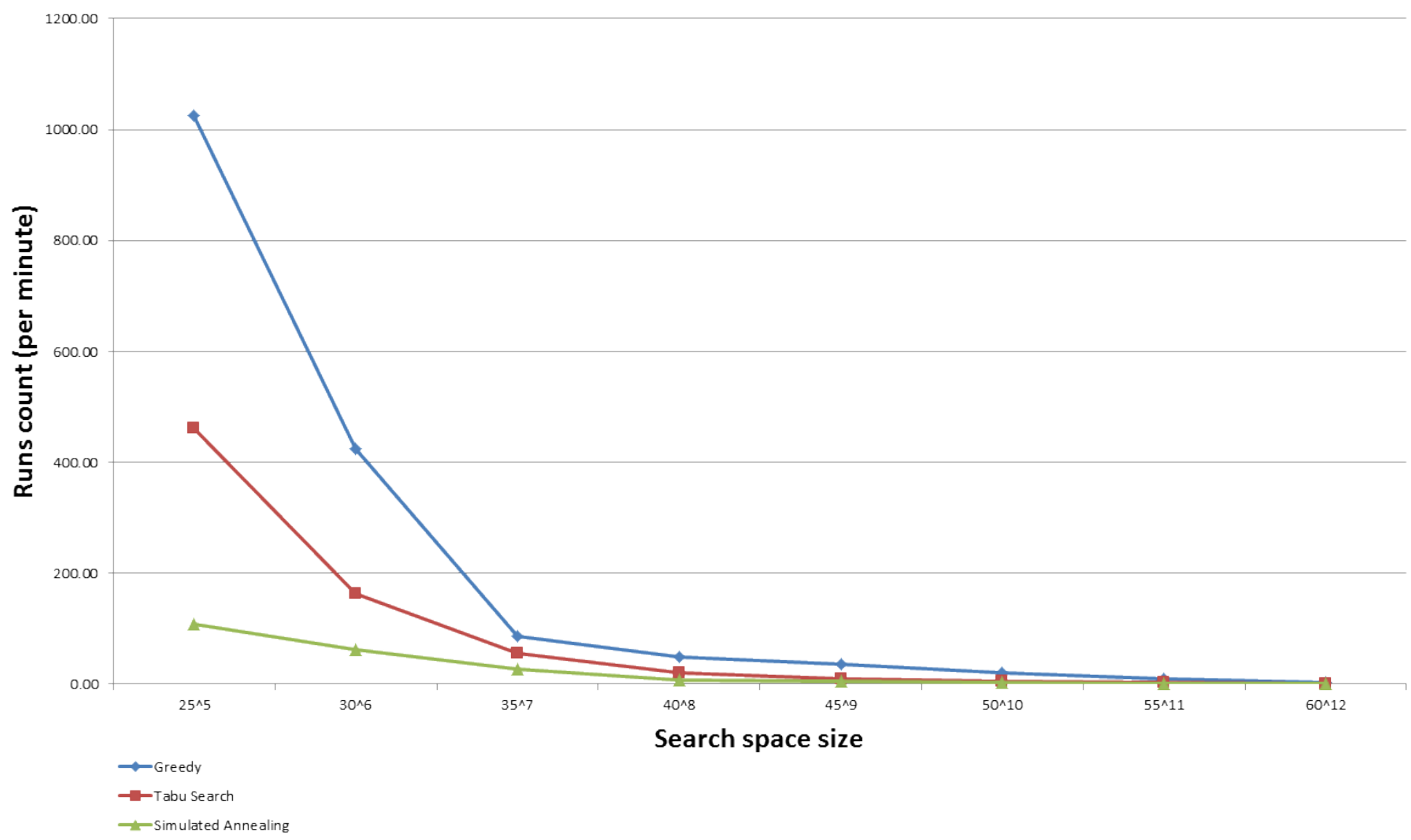

Figure 27: Runs count (per minute)

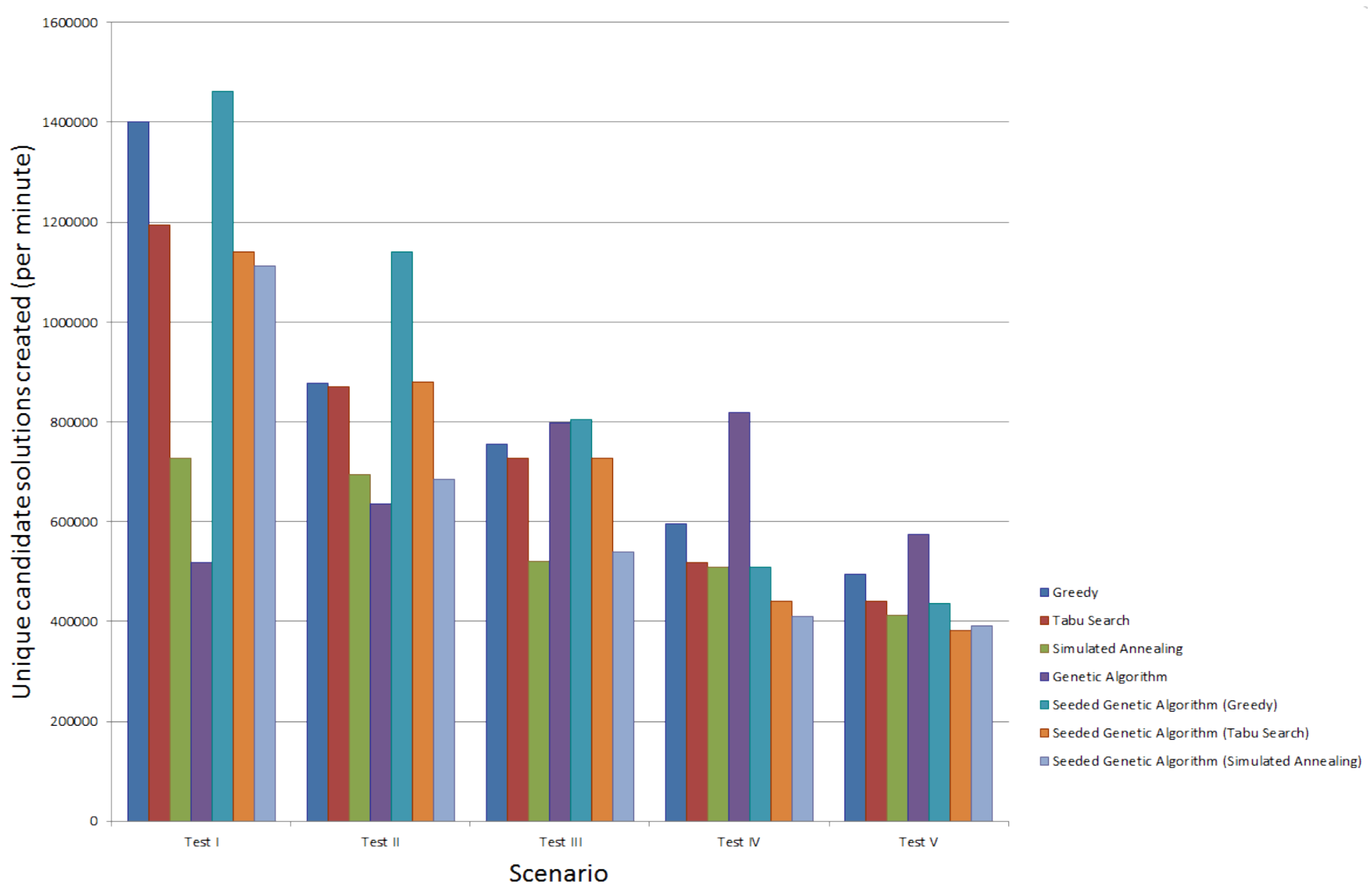

Figure 28: Unique candidate solutions created (per minute)

Five testing scenarios were designed to test how each strategy copes with the increasing complexity of the problem. An assumption was made that new nodes are added only when new tasks are deployed and the demand for computing resources increases. This scenario is simulated by enabling additional nodes, and in each test two additional nodes and ten more tasks are added. It is assumed that the load balancer will be run periodically, thus the selection of an arbitrary computation time, after which the best-found solution was selected as the output result.

| Scenario | Deployed tasks | Enabled nodes | Computation time | Search space size |
|---|---|---|---|---|
| Test I | 1-20 | A-D | 30 seconds | $20^4$ |
| Test II | 1-30 | A-F | 1 minute | $30^6$ |
| Test III | 1-40 | A-H | 2 minutes | $40^8$ |
| Test IV | 1-50 | A-J | 4 minutes | $50^{10}$ |
| Test V | 1-60 | A-L | 8 minutes | $60^{12}$ |

Table 12: Experiment data – Tests I, II, III, IV and V

The Full Scan strategy was used only as a benchmark if a global optimal solution was found and such a limit was not imposed. The Full Scan strategy was unable to finish scenarios Test IV and Test V in reasonable time, taking 24 hours and five days respectively. The results of all other strategies were plotted on the chart above. It should be noted that lower STCs are preferable.

## 6.3. EXPERIMENTAL RESULTS

As demonstrated in previous research (Józefowska et al., 2002; Leung, 2004; Sliwko, 2008), when solving RCPSP and its variants, more complex metaheuristics, such as TS, SA and GA, perform significantly better than simple algorithms such as Greedy. This was confirmed in the test results presented in Figure 29, where more sophisticated algorithms generally had better results, i.e. lower STC. A discussion of the outcomes of each strategy follows below.

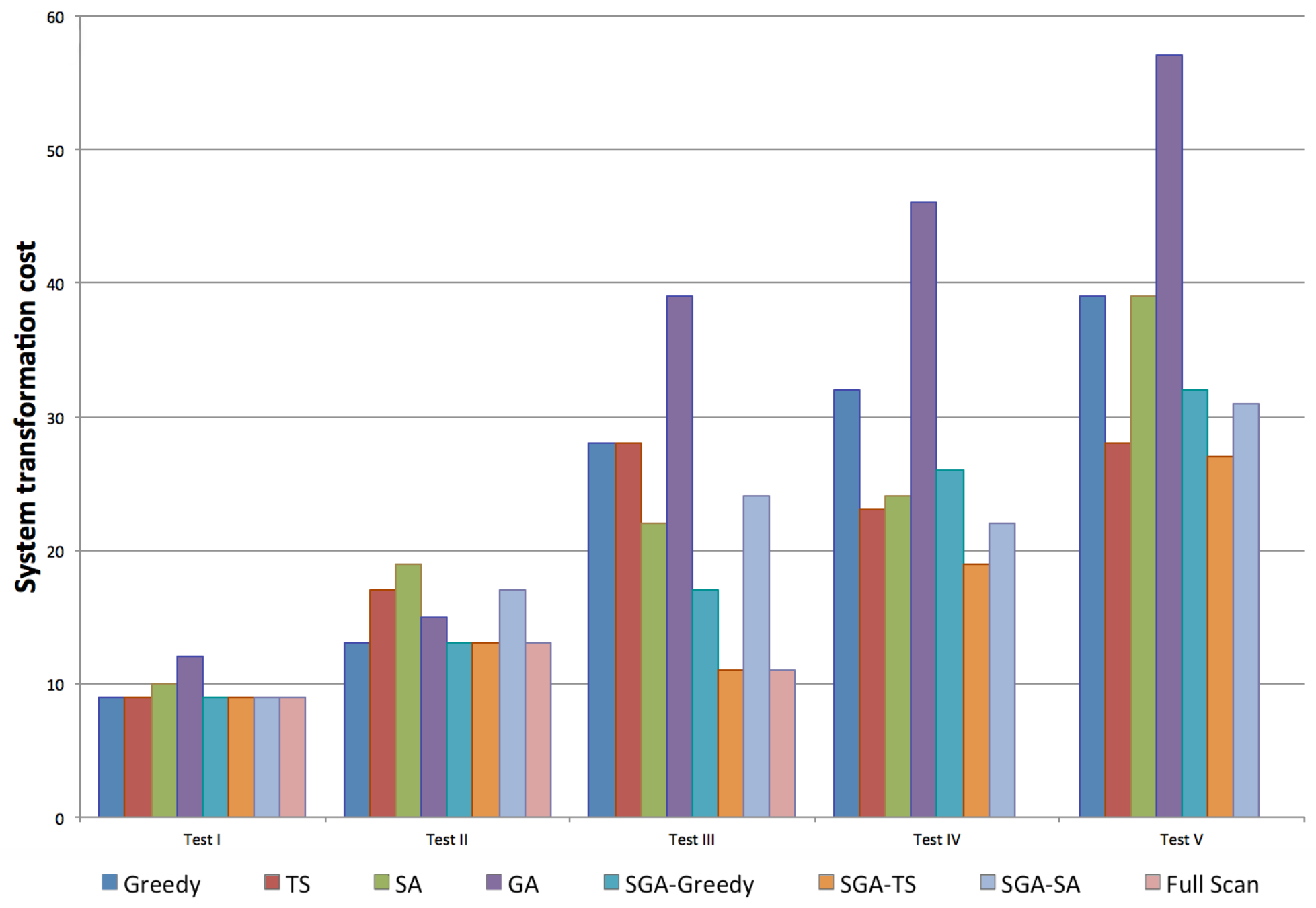

Figure 29: Simulation results

### 6.3.1. GREEDY

A very short execution time allowed this strategy to be repeatedly run and therefore a few stable solutions were found in each test. Result solutions were of average quality; the most time-consuming step was the generation of solution's neighbours, for example during the Test V scenario, each step required up to 60 x 12 = 720 solutions to be examined.

### 6.3.2. TABU SEARCH

The main bottleneck in this approach was the last step where all the same-value solutions had to be visited and marked as Tabu. Therefore, it was decided to introduce a maximum limit of dull moves (i.e. without bettering solution) the strategy will perform before it gives up and returns the actual solution. Overall, the TS algorithm worked very well in small instances of a problem, which confirms the results documented in Józefowska et al. (2002).

### 6.3.3. SIMULATED ANNEALING

SA strategy did require a much larger number of computations, often reaching only a fraction of runs in the same time as Greedy or TS. However, it did not require costly generation of all the solution neighbours, therefore the re-runs count decreased at a much slower pace than when the above strategies were deployed. This strategy benefited the most from introducing the solution cache.

### 6.3.4. GENETIC ALGORITHM

GA variant has been previously examined (Sliwko, 2008), where its main drawback was identified as being the costly generation of random solutions in the Genetic Drift step, especially when more types of resources are considered, and a solution space grows in size. Performance was shown to be sufficient when examining two kinds of resources. However, due to the number of random generations required in order to create the initial population, the strategy performed quite poorly when four resources were introduced. As in Józefowska et al. (1998), the larger the problem size, the lower the quality of the found solution. However, the performance of simpler algorithms, such as Greedy, TS and SA, was not impacted that much. Upon detailed examination it was found that the randomised solutions pool often contained a significant number of poor quality solutions. They were often eliminated in the next step; however, this process had a computation cost. This became apparent in instances of a larger problem, where ten or more nodes were involved.

### 6.3.5. SEEDED GENETIC ALGORITHM

SGA was the most interesting strategy in the experiment. As mentioned in GA, the randomised solutions pool contains low quality solutions, and eliminating those is costly. Therefore, solutions seeding replaced the previously designed

Genetic Drift step in the GA, which allowed for the downsize of the available genetic pool to 25% of its original size, thereby greatly reducing the computation time (ca. 50-70%) required to find good solutions without a reduction in quality. SGA returned the best results within the set time frame. In each case Greedy vs. SGA-Greedy, TS vs. SGA-TS and SA vs. SGA-SA, the found solution was improved, and generally less candidate solutions were examined. In Test V ca. 14% less candidates were visited. In this experiment the variant with TS strategy returned the best results.

### 6.3.6. FULL SCAN

Full Scan strategy guarantees that a globally optimum solution is found. Over the course of the research, this strategy has been heavily optimised. Currently, only ca. 9% of a solutions tree is traversed; the strategy starts moving tasks with the highest migration costs first, as the algorithm cuts solution tree's leaves as soon as partial solution is deemed unstable. However, this still cannot be considered an efficient strategy due to a large number of computations required. In this experiment, Full Scan strategy was used to produce a global optima solution only in minor instances of a problem.

## 6.4. SYSTEM OPTIMISATIONS

System wide enhancements and optimisations can dramatically increase the performance of certain algorithms. However, based on personal professional experience, hot spots – areas of a program's code where a high proportion of CPU-cycles is spent during the program's execution – can be found in very surprising places, especially in complex real-time systems. Generally speaking, this makes dry source code analysis pointless. A system developer needs to see a detailed and full application performance and memory profile before attempting to improve it.

Therefore, in the course of the research, the experiment routines have been profiled with YourKit Java Profiler (YKJP), with several bottlenecks being identified. Aside from programming optimisations, such as refactoring loops into tail recursive function, marking values and methods for lazy initialisation, and converting all objects as immutable case classes, there has been the identification and implementation of several system-wide optimisations. The majority of code optimisations and refactors were focused on improving the parallelism of the implemented prototype as to fully use the available HPC cluster machines (Appendix C) – the detailed explanations of profiling exercises and testability can be found in subsections 7.6.1 and 7.6.2 respectively.

### 6.4.1. ENHANCED RANDOM SOLUTION GENERATION

The starting point of many metaheuristic algorithms is the initial random state (or a pool of states), which the algorithm then revises into a better and improved result in each step. In this case, generating the initial candidate solution is expensive since it needs to be verified as stable - i.e. that no nodes are overloaded. The verification process is computation-intensive since all tasks on each node need to be iterated and their resource vectors need to be added to check if they exceed the available resources on this node.

The optimisation generates an initial random solution, which it then attempts to convert into a stable one through randomly moving tasks only from unstable nodes. This reduced the routine execution time by an order of magnitude. Previously, only one task had been randomly moved at a time (see definition of the neighbour solution (6) in section 3.3); however, further experimentation showed that an additional half of time could be shaved off by moving several tasks in each step. The fastest convergence was achieved by moving 10% of all tasks (but no less than 1) from the unstable nodes in one step. The pseudo-code is presented below:

**ALGORITHM**: Find stable random solution via multiple mutations on unstable nodes

**INPUT**:

A set of tasks where $\tau = \{t_1, t_2, \dots, t_p\}$

A set of nodes where $\eta = \{n_1, n_2, \dots, n_m\}$

A set of resource types $\psi$ (see definition of resource types in section 3.3)

**OUTPUT**:

$\mu_{out}: \tau \rightarrow \eta$ as output tasks assignment function

**BEGIN**

1 Randomly initialise task assignment function $\mu_{step}: \tau \rightarrow \eta$ to random

*(initially all tasks are assigned to random nodes)*

2 **WHILE** $\mu_{step}$ is not stable (as per definition (2) from 3.3)

*(repeat until current tasks assignment $\mu_{step}$ is not stable)*

2.1 Select a set of overloaded nodes

$\eta_{over} = \{n \in \eta : \exists i \in \psi : f_{step_i}(n) < 0\}$, where $f_{step_i}(n)$ is the available resources levels function for resource $i$ on node $n$ for task assignment $\mu_{step}$ (see (1) in section 3.3)

2.2 Select a set $\tau_{over} = \{t \in \tau : \mu_{step}(t) \in \eta_{over}\}$

*(select a set of all tasks on all overloaded nodes)*

2.3 Calculate migrations count $x = 10\% \cdot |\tau_{over}|$ (but no less than 1)

*(for higher number of overloading tasks, swap multiple nodes)*

2.4 Randomly select a set $\tau_{over_x} = \{t \in \tau\}$, where $|\tau_{over_x}| = x$

*(randomly select a subset of x tasks from $\tau_{over}$)*

2.5 Create a new task assignment function

*(re-assign tasks from $\tau_{over_x}$ to random different nodes)*

$$\mu_{next}(t) = \begin{cases} random(\eta - \{\mu_{step}(t)\}), & t \in \tau_{over_x} \\ \mu_{step}(t), & t \notin \tau_{over_x} \end{cases}$$

2.6 $\mu_{step} = \mu_{next}$

*(repeat loop with new task assignment)*

3 **RETURN** $\mu_{step}$

**END**

Additionally, to take advantage of the multi-core architecture, the candidate stable solutions were created in parallel. Initially, the implementation used Scala Future objects running on Executor from default ExecutionContext (Odersky et al., 2016); however, later Futures were replaced with Akka framework.

As a result of all the optimisations above, the CPU time spent in seeding solutions step was reduced from ca. 40% to 3% for GA and SGA. For other tested strategies (Greedy, SA and TS) the total CPU time spent in searching for starting solution was reduced from ca. 20% to 7%.

### 6.4.2. SOLUTION CANDIDATES CACHE

During execution, the strategies generated and tested a number of candidate solutions to compare them with the solution in the current step. Operations on the solution object, such as verifying whether the solution was stable, iterating unstable nodes and computing the STC, cost CPU cycles. To remedy this and to save CPU-cycles, a cache of created solutions was created. This meant that every newly created solution, from random generation, mutation, crossover and so on, were added to the cache if a solution was generated before the object in cache was used. This helps to avoid duplicate computations as solutions in cache might have had their methods executed once already. For example, unstable nodes may have been filtered before, and their results stored inside the object's private fields ('lazy' pattern).

The trade-off of this approach is the cache amount of memory that needs to be allocated to store all cached objects. Early experiments used default Scala mutable map implementation, which has been repeatedly cleared upon reaching a set size limit. However, further experimentation shows much enhanced results with CacheBuilder from Google Guava library. The base idea behind expiring cache is to evict entries that have not been used either recently or very often (‘sinking cache’ pattern). In this implementation, Google’s CacheBuilder starts evicting items when approaching a size limit of memory, which is specified upon cache initialisation (here: five hundred thousand).

It is difficult to measure the exact impact that the use of cache made on test performance, since cache is used in multiple of areas in the test algorithms. Test implementations of Greedy and plain GA were especially prone to testing a huge number of duplicate candidate solutions. In the experiments it was estimated that enabling cache speeds-up executions of tested strategies by around 25-45%, meaning that more algorithm’s steps were executed in the same amount of time.

## 6.5. SCALABILITY TESTS

The initial trials on a static dataset were promising, and clearly showed the potential for improving the quality of tasks' allocations. Given this, the next step was to test how the designed strategy would perform under a real-world workload.

Although the prototype centralised load balancer had been designed and implemented in the early stages of this research, the scalability tests were delayed until the AGOCS framework was ready. Eventually, the simulation framework was implemented, with the metaheuristic algorithms back-ported to it. Initially, only a fraction of the original GCD workload was used to find out how algorithms would perform on it.

Table 13 details the time required to compute a single load balancing sequence during simulation. All below test simulations were run on compute nodes from the University of Westminster HPC Cluster (see Appendix C).

| Test and size | Test I | Test II | Test III |
|---|---|---|---|
| | **62 nodes and 199 tasks** | **62 nodes and 326 tasks** | **62 nodes and 503 tasks** |
| Greedy[1] | 1 hour 9 minutes | 16 hours 59 minutes | 3 days 22 hours 9 minutes |
| TS[1] | 4 hours 35 minutes | 18 hours 18 minutes | 5 days 9 hours 34 minutes |
| SA[1] | 5 hours 27 minutes | 13 hours 20 minutes | 2 days 23 hours 51 minutes |
| GA[1,3] | 3 hours 12 minutes | 22 hours 13 minutes | 1 day 22 hours 48 minutes |
| SGA-Greedy[2,3] | 2 hours 48 minutes | 19 hours 33 minutes | 3 days 1 hour 1 minute |
| SGA-TS[2,3] | 2 hours 27 minutes | 12 hours 6 minutes | 1 day 21 hours 23 minutes |
| SGA-SA[2,3] | 4 hours 49 minutes | 14 hours 33 minutes | 2 days 16 hours 46 minutes |

1. Greedy, TS and SA strategies were constrained to run for a maximum of ten minutes and then restarted. This process was continued until a stable solution (i.e. no overloaded nodes) was found.
2. GA, SGA-Greedy, SGA-TS and SGA-TA strategies were run continuously until a stable solution was found, but for no less than ten minutes.
3. Due to high memory demand, the solution candidates' cache size (as detailed in 6.4.2) was limited to 500k of items.

Table 13: Time required to compute a single load balancing sequence

The above simulations were run on 62 nodes, which is roughly 0.5% of all GCD nodes. To test performance of solution, the total number of tasks was set to 0.15%, 0.25% and 0.4% of all GCD tasks in Tests I, II and III respectively.

Although previously presented tests have shown that a proposed strategy can indeed manage a small cluster with a limited number of running tasks, because of the lengthy computation time required to load balance a given instance, it cannot be considered as a feasible solution for tasks orchestration in Cloud. In the original GCD workload traces, the new tasks were scheduled with an average frequency of more than ca. 1.1k tasks per minute; this solution could not handle such a throughput.

Additionally, the proposed design employs VM-LM feature, where running tasks can be offloaded to alternative nodes and, therefore, a centralised load balancer must track all tasks existing in Cloud systems. This additional logic needed to handle running tasks and their migrations multiples the complexity of the load balancing algorithm, resulting in even higher computation power requirements. A number of optimisations were implemented during the code iterations which were focused especially on parallelisation, caching and non-blocking processing. Even so, the processing speed did not improve to an acceptable level.

## 6.6. SUMMARY AND CONCLUSIONS

After analysing the performance of the algorithms, the following conclusions were reached, which might assist in the design of new algorithms in the future and/or enhance algorithms which already exist:

- Metaheuristic algorithms rely on traversing a search space using small steps, meaning that the next selected solution is usually similar to the current one, and is also usually better. It might be beneficial to give higher priority to moving already-migrated tasks since they have already increased their migration cost, as well as to prioritise moving tasks with a smaller migration cost due to the reduced impact on the STC (introduced in Chapter 3). However, this step requires building problem-specific knowledge into the algorithms. This conclusion is very important for the design of the decentralised agent-based load balancer prototype detailed in Chapter 7, in which the chance of selecting a task to re-allocate is inversely proportional to its migration cost (see (12) in subsection 7.4.1).
- The initial random generation of candidate solutions is expensive. This behaviour is clearly visible in the upward trend in the number of candidate solutions created and tested using the GA strategy. The number of tested solutions does not correlate with the quality of solutions, and better results can be achieved if the solutions pool is initially created from an already precomputed set.

- Whilst a few strategies succeed in reaching a certain solution level, they face difficulties in moving out from this or in recognising a last state, for example when only one neighbouring solution is better. The TS algorithm, in particular, is prone to this and higher numbers of steps didn't increase the quality of the solution. However, the further experiments with TS variants have shown it to be a good candidate for selecting a set of tasks based on arbitrary criteria (as detailed in subsection 7.4.1).

While the proposed solution was able to efficiently schedule tasks on twelve nodes, further experiments have shown that the scalability of this approach is insufficient for supporting huge Clusters, such as 12.5k nodes in GCD traces. During scalability tests, the metaheuristic algorithm required several hours to execute a single load balancing sequence. When more nodes and tasks were added, the search space size grew exponentially. Whilst metaheuristics were still able to greatly reduce examined search space, the solutions found were either of decreasing quality or they consumed too much computation time to be viable.

Therefore, under the described Goals (I) and (II) from Chapter 3, and (III) and (IV) from Chapter 5, this method was recognised as being an inadequate foundation for the Cloud load balancer, and therefore an alternative approach was needed. Nevertheless, although it is unlikely that metaheuristic algorithms by themselves could orchestrate tasks allocation in a large computing cell, metaheuristic algorithms can still play a supporting role. For example, in the Czech National Grid Infrastructure MetaCentrum, experimental extensions based on TS are being used to improve the tasks queue in TORQUE Resource Manager (Klusáček et al., 2013). The following chapter will demonstrate that metaheuristic algorithms can indeed be efficiently used to form a local AI, which can locally manage a set of tasks on a node.

# 7. DECENTRALISED AGENT-BASED LOAD BALANCER

With a centralised load balancer prototype failing to scale well enough to yield satisfactory results (Chapter 6), the project shifted its focus to decentralised strategies such as agent-based systems.

The experiments in Chapter 6 have shown that improving the tasks' allocations quality requires higher computation time, and that a processing scheduling logic on a single head node machine would be the main holdup in scaling Cloud systems to larger sizes. The reasons behind this are, firstly, the higher rate of incoming tasks reduces the time window allowed for making the allocation decision, and secondly, as was observed, that the larger number of nodes increases the solution search space of feasible allocations.

In the subsequent design, the core strategy for developing the Cloud load balancer prototype was to offload the scheduling logic's processing to nodes themselves and to execute complex strategies locally. The principle of this approach is that when new nodes are added, the available processing capacity simultaneously grows. Early experiments have demonstrated that this strategy is not only viable, but it also allows the implementation of more sophisticated decision-making routines in the form of a software agent's AI. The sections below introduce a working prototype of a decentralised Cloud load balancer – Multi-Agent System Balancer (MASB).

It should be noted the solution presented in this chapter is partially based on work published in Sliwko and Zgrzywa (2009), Sliwko (2010), Sliwko et al. (2015) and Sliwko (2018).

## 7.1. LOAD BALANCING WITH AGENTS

Agent technologies can be dated back to 1992 (Sargent, 1992), at which point it was predicted that intelligent agent would become the next mainstream computing paradigm. Agents were described as the most important step in software engineering, representing a revolution in software (Guilfoyle and Warner, 1994). Since its inception, the field of multi-agent systems has experienced an impressive evolution, and today it is an established and vibrant field in computer studies. The software agents research field spans many disciplines, including mathematics, logic, game theory, cognitive psychology, sociology, organisational science, economics, philosophy, and so on (Weiss, 2013). Agents are considered to be a viable solution for large-scale systems, for example through spam-filtering and traffic light control (Brenner et al., 2012), or by managing an electricity gird (Brazier et al., 2002).

It is difficult to argue for any precise definition of an agent, with the research literature seeming to suggest that there are four key properties of an Agent (Castelfranchi, 1994; Gensereth and Ketchpel, 1994; Wooldridge and Jennings, 1995), namely:

- Autonomy when allowing agents to operate without direct human intervention;
- Social ability when agents communicate and interact with other agents;
- Reactivity when agents actively perceive their environment (physical or digital) and act on its changes;
- Proactiveness when agents not only dynamically respond to changes in environments but are also able to take initiative and exhibit goal-oriented behaviour as well as real-time communications.

A software agent it is generally defined as being of acting independently of its user in order to accomplish tasks on behalf of its user (Nwana, 1996). An agent can be described as a being which is supposed to act intelligently according to environmental changes and the user's input (Goodwin, 1995).

Software agents are found across many computer science disciplines, including AI, decentralised systems, self-organising systems, load balancing and expert systems (Guilfoyle and Warner, 1994; Milano and Roli, 2004; Cabri et al. 2006). Previous research has also shown that by deploying agents it is possible to achieve good global system performance (Nguyen et al., 2006), improve system stability and reduce downtime (Corsava and Getov, 2003), attain dynamic adaptation capability (Kim et al., 2004) and to realise robustness and fault-tolerance (Xu and Wims, 2000).

Agents were also found to be useful for the performance monitoring of distributed systems (Brooks et al., 1997). Several additional benefits may also be achieved, including more cost-effective resource planning (Buyya, 1999), a reduction of network traffic (Montresor et al., 2002), the autonomous activities of the agents (Goodwin, 1995), and decentralised network management (Yang et al., 2005). Multi-agent systems were also successfully used for forecasting demand and then adapting the charging schedule for electric cars (Xydas et al., 2016), and also to effectively coordinate emergency services during crisis (Othman et al., 2017). Reddy et al. (2017) presents an agent-based framework to model procurement operations in India. The most state-of-art research generally focuses on negotiation protocols and communications (Wang et al., 2014; Marey et al., 2015; Monteserin et al., 2017; Wyai et al., 2018).

Agent-based systems generally rely on decentralised architecture (Jones and Brickell, 1997; Shi et al., 2005; Wang et al., 2014; Monteserin et al., 2017), considering it to be more reliable. However, those schemas require complex communication algorithms, with negotiation protocols often being required for distributed architecture to attain a good level of performance (Bigham and Du., 2003; Yang, 2005; Wyai et al., 2018).

The idea of job scheduling with agents is not new; a single-machine multi-agent scheduling problem was introduced in 2003 (Baker and Smith, 2003; Agnetis et al., 2004). Since this time, the problem has been extended and exists in several variations, such as deteriorating jobs (Liu and Tang, 2008), the introduction of weighted importance (Nong et al., 2011), scheduling with partial information (Long et al., 2011), global objective functions (Tuong et al., 2012), and adding jobs' release times and deadlines (Yin et al., 2013). A suitable taxonomy of multi-agent scheduling problems in presented in Perez-Gonzalez et al. (2014).

The research on workload sharing via agents has a long history, with the papers below in particular having influenced the design of the MASB:

- Schaerf et al. (1995) presents a study concerning a multi-agent system in which all decision making is performed by a learning AI. The likeness of selection of a particular node for the processing of a given task depends on the past capacity of this node. The Agent’s AI uses only locally-accessible knowledge, meaning that it does not rely on information shared by other agents.
- Chavez et al. (1997) introduces Challenger, a multi-agent system, in which agents communicate with each other to share their available resources in an attempt to utilise them more fully. In Challenger, agents act as buyers and sellers in a resources marketplace, always trying to maximise their own utility. MASB follows a similar pattern, where nodes try to maximise their utilisation (via score system).
- Bigham and Du (2003) shows that cooperative negotiation between agents representing base stations in a mobile cellular network can lead to a near global optimal coverage agreement within the context of the whole cellular network. Instead of using a negotiation model of alternating offers, several possible local hypotheses are created, based on which parallel negotiations are initiated. The system commits to the best agreement found within a defined timeline. The cooperative model in which agents negotiate between themselves is the base of the distributed scheduling presented in this research.

- Kim et al. (2004) proposes a load-balancing scheme in which a mobile agent pre-reserves resources on a target machine prior to the occurrence of the actual migration. The system also prevents excessive centralisation through the implementation of a mechanism whereby when the workload processed on a particular machine exceeds a certain threshold, this machine will attempt to offload its agents to neighbouring machines.
- Cao et al. (2005) describes a solution in which agents representing a local grid resource uses past application performance data and iterative heuristic algorithms to predict the application's resource usage. In order to achieve a globally-balanced workload, agents cooperate with each other using a Point-to-Point (P2P) service advertisement and discovery mechanism. Agents are organised into a hierarchy consisting of agents, coordinators and brokers, who are at the top of the entire agent hierarchy. The authors conclude that for local grid load balancing, the iterative metaheuristic algorithm is more efficient than simple algorithms such as FCFS.
- Ilie and Bădică (2013) details a solution built on top of the ant colony algorithm, a solution which takes its inspiration from the metaphor of real ants searching for food. 'Ants' are software objects that can move between nodes managed by agents. A move between nodes which is managed by the same agent is less costly. Ants explore paths between nodes, marking them with different pheromone strength. Whenever an Ant visits a node, the agent managing it saves the recorded tour and updates its own database. Ants who subsequently visit this node read its current knowledge, meaning they have the potential to exchange information in this environment, which adds to the predictability of the whole solution.

- Eddy et al. (2015) presents a prototype in which agents operate an electricity market. Agents exchange ‘offers’ and ‘bids’ for those offers via a custom-designed communication protocol based on TCP/IP. Among other specialised agents, the system implements a short-lived coordinating agent to facilitate those exchanges, ensuring that the supply of electricity is managed. A comparable schema is implemented in MASB, in which the BA initially advises candidate target nodes where an overloading task can be re-allocated.

## 7.2. MASB DESIGN PRINCIPLES

The MASB project has been developed over several years, during which time it has undergone many changes in terms of both the technology used and the design of the architecture. This has included, for example, migration from Java to Scala, the change from thread pools to an Akka Actors/Streams framework, and the introduction and use of concurrency packages and non-locking object structures. However, the main design principles have not been altered and are presented below:

- To provide a stable and robust (i.e. no single point of failure) load balancer and scheduler for a Cloud-class system;
- To efficiently reduce the cost of scaling a Cloud-class system so that it can perform in an acceptable manner on smaller clusters (where there are tens of nodes) as well on huge installations (where there are thousands of nodes);

- To provide an easy way of tuning the behaviours of a load balancer where the distribution of tasks across system nodes can be controlled.

Many other Cluster managing systems, such as Google's Borg (Verma et al., 2015), Microsoft's Apollo (Boutin et al., 2014) and Alibaba's Fuxi (Zhang et al., 2014b), were built around the concept of the immovability and unstoppability of a task's execution. This means that once a task is started it cannot be re-allocated: it can only be stopped/killed and restarted on an alternative node. This design is particularly well suited when there is a high task churn, as observed in Apollo or Fuxi where tasks are generally short-lived, meaning that the system's scheduling decisions do not have a lasting impact. However, in order to support a mixed workload which features both short-lived batch jobs and long-running services, alternative solutions needed to be developed. One such solution is the resource recycling routines present in Borg wherein resources allocated to production tasks but not currently employed are used to run non-production applications (Verma et al., 2015).

MASB takes advantage of virtualisation technology features, namely VM-LM, to dynamically re-allocate overloading tasks. VM-LM allows programs which are running to be moved to an alternative machine without stopping their execution. As a result, a new type of scheduling strategy can be created which allows for the continuous re-balancing of the cluster's load. This feature is especially useful for long-term services which initially might not be fitted to the most suitable node, or where their required resources or constraints change.

Nevertheless, this design creates a very dynamic environment in which it is insufficient to schedule a task only once. Instead, a running task has to be continuously monitored and re-allocated if the task's current node cannot support its execution any longer.

The design of MASB relies on a number of existing tools and frameworks. The main technologies used are listed below:

- Decentralised software agents – a network of independent AI entities that can negotiate between each other and allocate Cloud workload between them. In MASB, specialised agents control nodes and manage the system workload. Due to the decentralised nature of MASB, there is no complete up-to-date system state. Instead, yet another type of agent is responsible for caching the nodes' statistics and providing an interface whereby a set of candidate nodes which a particular task can be migrated to can be requested.
- Metaheuristic selection algorithms – while the majority of the processing of load balancing logic is done via negotiation between agents, a few system processes are handled locally. One such example is that when an agent discovers its node is overloaded, it will select a subset of its tasks which it will attempt to migrate out. This selection is performed by TS algorithm.
- VM-LM which allows the transfer of a running application within the VM instance to an alternative node without stopping its execution. The vendors' strategy is to implement mixed production and low-priority jobs on a single machine. While production jobs are idler, low-priority jobs consume the nodes' resources. However, when production job resources need to be increased, the low-priority jobs are killed. The non-production jobs in Google Cluster (Verma et al., 2015) and the spot-instances in Amazon EC2 (Wang et al., 2018) use such an approach. MASB takes advantage of VM-LM to offload tasks without stopping their execution, collecting information about tasks in order to estimate the VM-LM cost of such a task.
- Functional programming language Scala and accompanying libraries (see Appendix B) – due to the decentralised design and loose coupling

between the system's components, the implementation language is of secondary importance. However, load balancing algorithms require a significant amount of tuning, especially if the Cloud is designed to have a high utilisation of available resources. This would mean that resource waste is low, and therefore the cost-per-job execution is also low. Due to the complexity of inner-system relations and dependencies, a high-fidelity simulation environment is necessary to evaluate the expected performance of a given configuration and implemented changes before is deployed to a production system, e.g. the FauxMaster simulator used by Google Engineers (Verma et al., 2015). In this implementation, Akka Actors framework was selected as the core parallelisation technology.

## 7.3. MASB ARCHITECTURE

The experiments in Chapter 6 that used a centralised load balancer based on metaheuristic algorithms demonstrated that, due to the high overheads of these algorithms, a scheduling strategy implemented on a single machine is highly unlikely to efficiently manage a large number of tasks. Therefore, MASB has been built around the concept of a decentralised load balancing architecture, an architecture which could scale well beyond the limits of a centralised scheduler.

The prototype has been built on top of an AGOCS framework (detailed in Chapter 5), meaning that the entire research and development process took advantage of the continuous testing on a real-world workload traces from the GCD project (Hellerstein et al., 2010).

MASB relies on a network of software agents to organically distribute and manage the sizeable system load. All communication between the agents is performed via a specialised stateless P2P protocol which promotes loose coupling. Figure 30 visualises the communications' flow within MASB system:

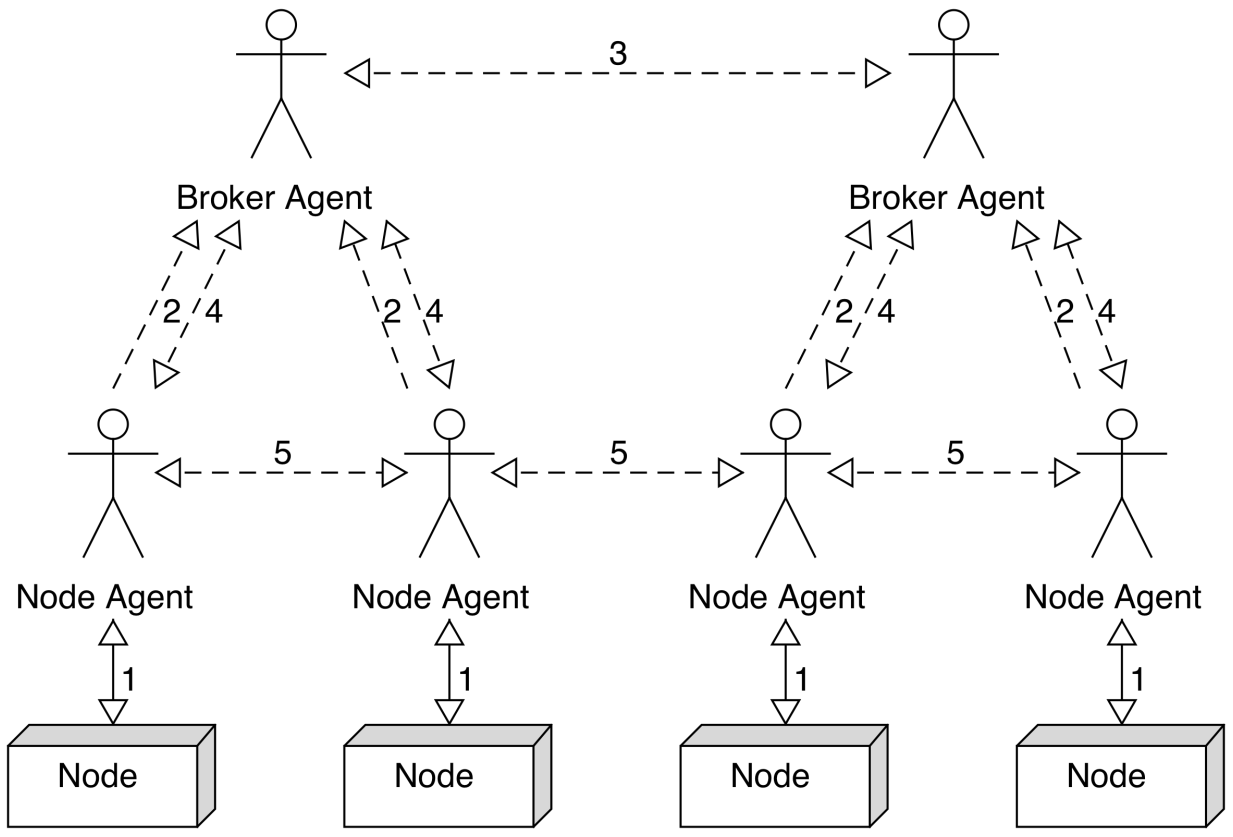

Figure 30: MASB communications' flow

Two types of agents are deployed: NA and BA. NAs are supervising system nodes, are responsible for keeping those nodes stable. NAs actively monitor the used resources on their nodes (1) and periodically forward this information to the subnetwork of BAs (2). BAs continuously exchange nodes' load information between themselves (3) and, therefore, effectively cache the state of the computing cell.

NA contains an AI module which is based on a metaheuristic algorithm TS. It manages a workload on a node. When an NA detects that its node is overloaded, it will attempt to find an alternative node for overloading tasks with the help of SAN protocol (the details can be found in section 7.4). The first step of SAN communication is to retrieve alternative nodes from BA (4). BAs provide a query-mechanism for NAs, which returns a set of candidate nodes for the migrations of tasks. However, because the information found in BAs is assumed to be outdated, once the NA completes this step, it communicates directly with their NAs so as to re-allocate this task (5).

The following two subsections describe the types of agents noted and detail their responsibilities. The annotated arrows 2 to 5 in Figure 30 correspond to inter-agent communications – messages that are exchanged within the system are detailed in subsection 7.3.3.

### 7.3.1. NODE AGENT

Every node in the system has a dedicated instance of NA. NA continuously monitors the levels of defined resources and periodically reports the state of its node and levels of utilised resources to BAs. Should any of the monitored resources be over-allocated, NA will initialise SAN process. In addition, NA performs the following functions:

- Accept/deny task migration requests – NA listens to task migration requests, and accepts or denies them. This routine is simple, with NA projecting its resource availability with that task as follows: projected allocation of resources = current allocation of resources (existing tasks which also includes tasks being migrated out from this node) + all tasks being migrated to this node + requested task (from request). If the projected resources do not overflow the node, the task is accepted and the migration process is initiated. The source node does not relinquish ownership of the task while it is being re-allocated, meaning that source node is regarded as a primary supplier of the service until the migration process successfully completes. It should be noted that during task migration, its required resources are allocated twice, to both the source node and the target node.
- Task migration – after accepting the task migration request, NA immediately starts listening for incoming VM-LM. In order to perform task migration, NA must have access to the administrative functions of VM and be able to initiate VM-LM to another node. This functionality can be either implemented by the calls of the VM manager API or by executing the command line command. This process may vary considerably per VM vendor.

### 7.3.2. BROKER AGENT

BA is responsible for storing and maintaining information about nodes' online status and their available resources. BA is a separate process which can coexist with NA on the same node since its operations are not computing-intensive. BA has two main purposes in the system. These are outlined below:

- Nodes resources utilisation database – NA periodically reports to its BA about the state of its node and available resources. BA stores all this data and can query them on demand. Every node entry is additionally stored with its timestamp, showing how long ago the data were updated. It has additional protection against the node silently going offline, for example through hardware malfunction or the network becoming unreachable, in that if this entry is not updated for five minutes, the node is assumed to be offline and entry is removed. This means that it will not be returned as the candidate node.
- Evaluating candidate nodes for a task migration – BA listens for GetCandidateNodesRequest and computes a list of candidate nodes for a task migration. In order to create a list of candidate nodes, BA retrieves nodal data from the local cache and then scores them using Allocation Scoring Function. BA scores the future state of the system as if task migration were being carried out. After scoring all the cached nodes, BA selects a configured number of candidate nodes with the highest score and sends them back to the asking node. In this research this number was set to fifteen candidate nodes, wherein higher numbers failed to yield superior results.

### 7.3.3. MESSAGE TYPES

In order to avoid costly broadcasts, since broadcast packages need to be rerouted through a whole network infrastructure consuming the available bandwidth, both NA and BA always communicate P2P. In the system there are several types of requests and responses between agents, outlined in Table 14 below:

| Request Type | Description |
| --- | --- |
| GetCandidateNodesRequest | Requests a number of candidate nodes for the migration of a specified tasks set. Send from NA to BA. |
| GetCandidateNodesResponse | Reply with a set of candidate nodes for task migration, together with their resource statistics. |
| TaskMigrationRequest | Request from source NA to candidate NA as to whether task migration is accepted. |
| TaskMigrationAcceptanceResponse | Replay from target candidate NA that task migration will be accepted.<br>Note: No resource allocation takes place after this request. |
| TaskMigrationRejectionResponse | Replay from target node’s NA that task migration will not be accepted. |
| TaskMigrationProcessRequest | Request to selected target node’s NA to start task migration.<br>Note: this request has an optional forced flag, requesting the target NA to skip the currently available resources check. The total node’s resources check and constraints check will be still performed. |
| TaskMigrationProcessConfirmationResponse | Confirmation from the target node’s NA that the task migration process can start.<br>Note: Resources are allocated for the migrated task and the live migration process starts. |
| TaskMigrationProcessErrorResponse | Denial of task migration process. This reply is generated if the NA can no longer accommodate the migrated task. |

Table 14: Message types

Agent-to-agent communications follow the ‘request-response’ pattern, in which each request object has one or more matching response objects. The message objects carry additional metadata such as fitness value (as explained in (12) in subsection 7.4.1), forced migration flag, and detailed node and task information.

Section 7.4 below explains the process in which messages are exchanged, while the subsections 7.4.1 to 7.4.4 show detailed samples of such objects.

## 7.4. SERVICE ALLOCATION NEGOTIATION PROTOCOL

When NA detects its node is overloaded, it will select a task (or a set of tasks) and attempt to migrate them to an alternative node or nodes. Since SAN is asynchronous, this means a single NA can run several SAN processes in parallel. In the current implementation, NA selects a number of tasks in the first step – Select Candidate Services (SCS) – and processes their allocation in parallel. Figure 31 visualises this process – for simplicity, the chart presents the allocation negotiation of one task only:

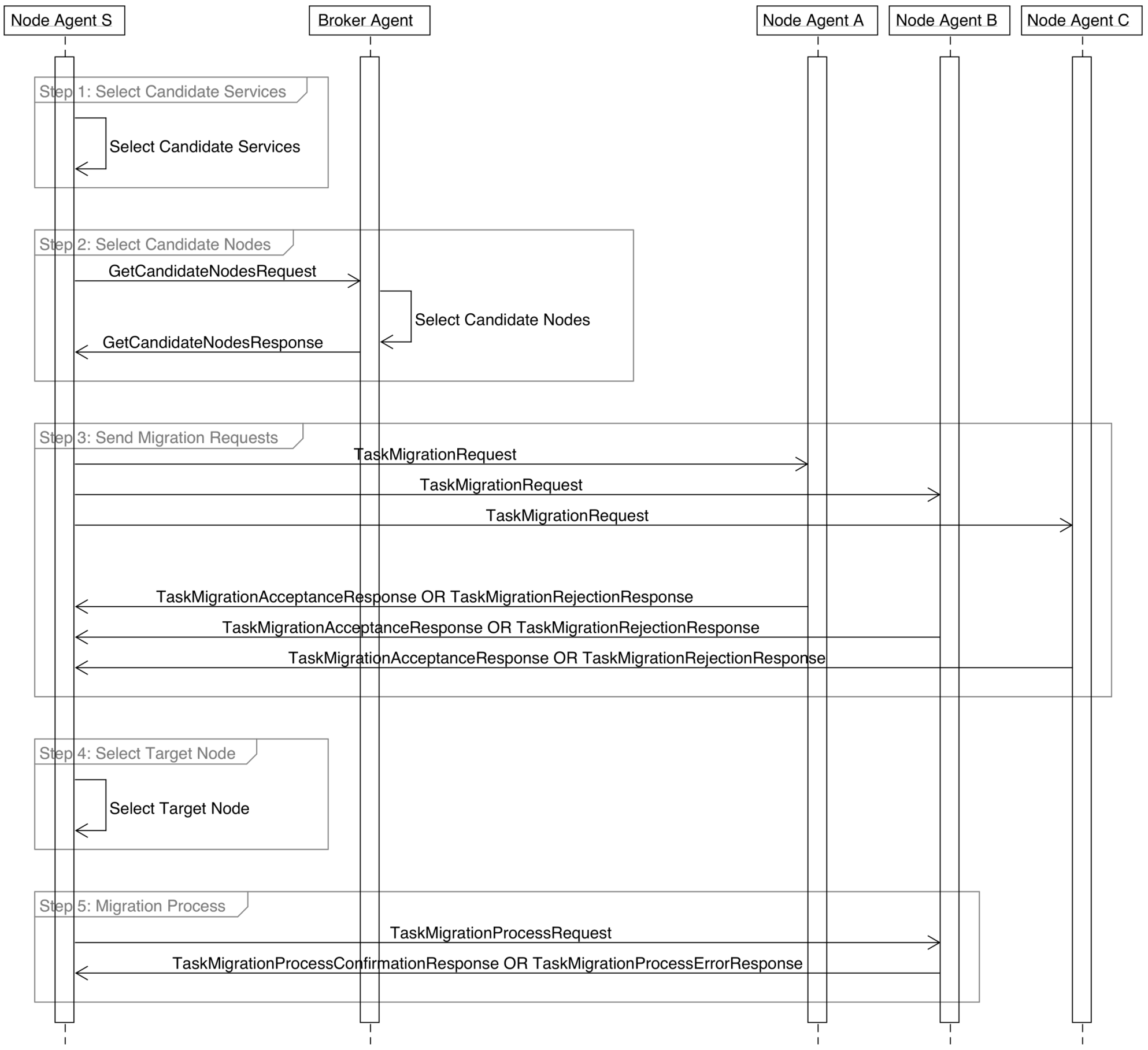

Figure 31: Service Allocation Negotiation

SAN is a five-stage process, involving a single source node (Node Agent S), one of the system BAs and several of other nodes in the system (Node Agent A, Node Agent B and Node Agent C).

When migrating-out a given task, NA at first sends a GetCandidateNodesRequest to BA to get with a set of candidate nodes where the task can potentially be migrated to. BA scores all its cached nodes and sends back the top fifteen to NA. Additionally, in order to help to avoid collisions, BA does not directly select only top candidate nodes, but instead selects them randomly from a node pool, where candidate node score is a weight, wherein higher scored nodes are selected more frequently. This design helps to avoid a situation where an identical subset of candidate nodes is repeatedly selected for a number of tasks with the same resource requirements.

Upon receiving this list, NA sends task migration requests to all of those candidate nodes (Step 3), and waits for a given time (in this case for thirty seconds) for all replies. After this time, NA evaluates all accepted task migration responses (Step 4) and orders them in relevance order (nodes with the highest score first) and then attempts to migrate a task to a target node with top score (Step 5). If target node returns an error, the source NA will pick the next target node and attempt to migrate a task there.

At each of these stages, the target node's NA might reject task migration or return an error, for example when task migration is no longer possible because the current node's resource utilisation levels have increased or because the node attributes no longer match the task's constraints. Depending on a system utilisation level, such collisions might be more or less frequent. However, they are resolved at node-to-node communication level and do not impact the system performance as a whole.

In a situation where there are insufficient candidate nodes available due to the lack of free resource levels, the BA will return candidate nodes with the 'forced migration' flag set to true.

The algorithm's five steps are explained in the following subsections, while the forced migrations feature is detailed in supplementary subsection 7.4.6 below.

### 7.4.1. STEP 1: SELECT CANDIDATE SERVICES

SCS routine is executed when the NA detects that the currently existing tasks are overloading its node. This step is processed on the node wholly locally. The purpose of this routine is to select the task (or set of tasks) that NA will attempt to migrate out and become stable (i.e. non-overloaded) during that process. All tasks currently running on this node are evaluated, taking into consideration various aspects, namely:

- The cost of running a task on this particular node. NA will aim to have the highest node score for its own node. If removing this particular task will cause its AS (calculated by SAS functions – see subsection 7.5 for details) to be higher, then this task is more likely to be selected.
- The cost of migration of a task – VM migrations cause disruptions on the Cloud system. In this research, cost is estimated by LMDT formula (Chapter 4) as the additional network traffic required to migrate the running VM instance to an alternative node. Additional notes are provided in subsection 5.8.
- The likeness to find an alternative node – the majority of tasks do not have major constraints and can be executed on a wide range of nodes. However, there are a small number of tasks with very restrictive constraints that significantly limit the number of nodes that the task can be executed on. If such a task can only be executed locally, i.e. the node has enough total resources capacity and task constraints are matched, then NA is unlikely to migrate out those tasks.

- Any task which cannot be executed on a local node is compulsory selected as a candidate task. This scenario could occur if the task constraints or node attributes were updated.

NA first computes a list of compulsory candidate tasks, i.e. tasks that can no longer be executed on this node. Following this, if the remaining tasks are still overloading the node, it will select a subset of tasks to be migrated out.

The candidate tasks selection algorithm tries to minimise the total migration cost of selected tasks, and also to achieve the highest AS for a node, under the assumption that the selected subset of candidate tasks is successfully migrated to the alternative node. In order to achieve this, the algorithm defines the Fitness Function as coded inside SCS:

$$Fitness\ Function = \frac{Node\ allocation\ score}{Total\ migration\ cost} \quad (12)$$

For the above in a NP-Hard problem with a substantial search space, e.g. twenty tasks on a node, the search space size is over one million combinations. Given this, the Full Scan approach (as detailed in subsection 6.1.6) will be substantially computation-intensive. Therefore, the use of metaheuristic algorithms is justified. In previously researched scheduling concept, a variant of TS has been successfully applied to solve a similar class of problems (subsection 6.3.2). The TS algorithm has the following properties:

- It has a small memory imprint since only the list of visited solutions is maintained thorough execution;
- It can be easily parallelised as a variant which is restarted multiple times;

- It is very controllable through setting up a limited number of steps and number of runs;
- It is stoppable, and the best-found result can be retrieved immediately;
- It generally returns good results.

It was found that multiple restarts (herein a twenty five re-run limit) with a shallow limit of steps (herein five) yield very good results, with only about 2-7% of solutions in the whole search space (i.e. selecting a subset of tasks being run on a node) being examined in each invocation. Additionally, instead of restarting the algorithm an arbitrary number of times, a stop condition for this algorithm has been implemented when the best-found solution has not been improved in a certain number of the last steps (herein six).

A sample log entry is presented below, wherein the subset of candidate tasks is being computed:

```
12:44:22.016 NodeAgentActor (node=2274790707) INFO
SAMPLE:
 Selected overloading tasks for node [2274790707]
 Node total resources = [0.5000000000,0.2493000000]
 Node used resources (all tasks) = [0.5598619000,0.2060380000]
 Node used prod resources (all tasks) = [0.4812960000,0.2190280000]
 All tasks (* Selected):
  Task [2902878580-1081] (PROD) Priority=11  Required resources=[0.0062480000,0.0014570000]
   Used resources=[0.0149800000,0.0269200000] Migration cost = 6876.02 [MB]
  Task [2902878580-3147] (PROD) Priority=11  Required resources=[0.0062480000,0.0014570000]
   Used resources=[0.0105300000,0.0250900000] Migration cost = 3820.05 [MB]
  Task [3998352223-38] (PROD) Priority=9  Required resources=[0.3125000000,0.1592000000]
   Used resources=[0.1680000000,0.0761700000] Migration cost = 69139054863.11 [MB]
  Task [5726057648-7] (PROD) Priority=9  Required resources=[0.0625000000,0.0077670000]
   Used resources=[0.0168200000,0.0058140000] Migration cost = 106.72 [MB]
* Task [6218406404-243] (PROD) Priority=0  Required resources=[0.0406500000,0.0206900000]
   Used resources=[0.0056840000,0.0057980000] Migration cost = 106.69 [MB]
  Task [6218406404-959] (PROD) Priority=0  Required resources=[0.0406500000,0.0206900000]
   Used resources=[0.0082550000,0.0057910000] Migration cost = 106.67 [MB]
* Task [6251414911-1447] Priority=1  Required resources=[0.0625000000,0.0318000000]
   Used resources=[0.0007629000,0.0076750000] Migration cost = 112.37 [MB]
  Task [6251664479-137] (PROD) Priority=2  Required resources=[0.0125000000,0.0077670000]
   Used resources=[0.0422400000,0.0055920000] Migration cost = 106.25 [MB]
  Task [6251784940-1615] Priority=2  Required resources=[0.0249900000,0.0254500000]
   Used resources=[0.0291700000,0.0135000000] Migration cost = 183.40 [MB]
  Task [6251787910-686] Priority=2  Required resources=[0.0249900000,0.0333900000]
   Used resources=[0.0321000000,0.0150100000] Migration cost = 236.79 [MB]
* Task [6251803864-88] Priority=2  Required resources=[0.0249900000,0.0254500000]
   Used resources=[0.1665000000,0.0102700000] Migration cost = 128.94 [MB]
* Task [6251812952-159] Priority=2  Required resources=[0.0249900000,0.0795900000]
   Used resources=[0.0648200000,0.0084080000] Migration cost = 115.72 [MB]
  Task [6251812952-2072] (unstarted) Priority=2  Required resources=[0.0249900000,0.0795900000]
   Used resources=[0.0000000000,0.0000000000] Migration cost = 101.00 [MB]
 Node used resources (remaining tasks) = [0.3220950000,0.1738870000]
 Node used prod resources (remaining tasks) = [0.4406460000,0.1983380000]
 Total migration cost (selected tasks) = 463.71561966381125 [MB]
```

Here, the thirteen tasks are being executed on node ‘2274790707’. However, the used resources exceed the node’s total resources, i.e. all tasks are utilising 0.5598619 CPU, while the node can provide only 0.5 CPU (values are normalised). The node's NA detects the node is overloaded and triggers the SCS routine. The SCS routine selects four tasks (here: the production task '6218406404-243' and non-production tasks: '6251414911-1447', '6251803864-88' and '6251812952-159'; marked with *) which are then added to candidate tasks, and NA will attempt to migrate out this set in the next step. The potential reduction of used resources is an effect of removing a subset of tasks from this node: (i) CPU reserved for production tasks is potentially reduced from 0.481296 to 0.440646 which is ca. 88% utilisation of total 0.5 CPU available on this node, and (ii) memory reserved for production tasks is potentially reduced from 0.219028 to 0.198338 which is ca. 80% utilisation of the total 0.2493 memory available on this node. The total migration cost for this set of migrations is ca. 463.72MB.

### 7.4.2. STEP 2: SELECT CANDIDATE NODES

After selecting candidate nodes, NA sends a GetCandidateNodes request to BA. A part of this request, task information data, such as currently used resources and constraints, are sent. BA also itself caches a list of all nodes in system with their available resources and attributes. Based on this information, BA prepares a list of alternative candidate nodes for a task in request. The main objective of this process is to find alternative nodes which have the potentially highest node AS, under the assumption that the task will be migrated to a scored node. The size of this list is limited to an arbitrary value to avoid network congestion when NA will send actual migration requests query in the next step. In this implementation, it is set to fifteen candidate nodes returned in each response.

This step is the most computing intensive of all, and represents a potential bottleneck for negotiating logic processing. BA needs to examine all system nodes, check their availability for a given task and score them accordingly. The request processing is self-contained and highly concurrent, meaning that the node scoring can be run in parallel and the final selection of top candidate nodes is run in sequence. Originally, this code was extensively profiled and improved, and designed BA to be able to run in a multi-instance mode if needed and to handle heavy usage. However, in experiments, the quoting mechanism proved to be very lightweight and the demand not that high, meaning that a single BA was sufficient to handle 12.5k nodes in the system. Below, a sample log entry is presented when such a list is computed and returned to a NA:

```
17:53:28.516 NodeAgentActor (node=97967489) INFO
SAMPLE:
 Candidate nodes recommendations for migration-out of task:
  Task [6251414911-740] Priority=1  Required resources=[0.0625000000,0.0318000000]
  Used resources=[0.0476100000,0.0097350000] Migration cost = 124.29 [MB]
 Source node: Node [97967489] [0.5000000000,0.4995000000]:
  CandidateNodeRecommendation[nodeId=2110696959,nodeAvailableResources=[0.1167000800,0.0578920000],
   fitnessValue=5.027797352070,forceMigration=false]
  CandidateNodeRecommendation[nodeId=2274669582,nodeAvailableResources=[0.0846342000,0.1311130000],
   fitnessValue=4.351440488446,forceMigration=false]
  CandidateNodeRecommendation[nodeId=294847211,nodeAvailableResources=[0.2073230300,0.0232970000],
   fitnessValue=3.990484728735,forceMigration=false]
  CandidateNodeRecommendation[nodeId=1302354,nodeAvailableResources=[0.2147855300,0.0715080000],
   fitnessValue=3.368267142480,forceMigration=false]
  CandidateNodeRecommendation[nodeId=7246234,nodeAvailableResources=[0.3283863000,0.0205610000],
   fitnessValue=2.444197290198,forceMigration=false]
  CandidateNodeRecommendation[nodeId=2887932822,nodeAvailableResources=[0.3051645700,0.1098830000],
   fitnessValue=2.147161970183,forceMigration=false]
  CandidateNodeRecommendation[nodeId=38743543,nodeAvailableResources=[0.3583948000,0.1018340000],
   fitnessValue=1.769829840087,forceMigration=false]
  CandidateNodeRecommendation[nodeId=6568110,nodeAvailableResources=[0.2394051100,0.2629800000],
   fitnessValue=1.711800790297,forceMigration=false]
  CandidateNodeRecommendation[nodeId=38709566,nodeAvailableResources=[0.3584505000,0.1189080000],
   fitnessValue=1.697701710745,forceMigration=false]
  CandidateNodeRecommendation[nodeId=3739348304,nodeAvailableResources=[0.2367339800,0.2681720000],
   fitnessValue=1.696017579836,forceMigration=false]
  CandidateNodeRecommendation[nodeId=1093461,nodeAvailableResources=[0.3801150000,0.0841200000],
   fitnessValue=1.681960083254,forceMigration=false]
  CandidateNodeRecommendation[nodeId=4217347623,nodeAvailableResources=[0.3635202600,0.1467840000],
   fitnessValue=1.553194840995,forceMigration=false]
  CandidateNodeRecommendation[nodeId=16918689,nodeAvailableResources=[0.3916948000,0.1250880000],
   fitnessValue=1.456396783346,forceMigration=false]
  CandidateNodeRecommendation[nodeId=257495090,nodeAvailableResources=[0.0367722000,0.0736920000],
   fitnessValue=0.000000000001,forceMigration=true]
  CandidateNodeRecommendation[nodeId=38679534,nodeAvailableResources=[0.3811526500,0.0066530000],
   fitnessValue=0.000000000001,forceMigration=true]
```

Here, NA on node ‘97967489’ requested candidate nodes for the migration of the task ‘6251414911-740’. BA returned top candidate nodes for a given task ordered by their suitability score, i.e. fitness value. Here values returned are: 5.02779735207 for node ‘2110696959’, 4.351440488446 for node ‘2274669582’, 3.990484728735 for node ‘294847211’, 3.36826714248 for node ‘1302354’, and so on. Additionally, the last recommendations for nodes ‘257495090’ and ‘38679534’ are forced-migrations (forceMigration is set to true).

Within the node recommendation there is additional information, such as node available resources and other metadata (not shown in listing). It is not necessary to return this extra information, but it was found to be very useful for logging and sampling purposes, and then efficient tuning of the system (for details see subsection 7.6.3).

### 7.4.3. STEP 3: SEND MIGRATION REQUESTS

Forced migration candidates will be always added to the list of accepted candidate nodes in the next step but with minimal scores. Each NA analyses its own node availability for a given task, i.e. both the available resources and the node's attributes, and responds with TaskMigrationAcceptanceResponse or TaskMigrationRejectionResponse.

Acceptance response only implies the readiness to accept a task with NA not yet allocating any resources. Additionally, TaskMigrationAcceptanceResponse message contains this node's current resources usage levels, which are used in the next step to rescore this node, since the data from BA are less recent.

### 7.4.4. STEP 4: SELECT TARGET NODE

NA waits for a defined time, or until all candidate nodes have responded by either the acceptance or rejection of a migrated task, and computes a list of nodes that accepted this task. NA evaluates each of the accepting nodes using the SRAS function, with the assumption that the task will be re-allocated to a scored node. From this pool, a target node is then selected. The selection is weighted with node scores but still randomised, which helps to avoid conflicts when many task migrations compete for the same node.

As noted above, all forced migration candidate nodes will be added to this list but will be selected only in last place, once all other alternative migrations attempts fail. This strategy ensures that NA always has an alternative node to offload the task. A scenario in which only one node is capable of running a given task is considered to be an error, and is reported to the system administrator. For fault-tolerance reasons, the system should always have multiple nodes able to run any given task.

A sample log entry is presented below:

```
17:48:51.541 NodeAgentActor (node=30790115) INFO
SAMPLE:
 Accepted recommendations for migration-out of task:
  Task [4844000327-3] (PROD) Priority=10  Required resources=[0.0625000000,0.0031090000] Used
resources=[0.0037420000,0.0018860000] Migration cost = 101.86 [MB]
 Source node: Node [30790115] [0.5000000000,0.2493000000]
 All non-expired recommendations (* selected):
  CandidateNodeRecommendation[nodeId=72,nodeAvailableResources=[0.2409250200,0.0802350000],
   fitnessValue=2.737788312063,forceMigration=false]
  CandidateNodeRecommendation[nodeId=4995304750,nodeAvailableResources=[0.2017312000,0.1336360000],
   fitnessValue=2.704122369764,forceMigration=false]
  CandidateNodeRecommendation[nodeId=6608641,nodeAvailableResources=[0.1552938500,0.1871200000],
   fitnessValue=2.657728011619,forceMigration=false]
  CandidateNodeRecommendation[nodeId=336053478,nodeAvailableResources=[0.2798536000,0.0479556000],
   fitnessValue=2.558664832112,forceMigration=false]
  CandidateNodeRecommendation[nodeId=351638129,nodeAvailableResources=[0.2121407240,0.1468220000],
   fitnessValue=2.505822852307,forceMigration=false]
  CandidateNodeRecommendation[nodeId=431038304,nodeAvailableResources=[0.3267638000,0.0397480000],
   fitnessValue=2.142784872639,forceMigration=false]
* CandidateNodeRecommendation[nodeId=3650320528,nodeAvailableResources=[0.3118476200,0.0739690000],
   fitnessValue=2.101080438228,forceMigration=false]
  CandidateNodeRecommendation[nodeId=351664198,nodeAvailableResources=[0.3099791000,0.1114180000],
   fitnessValue=1.926755718413,forceMigration=false]
  CandidateNodeRecommendation[nodeId=6565510,nodeAvailableResources=[0.3613202000,0.1106346000],
   fitnessValue=1.564594925411,forceMigration=false]
  CandidateNodeRecommendation[nodeId=1273895,nodeAvailableResources=[0.3402396000,0.1485660000],
   fitnessValue=1.556209187067,forceMigration=false]
  CandidateNodeRecommendation[nodeId=662212,nodeAvailableResources=[0.4032697100,0.0658030000],
   fitnessValue=1.431851646113,forceMigration=false]
  CandidateNodeRecommendation[nodeId=1272936,nodeAvailableResources=[0.3443891900,0.2676010000],
   fitnessValue=1.119583713082,forceMigration=false]
  CandidateNodeRecommendation[nodeId=2594787,nodeAvailableResources=[0.3313337000,0.3637160000],
   fitnessValue=0.874210901828,forceMigration=false]
  CandidateNodeRecommendation[nodeId=2098371268,nodeAvailableResources=[0.3326411500,0.0528420000],
   fitnessValue=0.000000000001,forceMigration=true]
  CandidateNodeRecommendation[nodeId=1332336,nodeAvailableResources=[0.2588359500,0.3254990000],
   fitnessValue=0.000000000001,forceMigration=true]
```

Here, NA on a node ‘30790115’ is selecting a target node for the migration of task ‘4844000327-3’ (with the migration cost of 101.86MB). All accepted recommendations from previous step (within thirty seconds) or forced recommendations (forceMigration is set to true) are re-scored and a single node is selected (here: node ‘3650320528’; marked with *). Then, NA sends TaskMigrationProcessRequest to initiate a task migration process itself. NA stores received candidate node recommendations in its memory in case the task migration fails, and the next target node has to be selected.

Once the task is removed from a node, meaning it is re-allocated, and has finished its execution, is killed or crashes, all its candidate node recommendations are automatically invalidated and deleted. Additionally, candidate node recommendations expire after an arbitrary defined time, in this case three minutes. This mechanism exists in order to remove recommendations with out-dated node data. If no candidate node recommendations are left (or expire), and the node is still overloaded, the SAN process restarts from Step 1.

### 7.4.5. STEP 5: MIGRATION PROCESS

When NA receives TaskMigrationProcessRequest, it performs a final suitability check, wherein both node's available resources and task constraints are validated. If the forced-migration flag is set, NA ignores the existing tasks and validates the required resources against total node resources. Occasionally, the target NA can reject task migration process or migration fails. In such a scenario the algorithm returns to Step 4 and selects the next candidate node (via weighted randomised selection).

In practice, this happens only for 6-8% of all task migration attempts (in simulated GCD workload), the majority being the result of task migration collisions where two or more tasks are being migrated to the same node. The first-to-arrive TaskMigrationProcessRequest is generally successful, meaning that Steps 4 and 5 are repeated only for the rejected migrations. There have been no observations of an increase in collisions when the larger Cloud system is simulated (up to 100k nodes, as detailed in section 7.6.10). This is because a single NA communicates with only a limited set of other agents, and the P2P communication model is used exclusively. This means that the communication overhead does not go up when the system size is increased.

### 7.4.6. FORCED MIGRATION

In rare circumstances, approximately 10-15 out of 10k tasks present constraints which restrict the execution of a task to a very limited number of nodes.

Considering this, there is a scenario in which NA wants to migrate out a given task but is unable to find an alternative node because all suitable nodes have already been allocated to other tasks, and the majority of their resources have been utilised. In such a scenario, BA returns candidate node recommendations with a forced-migration flag set. In response, the BA can also mix non-forced migrations and forced migrations. In a worst-case scenario, all returned recommendations would be forced, but this approach ensures there is always an acceptable node to run a given task on. This prevents a starvation of the task resources, where the task is never executed.

A forced migration flag signals that a node is capable of executing a task but that its current resources utilisation levels do not allow it to allocate additional tasks, since this will cause the node to be overloaded. Forced migration forces the node to accept the task migration request while skipping the available resources check. However, task constraints are still validated, including the check if the node's total resources are sufficient to run the task. This design helps to avoid a situation where a task has very limiting constraints and only a few nodes in the system can execute it. If those nodes have no available resources then it will not be possible to allocate a task to them, and therefore tasks will not run. As such, the nodes are forced to accept this task, which then many trigger the target node's NA to migrate out some of its existing tasks to alternative nodes.

## 7.5. SERVICE ALLOCATION SCORE FUNCTIONS

SAS functions are a crucial part of the system, which greatly impacts global resource usage level. That is, they determine how well nodes' resources are utilised. They are used when a new task is allocated or when a system needs to re-allocate an existing task to an alternative node.

SAS functions evaluate how well a given task will fit a scored node system-wise by returning AS value. In this implementation, SAS input is constructed from the total node resources, the currently available node resources and the currently required resources for a given task. SAS function returns a value when a task fits the available resources on a node, and also when a node is overloaded by a task. If a node cannot fulfil a task's constraints, the node is deemed non-suitable and the scoring function is undefined.

This research concludes that node AS are failing in six separate areas:

- Idle Node – a completely idle node is a special case of allocation, in which no task has been allocated to this node. Such a node could be completely shut down, resulting in lower power usage for a cluster. In this research, idle nodes are scored most highly when determining a suitable node for initial task allocation.
- Super Tight Allocation (STA) – where some of the node's resources are utilised in the 90%-100% range. STA is regarded as stable allocation; however, due to the dynamic resource usage, this is actually not a desirable scenario. Complete, or almost complete, resource usage can frequently lead to resource over-allocation, whereby one or more tasks increase their resource utilisation. This experimentation has determined that leaving 10% of any given resource unutilised gives the best results since it reduces task migration but still ensures the efficient use of the system resources (see discussion in subsection 7.5.4).
- Tight Allocation (TA) – where all node resources are utilised in the 70-90% range. This is the most desirable outcome as it promotes the best fitting allocation of tasks and, therefore, low resource wastage.
- Proportional Allocation (PA) – while tight-fit is the most desirable outcome, the majority of tasks in this research consumed a small amount of each resource. Most scheduled tasks are short batch jobs which have a very short execution time. In such a scenario, it is desirable to keep proportional resources' usage ratios on all nodes which would, therefore, generally enable nodes to fit more tasks with ease.

- Disproportional Allocation (DA) – where the node's resources are not proportionally utilised, thereby making it difficult to allocate additional tasks if required. For example, a setup where tasks on a node allocate 75% of CPU but only 20% of memory is not desirable.
- Overloaded Node – when allocated resources overload the total available resources on the node. Naturally, this is an unwanted situation, and such a node is given a score of zero.

Figure 32 visualises AS types for the two resources (CPU and memory):

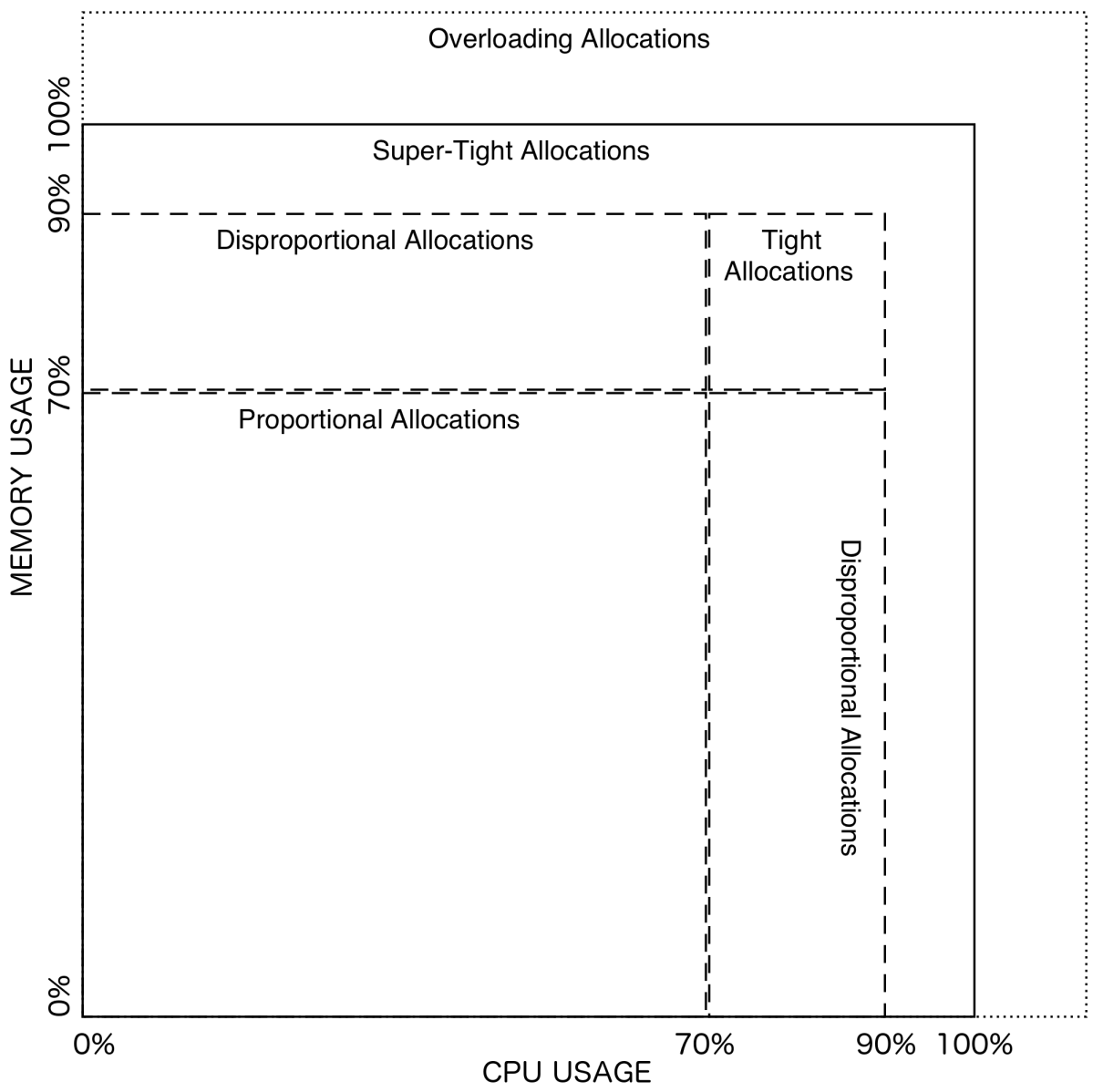

Figure 32: Allocation Score types (two resources)

SAS function should never allow overloading allocations to take place in order to prevent a scored node to become overloaded and unstable. Additionally, during the research it was determined that STAs are very prone to over-allocate nodes and are damaging to overall system stability. Therefore, they are also accorded a score of zero. DAs increase global resource wastage and should be avoided;

nevertheless, they are acceptable if none of the more desired types of AS are possible. The desirability order varies and depends on the task's state, as discussed in subsections 7.5.2 and 7.5.3 below, while the following subsection introduces the concept of Service Allocation Lifecycle (SAL).

### 7.5.1. SERVICE ALLOCATION LIFECYCLE

Tightly fitting tasks on as few nodes as possible are beneficial for global system throughput. However, during this research the following facts were observed:

- Initially, a Cloud user specifies the task's required resources. Users tend to overestimate the amount of resources required, wasting in some cases close to 98% of the requested resource (Moreno et al., 2013). Therefore, only after the task is executed could realistic resource utilisation values be expected. Allocating new tasks in a tight-fit way (i.e. TA and STA areas in Figure 32) does result in turmoil when the task is actually executed and the exact resource usages levels are logged. Therefore, the initial allocation should rather aim to distribute tasks across nodes and keep the resource utilisation levels on individual nodes low (i.e. PA area in Figure 32), than pile them on the lowest possible number of nodes.
- In GCD, only about 20-40% of tasks qualify as long-running tasks, meaning that they run for longer than twenty minutes (Schwarzkopf et al., 2013). The remaining scheduled tasks consisted of short-term jobs which generally have much lower resource requirements than long-running tasks. The majority of tasks are short and will not exist for long at all in the system. Therefore, it is important for an initial allocation not to spend too much time in trying to tightly fit them into available nodes.
- While the majority of tasks are short-lived (up to twenty minutes), there exists a number of long-running tasks that have more demanding resource requirements, meaning that the majority of resources (55–80%) are allocated to long-lived services (ibid.). Therefore, it is more difficult to fit them into nodes, and these allocations should be much tighter to minimise global system resource waste. More nodes need to be scored which therefore consumes more CPU time when allocating a single task.

Given the above reasons, the ideal scenario for a task is to be initially allocated on a lowly-utilised node, before it is gradually migrated towards more tightly-fitted allocations with other tasks. Figure 33 represents resulting SAL:

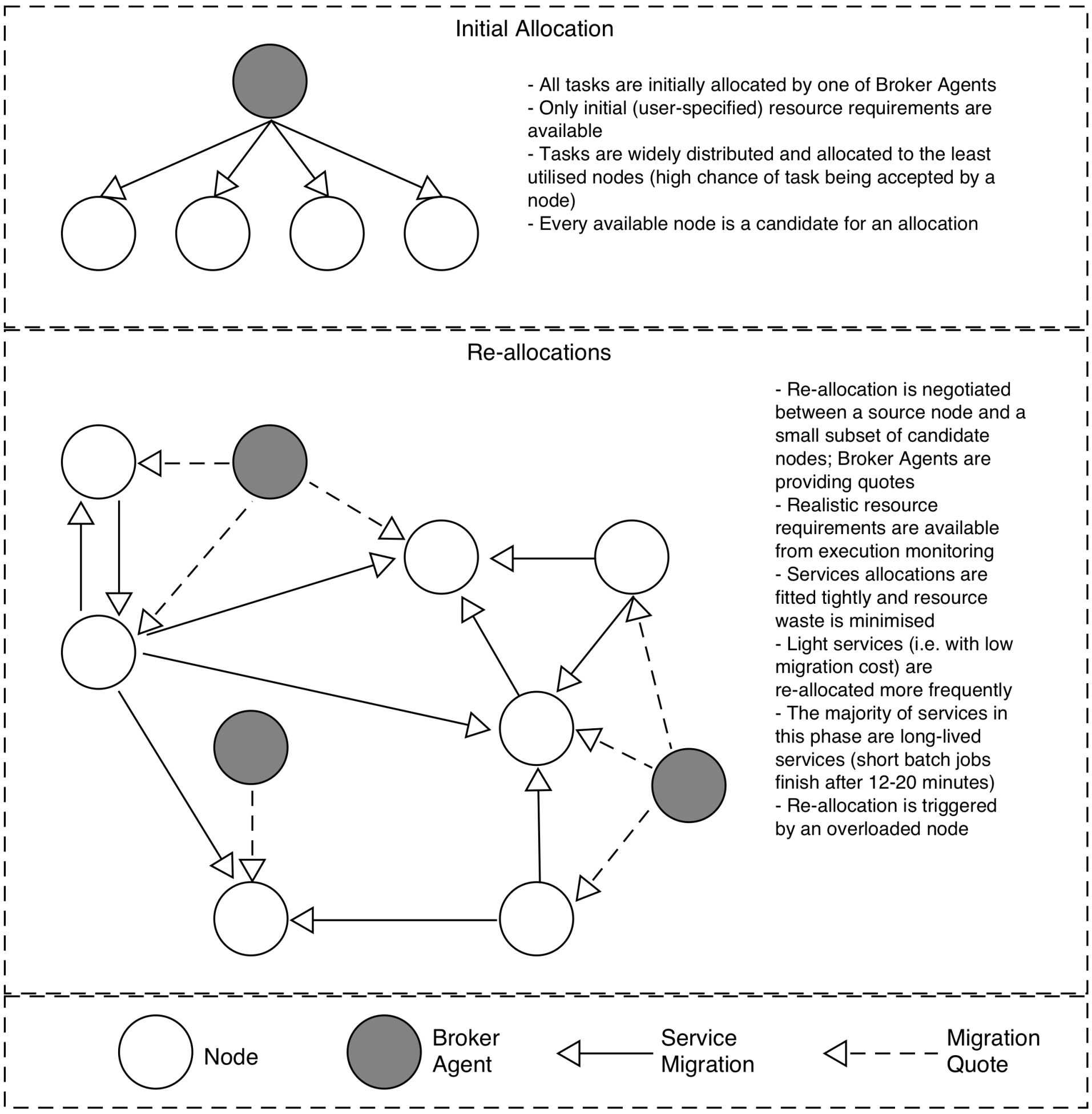

Figure 33: Service Allocation Lifecycle

Originally, the MASB framework did not have distinct scoring functions for SIAS and SRAS; a single SAS function, with the same scoring model as SRAS, was used for all allocations which resulted in lowered performance. The design was ultimately altered, and SAS function was split.

During Initial Allocation, a randomly selected BA is responsible for allocating a newly arrived task to a worker node. BA uses SIAS function (detailed in subsection 7.5.2) to score nodes. Only a limited number of candidate node recommendations are calculated (here: 200) before selecting the top recommendations. This is to prevent scoring routine calculations from processing for too long. The limit of 200 applies only to non-forced recommendations for matching nodes.

MASB uses a network of BAs to provide a set of the best candidate nodes (nodes with the highest AS) to allocate the task. However, some applications such as Big Data frameworks often send multiples of an identical task in a batch. Those tasks execute the same program and have the same (or very similar) resource requirements. As such, a limited set of nodes will be highly scored and may result in a multiple repeated allocations requests to the same node over a very short period of time. To prevent this phenomenon, the pool of candidate nodes is randomly shuffled each time BA receives a request.

A once allocated (and running) task can be re-allocated to an alternative node if necessary. In such a scenario NA of a node which the task is being executed is responsible for finding a candidate node. Both NA and BA use SRAS function (detailed in subsection 7.5.3) to score candidate nodes. Similar to calculating recommendations for new tasks, as an additional optimisation, only a limited number of candidate node recommendations are calculated before selecting the top recommendations. However, because this routine is invoked much less frequently, two thousand nodes are analysed. The two thousand limit applies only to non-forced recommendations for matching nodes.

### 7.5.2. SERVICE INITIAL ALLOCATION SCORE

As explained in the subsection above, in order to minimise the impact of Cluster user's overestimating resource requirements, the initial allocation should attempt to spread tasks widely across all system nodes. Therefore, when initially allocating existing tasks, candidate nodes should be scored in the following order: PA, TA and finally DA.

In this implementation, the SIAS function for two resource types (CPU and memory) was used. Figure 34 is a graphical representation of SIAS function:

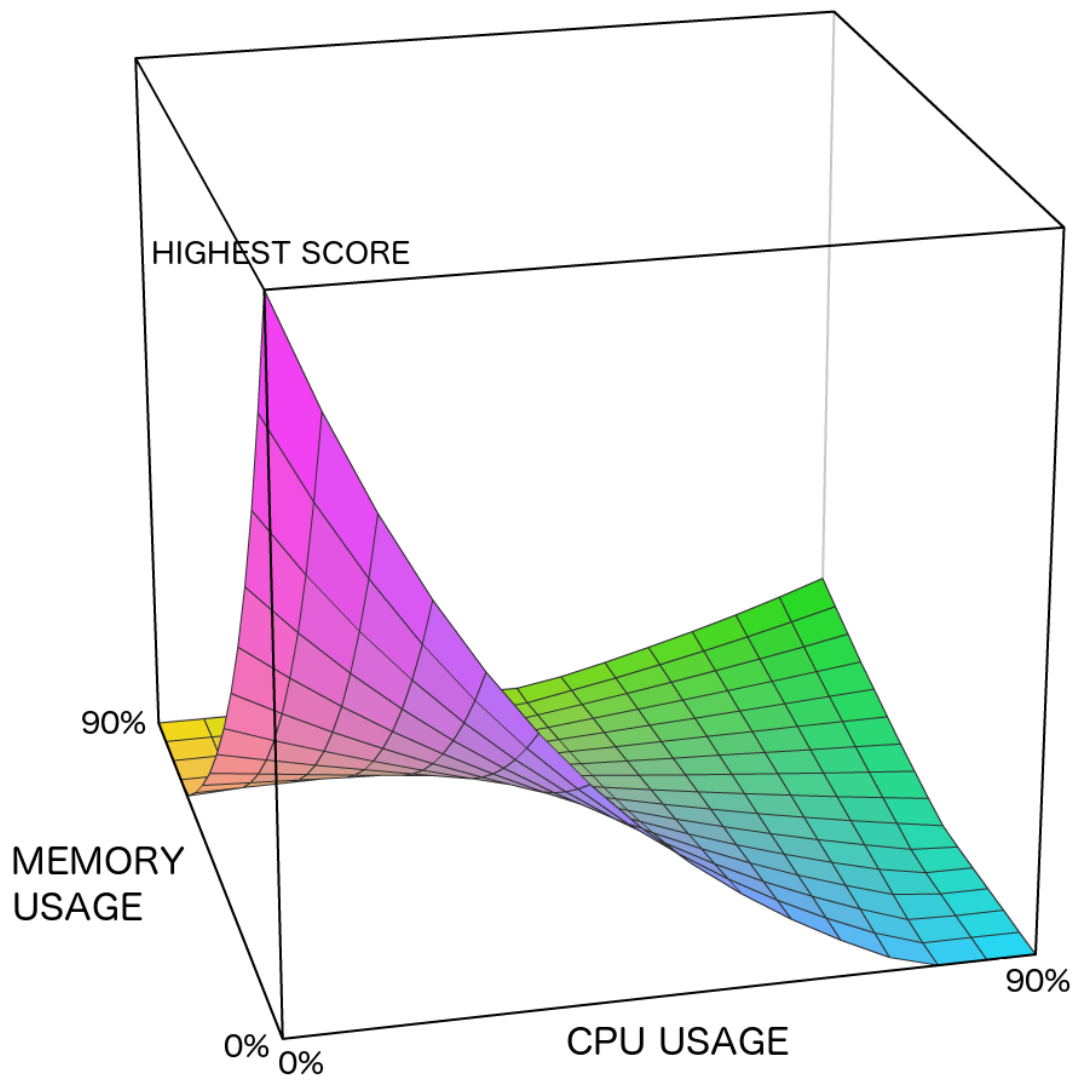

Figure 34: Service Initial Allocation Score (two resources)

Three separate areas can be noticed:

- Lower-left (the highest score) – this promotes PA, which will leave resource utilisation at a low level or proportionately used.
- Upper-right corner (the medium score) – this promotes TA, where tasks on this node will closely utilise all its resources.
- The upper-left and lower-right corners (the lowest score) – these DAs will leave one resource utilised almost fully and the other resource wasted.

It should be noted that the maximum resource usage is 90%, and that values above this level are in an undesired STA's area (and have zero AS). The following SIAS function was used:

$$SCORE = F_STEEP^{(r_{CPU} - F_BIAS \cdot r_{CPU_MAX}) \cdot (r_{MEM} - F_BIAS \cdot r_{MEM_MAX})} - F_FLOOR \quad (13)$$

- $r_{CPU}, r_{MEM}$ – current resources utilisation levels on a node (values are normalised to between 0 and 1);
- $r_{CPU_MAX}, r_{MEM_MAX}$ – total resources available on a node (values are normalised to between 0 and 1);
- $F_BIAS$ – score factor which sets the bias towards low (i.e. SIAS function) or high (i.e. SRAS function) utilisation of resources on a node. Here, a value of 0.3 was used;
- $F_STEEP$ – parameter describing how aggressively the system should increase scores of the more desired AS-es (which impacts the probability of a node selection). Here, a value of 350 was used;
- $F_FLOOR$ – parameter describing how aggressively the system should reduce scores of less desired AS-es (which impacts the probability of skipping a node). Here, a value of 0.8 was used;
- Additionally, negative score values are adjusted to zero (to prevent the selection of a node).

It should be noted that the SIAS is calculated exclusively from user-defined resource requirements since the actually-used resource requirements are unknown before the task execution actually starts.

### 7.5.3. SERVICE RE-ALLOCATION SCORE

This research has found that the best throughput results are achieved when tasks are packed tightly into available nodes, i.e. where global resource utilisation is the highest. The best fit scenario, where the task fully utilises 90% of all available resources on a node, is scored the highest. Therefore, when migrating existing tasks, candidate nodes should be scored in the following order: TA, PA, then DA.

Like the SIAS function presented in 7.5.2, the SRAS function for two resource types (CPU and memory) was used. Figure 35 is a graphical representation of SRAS function:

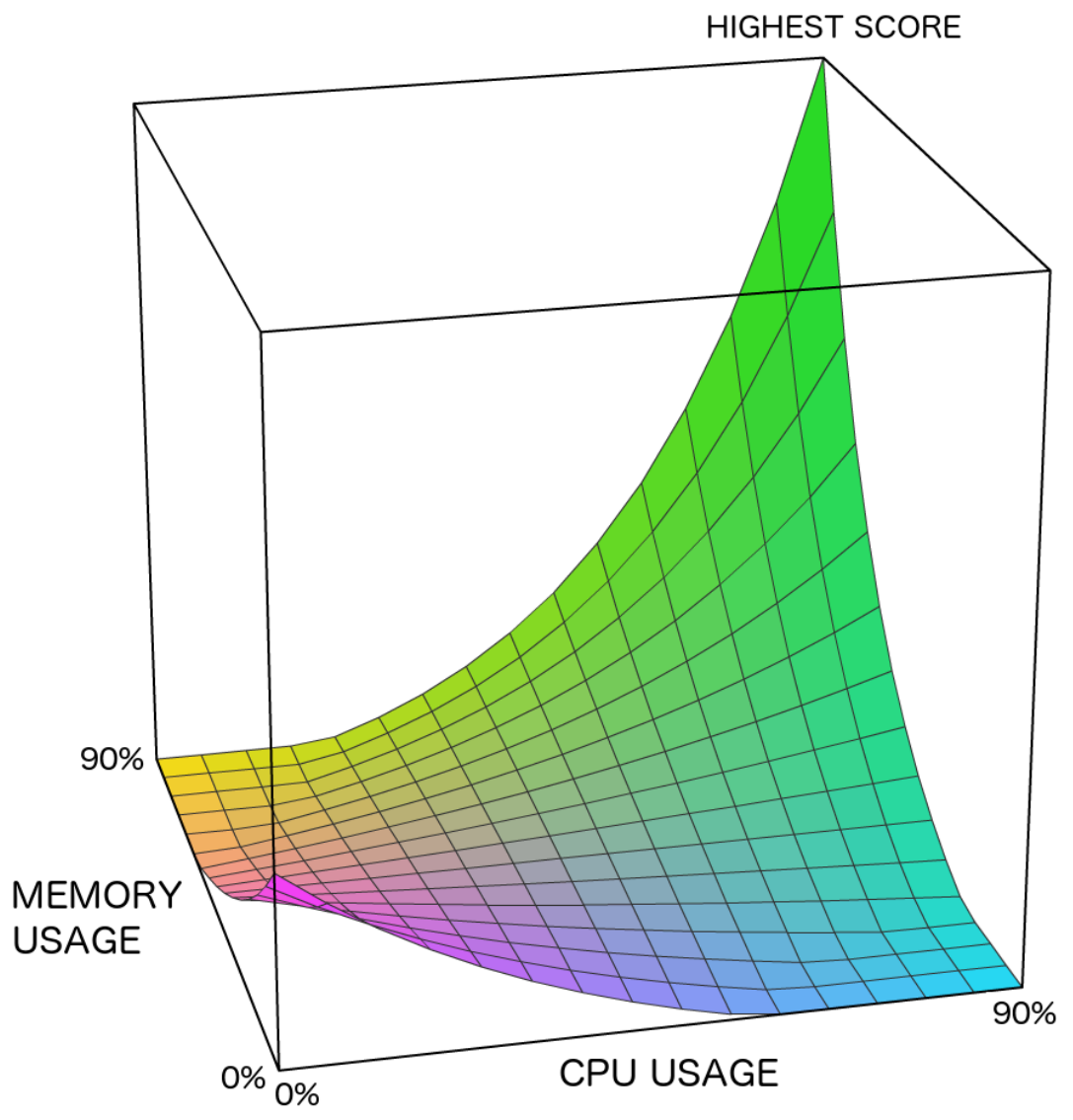

Figure 35: Service Re-allocation Score (two resources)

Three separate areas can be noticed:

- Upper-right corner (the highest score) – this promotes TA, where tasks on this node will closely utilise all its resources.
- Lower-left (the medium score) – this promotes PA that will leave resource utilisation at a low level or proportionately used.
- The upper-left and lower-right corners (the lowest score) – these DAs will leave one resource utilised almost fully and the other resource wasted.

In this implementation, the following SRAS was used:

$$SCORE = F_STEEP^{(r_{CPU} - F_BIAS \cdot r_{CPU_MAX}) \cdot (r_{MEM} - F_BIAS \cdot r_{MEM_MAX})} - F_FLOOR \qquad (14)$$

(with the exceptions of $F_STEEP$ where a value of 500 was used and $F_BIAS$ where a value of 0.6 was used; the parameter definitions are the same as in (13) in subsection 7.5.2)

As can be observed visually, SRAS is a mirror image to the SIAS function (presented in Figure 34). The main difference is changing the score bias (i.e. $F_BIAS$ parameter) which shifts the peak score point from (0,0) to (90,90) (percentage of utilised resources), and which relates to the change in the most desirable AS from PA to TA.

It should be noted that the SRAS is calculated exclusively from actually-allocated resource requirements. User-defined resource requirements are evaluated as part of the RUS routine, explained in detail below.

### 7.5.4. RESOURCE USAGE SPIKES

Occasionally, a task might instantly increase its resource usage as the result of sudden increase of a demand for a task; at such times, a node should have the capacity to immediately accommodate this request, without needing to migrate the task to an alternative node (since this takes time). In such a situation, other VMs running on this machine can be paused or killed to let the VM instance executing this task instantly allocate more resources.

As such, an additional feature was implemented in MASB to handle RUS. Aside from checking the actually-used resources for tasks and ensuring that the node has the capacity to support it, the system also calculates the maximum possible resource usage of all production tasks based on user-defined resource requirements, as well as making sure that the node has the capacity to support all production tasks at their full resource utilisation. This constraint is limited only to production jobs since VMs running non-production jobs can be suspended without disturbing business operations. The continuously fulfilment of this constraint is referred to as Goal (IV).

The introduction of RUS constraint adds another dimension to the tasks allocations' logic. Figure 36 visualises how user-defined resource requirements for production tasks and actually-used resources for all tasks are integrated:

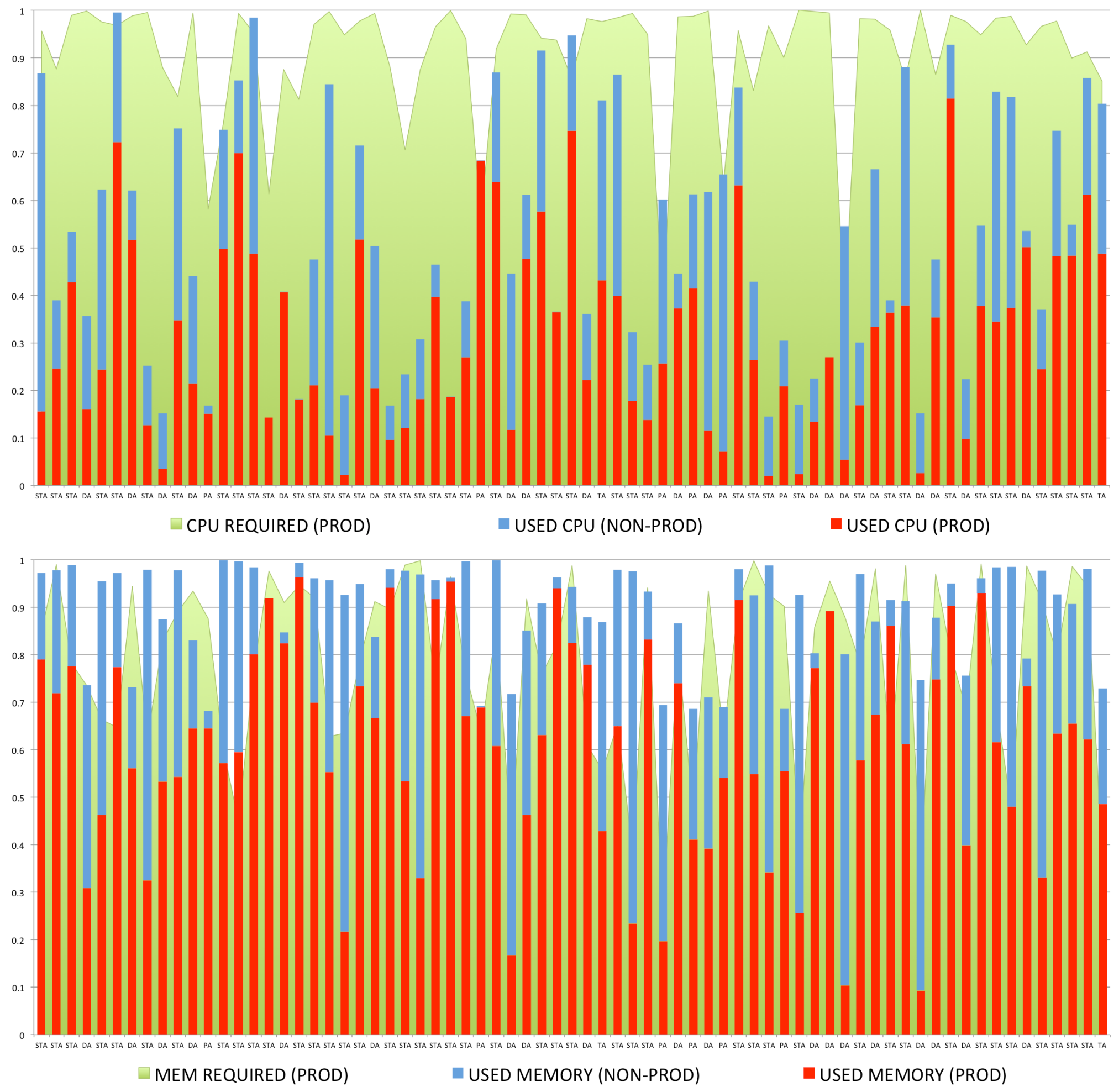

Figure 36: Production vs. non-production allocated resources

In this 60 node sample (a single bar represents one node), approximately half the nodes have a very high CPU user-defined allocation for production tasks, while the real usage is much lower. It should be noted that while memory usage stays proportionally high thorough the GCD workload, the gaps between the requested and the actually-used memory are much smaller. This is a relatively common pattern for GCD workload. Additionally, the chart marks the allocation type (STA, TA, DA, PA) for each node in this sample on the horizontal axis.

Whilst RUS do not occur frequently, they do have the significant potential to destabilise an affected node. Table 15 represents the average frequency of RUS in examined GCD workload traces with ca. 12.5k nodes and ca. 140k tasks being continuously executed by them (with different RUS thresholds examined):

| RUS threshold | Average RUS (count per minute) | Peak RUS (count per minute) |
|---|---|---|
| 5% | 659 | 7538 |
| 10% | 212 | 4362 |
| 15% | 66 | 2390 |
| 20% | 47 | 1925 |
| 25% | 26 | 1135 |

Table 15: Resource Usage Spike frequencies (GCD)

Here, while running a simulation based on replaying the original Borg's allocation decisions (as detailed in 7.6.7), the RUS threshold of 10%, i.e. whenever there was a greater than 10% increase in the overall node resource utilisations levels in any of the monitored resources, was breached 212 times per minute on average, with a peak of 4362 breaches.

In this research, a threshold of 10% was selected for the experimental simulations as an overall good balance between efficiently allocating nodes' resources and, at the same time, leaving the running tasks enough headroom for occasional activity spikes. Generally, lower thresholds resulted in many task migrations (and thus incurred additional task migration costs), and the thresholds above 10% were not utilising resources effectively (the system throughput was lowered). Consequently, the SAS functions were tuned to allocate up to 90% of all available resources on the node (as seen in Figure 32) which seem to give the best overall results.

RUS are a significant design consideration, and a misconfiguration might lead to multiple premature terminations of the tasks and suboptimal performance of the system. Google's engineers implemented a custom resource reservation strategy using a variant of step moving average, as detailed by John Wilkes in a presentation during the GOTO 2016 conference in Berlin (Wilkes, 2016). To cite an alternative solution to handle RUS, El-Sayed et al. (2017) proposes a Machine Learning framework for predicting task terminations, with the resulting task-cloning policy mitigating the effect.

## 7.6. EXPERIMENTAL RESULTS

The previously developed AGOCS framework (Chapter 5) was used as the base of the experimental simulation. AGOCS is a very detailed simulator which provides a multiple of parameters and logical constraints for simulated jobs. The scope of the available variables is very broad, including memory page cache hit and instructions per CPU cycle; however, in this project simulations were based on the following assumptions:

- Requested (by user) and realistic (monitored) resources' utilisation levels for memory and CPU;
- Detailed timing of incoming tasks and any changes in available nodes (within one-minute cycles);
- Nodes attributes and attributes' constraints defined for tasks (as specified in GCD workload traces).

This level of detail comes at the price of extensive computing power requirements. While dry simulation itself can run on a typical desktop machine (see Appendix A) in ca. nine hours, adding layers of scheduling logic, agents' states and inter-system communication requires a significant increase in processing time. In order to realistically and correctly simulate scheduling processes on a Cloud system, the Westminster University HPC Cluster was used.

### 7.6.1. TEST ENVIRONMENT AND CODE PROFILING

The MASB prototype was initially developed on a personal desktop, but as the size and level of detail of the simulations grew, it was necessary to move to a Cluster environment where more computing power was available. All the experiments were executed on the Westminster University HPC Cluster, regarding which more details concerning the software and hardware specifications can be found in Appendix C.

While this cluster offered a sizable array of GPUs, the simulations did not take advantage of that computing power, and instead all processing took place on CPUs. Although it would have been possible to achieve higher throughput when using GPU with frameworks such as ScalaCL or Rootbeer, JVM does not natively support GPU processing. Having as few external dependencies as possible was therefore preferred, since they make maintaining the project more time-consuming. Interestingly, Google's BorgMaster process, which manages a single cell in the production environment for one computing cell, uses 10–14 CPU cores and up to 50GB of memory. The statistics presented are valid for an intensely utilised computing cell, for example one which completes more than 10k tasks per minute on average (Verma et al., 2015).

In experiments, all available forty CPU cores on HPC machine and used them continuously at 60% to 80%. The MASB process allocated ca. 7GB of memory. It is difficult to measure exactly how much computing power was spent on supporting activities such as simulating messaging interactions between agents, i.e. enqueuing and dequeuing messages to and from Akka actors. However, after tuning exercises of the default configuration, the Akka Actors framework proved to be quite resilient. It is estimated that the framework's processing did not take up more than 10-15% of the total CPU time, with the relatively lightweight AGOCS simulator framework consuming about 15-25% of all CPU time. As an interesting note, Akka's optional Thread-pool executor performed noticeably better on the test HPC machines (Appendix C) than on the default Fork-join-pool executor, which is based on a work-stealing pattern. This phenomenon, as well as other experiences of running computation-intensive applications on HPC machines, were discussed during in two presentations (Sliwko, 2018a; Sliwko, 2018b). All the profiling and the above estimation were completed with help of YKJP, similar to the profiling exercises detailed in section 6.4. Figure 37 presents a sample screenshot from the code profiling exercises:

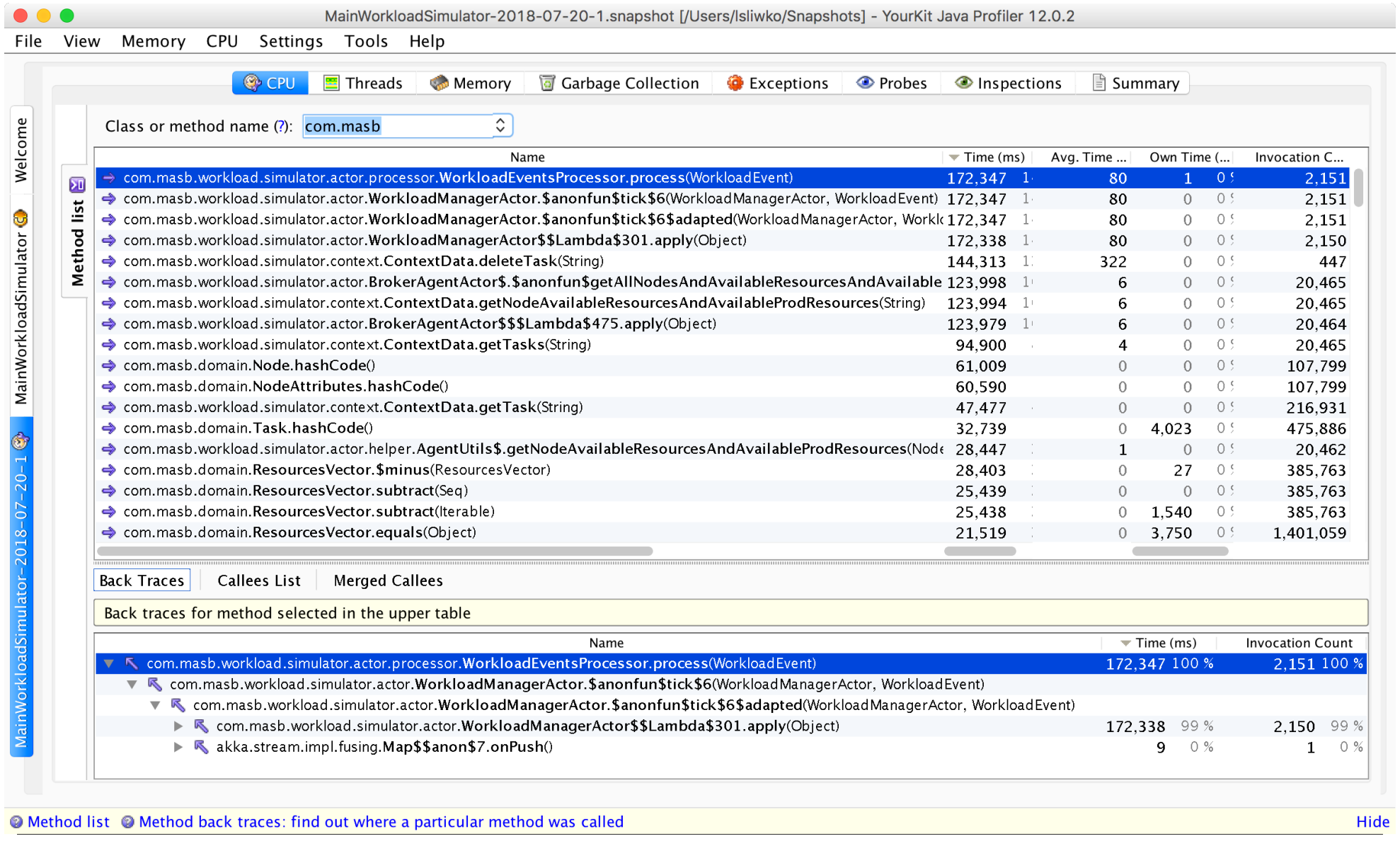

Figure 37: YourKit Java Profiler exercise

YKJP was an excellent tool which helped to optimise the code execution time. However, in a truly multi-core environment, a different approach was required – one which focused on minimising context switches frequency and average CPU idle time across all available cores. Once the MASB framework was moved into the Cluster environment, the 'pidstat' command tool was used to gather statistics, before the refactor and fine-tune framework so as to achieve better parallelism.

During MASB simulations, the typical observed context switches frequency was ca. 500-700 per second per thread, which is comparable with a fully loaded webserver (Mechalas, 2012).

### 7.6.2. TESTABLE DESIGN

Building a framework which fully simulates the Google computing cell from GCD traces has been previously recognised as a challenging task, where there are many aspects to consider (Sharma et al., 2011; Abdul-Rahman, et al., 2014; Zhu et al., 2015). GCD traces contain details of nodes, including their resources, attributes and historical changes in their values. Traces also contain corresponding parameters for tasks, such as user-defined and actually-used resources, as well as attributes' constraints. This has created a multi-dimensional domain with a range of relations which has resulted in complex error-prone implementation. In order to mitigate the risk of coding errors, especially during rapid iterations, a number of programming practices were used:

- A comprehensive test units suite was developed (see Appendix E), along with prototype code. Test units were executed upon every build to catch errors before being deployed to production. This software engineering pattern allowed for a rapid development of prototype and helped to maintain the high code quality;
- A number of sanity checks were built into the runtime logic, such as checking whether the task's constraints could be matched to any node's attributes within the system and checking whether the total of all scheduled tasks' resources exceeded the computing cell compatibilities;
- Recoverable logic flow was implemented for both NA and BA. In the case of various errors such as division by zero or null pointer exceptions, the error is logged but the agent continues to run;

- Keeping a separate error log file with the output of all warnings and errors was a considerable help in terms of resolving bugs.

The implementation of the above features gave high confidence in terms of realising a good quality and reasonably bug-free code.

### 7.6.3. PLATFORM OUTPUTS

Adding detailed logging features to MASB has proved surprisingly difficult. Due to the highly parallel nature of the simulated Cloud environment, an enormous number of log messages were generated upon each simulation, making it difficult to analyse the behaviour of tested algorithms. In addition, writing and flushing log streams caused pauses in simulation. Switching to a Logback framework designed with a focus on concurrent writes provided a solution to this problem, although it was necessary to split the data into distinct log files in order to improve readability, e.g. separate errors from algorithms' output data.

#### 7.6.3.1. LOGGING

In order to fine-tune MASB, excessive logging routines were implemented. All messages, counters and errors are logged to four types of log-files:

- /logs/*.log files – standard log outputs containing all logs messages and also samples;
- /logs/*-error.log – errors and corrupted data exceptions are written to separate files to help with debugging and troubleshooting;
- /logs/*-ticks.csv – CSV files with periodically generated overall system stats, such as the number of idle and overloaded nodes, number of migration attempts, global resources-allocation ratio, and so on;
- /usage/*.csv – detailed node usage stats and task allocations are written periodically to a file, that is, every hundred minutes of simulation time.

#### 7.6.3.2. SAMPLING

Sampling proved to be one of the most important logging features implemented. While examining every decision process in MASB simulation is virtually impossible, frequent and recurrent analysis of the details and values was useful for fine-tuning the system and the scoring functions. Not all the details of every single decision process were logged, rather just a small percentage of all invocations. In the current implementation, the following items are sampled:

- The selection of overloading tasks by the NA, ca. 1 sample per 50 invocations (a sample is presented in 7.4.1);
- The scoring and selection of candidate nodes by the BA, ca. 1 sample per 5k invocations (see log entry in 7.4.2);
- The selection of the target node from the candidate node list, ca. 1 sample per 5k invocations (as listed in subsection 7.4.4).

### 7.6.4. SYSTEM EVOLUTIONS AND OPTIMISATIONS

In order to achieve high resources utilisation and low resources waste, several enhancements were implemented and then fine-tuned, including:

- Limiting the number of candidate nodes returned from BA to fifteen, and introducing the forced migrations feature (subsection 7.4.6);
- Fine-tuning SCS routine to maintain the balance between migration cost and the node allocation score as specified in subsection 7.4.1, which refers to finding the right combination of steps of the TS algorithm, as well as its termination depth;
- Splitting the SAS function into SIAS and SRAS and then limiting the number of candidate nodes examined in those functions (200 and 2k respectively);

- Adjusting input parameters for SIAS and SRAS functions, namely values for $F_BIAS$, $F_STEEP$ and $F_FLOOR$ for the best results based on samples logged (subsections 7.5.2 and 7.5.3);
- Adding the timestamp parameter to the candidate node recommendations, and regularly removing those which have expired. In scenarios where the task migration request is repeatedly refused, this mechanism forces NA to disregard the results of old calculations and request newly scored recommendations from BAs. In this implementation, the recommendation's age threshold was set to three minutes (simulation time) with lower values not yielding better results (see subsection 7.4.4).

### 7.6.5. TEST SIMULATIONS SETUP

During the later stages of the development of the MASB prototype, several simulations were continuously run. They were frequently paused, tuned and then resumed to see whether a given tweak would improves the results. This methodology allowed the research to progress at a good speed while simultaneously iterating a number of ideas and tweaks. Therefore, the testing process did not have noticeable stages, but instead the stages blended into each other. This said, it is possible to logically split the testing into four main areas:

- Benchmarking – GCD workload traces also contain actual Google's Borg scheduler task allocations. In the Borg's simulation, MASB will replay all recorded events, mirroring tasks allocations as per the Google scheduler, i.e. not using its own scheduling logic. This simulation was used as a controlling run in order to test the system, and also as a benchmark to compare results with the original allocations.
- Throughput – secondly, MASB was tested to identify whether it was capable of allocating the same workload as Borg system. The size of the

workload was then increased gradually in 2% steps while preserving the configuration of the system nodes. To ensure the correctness of results, another technique, called 'cell compaction' (Verma et al., 2015) was used in which, instead of adding additional tasks, the system nodes were removed. The results were then compared to the original GCD workload.

- Migration Cost – thirdly, this batch of experiments focused on migration costs incurred via use of VM-LM. A collection of different SAS functions and their variants were tried in order to research their impact on total migration cost while allocating the given workload.
- Scalability – finally, the MASB simulation was run with multiplies of GCD workload in order to test the scalability limits of the designed solution. Although this step was the least work-intensive, it took the longest time to perform.

As noted in Zhu et al. (2015), simulating GCD workload is not a trivial task. The main challenge when running such large and complex simulations is the demand for computation power and the continuous processing. During this experiment, the AGOCS framework was modified to also allow the testing of computing cells larger than 12.5k. This was achieved by duplicating randomly selected existing tasks and their events, for example 'Create Task A event from GCD workload trace files' will create events AddTaskWorkloadEvent events for task A and A'. This feature is based on the hashcode of object's ID, which is a constant value.

The largest experiments simulated a single Cloud computing cell with 100k nodes and required nine months of uninterrupted processing on one of the University of Westminster HPC cluster's nodes. At this juncture, it should be noted that early simulations often fail due to unforeseen circumstances, such as NAS detachment or network failure. One solution to this was to frequently save snapshots of the state of the simulation and to keep a number of previous snapshots in case of write file failure.

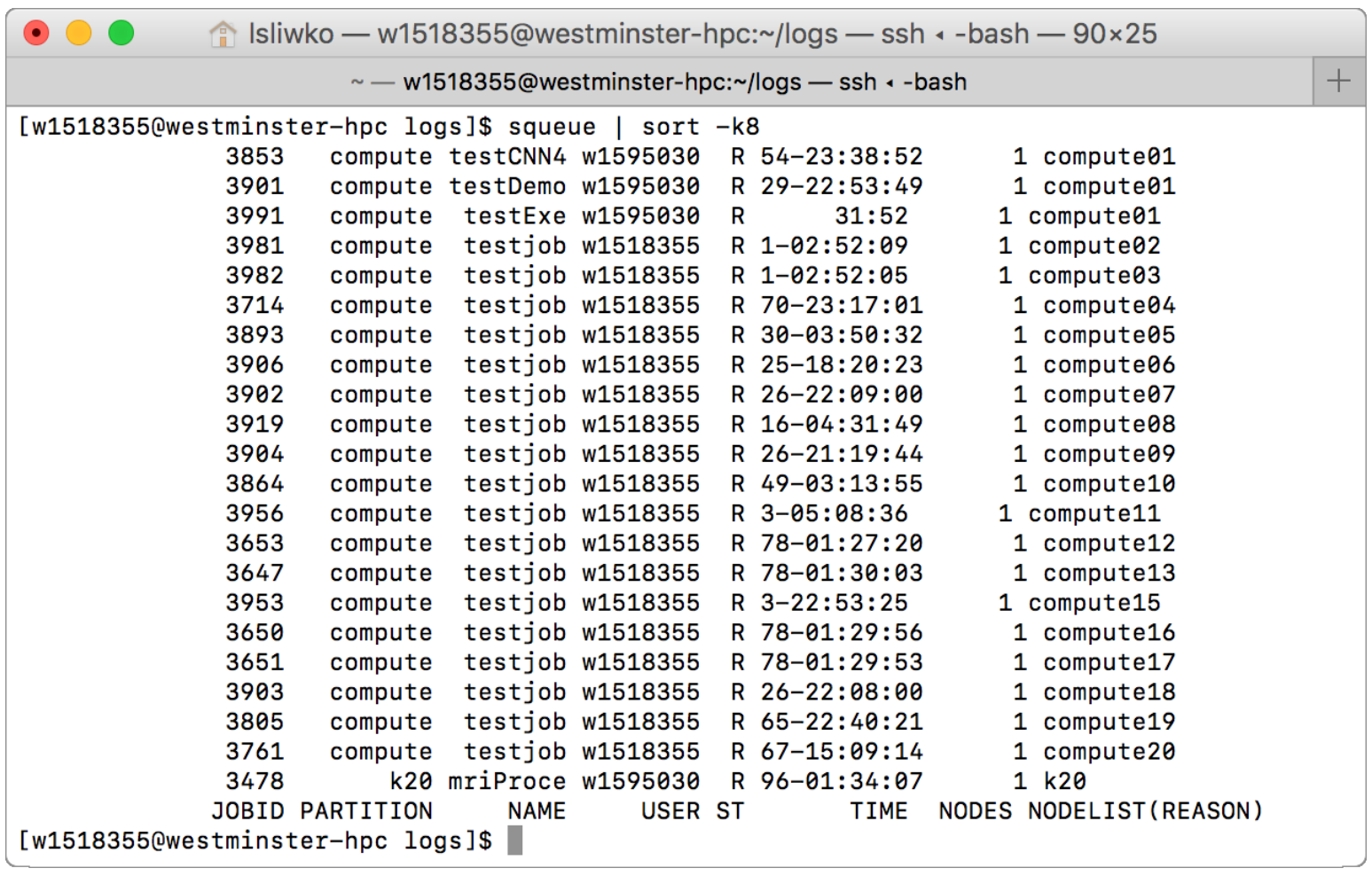

lsliwko — w1518355@westminster-hpc:~/logs — ssh ‹ -bash — 90×25

~ — w1518355@westminster-hpc:~/logs — ssh ‹ -bash

```
[w1518355@westminster-hpc logs]$ squeue | sort -k8
               3853   compute testCNN4 w1595030  R 54-23:38:52      1 compute01
               3901   compute testDemo w1595030  R 29-22:53:49      1 compute01
               3991   compute  testExe w1595030  R       31:52      1 compute01
               3981   compute  testjob w1518355  R 1-02:52:09      1 compute02
               3982   compute  testjob w1518355  R 1-02:52:05      1 compute03
               3714   compute  testjob w1518355  R 70-23:17:01      1 compute04
               3893   compute  testjob w1518355  R 30-03:50:32      1 compute05
               3906   compute  testjob w1518355  R 25-18:20:23      1 compute06
               3902   compute  testjob w1518355  R 26-22:09:00      1 compute07
               3919   compute  testjob w1518355  R 16-04:31:49      1 compute08
               3904   compute  testjob w1518355  R 26-21:19:44      1 compute09
               3864   compute  testjob w1518355  R 49-03:13:55      1 compute10
               3956   compute  testjob w1518355  R 3-05:08:36      1 compute11
               3653   compute  testjob w1518355  R 78-01:27:20      1 compute12
               3647   compute  testjob w1518355  R 78-01:30:03      1 compute13
               3953   compute  testjob w1518355  R 3-22:53:25      1 compute15
               3650   compute  testjob w1518355  R 78-01:29:56      1 compute16
               3651   compute  testjob w1518355  R 78-01:29:53      1 compute17
               3903   compute  testjob w1518355  R 26-22:08:00      1 compute18
               3805   compute  testjob w1518355  R 65-22:40:21      1 compute19
               3761   compute  testjob w1518355  R 67-15:09:14      1 compute20
               3478       k20 mriProce w1595030  R 96-01:34:07      1 k20
              JOBID PARTITION     NAME     USER ST       TIME  NODES NODELIST(REASON)
[w1518355@westminster-hpc logs]$
```

Figure 38: University of Westminster HPC Cluster utilisation

At the peak of the experiment, eighteen out of twenty computing nodes were committed to running MASB simulations, as can be seen in Figure 38.

### 7.6.6. ALLOCATION SCORE RATIOS

Clearly, when examining the suitability of load balancing, the key parameter is the number of overloaded nodes, which should be kept to minimum. It was found that replaying GCD traces using Google's original Borg's allocation decisions results in up to 0.5% of nodes being overloaded in a simulated one-minute period. It was assumed that this phenomenon was the result of delayed and compacted resource usage statistics, which were recorded and averaged over ten-minute periods. As such, in further experiments this ratio was used as an acceptable error margin.

The second researched property was how nodes were distributed amongst allocation score types during simulations. Therefore, each experiment recorded a number of nodes with each allocation score type, and averaged them out over the simulation period. The set of normalised values for STA, TA, PA and DA are referred to as Allocation Score Ratios (ASR). Idle Nodes and Overloaded Nodes are discussed separately, and they are excluded from the ASR. The ASR values describe how well the Cluster is balanced, that is, how well nodes are balanced as a whole group.

The ASR values are used to describe the experimental results presented in the subsections below to highlight the differences in how various load balancing strategies perform under a GCD workload. Figure 39 chart visualises the AS distribution during a month-long simulation. The horizontal axis is the measure of time and the vertical axis represents the number of nodes having a particular allocation type (as per coloured legend):

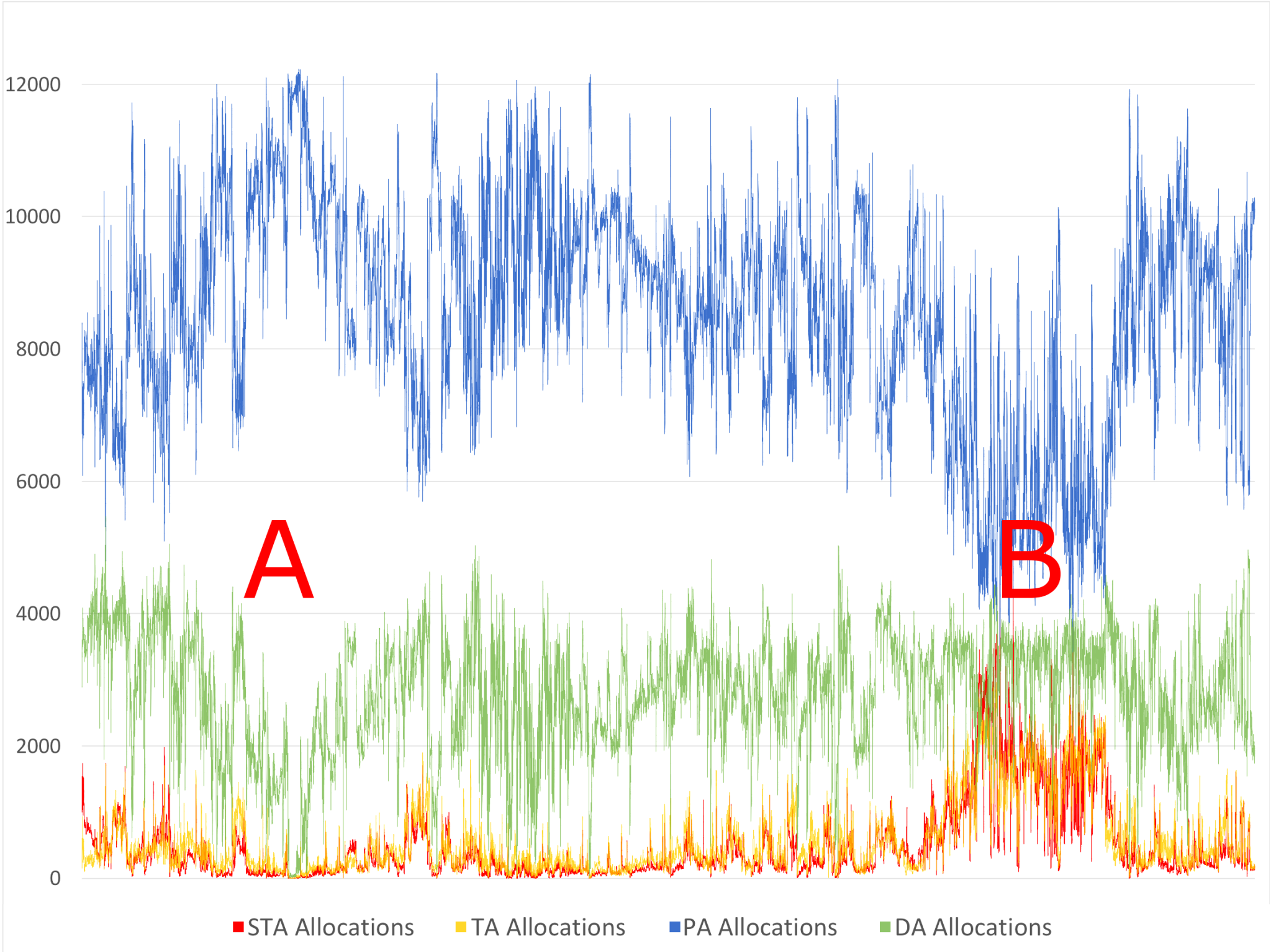

Figure 39: MASB – Allocation Scores distribution (12.5k nodes)

The most dominant AS was PA, meaning that each of the node's resources is utilised between 0% and 70%. Ca. 68% of all the cluster's nodes are found within these parameters, which is the direct result of their initial allocation using SIAS function. The second biggest group, ca. 22% of all servers, are nodes allocated disproportionally in which one or more resources are highly used but the other resources are relatively idle. The remainder of the nodes have either an STA or TA allocation score type. The PA to DA ratio of roughly 3:1 is characteristic for a typical workload as recorded in GCD traces and processed by MASB.

The chart also highlights two periods of low and elevated workload, marked A and B respectively:

- During the low workload period (A), SIAS function can schedule most newly-arriving tasks to relatively unused nodes, thereby successfully preserving their resource usage proportions. As such, the number of PAs increases while the number of DAs decreases. Existing long-running services continue to run uninterrupted on their nodes, and so the ratio of STA to TA remains flat.
- During an elevated workload period (B), SIAS function is unable to find relatively unused nodes anymore. It thus selects lower quality allocations, resulting in a decrease in PAs. Due to the scarcity of resources, tasks are also re-allocated more frequently by SRAS function. This results in tighter fit allocations, which is seen as an increase in STAs and TAs counts.

This cycle is repeated thorough cluster activity, wherein MASB balances the workload. The subsections which follow describe several implemented optimisations and their rationales, as well as the experimental results and a commentary on them.

### 7.6.7. BENCHMARK

Given that GCD traces have a complicated structure and contain a vast amount of data, only rarely are they analysed to the full extent of their complexity. MASB design shares similarities with BorgMaster in areas such as constraining tasks, defining memory and CPU cores as resources, using scoring functions for candidate node selection, and handling RUS. It also closely follows the lifecycle of tasks as presented in subsection 5.5.1. As things stand, there is no publicly available literature which contains descriptions of similar experiments which could be compared with the simulation results of MASB. Therefore, the closest comparable results are the original Borg's allocation decisions that were recorded in GCD traces. For the purposes of this research, it was decided that they be used as a benchmark for the results from MASB's experiments.

Both simulations processed full month-long GCD traces. The average values were used because MASB simulation works in one-minute intervals whilst GCD traces provide usage statistics in ten-minute windows that occasionally overlap. Given this, peak or median values were not accurate. To highlight differences in workings between the MASB and Google Borg algorithms, Figure 40 presents the AS distribution during the period recorded in GCD (replayed Google's Borg allocation events):

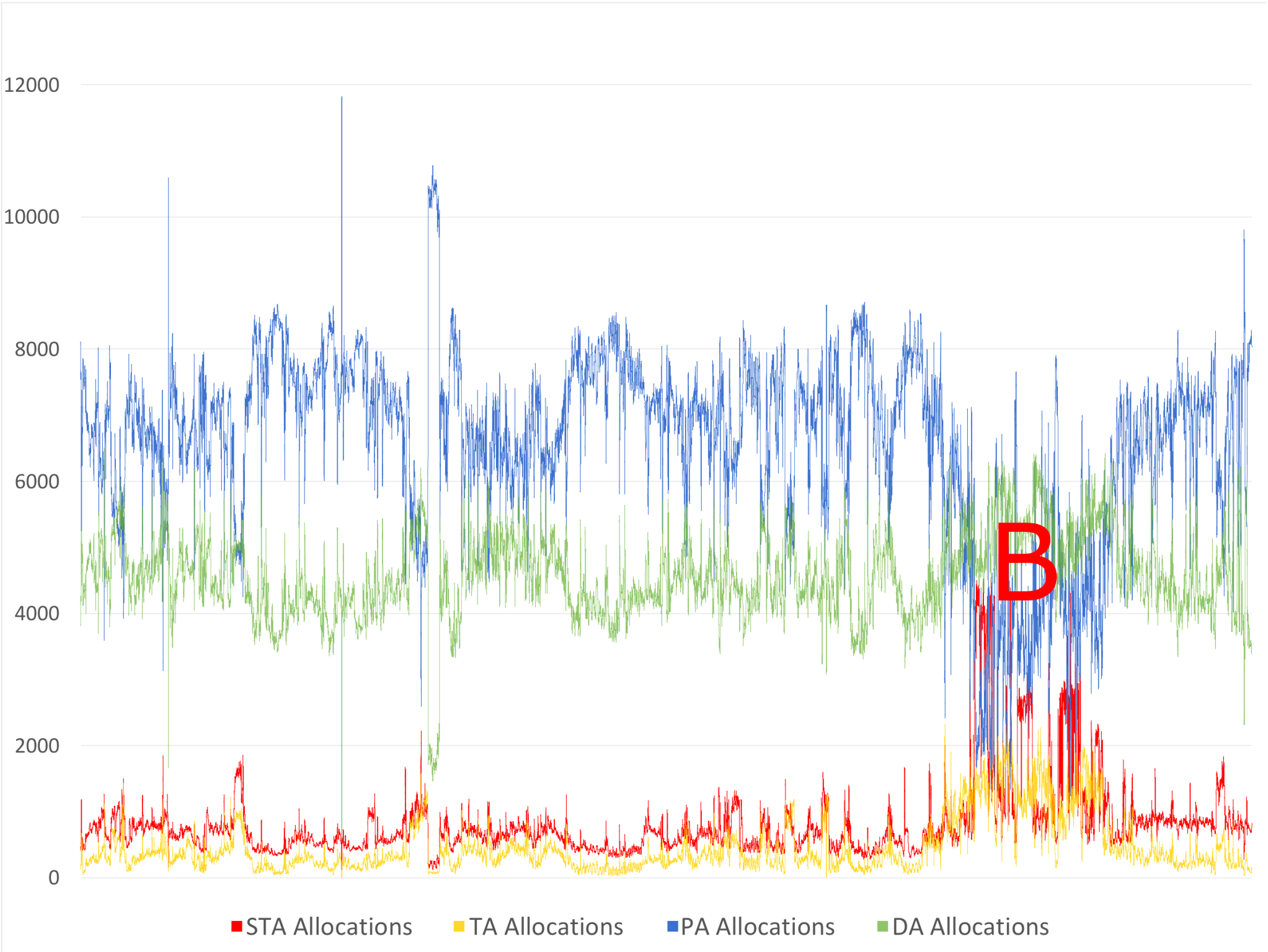

Figure 40: Borg – Allocation Scores distribution (12.5k nodes)

In comparison to the experimental data presented in Figure 39, MASB behaves more organically during periods of low and elevated workload. This is especially visible during the period of elevated workload (B) where MASB managed to preserve a better ratio of PA to DA Nodes than Google's Borg. This behaviour is the result of allowing a given task to be re-allocated during its execution, meaning that MASB can dynamically shape its workload and improve the health of its allocations. This feature also allows greater flexibility in altering the requirements of running tasks, in which the load balancer attempts to offload an alternative node.

Table 16 directly compares ASR parameters of both pre-recorded Google's Borg and MASB simulations.

| **Parameter (average, one-minute interval)** | **Framework** | |
|---|---|---|
| | **Borg** (Figure 40) | **MASB** (Figure 39) |
| Idle Nodes | 1.01 (0.01%[2]) | 78.10 (0.63%[2]) |
| STA[1] Nodes | 820.49 (6.58%[2]) | 487.01 (3.91%[2]) |
| TA[1] Nodes | 459.57 (3.69%[2]) | 564.18 (4.53%[2]) |
| PA[1] Nodes | 6597.14 (52.94%[2]) | 8508.08 (68.28%[2]) |
| DA[1] Nodes | 4578.49 (36.74%[2]) | 2810.69 (22.56%[2]) |
| Overloaded Nodes | 4.04 (0.03%[2]) | 12.62 (0.10%[2]) |

1. STA, TA, PA and DA as defined in section 7.5.
2. Totals do not sum to 100 percent due to rounding.

Table 16: Benchmark results – Borg and MASB

The listed ASR values highlight the differences in Borg and MASB workings:

- Idle Nodes – Borg's design has a definite advantage over MASB because Borg's schedulers can access the shared cluster's state and iterate over the complete set of system nodes. MASB relies on a network of BAs, each of which has only partial information about the cluster's state. Therefore, a subset of idle nodes might never be scored, even if they represent the best allocation for a given task.
- STA and TA Nodes – in both systems, under normal workload conditions, incoming tasks are reasonably well distributed between the nodes. Only ca. 10% of all system nodes register higher resource usage scores, when at least one of resource utilisation levels crosses 90%. The exact scoring algorithm of Google's Borg has not been disclosed, but the results suggest a degree of similarity to the SIAS function.
- PA and DA Nodes – the ratio of PAs to DAs is visibly different in Borg and MASB. Borg's original scheduling decisions had a ratio of roughly 3:2, meaning that for every three proportionally allocated nodes in the system, there were two nodes that were disproportionately allocated. MASB managed to achieve a better ratio of 3:1, suggesting that the use of SIAS and SRAS scoring functions together with VM-LM feature can potentially create a more balanced scheduling system.

Given the superior ratio of PA to DA nodes as measured, and the possibility of increased throughput, the next experiment focused on processing increased workload.

### 7.6.8. THROUGHPUT TESTS

The MASB framework has been designed as a general solution for balancing workload in a decentralised computing system. After numerous iterations, MASB was eventually able to schedule the entire GCD workload, with additional tasks also added. Table 17 presents a comparison of the results:

| Parameter (average per minute) | Workload Size (tasks) | | | |
|---|---|---|---|---|
| | 100% (original) | 102% | 104% | 106% |
| Nodes Count | 12460.39[1] | 12460.36[1] | 12460.68[1] | 12460.35[1] |
| Tasks Count | 132061.15[1] | 134738.92[1] | 137399.93[1] | 142936.05[1] |
| Global CPU Usage Ratio | 43.64% | 44.54% | 45.42% | 46.89% |
| Global Memory Usage Ratio | 62.05% | 63.33% | 64.58% | 66.57% |
| Idle Nodes | 76.41 (0.61%) | 73.08 (0.59%) | 72.75 (0.58%) | 52.18 (0.42%) |
| STA Nodes | 479.91 (3.85%) | 480.22 (3.85%) | 423.51 (3.40%) | 447.73 (3.59%) |
| TA Nodes | 566.20 (4.54%) | 545.74 (4.38%) | 447.75 (3.59%) | 355.88 (2.86%) |
| PA Nodes | 8507.49 (68.28%) | 8718.76 (69.97%) | 9316.35 (74.77%) | 9576.67 (76.86%) |
| DA Nodes | 2818.11 (22.62%) | 2610.04 (20.95%) | 2084.06 (16.73%) | 1718.08 (13.79%) |
| Overloaded Nodes | 12.28 (0.10%) | 32.53 (0.26%) | 116.25 (0.93%) | 309.79 (2.49%) |

1. The AGOCS framework itself records minuscule variances in node and task counts that are the result of concurrent update operations, while modifying shared context data objects. See section 5.8 for details.

Table 17: Throughput results (100%-106% workload size)

As demonstrated above, MASB was able to schedule, on average, an additional ca. 2.6k tasks per minute (ca. 2% more tasks). Further tuning was unable to improve those results, with workload sizes greater than 102% increasing the number of overloaded nodes above the defined threshold of 0.5%.

To further ensure the correctness of the attained results, another set of experiments was run in parallel. Here, instead of multiplying the original GCD workload, the random machines were removed from the cluster until the workload could no longer be fitted. This method, known as 'cell compaction', is suggested in Verma et al. (2015) for simulations with GCD traces.

Similar to the previously detailed experiments which had augmented workload, even when the cluster size was reduced to ca. 98% of its original size (242 nodes being removed), the original GCD workload could still be fitted without breaching the 0.5% limit of overloaded nodes. Table 18 details the experimental results:

| **Parameter (average per minute)** | **Cluster Size (nodes)** | | | |
|---|---|---|---|---|
| | **100% (original)** | **99%** | **98%** | **97%** |
| Nodes Count | 12460.39 | 12332.92 | 12218.61 | 12081.30 |
| Tasks Count | 132061.15 | 132057.96 | 132057.54 | 132055.86 |
| Global CPU Usage Ratio | 43.64% | 44.09% | 44.52% | 45.05% |
| Global Memory Usage Ratio | 62.05% | 62.72% | 63.39% | 64.08% |
| Idle Nodes | 76.41 (0.61%) | 53.29 (0.43%) | 58.96 (0.48%) | 75.87 (0.63%) |
| STA Nodes | 479.91 (3.85%) | 480.28 (3.89%) | 404.13 (3.31%) | 448.67 (3.71%) |
| TA Nodes | 566.20 (4.54%) | 572.24 (4.64%) | 485.55 (3.97%) | 500.75 (4.14%) |
| PA Nodes | 8507.49 (68.28%) | 8412.71 (68.21%) | 8866.13 (72.56%) | 8663.88 (71.66%) |
| DA Nodes | 2818.11 (22.62%) | 2800.30 (22.71%) | 2361.64 (19.33%) | 2339.31 (19.35%) |
| Overloaded Nodes | 12.28 (0.10%) | 14.11 (0.11%) | 42.20 (0.35%) | 62.07 (0.51%) |

Table 18: Throughput results (97%-100% cluster size)

On average, GCD traces utilise ca. 40-50% of the globally available CPUs and ca. 60-70% of globally available memory while continuously guaranteeing ca. 85% of CPUs and ca. 70% of memory to production tasks to handle RUS. It should be noted that Borg's scheduling routines have been perfected following decades of work by a team of brilliant Google engineers. The conclusion of this research is that, it is hard to substantially improve this impressive result given those constraints.

Although the throughput of the original Google Scheduler could not be significantly improved, the results from both methods of evaluation show the benefits of using VM-LM to fit additional tasks in an already very tightly-fitted cluster.

### 7.6.9. MIGRATION COST

The MASB framework relies on a VM-LM feature to balance workload by moving running tasks across Cloud nodes. While the VM-LM process is reasonably cheap in terms of the computing power, it does incur a non-trivial cost on the Cloud's infrastructure. In order to avoid excessive networks transfers, NAs carefully decide which tasks will be migrated out from a given node. To score candidate tasks, the SCS function is used which takes the task's estimated migration cost into consideration as well as released resources (see 7.4.1 for more details).

Unexpectedly, when searching for ways to lower the total migration cost, although modifications of SCS function seemed to be the most palpable place to start, significantly better results were not obtained. Based on experience from previous experiments, it was discovered that the biggest reduction in task migrations was achieved by improving the quality of the initial task allocation. Therefore, further experimentation focused on testing variants and combinations of the score functions.

Figure 41 presents the evolutions of scoring functions:

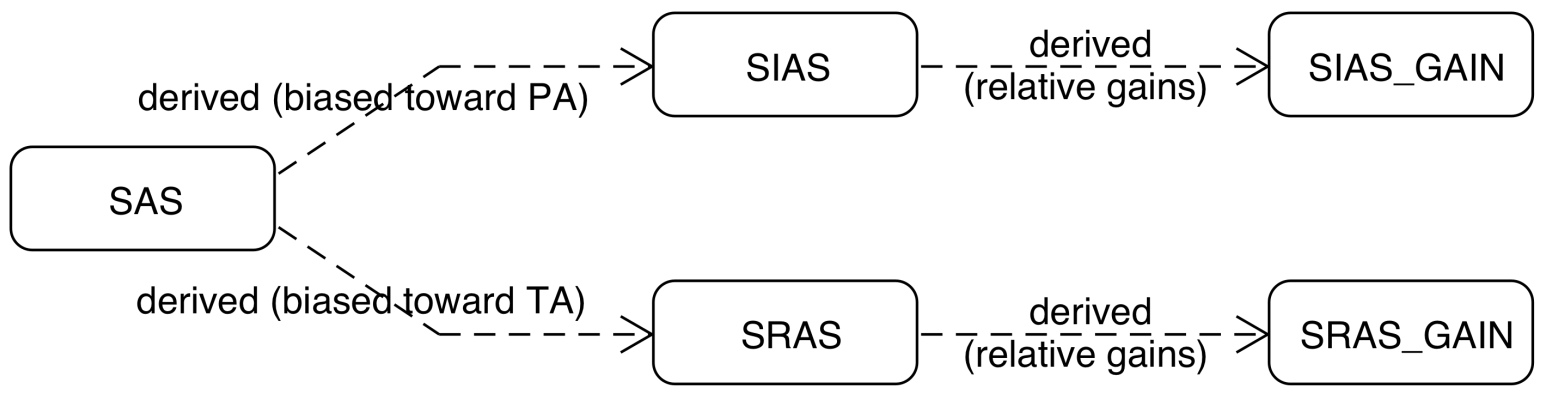

Figure 41: Scoring functions evolution

As previously mentioned, initially MASB implemented a single SAS function which prioritised the scattering of tasks amongst nodes. With introduction of SAL (detailed in subsection 7.5.1), the SAS function was split into SIAS and SRAS functions biased towards opposite allocation types, namely PA and TA. However, during the study of the impact of frequent re-allocations on STC, it was found that those scoring functions can be further improved by introducing GAIN variants.

The GAIN variants of SIAS and SRAS functions are defined here as SIAS_GAIN and SRAS_GAIN respectively:

- $SIAS_GAIN = SIAS(T') - SIAS(T)$
- $SRAS_GAIN = SRAS(T') - SRAS(T)$ (15)

where $T$ is the current set of allocated tasks, and $T'$ is the candidate set of allocated tasks on a given node. Additionally, cases when a node would lower its AS as a result of migrations have a zero score.

In the GAIN variants of scoring functions, the relative AS gains are prioritised over the absolute AS values for an individual node. For example, given the scenario in which the task migration to node A would change its AS from 0.1 to 0.4 (a 300% gain), while the same task could also be migrated to node B, changing its AS from 0.4 to 0.6 (a 50% gain), the former option will be selected as yielding a higher gain (since 300% is greater than 50%) regardless of the potentially higher absolute score value of node B.

Table 19 presents the results under the variants of the scoring functions:

| Parameter (average per minute) | Scoring Functions | | | |
|---|---|---|---|---|
| | SIAS SRAS | SIAS SRAS_GAIN | SIAS_GAIN SRAS | SIAS_GAIN SRAS_GAIN |
| Total Migration Cost [GB] | 1490.65 | 7008.30 | 1252.50 | 5925.41 |
| Cost per Task Migration [MB] | 338.09 | 795.90 | 339.02 | 954.21 |
| Idle Nodes | 83.54 (0.67%) | 105.80 (0.85%) | 76.14 (0.61%) | 79.33 (0.64%) |
| STA Nodes | 495.09 (3.97%) | 687.77 (5.52%) | 490.42 (3.94%) | 654.18 (5.25%) |
| TA Nodes | 656.67 (4.54%) | 560.65 (4.50%) | 558.57 (4.48%) | 547.38 (4.39%) |
| PA Nodes | 8515.95 (68.34%) | 8492.21 (68.15%) | 8511.61 (68.31%) | 8451.12 (67.82%) |
| DA Nodes | 2785.23 (22.35%) | 2586.43 (20.76%) | 2810.95 (22.56%) | 2707.73 (21.73%) |
| Overloaded Nodes | 14.88 (0.12%) | 27.54 (0.22%) | 12.67 (0.10%) | 20.61 (0.17%) |

Table 19: Results comparison of SAS, SIAS and SRAS (migration cost)

The combination of SIAS_GAIN and SRAS functions was most efficient, i.e. the total average migration cost as well as the average cost per task migration were lowest, while ASR remained virtually unchanged. Nonetheless, the good results were also yielded with the combination of SIAS and SRAS.

The experiment showed that focusing on the node AS's absolute value as well as value gain are both viable strategies during the initial task allocation (with the former being relatively better). However, it is the selection of the task re-allocation strategy that is crucial and should be dedicated to maximising the absolute value of the node's allocation score. As mentioned previously, the majority of tasks scheduled on the GCD cluster are short-lived batch jobs which tend not to have high resource requirements (see section 5.2). As such, there is no need to carefully fit them to a node. As a result of their limited time on the cluster, the chance of re-allocation is low. Long-running services, however, should be fitted tightly onto available nodes and continue to run there due to the additional cost of further re-allocations because of the typically large amounts of used memory.

### 7.6.10. SCALABILITY STUDY

The final step in the experiments was to examine the scalability of the MASB framework. Due to the simulation's high computational requirements, its one-minute time slices were split into 'rounds', in which every NA could both respond to migration requests as well as send its own requests, although sent requests would be unanswered until the next 'round'. This meant that the simulated scenarios were as realistic as possible whilst also emulating massive Cloud installations.

Such a long simulation was necessary in order to achieve reliable and quality results. The month-long GCD workload traces were produced by an actual Cluster system and contain many real-world scenarios which would not be possible to synthesise in any other way. Special thanks are due to University of Westminster IT staff which provided a massive help and support during those experiments.

Table 20 demonstrates the results achieved through the multiplication (here: two, four and eight times) of the original GCD workload; it also highlights the lack of changes in ASR values:

| Parameter (average per minute) | Cluster Size (nodes) | | | |
|---|---|---|---|---|
| | 12.5k (original) | 25k (2x) | 50k (4x) | 100k (8x) |
| Nodes Count | 12460.70 | 24921.49 | 49842.99 | 99685.97 |
| Tasks Count | 132061.35 | 264155.80 | 528336.38 | 1056645.92 |
| Idle Nodes | 71.61 (0.57%) | 95.82 (0.38%) | 226.42 (0.45%) | 413.03 (0.41%) |
| STA Nodes | 492.67 (3.95%) | 805.60 (3.23%) | 1920.99 (3.85%) | 3868.22 (3.88%) |
| TA Nodes | 570.37 (4.58%) | 962.14 (3.86%) | 2232.10 (4.48%) | 4300.70 (4.31%) |
| PA Nodes | 8502.06 (68.24%) | 18118.11 (72.71%) | 34102.21 (68.42%) | 68999.49 (69.22%) |
| DA Nodes | 2812.74 (22.57%) | 4914.55 (19.72%) | 11324.77 (22.72%) | 22031.79 (22.10%) |
| Overloaded Nodes | 11.26 (0.09%) | 25.25 (0.10%) | 36.49 (0.07%) | 71.83 (0.07%) |

Table 20: Scalability tests – 12.5k, 25k, 50k and 100k nodes

MASB was able to orchestrate a cell size of 100k without a noticeable scalability cost and without crossing the limit of 0.5% overloaded nodes. With the current MASB framework implementation, the simulation of this size took around nine months on a single node of the University of Westminster HPC (see Appendix C for specifications).

Google has never disclosed the size of their largest cluster, but it has been noted in Verma et al. (2015) that Borg computing cells are similarly sized to the clusters managed by Microsoft's Apollo system, which have in excess of 20k nodes (Boutin et al., 2014). A 12.5k node cells in GCD traces have been described as 'average' or 'median', cells with fewer than 5k nodes have been called 'small' or 'test' (Verma et al., 2015). Additionally, (ibid.) gives an example of a larger cell C, which is 150% the size of cell A and therefore also approximately 20k nodes. As such, in this research it is assumed that the computing cell of the large Borg is around 20-25k nodes.

Therefore, as demonstrated, the designed multi-agent load balancing strategy scaled beyond the original GCD workload without incurring noticeable scalability costs. The paradigm of offloading the scheduling logic onto nodes themselves has the following benefits: (i) it enables the implementation of more complex scheduling schemas as the nodes resources can be used for that purpose; (ii) the computing power dedicated to cluster orchestration increases together with the Cluster size (so allowing for greater scalability); and, (iii) limits the amount of communications required to maintain up-to-date Cluster state information. The result of such a schema is the ability to enlarge the computing cells to the sizes of 100k nodes while preserving a good throughput and performance.

## 7.7. COMPETITIVE SOLUTIONS

During work on the Cloud load balancer prototype, a number of publications were examined and later compared with the proposed MASB design. Aside from the solutions presented in section 7.1, the following three systems listed in the subsections below have been found to share a degree of similarity with MASB.

### 7.7.1. ANGEL SYSTEM

The ANGEL system (Zhu et al., 2015) is based on a concept wherein a multi-agent system manages its workload in a virtualised Cloud environment. This solution also takes advantage of the VM-LM feature to re-allocate running tasks to an alternative node if necessary. While the basic concept of ANGEL and the MASB system is similar, the design of the architecture and features differ substantially:

- Within ANGEL each task is represented by Task Agent created upon task arrival and destroyed when the task is complete. VM Agent represents a VM hypervisor running on a physical node and accepting/rejecting tasks. In comparison, during the development of MASB, it was found that the sheer number of tasks made it impractical to create an entity for each task responsible for its allocation; given this, the responsibility was assigned to NAs. In MASB, NAs themselves are responsible for keeping their node stable and offloading overloading tasks to alternative nodes. Therefore, MASB can potentially support very larger number of tasks. Indeed, during simulations one million tasks were continuously managed.
- In ANGEL, Manager Agent acts as a leader for this computing cell and stores the complete system state in a 'VM Information Board'. VM Agents are constantly updating Manager Agent as to changes in their state, such as available resource (CPU and memory) changes, VM creations and cancellations. The ANGEL system assumes that the stored system state is always current, and Manager Agent this information to match Task Agents with VM Agents. In MASB a subnetwork of BAs has responsibility of caching the global system; however, this information is accepted as outdated by design, and so system uses it only for building initial candidate nodes list which is then sent to NAs. Therefore, MASB doesn't rely on accurate and timed updates from system nodes and the actual task allocation is resolved later between NAs themselves.
- MASB is focused on a Cluster throughput and scalability whereby resource usages gaps are reduced, and tasks are fitted into available nodes. The focus of the project was to achieve tightness of task allocations no worse than in the GCD traces while improving scalability. The aim of ANGEL is to guarantee the ratio of tasks guaranteed to meet their deadlines which are also priority-adjusted. Therefore, ANGEL seems to be more aimed at high churn of short-term tasks, while MASB is designed to support mixed-workload consisting of batch jobs as well as long-lived services.

The authors of ANGEL also tested their solution on GCD traces. In so doing, they acknowledged the difficulty of conducting experiments on the whole month-long traces because of the enormous count of tasks in the trace logs. As such, they performed their experiments exclusively on the 18$^{th}$ day of traces, which has been recognised as being the most representative time period in GCD traces (Moreno et al., 2013). However, the results presented use different metrics and do not specify further details of the experiments, such as whether authors also matched task constraints and whether tasks were allocated with regards to handing RUS.

### 7.7.2. US PATENT 5,031,089

Liu and Silvester (1991) filed a patent which described a set of routines that could be deployed on nodes in order to balance system-wide workload. The first routine periodically examines a number of jobs on the node's queue and computes the 'workload value', which is then provided on request to other nodes by the second routine. The third routine, meanwhile, is triggered periodically when the node is idle, and at the end of each job completion. This routine contains the main bulk of load balancing logic and evaluates whether the node's 'workload value' is below a pre-established value that would indicate that the node is relatively idle. If the node is recognised as being under-utilised and available for more jobs, then the routine will poll all the other nodes for their 'workload value', and transfer jobs from the node with the highest 'workload value' to its own queue.

The feasibility of this invention was validated via several simulations, although those results are not shared in the cited patent. The authors list several assumptions made during the performance testing of this study, such as the homogeneity of all the tasks and their resource requirements, as well as the assumption that the job's transfer cost is negligible. The main criticism of this solution is that it oversimplifies the Cluster workload's model, and it omits the continuous changes of resources used by jobs. Only the job’s queue length was used as 'workload value'. Furthermore, only non-started jobs can be transferred to alternative nodes. The solution relies on polling all nodes in the cluster for their utilisation levels, which in a large cluster might be not feasible and may create a bottleneck.

### 7.7.3. US PATENT 8,645,745

Barsness et al. (2014) notes that there is a problem when a centralised job scheduler needs to pass through a large number of nodes in order to find one which can be used to run the task, and proposed a solution whereby each node is continuously scanning a shared-file to determine which job could be executed on this node. When a job requires multiple nodes, the one on the nodes becomes a primary node, which then assigns and monitors the job execution on the multiple nodes.

In comparison to MASB, the main similarity is that there is no centralised manager to assign tasks to nodes. This means that nodes are themselves responsible for selecting and then running the accepted tasks. However, the main difference is that proposed patented strategy doesn’t examine all nodes, and the task is allocated to the first (quickest) scheduler that picks the task. In MASB a task allocation is a multi-step process in which each node tries to increase its AS by selecting the best-matching tasks. Moreover, MASB dynamically manages workload by offloading currently running tasks to the best candidate nodes (with the highest AS score), and, by doing that, the overall system efficiency is increased.

Given that the patent paper provides no results from experiments, it is difficult to directly compare systems' performances.

## 7.8. SUMMARY AND CONCLUSIONS

The primary challenge when sequencing a queue of tasks on a cluster is to fit them tightly so as to reduce resource usage gaps. The scheduling algorithm attempts to reduce the situations where a resource on a given node is overly un-utilised at the same time that other resources on that node are mostly allocated. It is extremely important to shrink the gaps in resource utilisation and to allocate them proportionally, especially when initially scheduling new tasks which tend to have balanced resource requirements.

Fitting objects of different volumes into a finite number of containers is known as a 'bin-packing' problem, and belongs to class of NP-Hard problems. The traditional way of solving NP-Hard problems are metaheuristic algorithms. However, experiments in Chapter 6 demonstrated that although metaheuristic algorithms yield good solutions, they do not scale well to the required number of nodes in a Cloud system.

Alternative solutions and a large number of optimisations can be devised, such as caching computed solutions and then retrieving them based on task similarity, multiple concurrent schedulers working on a single data store, and pre-allocating resources for the whole task batches (Verma et al., 2015). However, these solutions and optimisations still incur substantial computational costs, and it is inevitable that any model where the head node processes all scheduling logic by itself will eventually work less effectively when the cluster size grows and the frequency of incoming tasks increases.

The MASB framework offers an alternative approach to task allocations in that all the actual processing of scheduling logic is offloaded to nodes themselves. This framework uses loose coupling at every stage of its scheduling flow, meaning that scheduling decisions are made only on locally-cached knowledge and all communication between nodes is kept to minimum. Each node tries to increase its AS by selecting and offloading tasks, with the assumption being that by bettering individual ASs, the global system performance will be improved. This design also takes advantage of the VM-LM feature, where a running program within a VM instance can be migrated on the fly to an alternative node without stopping a program execution.

Design of this schema created a set of new challenges, such as selecting alternative nodes with limited and non-current knowledge about the state of other nodes, estimating the VM-LM cost of migrating a running program, understanding the classifying and scoring functions of the allocation type of a node, and designing the stateless node-to-node communication protocol, to identify just a few.

In this research, realistic (i.e. pre-recorded) workload traces from GCD were used and were run on the AGOCS framework described above as a very detailed simulation. The costs involved were the substantial computing power required to run experiments as well as time, in that a single simulation run took about a month on a forty-core HPC machine. In order to benchmark the research results, original scheduling decisions made by Google’s Borg scheduler are examined which are also part of GCD traces. This generated statistics such as total resource usage, the number of idle nodes and production-allocated resources.

When examining GCD traces, it is important to note that Google’s engineers did a phenomenal job in first designing and then iteratively improving the Borg system. Incoming tasks are packed very tightly and, although production jobs always have additional resources available to them within defined requirements' limits, the spare resources are efficiently recycled for low priority jobs. Google Cluster has been built upon hardware without direct support for virtualisation, meaning that its orchestrating software design had to accommodate this limitation. This research should be considered an as-if scenario and assumes the availability of the VM-LM feature to shuffle running tasks within a Cluster.

In this research, there was only limited success in terms of improving the throughput of executed tasks on a simulated computing cell. This was mainly due to the constraints arising from handling RUS. During throughput tests, the MASB achieved a similar level as Google's Borg, understood here as the total number of executed tasks. During the progressively more intensive workload, ASR values indicated a degradation in the quality of allocations so that eventually the throughput could be improved by a margin of 2%. However, MASB could achieve higher scalability and run multiple sizes of examined computing cell without noticeable scalability costs. Simulations up to 100k nodes from GCD were tested, yielding relatively comparable results when run with smaller instances of simulations.

Although the experimental results prove that it is feasible to deploy the presented decentralised architecture in a live environment, there are several possible other improvements, as listed below:

- During experimentations, several nodes remained idle. This effect was a result of iterating only a limited number of nodes while computing a candidate node's set for a given task migration. A potential solution to this issue is a separate size-limited list of relatively under-utilised nodes which would be compulsorily scored each time a BA is issued a GetCandidateNodesRequest request. Such a list could be exchanged separately between BAs;

- The SCS routine (Step 1 in the SAN protocol) is triggered only when the NA detects that its node is overloaded. However, the system could employ a more proactive approach in which the NA would periodically try to offload its tasks in order to improve its AS, even if the node is stable. This would create a secondary mechanism to distribute the load, which would potentially reduce resource utilisation gaps even further. However, this feature would also place additional pressure on BAs and, as such, needs to be carefully balanced;
- In a real-world system it is expected that a number of nodes will experience failure. NA's AI module could maintain a set of blacklisted nodes which repeatedly did not respond to requests. Such a set could be shared with BAs, similar to the way it is implemented in Fuxi (Zhang et al., 2014b), and presented to system administrators.

These suggestions deal with algorithm- and protocol-level details. A list of high-level propositions is presented in the research summary in Chapter 8.

# 8. SUMMARY AND CONCLUSIONS

The chapters above detail a journey from an initial concept, through the research process, the multiple iterations of implementations and experiments and, finally, to achieving a working prototype for the Cloud load balancer. The initial assumption of the project was that existing Cloud management software could be improved by deploying intelligent load balancing routines such as dynamic re-allocations of running tasks, and that task allocation quality could be bettered by adding more refined strategies, such as metaheuristic algorithms. The research began with a presentation in Chapter 2 of a review of existing scheduling software and strategies which helped to define the CRUM in Chapter 3.

CRUM assumed the mobility of tasks being executed on nodes. It also defined the cost incurred on Cloud's infrastructure when a task is re-allocated to an alternative node. There are limitless potential scenarios as to how such procedures could be performed, ranging from simple stop-copy-restart, to snapshotting processes' memory and then restoring it on another machine, or indeed moving the program state using custom routines, to name but a few. This research assumes that Cloud environments are substantially virtualised and that its applications are run within VM instances and, thus, can be migrated by VM-LM process.

Chapter 4 investigated a VM-LM feature which allows the migration of a running VM instance to an alternative machine on the fly, without stopping its execution. The chapter also presents the LMDT formula which can be used to estimate VM-LM migration cost. Since Public Cloud companies are selling their platforms to a range of business customers, they must prioritise the availability of applications run on their platform. The VM-LM feature, although incurring additional costs, ensures the continuous and uninterrupted execution of tasks. The VM-LM process cost definition could be interpreted in many ways, such as performance drop, the extra resources needed for the migration process itself, or the additional expense of energy. In this research, the task migration cost is a direct cost inflicted by VM-LM on a global Cloud network infrastructure that is the size of the data transferred over the network during VM-LM.

Chapter 5 details the design of AGOCS, a high-fidelity Cloud workload simulator that has been developed to test load balancer prototypes using a realistic workload. AGOCS is based on the notion of replaying the workload data available from month-long GCD traces. GCD traces are very rich because they contain many monitored parameters such as the number of CPU cores requested and used, canonical (kernel) and assigned (application) memory requested and used, page cache memory stats, disk I/O time, average cycles per instruction, average memory access per instruction, task priority and local scheduler process priority, to name a few. Additionally, GCD traces provide task constraints and matching attributes on nodes. Those qualities make GCD traces a remarkable source for real-world Cloud computing workload.

## 8.1. RESEARCH SUMMARY

Chapters 2 to 5 discussed research in order to establish a theoretical as well as a practical base for further experimentations. AGOCS provided a solid base for further experiments and simulations with the load balancing prototype. Chapters 6 and 7 described the proposed load balancing designs and the way in which they have evolved during the project. The main consideration of this project was to research the possible improvements to load balancing strategies whilst also maintaining scalability. The main novelty has been the use of the VM-LM feature on a large scale to re-allocate running applications on a Cloud to alternative nodes without stopping their execution.

Orchestrating workload on thousands of machines is a surprisingly complex challenge which has many different dimensions, including:

- the cluster's operations need to be continuous, and applications and programs must be able to recover in the case of failure;
- the cluster's resources should be utilised to their highest extent because every resource utilisation gap potentially blocks a number of tasks from being run, and lowers the overall cluster throughput;
- tasks scheduled on the cluster should be run with minimum delay, and the fairness of cluster utilisation should be maintained;
- the load balancer should be able to handle a variety of tasks, each with unique requirements and rapidly changing resource utilisation levels;
- the load balancer should also be able to proactively adapt to the cluster's configuration changes as new nodes are added, or when existing nodes are taken offline for maintenance or removed.

Those considerations lead to an interesting challenge in which all the parameters identified above needed to be balanced against each other and where, based on business requirements, it was necessary for the Cloud architecture to be able to adapt to variable workloads.

The first prototype of a centralised metaheuristic load balancer was developed as a proof of concept very early in this research. A number of metaheuristic algorithms and their variants were implemented and experimented with, their results being subsequently discussed. Although early trials were promising, the subsequent full-scale experiments using the AGOCS framework demonstrated that while there are scenarios (i.e. small-size clusters) that could benefit from the above approach, it is not a solution that is likely to scale well enough to support large distributed environments with thousands of nodes, such as 12.5k-nodes computing cell from GCD traces.

Given this, an alternative approach was required. Experiences from the first design suggested that metaheuristic algorithms could indeed improve allocation decisions when compared to traditional methods such as Round Robin, FCFS, JSF, 'best-fit' and so on; however, those results are isolated to smaller sets of examined entries. Nonetheless, the literature suggested that a network of software agents could be deployed to offload heavy logic processing to remote machines whilst simultaneously communicating with each other via P2P protocol. Hence, the focus of the research shifted to a decentralised agent-based scheduler in which an agent represents each node and communicates with other agents, trying to keep its node stable.

Subsequently, AGOCS was refactored into a multi-agent system based on the Akka Actors/Streams framework. A new scheduling logic layer was added in the form of a SAN protocol, together with NA's AI module. The resulting system was named MASB, which has been iterated multiple times with many optimisations having been progressively folded into its source code. These include: (i) a subnetwork of BAs, caching the global system state; (ii) original SAS functions were split into SIAS and SRAS functions, with GAIN variants of scoring functions then created; (iii) planning for RUS was introduced; and (iv) NA's AI module was refactored into using metaheuristic algorithms, namely TS variant.

While replaying GCD traces via the AGOCS framework is a reasonably un-obstructing process which consumes little CPU time, the scheduling logic added in MASB extension proved to be a process with excessive computational demands, especially during the development of NA's AI module. As such, MASB instances were deployed on University of Westminster's HPC Cluster (see Appendix C for software and hardware specifications). During peak periods, eighteen out of the twenty available computing nodes were committed to running simulations.

As a benchmark, the project utilised the workload traces of Google's Borg scheduler as recorded in May 2011. These are freely available from the GCD repository. Comparing the results between the original Borg allocations and the MASB simulations suggested a degree of similarity between implemented algorithms, albeit with a few noteworthy differences:

- The centrally managed scheduling mechanism which relied on shared Cluster state information (as implemented in Google's Borg) provided a better overall view concerning the state of all Cluster nodes. The idle nodes were swiftly identified and used, while the decentralised solution had a larger ratio of unutilised nodes.
- The VM-LM feature, together with different scoring schemas for initial and secondary task migrations (i.e. SIAS and SRAS functions), improved the load balance of the cluster by keeping the nodes' resources more proportionally utilised. The ASR vectors, derived from the counts of nodes with a given AS type, were used to monitor the health state of the Cluster. In comparison to Borg, MASB behaved more organically, fluently moving between the higher concentration of tasks under elevated workload and the wider distribution of allocations during low workload periods.
- The experiments resulted in the Cluster performing comparably, with MASB having a small edge and being able to additionally schedule ca. 2.6k tasks per minute in comparison to original GCD worktraces. However, this was at the expense of the increased infrastructure utilisation required to perform task migrations. Additional experiments were performed where the focus was on lowering the total cost of task migrations, and the resulting combination of scoring function variants could reduce this property to a fraction of its original value.
- Most importantly, although throughput could not be substantially improved in this research, the final version of the system could scale multiple times of the original GCD size without a noticeable scalability cost.

The MASB was able to execute its role well even when the experiments simulated massive 100k nodes computing cells. As in throughput testing, the key indicator of the health of a Cluster workload balance was the ASR vectors. Despite increases in the simulated Cluster size, the ASR values remained the same, which demonstrates the greater scalability of proposed solution.

The following sections will: (i) summarise the key findings made during this research; (ii) list the potential applications of the developed technology; and (iii) conclude the project with recommendations for future directions.

## 8.2. KEY FINDINGS

In this section, the most important achievements and key findings of this research are detailed:

- Running a detailed simulation of a Cloud environment is no simple challenge. The sheer number of tasks, the complexity and dynamicity of the requirements, the split between production and non-production applications, the dependencies on other tasks and timings are some of the factors which create a very multifaceted system. AGOCS, a custom fine-grained Cloud workload simulation framework created within this research, is a unique creation on its own;
- VM-LM technology can seamlessly migrate a running task within the VM instance. This technology can be efficiently used to dynamically load balance a Cloud system whilst not inflicting a cost on Cloud network infrastructure. This research provided a LMDT formula which can be used to closely estimate this cost as an alternative to historical data;
- Whilst metaheuristic algorithms can indeed improve task allocation quality, their scalability is insufficient for managing workload on the larger clusters, such as 12.5k-nodes cell from GCD traces. Metaheuristic algorithms can efficiently manage workload on smaller scale and are a good candidate for managing workload on a single node. In the presented solution, metaheuristic algorithms are part of NA's AI routines;

- Decentralised load balancing is a viable approach and, while allocation quality could not be substantially improved in this research, the prototype load balancer was able to manage multiplies of the original GCD workload without a noticeable scalability cost. Under this approach, nodes are represented by NAs, continuously negotiating tasks' allocations between themselves using P2P communications model. This design is supported by a subnetwork of BAs caching the global state of cluster;
- The proposed approach did not eliminate the centralised cluster's state knowledge store; instead, a network of BAs was created which was able to cache the cluster's state knowledge and provide an interface to query it. By design, this knowledge is expected to be outdated, and so is required only during the initial step of the SAN protocol. During the subsequent steps, however, NAs exchange load information between themselves.

## 8.3. APPLICATIONS OF TECHNOLOGY

In late 2017, a team of marketing experts from IBM estimated that the world generates roughly 2.5 million TBs of data per day, with 90% of all data having been created in the past few years alone (IBM, 2017). With novel technologies emerging, new devices and sensors being connected, the data growth rate will accelerate even more. To process such vast data streams, new distributed computing models are being designed. In recent years, the trend for software development has been towards Big Data systems and Machine Learning algorithms, specifically:

- Big Data systems are characterised by a high degree of parallelism. A typical Big Data system design is based on a distributed file system, where nodes have the dual function of storing data as well as processing it. One program in such a system might need to crunch tens of TBs of data split across thousands of nodes. Even with the ideal allocation of Big Data tasks, where every node is processing data from local storage, a single machine would still need to process GBs of data. In order to speed up this time-consuming process, the partial datasets can be split even further and processed on more nodes;
- Machine Learning is yet another rapidly developing area where there is high demand for computing power. The training algorithms for deep neural networks require multiple iterations over datasets, and the recent research is shifting towards greater parallelism (Chung et al., 2017; Sun and Liu, 2018). However, important algorithms such as k-means clustering, alternating least squares, and logistic regression are already very suited to run in parallel (Abadi et al., 2016). Open Source libraries such as Google's TensorFlow and Spark's MLlib, and the affordability of specialised clusters (e.g. Google's Cloud TPU) makes it easy for businesses as well as researchers to utilise those technologies to enhance their offerings. It can certainly be argued that industries will be adopting Machine Learning in order to increase competitiveness.

Therefore, the organisations which employ those modern technologies are highly likely to build computing cells with even more inter-connected nodes in the near future. To manage larger computing cells, more scalable workload orchestration technologies are required, such as the presented MASB prototype. Experiments have shown that MASB design can run a workload on a large Cloud system (100k nodes) with a throughput comparable to Google's Borg system. It should be noted that larger computing cells are also more economical – Google's Borg demonstrated (Verma et al., 2015) that running a mixed-workload consisting of short-lived batch jobs and long-running services as well as production and non-production jobs on the same cluster is not only possible, but allows to utilise of available resources more efficiently. Essentially, resource usage gaps are reduced. Therefore, industries such as financials, health or even government, could make monetary savings if their processing centres were joined and more heterogenous workload was introduced in those clusters. MASB is a good candidate for such an integration.

## 8.4. FUTURE DIRECTIONS

The project was challenging as well as very satisfying. However, no research is ever complete, and this document is by no means a final blueprint for a Cloud load balancer. This research has tackled the problem of load balancing within large Cloud systems, proving that the presented decentralised load balancing solution is feasible and can improve certain aspects of currently used strategies. The prototype has been experimented with on real-world data from GCD traces, and the experimental results demonstrate that the selected strategy is feasible. It has given a strong indication that it would be a viable approach were it implemented on a real Cloud system.

Nevertheless, the list below provides a series of possible next directions and areas that could be further developed in order of perceived importance:

- The MASB prototype does not address fault-tolerance, which is an important aspect of Cloud design. This feature could be realised in multiple ways, such as running cloned instances of tasks, periodically saving process checkpoints, and ensuring the applications' state is synchronised across all its instances. The fault tolerance could also be improved by implementing service/node anti-affinity scheduling strategies where a scheduler tries to allocate replicas of a given service to possibly distanced nodes. In critical failure scenarios, such programs have a greater chance to survive and continue operations. For example, the Kubernetes scheduler implements anti-affinity scoring functions, which gives higher priority to nodes not running services from the same application (Lewis and Oppenheimer, 2017);

- Resource usage quotas per user, group or other entity, would make another welcome feature. This is something which is often present in commercial Cluster schedulers. However, it would also require adding an accountancy module with a decentralised dataset in order to maintain scalability. The same mechanism could be used to throttle the submissions of new tasks so as to not extend the Cluster's capabilities;
- The proposed design does not account for task priorities, meaning that tasks are only split into production and non-production groups. Production tasks have committed resources which, under normal circumstances, are guaranteed to be available. However, during critical system-wide failures, such as a power failure or network infrastructure collapse, the system should degrade gracefully (as opposed to an uncontrolled crash). In scenarios where the current workload cannot be sustained, the system should shut down lower priority tasks first and use the remaining available nodes to offload high-priority tasks;
- In this project, it is assumed that NAs and BAs agents are continuously running without breakdowns. Nevertheless, agents are also a piece of software, meaning that they are prone to bugs and errors. As a possible improvement to detect and restore hung agents, a hierarchy model could be introduced in which an agent supervises a number of other agents and restarts them if necessary. This concept is similar to the Akka Actors implementation (Roestenburg et al., 2015) in which a parent actor manages the failures of its children. Additionally, a hierarchy of BAs could be used to propagate the cluster's state knowledge in a more efficient manner;

- MASB does not attempt to implement locality optimisation when the task's part of a distributed file is processed faster if accessed locally. Currently, GCD task descriptions contain only restrictions which disallow nodes that the task could be executed on. However, adding optional metadata, such as the ID of the distributed file's part, could prioritise a set of nodes and improve the overall cluster performance. This functionality is featured in some of the Big Data frameworks;
- Even though the experimental results presented are of good quality, they suggest a number of potential improvements, especially in locating and then scheduling tasks to idle nodes. One possible improvement could be sharing vector idle nodes between all BAs, and then compulsory prioritising them over utilised nodes;
- The LMDT formula presented in Chapter 4 specifically addresses VM-LM impact on Cloud network infrastructure. Other migration costs of tasks were considered marginal since they impacted only individual nodes rather than the Cloud system as a whole. However, given advances in virtualisation technology, such as common VDI standard in addition to progressively better hardware support for virtualisation, more comprehensive future research in that area might be advantageous. It should be noted that presented input parameters are for a particular test configuration, and so might need to be re-adjusted for the configuration of a specific Cluster, i.e. hardware, network infrastructure, VM vendor and version, and deployed applications;
- The Cloud architectures' design is moving towards greater use of VCs such as Docker. At the time of writing, Docker does not fully support LM – the integration with CRIU does not allow the migration of a running application to the alternative container on the fly. Instead, the user must copy checkpoint files and restore them on an alternative node (cold migration). However, the available literature describes early experiments with LM feature (Yu and Huan, 2015) and the working prototype was demonstrated in a presentation during the OpenStack Summit 2016 conference in Barcelona (Estes and Murakami, 2016). Once LM becomes the part of mainstream technology, the load balancing strategy presented in this research could be adapted to use VCs;

- MASB estimates the task migration cost, and considers this value when selecting which tasks to migrate out from a node. However, it does not calculate the fact that neighbouring nodes (e.g. those in the same server rack) might offer much faster transfer rates than more remote nodes. Therefore, adjusting the task migration cost by the nodes' distances could improve the overall cluster performance;
- Energy efficiency is the next possible area to expand. In its current design, MASB focuses on reducing the cost of task re-allocations necessary to keep the Cloud system stable. However, this approach could be shifted to focus on completely offloading idle nodes, at which point the system would be able to power down those nodes in order to save energy.

The suggested directions of future study and possible expansions as listed above have the potential to improve the results of this research. Nevertheless, the ultimate aim of this work was to advance the ability to design a feasible strategy for managing and proactively balancing a workload within a virtualised Cloud system – an objective which has been achieved.

# APPENDICES

## A. DEVELOPMENT ENVIRONMENT

| | |
|---|---|
| **Model** | MacBook Pro11,1 |
| **Operating System** | OS X 10.13.5 (High Sierra) |
| **CPU** | 2.4GHz dual-core Intel Core i5 |
| **Memory** | 8GB 1600 MHz memory |
| **Storage** | 256GB PCIe-based flash storage |
| **Java Virtual Machine** | 1.8.0_111-b14 Oracle (previously Sun Microsystems) |
| **Scala IDE** | IntelliJ IDEA 2017.1.4 (Community Edition) |
| **YourKit Java Profiler** | 2017.02 build 75 |

Table A1: Development environment specifications

## B. SYSTEM DEPENDENCIES

| | |
|---|---|
| **JVM** | OpenJDK 64-Bit Server VM (build 25.91-b14, mixed mode) |
| **Scala** | Scala 2.12.4 |
| **Akka** | akka-actors (2.5.6), akka-streams (2.5.6) |
| **Google's Guava** | guava 23.0 |
| **Apache Commons** | commons-math3 (3.6.1), commons-lang3 (3.5), commons-csv (1.4) |
| **Logback** | logback-classic (1.2.3) |
| **Kyro** | kyro-shaded (4.0.1), chill (0.9.2) |

Table B1: Runtime libraries specifications

## C. UNIVERSITY OF WESTMINSTER HPC CLUSTER

| | |
|---|---|
| **Model** | Dell R730xd |
| **Operating System** | CentOS Linux release 7.2.1511 (Core) |
| **CPU** | 16x Intel E5-2630 v3 |
| **Memory** | 32GB memory |
| **Storage** | 11TB |
| **Networking** | 10Gb Ethernet |
| **Java Virtual Machine** | OpenJDK 64-Bit Server VM (build 25.91-b14, mixed mode) |

Table C1: Head node (March 2016)

| | |
|---|---|
| **Model** | Dell R630 |
| **Operating System** | CentOS Linux release 7.2.1511 (Core) |
| **CPU** | 20x 2.3GHz Intel E5-2650 v3 |
| **Memory** | 96GB memory |
| **Storage** | 1TB |
| **Networking** | 10Gb Ethernet |
| **Java Virtual Machine** | OpenJDK 64-Bit Server VM (build 25.91-b14, mixed mode) |

Table C2: Nodes compute01-20 (March 2016)

## D. VM ALLOCATOR SOURCE CODE

The code below is an application used to measure the impact of WWS size on data size transferred over the network during VM-LM.

```
#include <stdio.h>
#include <stdlib.h>
#include <ctime>

void writeRandomMemory(char* buffer, int memSizeBytes) {
   for (size_t i = 0; i < memSizeBytes; i++) {
       buffer[i] = rand() % 256;
   }
}

int main(int argc, char **argv) {
   printf("VM-Allocation Tester\n");

   if (argc != 4) {
       printf("Usage" \
              "[memory size MB] " \
              "[writable working set size MB] "\
              "[writable working set over-write interval ms] " \
              "[over-writing threads count]\n");
       exit(EXIT_SUCCESS);
   }

   //read arguments
   int memSize = atoi(argv[1]);
   int wwsSize = atoi(argv[2]);
   int wwsInterval = atoi(argv[3]);
   int wwsThreadsCount = atoi(argv[4]);

   if (memSize<=0) {
       printf("Memory Size must be positive!\n");
       exit(EXIT_FAILURE);
   }

   if (wwsSize<=0) {
       printf("Writable Working Set Size must be positive!\n");
       exit(EXIT_FAILURE);
   }

   if (wwsInterval<=0) {
       printf("Writable Working Set Over-Write Interval must be positive!\n");
       exit(EXIT_FAILURE);
   }

   if (wwsThreadsCount<=0) {
       printf("Over-Write Threads Count must be positive!\n");
       exit(EXIT_FAILURE);
   }

   printf("Allocating %i MB memory\n", memSize);
   printf("Writable Working Set Size is %i MB\n", wwsSize);
   printf("Writable Working Set Over-Write Interval %i ms\n", wwsInterval);

   int memSizeBytes = memSize * 1024 * 1024 / sizeof(char);
   int wwsSizeBytes = wwsSize * 1024 * 1024 / sizeof(char);

   char* buffer = (char*) malloc(memSizeBytes);

   if (buffer == 0) {
       printf("Cannot allocate memory!\n");
       exit(EXIT_FAILURE);
   }

   //write all memory
   printf("Over-Writting all memory (%i bytes)\n", memSizeBytes);
   writeRandomMemory(buffer, memSizeBytes);
   printf("Done\n");

   for (int i=0; i<wwsThreadsCount; i++) {
       int uid = fork();
```

```
        if (uid<0) {
            printf("Cannot fork!\n");
            exit(EXIT_FAILURE);
        }

        //create timespec structure for nanosleep
        struct timespec tim;
        tim.tv_sec = rand() % 10;
        tim.tv_nsec = 0;

        if (uid>0) {
            //random delay (threads won't start on the same memory)
            nanosleep(&tim, NULL);

            printf("%i: Over-writing Thread started\n", i);
            tim.tv_sec = wwsInterval / 1000;
            tim.tv_nsec = wwsInterval % 1000 * 1000000;

            //keep writing random values to wws buffer
            for (;;) {
                printf("%i: Over-writing Writable Working Set (%i bytes)\n", i, wwsSizeBytes);
                writeRandomMemory(buffer, wwsSizeBytes);

                //sleep
                nanosleep(&tim, NULL);
            }
        }
    }

    //endlessly wait for kill signal
    while (wait(1)>0) {
        /* no-op */;
    }

    return EXIT_SUCCESS;
}
```

## E. TEST UNITS SAMPLE

The code below is a sample of the MASB unit tests package which is used to ensure the correctness of the implemented solution.

```
package com.masb.domain

import org.scalatest.{FlatSpec,Matchers}

class TaskConstraintsTest extends FlatSpec with Matchers {
  "EqualAttributeConstraint" should "be implemented correctly" in {

      EqualAttributeConstraint("attribute 1", "value A").checkConstraint(
      NodeAttributes(Map(("attribute 1", "value A")))) should be(true)

      EqualAttributeConstraint("attribute 1", "value A").checkConstraint(
      NodeAttributes(Map(("attribute 1", "value B")))) should be(false)

      EqualAttributeConstraint("attribute 1", "value A").checkConstraint(
      NodeAttributes(Map(("attribute 1", "")))) should be(false)

      EqualAttributeConstraint("attribute 1", "value A").checkConstraint(
      NodeAttributes.NONE) should be(false)

      EqualAttributeConstraint("attribute 1", "value A").checkConstraint(
      NodeAttributes(Map(("attribute 1", "")))) should be(false)

      EqualAttributeConstraint("attribute 1", "").checkConstraint(
      NodeAttributes.NONE) should be(true)

      EqualAttributeConstraint("attribute 1", "").checkConstraint(
      NodeAttributes(Map(("attribute 1", "")))) should be(true)

      EqualAttributeConstraint("attribute 1", "").checkConstraint(
      NodeAttributes(Map(("attribute 2", "value A")))) should be(true)
  }
```

```
    "NotEqualAttributeConstraint" should "be implemented correctly" in {

        NotEqualAttributeConstraint("attribute 1", "value A").checkConstraint(
        NodeAttributes(Map(("attribute 1", "value A")))) should be(false)

        NotEqualAttributeConstraint("attribute 1", "value A").checkConstraint(
        NodeAttributes(Map(("attribute 1", "value B")))) should be(true)

        NotEqualAttributeConstraint("attribute 1", "value A").checkConstraint(
        NodeAttributes(Map(("attribute 1", "")))) should be(true)

        NotEqualAttributeConstraint("attribute 1", "value A").checkConstraint(
        NodeAttributes.NONE) should be(true)

        NotEqualAttributeConstraint("attribute 1", "value A").checkConstraint(
        NodeAttributes(Map(("attribute 1", "")))) should be(true)

        NotEqualAttributeConstraint("attribute 1", "").checkConstraint(
        NodeAttributes.NONE) should be(false)

        NotEqualAttributeConstraint("attribute 1", "").checkConstraint(
        NodeAttributes(Map(("attribute 1", "")))) should be(false)
    }

    "LessThanAttributeConstraint" should "be implemented correctly" in {

        LessThanAttributeConstraint("attribute 1", 10).checkConstraint(
        NodeAttributes(Map(("attribute 1", "10")))) should be(false)

        LessThanAttributeConstraint("attribute 1", 10).checkConstraint(
        NodeAttributes(Map(("attribute 1", "9")))) should be(true)

        LessThanAttributeConstraint("attribute 1", 10).checkConstraint(
        NodeAttributes(Map(("attribute 1", "99")))) should be(false)

        LessThanAttributeConstraint("attribute 1", 10).checkConstraint(
        NodeAttributes(Map(("attribute 1", "11")))) should be(false)

        LessThanAttributeConstraint("attribute 1", 10).checkConstraint(
        NodeAttributes.NONE) should be(true)
    }

    "GreaterThanOrEqualAttributeConstraint" should "be implemented correctly" in {

        GreaterThanAttributeConstraint("attribute 1", 10).checkConstraint(
        NodeAttributes(Map(("attribute 1", "10")))) should be(false)

        GreaterThanAttributeConstraint("attribute 1", 10).checkConstraint(
        NodeAttributes(Map(("attribute 1", "9")))) should be(false)

        GreaterThanAttributeConstraint("attribute 1", 10).checkConstraint(
        NodeAttributes(Map(("attribute 1", "99")))) should be(true)

        GreaterThanAttributeConstraint("attribute 1", 10).checkConstraint(
        NodeAttributes(Map(("attribute 1", "11")))) should be(true)

        GreaterThanAttributeConstraint("attribute 1", 10).checkConstraint(
        NodeAttributes.NONE) should be(false)
    }

   "BetweenAttributeConstraint" should "be implemented correctly" in {

       BetweenAttributeConstraint("attribute 1", 0, 10).checkConstraint(
       NodeAttributes(Map(("attribute 1", "10")))) should be(false)

       BetweenAttributeConstraint("attribute 1", 5, 10).checkConstraint(
       NodeAttributes(Map(("attribute 1", "9")))) should be(true)

       BetweenAttributeConstraint("attribute 1", 0, 10).checkConstraint(
       NodeAttributes(Map(("attribute 1", "99")))) should be(false)
  }
}
```

## F. CONCURRENT MAP UPDATE OPERATIONS

The code below is an implementation of concurrent map update operations in the MASB package. Although the operations are asynchronously executed and updates are non-blocking, there is no strict guarantee of the order of consecutive modifications on the same value object. In practice, during MASB simulations, less than 0.02% of all 'replaceWith' operations weren't immediately executed.

```
package com.masb.helper

object ConcurrentMapOps {
  implicit class ConcurrentMapOpsImpl[A,B](val map: collection.concurrent.Map[A, B]) {

    @inline
    def replaceWith(key: A, function: B => B): Option[B] = {
      //repeat till replace is successful
      while (true) map.get(key) match {
        case None => return None
        case Some(value) =>
          if (map.replace(key, value, function(value))) {
            return Some(value)  //replace success - return value (exit)
          } else {
            Thread.`yield`()    //replace failure - yield control
          }
      }
      None
    }
  }
}
```

# GLOSSARY

- AS – Allocation Score
- AGOCS – Accurate Google Cloud Simulator
- AMM – Automatic Memory Management
- ASR – Allocation Score Ratios
- BA – Broker Agent
- BNS – Borg Name Service
- CFS – Completely Fair Scheduler
- CIFS – Common Internet File System
- CQ – Circular Queue
- CRIU – Checkpoint/Restore In Userspace tool
- CRUM – Cloud Resource Utilisation Model
- CS – Cooperative Scheduling
- D-RSOP – D-Resource System Optimisation Problem
- DA – Disproportional Allocation
- DAG – Directed Acyclic Graph
- E-PVM – Enhanced Parallel Virtual Machine algorithm
- FCFS – First-Come-First-Serve
- GA – Genetic Algorithm
- GC – Java's Garbage Collector
- GCD – Google Cluster Data project
- GFS – Google File System
- GWP – Google Wide Profiling framework
- HT – Hyper-Threading
- IAAS – Infrastructure As A Service
- IOPS – I/O Operations Per Second
- JVM – Java Virtual Machine
- LMDT – Live Migration Data Transfer formula
- MASB – Multi-Agent System Balancer
- MLFQ – Multilevel Feedback Queue
- NA – Node Agent
- NAS – Network Attached Storage device
- NUMA – Non-Uniform Memory Access
- OS – Operating System
- P2P – Point-to-point communication
- PA – Proportional Allocation

- PAAS – Platform As A Service
- QA – Quantum Annealing
- RCPSP – Resource-Constrained Project Scheduling Problem
- RDD – Resilient Distributed Datasets
- RPC – Remote Procedure Call
- RUS – Resource Usage Spike
- SA – Simulated Annealing
- SAAS – Software As A Service
- SAL – Service Allocation Lifecycle
- SAN – Service Allocation Negotiation protocol
- SAS – Service Allocation Score function
- SCS – Select Candidate Services routine
- SGA – Seeded Genetic Algorithm
- SGA-Greedy – Genetic Algorithm seeded by Greedy
- SGA-SA – Genetic Algorithm seeded by Simulated Annealing
- SGA-TS – Genetic Algorithm seeded by Tabu Search
- SIAS – Service Initial Allocation Score function
- SJF – Shortest Job First
- SLURM – Simple Linux Utility for Resource Management
- SRAS – Service Re-allocation Score function
- STA – Super Tight Allocation
- STC – System Transformation Cost
- TA – Tight Allocation
- TPU – Tensor Processing Unit
- TS – Tabu Search
- UCAAS - Unified Communication as a Service
- VC – Virtual Container
- VDI – Virtual Disk Image
- VM – Virtual Machine
- VM-LM – Virtual Machine Live Migration
- WFQ – Weighted Fair Queueing
- WWS – Writable Working Set memory
- YARN – Yet Another Resource Negotiator
- YKJP – YourKit Java Profiler tool

# LIST OF REFERENCES

"10 Key Marketing Trends for 2017 and Ideas for Exceeding Customer Expectations." IBM Marketing Cloud. November 29, 2017. Available from: https://www-01.ibm.com/common/ssi/cgi-bin/ssialias?htmlfid=WRL12345USEN Retrieved June 22, 2018.

"Apache Aurora." Aurora. Available from: http://aurora.apache.org/ Retrieved December 5, 2018. Version 0.19.0.

"Developer Survey Results 2017." Stack Overflow. May 13, 2017. Available from: https://news.netcraft.com/archives/2017/11/21/november-2017-web-server-survey.html Retrieved November 11, 2017.

"Marathon: A container orchestration platform for Mesos and DC/OS." Mesosphere, Inc. January 10, 2018. Available from: https://mesosphere.github.io/marathon/ Retrieved February 7, 2018.

"Maui Administrator's Guide." Adaptive Computing Enterprises, Inc. May 16, 2002. Available from: http://docs.adaptivecomputing.com/maui/pdf/mauiadmin.pdf Retrieved November 5, 2014. Version 3.2.

"Nomad - Easily Deploy Applications at Any Scale", HashiCorp. Available from: https://www.nomadproject.io Retrieved March 19, 2018. Version 0.7.1.

"November 2017 Web Server Survey." Netcraft, Web Server Survey. November 21, 2017. Available from: https://news.netcraft.com/archives/2017/11/21/november-2017-web-server-survey.html Retrieved November 24, 2017.

"Top500 List - November 2017". TOP500 Project. November, 2017. Available from: https://www.top500.org/lists/2017/11/ Retrieved November 17, 2017.

"TORQUE Resource Manager. Administration Guide 5.1.2." Adaptive Computing Enterprises, Inc. November 2015. Available from: http://docs.adaptivecomputing.com/torque/5-1-2/torqueAdminGuide-5.1.2.pdf Retrieved November 15, 2016.

"VirtualBox User Manual." Oracle Corporation. Available from: http://download.virtualbox.org/virtualbox/UserManual.pdf Retrieved April 30, 2018. Version 5.2.12.

Abadi, Martín, Paul Barham, Jianmin Chen, Zhifeng Chen, Andy Davis, Jeffrey Dean, Matthieu Devin et al. "TensorFlow: A System for Large-Scale Machine Learning." In 12th USENIX Symposium on Operating Systems Design and Implementation, vol. 16, pp. 265-283. 2016.

Abali, Bulent, Canturk Isci, Jeffrey O. Kephart, Suzanne K. McIntosh, and Dipankar Sarma. "Live virtual machine migration quality of service." U.S. Patent 9,619,258, issued April 11, 2017.

Abdul-Rahman, Omar Arif, and Kento Aida. "Towards understanding the usage behavior of Google cloud users: the mice and elephants phenomenon." In Cloud Computing Technology and Science (CloudCom), 2014 IEEE 6th International Conference on, pp. 272-277. IEEE, 2014.

Agnetis, Allesandro, Pitu B. Mirchandani, Dario Pacciarelli, and Andrea Pacifici. "Scheduling problems with two competing agents." Operations research 52, no. 2 (2004): 229-242.

Akoush, Sherif, Ripduman Sohan, Andrew Rice, Andrew W. Moore, and Andy Hopper. "Predicting the performance of virtual machine migration." In Modeling, Analysis & Simulation of Computer and Telecommunication Systems (MASCOTS), 2010 IEEE International Symposium on, pp. 37-46. IEEE, 2010.

Amir, Yair, Baruch Awerbuch, Amnon Barak, R. Sean Borgstrom, and Arie Keren. "An opportunity cost approach for job assignment in a scalable computing cluster." IEEE Transactions on parallel and distributed Systems 11, no. 7 (2000): 760-768.

Anderson, Jennifer, Lance Berc, George Chrysos, Jeffrey Dean, Sanjay Ghemawat, Jamey Hicks, Shun-Tak Leung et al. "Transparent, Low-Overhead Profiling on Modern Processors." In Proceedings of the Workshop on Profile and Feedback-Directed Compilation. 1998.

Armbrust, Michael, Armando Fox, Rean Griffith, Anthony D. Joseph, Randy H. Katz, Andrew Konwinski, Gunho Lee et al. Above the clouds: A berkeley view of cloud computing. Vol. 17. Technical Report UCB/EECS-2009-28, Department of Electrical Engineering and Computer Sciences, University of California, Berkeley, 2009.

Arpaci-Dusseau, Remzi H., and Andrea C. Arpaci-Dusseau. "Operating systems: Three easy pieces." Arpaci-Dusseau Books, 2015.

Ausiello, Giorgio, Pierluigi Crescenzi, Giorgio Gambosi, Viggo Kann, Alberto Marchetti-Spaccamela, and Marco Protasi. Complexity and approximation: Combinatorial optimization problems and their approximability properties. Springer Science & Business Media, 2012.

Baker, Kenneth R., and J. Cole Smith. "A multiple-criterion model for machine scheduling." Journal of scheduling 6, no. 1 (2003): 7-16.

Barham, Paul, Boris Dragovic, Keir Fraser, Steven Hand, Tim Harris, Alex Ho, Rolf Neugebauer, Ian Pratt, and Andrew Warfield. "Xen and the art of virtualization." In ACM SIGOPS operating systems review, vol. 37, no. 5, pp. 164-177. ACM, 2003.

Barroso, Luiz André, Jeffrey Dean, and Urs Hölzle. "Web search for a planet: The Google cluster architecture." Micro, IEEE 23, no. 2 (2003): 22-28.

Barsness, Eric L., David L. Darrington, Ray L. Lucas, and John M. Santosuosso. "Distributed job scheduling in a multi-nodal environment." U.S. Patent 8,645,745, issued February 4, 2014.

Becchetti, L, Stefano Leonardi, Alberto Marchetti-Spaccamela, Guido Schäfer, and Tjark Vredeveld. (2006) "Average-case and smoothed competitive analysis of the multilevel feedback algorithm." Mathematics of Operations Research 31, no. 1: 85-108.

Bedra, Aaron. "Getting started with google app engine and clojure." IEEE Internet Computing 14, no. 4 (2010): 85.

Beitch, Aaron, Brandon Liu, Timothy Yung, Rean Griffith, Armando Fox, and David A. Patterson. "Rain: A workload generation toolkit for cloud computing applications." University of California, Tech. Rep. UCB/EECS-2010-14 (2010).

Bernstein, David, Erik Ludvigson, Krishna Sankar, Steve Diamond, and Monique Morrow. "Blueprint for the intercloud-protocols and formats for cloud computing interoperability." In Internet and Web Applications and Services, 2009. ICIW'09. Fourth International Conference on, pp. 328-336. IEEE, 2009.

Bigham, John, and Lin Du. "Cooperative negotiation in a multi-agent system for real-time load balancing of a mobile cellular network." In Proceedings of the second international joint conference on Autonomous agents and multiagent systems, pp. 568-575. ACM, 2003.

Blagodurov, Sergey, Sergey Zhuravlev, Alexandra Fedorova, and Ali Kamali. "A case for NUMA-aware contention management on multicore systems." In Proceedings of the 19th international conference on Parallel architectures and compilation techniques, pp. 557-558. ACM, 2010.

Boctor, Fayer F. "Some efficient multi-heuristic procedures for resource-constrained project scheduling." European journal of operational research 49, no. 1 (1990): 3-13.

Bode, Brett, David M. Halstead, Ricky Kendall, Zhou Lei, and David Jackson. "The Portable Batch Scheduler and the Maui Scheduler on Linux Clusters." In Annual Linux Showcase & Conference. 2000.

Bonald, Thomas, Laurent Massoulié, Alexandre Proutiere, and Jorma Virtamo. "A queueing analysis of max-min fairness, proportional fairness and balanced fairness." Queueing systems 53, no. 1 (2006): 65-84.

Bouleimen, K. and Lecocq, H. "A new efficient simulated annealing algorithm for the resource-constrained project scheduling problem and its multiple mode version." European Journal of Operational Research 149, no. 2 (2003): 268-281.

Boutin, Eric, Jaliya Ekanayake, Wei Lin, Bing Shi, Jingren Zhou, Zhengping Qian, Ming Wu, and Lidong Zhou. "Apollo: Scalable and Coordinated Scheduling for Cloud-Scale Computing." In *OSDI*, vol. 14, pp. 285-300. 2014.

Brazier, Frances MT, Frank Cornelissen, Rune Gustavsson, Catholijn M. Jonker, Olle Lindeberg, Bianca Polak, and Jan Treur. "A multi-agent system performing one-to-many negotiation for load balancing of electricity use." Electronic Commerce Research and Applications 1, no. 2 (2002): 208-224.

Brenner, Walter, Rüdiger Zarnekow, and Hartmut Wittig. "Intelligent software agents: foundations and applications." Springer Science & Business Media, 2012.

Brooks, Chris, Brian Tierney, and William Johnston. "JAVA agents for distributed system management." LBNL Report (1997).

Brucker, Peter, and Sigrid Knust. "Lower bounds for resource-constrained project scheduling problems." European Journal of Operational Research 149, no. 2 (2003): 302-313.

Bu, Yingyi, Bill Howe, Magdalena Balazinska, and Michael D. Ernst. "HaLoop: Efficient iterative data processing on large clusters." Proceedings of the VLDB Endowment 3, no. 1-2 (2010): 285-296.

Bulpin, James R. "Operating system support for simultaneous multithreaded processors." No. UCAM-CL-TR-619. University of Cambridge, Computer Laboratory, 2005.

Burnett, R. Michael Haines, Calvin Pettiecord, Bryan DiGiorgio, Darren Molz, and Scott Koerner. "Supporting domain variation within a cloud provided multitenant unified communications environment." U.S. Patent Application 13/180,773, filed July 12, 2011.

Burns, Brendan, Brian Grant, David Oppenheimer, Eric Brewer, and John Wilkes. "Borg, Omega, and Kubernetes." Communications of the ACM 59, no. 5 (2016): 50-57.

Buyya, Rajkumar. "High Performance Cluster Computing: Architectures and Systems, Volume I." Prentice Hall, Upper SaddleRiver, NJ, USA 1 (1999): 999.

Buyya, Rajkumar, Chee Shin Yeo, Srikumar Venugopal, James Broberg, and Ivona Brandic. "Cloud computing and emerging IT platforms: Vision, hype, and reality for delivering computing as the 5th utility." Future Generation computer systems 25, no. 6 (2009): 599-616.

Byrne, Debora Jean, John Mark McConaughy, Shaw-Ben Shi, Chin-Long Shu, and Trung Minh Tran. "Reverse string indexing in a relational database for wildcard searching." U.S. Patent 6,199,062, issued March 6, 2001.

Cabri, Giacomo, Luca Ferrari, Letizia Leonardi, and Raffaele Quitadamo. "Strong agent mobility for aglets based on the ibm jikesrvm." In Proceedings of the 2006 ACM symposium on Applied computing, pp. 90-95. ACM, 2006.

Calheiros, Rodrigo N., Rajiv Ranjan, Anton Beloglazov, César AF De Rose, and Rajkumar Buyya. "CloudSim: a toolkit for modeling and simulation of cloud computing environments and evaluation of resource provisioning algorithms." Software: Practice and experience 41, no. 1 (2011): 23-50.

Calheiros, Rodrigo N., Marco AS Netto, César AF De Rose, and Rajkumar Buyya. "EMUSIM: an integrated emulation and simulation environment for modeling, evaluation, and validation of performance of cloud computing applications." Software: Practice and Experience 43, no. 5 (2013): 595-612.

Callau-Zori, Mar, Lavinia Samoila, Anne-Cécile Orgerie, and Guillaume Pierre. "An experiment-driven energy consumption model for virtual machine management systems." Sustainable Computing: Informatics and Systems (2017).

Campbell, Matthew. "Distributed Scheduler Hell." DigitalOcean. SREcon17 Asia/Australia. May 24, 2017.

Cao, Junwei, Daniel P. Spooner, Stephen A. Jarvis, and Graham R. Nudd. "Grid load balancing using intelligent agents." Future generation computer systems 21, no. 1 (2005): 135-149.

Castelfranchi, Cristiano. "Guarantees for autonomy in cognitive agent architecture." In International Workshop on Agent Theories, Architectures, and Languages, pp. 56-70. Springer, Berlin, Heidelberg, 1994.

Chang, Fay, Jeffrey Dean, Sanjay Ghemawat, Wilson C. Hsieh, Deborah A. Wallach, Mike Burrows, Tushar Chandra, Andrew Fikes, and Robert E. Gruber. "Bigtable: A distributed storage system for structured data." ACM Transactions on Computer Systems (TOCS) 26, no. 2 (2008): 4.

Chavez, Anthony, Alexandros Moukas, and Pattie Maes. "Challenger: A multi-agent system for distributed resource allocation." In Proceedings of the first international conference on Autonomous agents, pp. 323-331. ACM, 1997.

Che, Jianhua, Yong Yu, Congcong Shi, and Weimin Lin. "A synthetical performance evaluation of openVZ, Xen and KVM." In Services Computing Conference (APSCC), 2010 IEEE Asia-Pacific, pp. 587-594. IEEE, 2010.

Chen, Yanpei, Sara Alspaugh, and Randy H. Katz. Design insights for MapReduce from diverse production workloads. No. UCB/EECS-2012-17. California Unversity Berkley, Department of Electrical Engineering and Computer Science, 2012.

Chierici, Andrea, and Riccardo Veraldi. "A quantitative comparison between Xen and KVM." In Journal of Physics: Conference Series, vol. 219, no. 4, p. 042005. IOP Publishing, 2010.

Chisnall, David. "The definitive guide to the Xen hypervisor." Pearson Education, 2008.

Chung, I-Hsin, Tara N. Sainath, Bhuvana Ramabhadran, Michael Picheny, John Gunnels, Vernon Austel, Upendra Chauhari, and Brian Kingsbury. "Parallel deep neural network training for big data on blue gene/q." IEEE Transactions on Parallel and Distributed Systems 28, no. 6 (2017): 1703-1714.

Chvatal, Vasek. "A greedy heuristic for the set-covering problem." Mathematics of operations research 4, no. 3 (1979): 233-235.

Clark, Christopher, Keir Fraser, Steven Hand, Jacob Gorm Hansen, Eric Jul, Christian Limpach, Ian Pratt, and Andrew Warfield. "Live migration of virtual machines." In Proceedings of the 2nd Conference on Symposium on Networked Systems Design & Implementation-Volume 2, pp. 273-286. USENIX Association, 2005.

Clark, Jack "5 Numbers That Illustrate the Mind-Bending Size of Amazon's Cloud." Bloomberg Technology. November 14, 2014.

Coffman Jr, Edward G., Michael R. Garey, and David S. Johnson. "Approximation algorithms for bin packing: A survey." In Approximation algorithms for NP-hard problems, pp. 46-93. PWS Publishing Co., 1996.

Cook, Stephen A. "The complexity of theorem-proving procedures." In Proceedings of the third annual ACM symposium on Theory of computing, pp. 151-158. ACM, 1971.

Cooper, Dale F. "Heuristics for scheduling resource-constrained projects: An experimental investigation." Management Science 22, no. 11 (1976): 1186-1194.

Corbató, Fernando J., Marjorie Merwin-Daggett, and Robert C. Daley. "An experimental time-sharing system." In Proceedings of the May 1-3, 1962, spring joint computer conference, pp. 335-344. ACM, 1962.

Corbet, Jonathan. "The staircase scheduler." LWN.net. June 2, 2004. Available from: https://lwn.net/Articles/87729/ Retrieved September 25, 2017.

Corbet, Jonathan. "The Rotating Staircase Deadline Scheduler." LWN.net. March 6, 2007. Available from: https://lwn.net/Articles/224865/ Retrieved September 25, 2017.

Corbett, James C., Jeffrey Dean, Michael Epstein, Andrew Fikes, Christopher Frost, Jeffrey John Furman, Sanjay Ghemawat et al. "Spanner: Google’s globally distributed database." ACM Transactions on Computer Systems (TOCS) 31, no. 3 (2013): 8.

Corsava, Sophia, and Vladimir Getov. "Intelligent architecture for automatic resource allocation in computer clusters." In Parallel and Distributed Processing Symposium, 2003. Proceedings. International, pp. 8-pp. IEEE, 2003.

Dargie, Waltenegus. "Estimation of the cost of vm migration." In Computer Communication and Networks (ICCCN), 2014 23rd International Conference on, pp. 1-8. IEEE, 2014.

Davis, Edward W., and James H. Patterson. "A comparison of heuristic and optimum solutions in resource-constrained project scheduling." Management science 21, no. 8 (1975): 944-955.

Dean, Jeffrey, and Sanjay Ghemawat. "MapReduce: a flexible data processing tool." Communications of the ACM 53, no. 1 (2010): 72-77.

Debels, Dieter, Bert De Reyck, Roel Leus, and Mario Vanhoucke. "A hybrid scatter search/electromagnetism meta-heuristic for project scheduling." European Journal of Operational Research 169, no. 2 (2006): 638-653.

Demassey, Sophie, Christian Artigues, and Philippe Michelon. "Constraint-propagation-based cutting planes: An application to the resource-constrained project scheduling problem." INFORMS Journal on computing 17, no. 1 (2005): 52-65.

Demeulemeester, Erik, and Willy Herroelen. "A branch-and-bound procedure for the multiple resource-constrained project scheduling problem." Management science 38, no. 12 (1992): 1803-1818.

Deshpande, Umesh, and Kate Keahey. "Traffic-sensitive live migration of virtual machines." Future Generation Computer Systems 72 (2017): 118-128.

Di, Sheng, Derrick Kondo, and Walfredo Cirne. "Characterization and comparison of cloud versus grid workloads." In Cluster Computing (CLUSTER), 2012 IEEE International Conference on, pp. 230-238. IEEE, 2012.

Drepper, Ulrich. "What every programmer should know about memory." Red Hat, Inc. 11 (2007): 2007.

Du, Yuyang, Hongliang Yu, Guangyu Shi, Jian Chen, and Weimin Zheng. "Microwiper: efficient memory propagation in live migration of virtual machines." In Parallel Processing (ICPP), 2010 39th International Conference on, pp. 141-149. IEEE, 2010.

Dua, Rajdeep, A. Reddy Raja, and Dharmesh Kakadia. "Virtualization vs Containerization to Support PaaS." In Cloud Engineering (IC2E), 2014 IEEE International Conference on, pp. 610-614. IEEE, 2014.

Eddy, YS Foo, Hoay Beng Gooi, and Shuai Xun Chen. "Multi-agent system for distributed management of microgrids." IEEE Transactions on power systems 30, no. 1 (2015): 24-34.

El-Sayed, Nosayba, Hongyu Zhu, and Bianca Schroeder. "Learning from Failure Across Multiple Clusters: A Trace-Driven Approach to Understanding, Predicting, and Mitigating Job Terminations." In Distributed Computing Systems (ICDCS), 2017 IEEE 37th International Conference on, pp. 1333-1344. IEEE, 2017.

Estes, Phil, and Shaun Murakami. "Live Container Migration on OpenStack." OpenStack Summit Barcelona 2016. October 25, 2016.

Etsion, Yoav, and Dan Tsafrir. "A short survey of commercial cluster batch schedulers." School of Computer Science and Engineering, The Hebrew University of Jerusalem 44221 (2005): 2005-13.

Fanjul-Peyro, Luis, Federico Perea, and Rubén Ruiz. "Models and matheuristics for the unrelated parallel machine scheduling problem with additional resources." European Journal of Operational Research 260, no. 2 (2017): 482-493.

Feitelson, Dror G., Dan Tsafrir, and David Krakov. "Experience with using the parallel workloads archive." Journal of Parallel and Distributed Computing 74, no. 10 (2014): 2967-2982.

Feng, Xiujie, Jianxiong Tang, Xuan Luo, and Yaohui Jin. "A performance study of live VM migration technologies: VMotion vs XenMotion." In Asia Communications and Photonics Conference and Exhibition, p. 83101B. Optical Society of America, 2011.

Fortnow, Lance. "The status of the P versus NP problem." Communications of the ACM 52, no. 9 (2009): 78-86.

Foster, Ian, and Carl Kesselman. "Globus: A metacomputing infrastructure toolkit." The International Journal of Supercomputer Applications and High Performance Computing 11, no. 2 (1997): 115-128.

Foster, Ian, Carl Kesselman, and Steven Tuecke. "The anatomy of the grid: Enabling scalable virtual organizations." The International Journal of High Performance Computing Applications 15, no. 3 (2001): 200-222.

Frieze, Alan M. "On the Lagarias-Odlyzko algorithm for the subset sum problem." SIAM Journal on Computing 15, no. 2 (1986): 536-539.

Gabriel, Edgar, Graham E. Fagg, George Bosilca, Thara Angskun, Jack J. Dongarra, Jeffrey M. Squyres, Vishal Sahay et al. "Open MPI: Goals, concept, and design of a next generation MPI implementation." In European Parallel Virtual Machine/Message Passing Interface Users' Group Meeting, pp. 97-104. Springer Berlin Heidelberg, 2004.

Ganapathi, Archana, Yanpei Chen, Armando Fox, Randy Katz, and David Patterson. "Statistics-driven workload modeling for the cloud." In Data Engineering

Workshops (ICDEW), 2010 IEEE 26th International Conference on, pp. 87-92. IEEE, 2010.

Garg, Saurabh Kumar, and Rajkumar Buyya. "Networkcloudsim: Modelling parallel applications in cloud simulations." In Utility and Cloud Computing (UCC), 2011 Fourth IEEE International Conference on, pp. 105-113. IEEE, 2011.

Gensereth, Michael R., and Steven P. Ketchpel. "Software agents." Communications of the ACM 37, no. 7 (1994): 48.

Gentzsch, Wolfgang. "Sun grid engine: Towards creating a compute power grid." In Cluster Computing and the Grid, 2001. Proceedings. First IEEE/ACM International Symposium on, pp. 35-36. IEEE, 2001.

Ghemawat, Sanjay, Howard Gobioff, and Shun-Tak Leung. "The Google file system." In ACM SIGOPS operating systems review, vol. 37, no. 5, pp. 29-43. ACM, 2003.

Ghemawat, Sanjay, and Paul Menage. "Tcmalloc: Thread-caching malloc." Google performance tools. November 15, 2005.

Glover, Fred. "Future paths for integer programming and links to artificial intelligence." Computers & operations research 13, no. 5 (1986): 533-549.

Glover, Fred. "Tabu search—part I." ORSA Journal on computing 1, no. 3 (1989): 190-206.

Gog, I. "Dron: An Integration Job Scheduler." Imperial College London (2012).

Goodwin, Richard. "Formalizing properties of agents." Journal of Logic and Computation 5, no. 6 (1995): 763-781.

Grimshaw, Andrew S. "The Mentat run-time system: support for medium grain parallel computation." In Distributed Memory Computing Conference, 1990., Proceedings of the Fifth, vol. 2, pp. 1064-1073. IEEE, 1990.

Grimshaw, Andrew S., William A. Wulf, James C. French, Alfred C. Weaver, and Paul Reynolds Jr. "Legion: The next logical step toward a nationwide virtual computer. " Technical Report CS-94-21, University of Virginia, 1994.

Groves, Taylor, Jeff Knockel, and Eric Schulte. "BFS vs. CFS - Scheduler Comparison." The University of New Mexico, 11 December 2009.

Guilfoyle, Christine, and Ellie Warner. "Intelligent agents: The new revolution in software." Ovum, 1994.

Hacker, T. "Toward a reliable cloud computing service." Cloud Computing and software services: Theory and Techniques 139 (2010).

Hamscher, Volker, Uwe Schwiegelshohn, Achim Streit, and Ramin Yahyapour. "Evaluation of job-scheduling strategies for grid computing." Grid Computing—GRID 2000 (2000): 191-202.

Hart, Johnson M. "Win32 systems programming." Addison-Wesley Longman Publishing Co., Inc., 1997.

Helland, Pat, and Harris Ed "Cosmos: Big Data and Big Challenges." Stanford University, October 26, 2011.

Hellerstein, Joseph L., W. Cirne, and J. Wilkes. "Google Cluster Data." Google Research Blog. January 7, 2010.

Hindman, Benjamin, Andy Konwinski, Matei Zaharia, Ali Ghodsi, Anthony D. Joseph, Randy H. Katz, Scott Shenker, and Ion Stoica. "Mesos: A Platform for Fine-Grained Resource Sharing in the Data Center." In NSDI, vol. 11, no. 2011, pp. 22-22. 2011.

Hou, Kai-Yuan, Kang G. Shin, and Jan-Lung Sung. "Application-assisted live migration of virtual machines with Java applications." In Proceedings of the Tenth European Conference on Computer Systems, p. 15. ACM, 2015.

Hu, Wenjin, Andrew Hicks, Long Zhang, Eli M. Dow, Vinay Soni, Hao Jiang, Ronny Bull, and Jeanna N. Matthews. "A quantitative study of virtual machine live migration." In Proceedings of the 2013 ACM cloud and autonomic computing conference, p. 11. ACM, 2013.

Huang, Qiang, Fengqian Gao, Rui Wang, and Zhengwei Qi. "Power consumption of virtual machine live migration in clouds." In Communications and Mobile

Computing (CMC), 2011 Third International Conference on, pp. 122-125. IEEE, 2011.

Ilie, Sorin, and Costin Bădică. "Multi-agent approach to distributed ant colony optimization." Science of Computer Programming 78, no. 6 (2013): 762-774.

Iosup, Alexandru, Hui Li, Mathieu Jan, Shanny Anoep, Catalin Dumitrescu, Lex Wolters, and Dick HJ Epema. "The grid workloads archive." Future Generation Computer Systems 24, no. 7 (2008): 672-686.

Iosup, Alexandru, Simon Ostermann, M. Nezih Yigitbasi, Radu Prodan, Thomas Fahringer, and Dick Epema. "Performance analysis of cloud computing services for many-tasks scientific computing." IEEE Transactions on Parallel and Distributed systems 22, no. 6 (2011): 931-945.

Isard, Michael, Mihai Budiu, Yuan Yu, Andrew Birrell, and Dennis Fetterly. "Dryad: distributed data-parallel programs from sequential building blocks." In ACM SIGOPS operating systems review, vol. 41, no. 3, pp. 59-72. ACM, 2007.

Isard, Michael, Vijayan Prabhakaran, Jon Currey, Udi Wieder, Kunal Talwar, and Andrew Goldberg. "Quincy: fair scheduling for distributed computing clusters." In Proceedings of the ACM SIGOPS 22nd symposium on Operating systems principles, pp. 261-276. ACM, 2009.

Jackson, David, Quinn Snell, and Mark Clement. "Core algorithms of the Maui scheduler." In Workshop on Job Scheduling Strategies for Parallel Processing, pp. 87-102. Springer, Berlin, Heidelberg, 2001.

Jackson, Keith R., Lavanya Ramakrishnan, Krishna Muriki, Shane Canon, Shreyas Cholia, John Shalf, Harvey J. Wasserman, and Nicholas J. Wright. "Performance analysis of high performance computing applications on the amazon web services cloud." In Cloud Computing Technology and Science (CloudCom), 2010 IEEE Second International Conference on, pp. 159-168. IEEE, 2010.

Jain, Abhinivesh, and Niraj Mahajan. "Introduction to Cloud Computing." In The Cloud DBA-Oracle, pp. 3-10. Apress, Berkeley, CA, 2017.

Jin, Hai, Li Deng, Song Wu, Xuanhua Shi, and Xiaodong Pan. "Live virtual machine migration with adaptive, memory compression." In Cluster Computing and Workshops, 2009. CLUSTER'09. IEEE International Conference on, pp. 1-10. IEEE, 2009.

Jin, Hai, Shadi Ibrahim, Tim Bell, Li Qi, Haijun Cao, Song Wu, and Xuanhua Shi. "Tools and technologies for building clouds." In Cloud Computing, pp. 3-20. Springer London, 2010.

Jones, James Patton, and Cristy Brickell. "Second evaluation of job queuing/scheduling software: Phase 1 report." Technical Report NAS-97-013, NASA Ames Research Center, 1997.

Jones, M. Tim. "Inside the Linux 2.6 Completely Fair Scheduler - Providing fair access to CPUs since 2.6.23" In IBM DeveloperWorks. December 15, 2009.

Jouppi, Norman P., Cliff Young, Nishant Patil, David Patterson, Gaurav Agrawal, Raminder Bajwa, Sarah Bates et al. "In-datacenter performance analysis of a tensor processing unit." In Proceedings of the 44th Annual International Symposium on Computer Architecture, pp. 1-12. ACM, 2017.

Józefowska, Joanna, Marek Mika, Rafał Różycki, Grzegorz Waligóra, and Jan Węglarz. "Local search metaheuristics for discrete–continuous scheduling problems." European Journal of Operational Research 107, no. 2: 354-370, 1998.

Józefowska, Joanna, Marek Mika, Rafał Różycki, Grzegorz Waligóra, and Jan Węglarz. "Simulated annealing for multi-mode resource-constrained project scheduling." Annals of Operations Research 102, no. 1-4: 137-155, 2001.

Józefowska, Joanna, Marek Mika, Rafał Różycki, Grzegorz Waligóra, and Jan Węglarz. "A heuristic approach to allocating the continuous resource in discrete–continuous scheduling problems to minimize the makespan." Journal of Scheduling 5, no. 6: 487-499, 2002.

Kalra, Mala, and Sarbjeet Singh. "A review of metaheuristic scheduling techniques in cloud computing." Egyptian informatics journal 16, no. 3 (2015): 275-295.

Kanev, Svilen, Juan Pablo Darago, Kim Hazelwood, Parthasarathy Ranganathan, Tipp Moseley, Gu-Yeon Wei, and David Brooks. "Profiling a warehouse-scale computer." In Computer Architecture (ISCA), 2015 ACM/IEEE 42nd Annual International Symposium on, pp. 158-169. IEEE, 2015.

Kannan, Subramanian, Mark Roberts, Peter Mayes, Dave Brelsford, and Joseph F. Skovira. "Workload management with loadleveler." IBM Redbooks 2, no. 2 (2001).

Karp, Richard M. "Reducibility among combinatorial problems." In Complexity of computer computations, pp. 85-103. Springer, Boston, MA, 1972.

Kavulya, Soila, Jiaqi Tan, Rajeev Gandhi, and Priya Narasimhan. "An analysis of traces from a production mapreduce cluster." In Proceedings of the 2010 10th IEEE/ACM International Conference on Cluster, Cloud and Grid Computing, pp. 94-103. IEEE Computer Society, 2010.

Kay, Judy, and Piers Lauder. "A fair share scheduler." Communications of the ACM 31, no. 1 (1988): 44-55.

Kim, Gu Su, Kyoung-in Kim, and Young Ik Eom. "Dynamic load balancing scheme based on Resource reservation for migration of agent in the pure P2P network environment." In International Conference on AI, Simulation, and Planning in High Autonomy Systems, pp. 538-546. Springer, Berlin, Heidelberg, 2004.

Kliazovich, Dzmitry, Pascal Bouvry, and Samee Ullah Khan. "GreenCloud: a packet-level simulator of energy-aware cloud computing data centers." The Journal of Supercomputing 62, no. 3 (2012): 1263-1283.

Klusáček, Dalibor, and Hana Rudová. "The Use of Incremental Schedule-based Approach for Efficient Job Scheduling." In Sixth Doctoral Workshop on Mathematical and Engineering Methods in Computer Science, 2010.

Klusáček, Dalibor, Václav Chlumský, and Hana Rudová. "Optimizing user oriented job scheduling within TORQUE." In SuperComputing The 25th International Conference for High Performance Computing, Networking, Storage and Analysis (SC'13). 2013.

Klusáček, Dalibor. "MetaCentrum Workload Log." Czech National Infrastructure Grid MetaCentrum. Available from: http://www.fi.muni.cz/~xklusac/index.php?page=meta2009 Retrieved November 15, 2014.

Klusáček, Dalibor, and Boris Parák. "Analysis of Mixed Workloads from Shared Cloud Infrastructure." In Workshop on Job Scheduling Strategies for Parallel Processing, pp. 25-42. Springer, Cham, 2017.

Kochut, Andrzej, Yu Deng, Michael R. Head, Jonathan Munson, Anca Sailer, Hidayatullah Shaikh, Chunqiang Tang et al. "Evolution of the IBM Cloud: Enabling an enterprise cloud services ecosystem." IBM Journal of Research and Development 55, no. 6 (2011): 7-1.

Kolisch, Rainer, and Arno Sprecher. "PSPLIB-a project scheduling problem library: OR software-ORSEP operations research software exchange program." European journal of operational research 96, no. 1 (1997): 205-216.

Kolisch, Rainer, and Sönke Hartmann. "Experimental investigation of heuristics for resource-constrained project scheduling: An update." European journal of operational research 174, no. 1 (2006): 23-37.

Kolivas, Con. "linux-4.8-ck2, MuQSS version 0.114." -ck hacking. October 21, 2016. Available from: https://ck-hack.blogspot.co.uk/2016/10/linux-48-ck2-muqss-version-0114.html Retrieved December 8, 2016.

Krauter, Klaus, Rajkumar Buyya, and Muthucumaru Maheswaran. "A taxonomy and survey of grid resource management systems for distributed computing." Software: Practice and Experience 32, no. 2 (2002): 135-164.

Kulkarni, Sanjeev, Nikunj Bhagat, Maosong Fu, Vikas Kedigehalli, Christopher Kellogg, Sailesh Mittal, Jignesh M. Patel, Karthik Ramasamy, and Siddarth Taneja. "Twitter Heron: Stream processing at scale." In Proceedings of the 2015 ACM SIGMOD International Conference on Management of Data, pp. 239-250. ACM, 2015.

Lamport, Leslie. "The part-time parliament." ACM Transactions on Computer Systems (TOCS) 16, no. 2 (1998): 133-169.

Lang, Willis, and Jignesh M. Patel. (2010) "Energy management for mapreduce clusters." Proceedings of the VLDB Endowment 3, no. 1-2: 129-139.

Lewis, Ian, and David Oppenheimer. "Advanced Scheduling in Kubernetes". Kubernetes.io. Google, Inc. March 31, 2017. Available

https://kubernetes.io/blog/2017/03/advanced-scheduling-in-kubernetes Retrieved January 4, 2018.

Leung, Joseph Y-T. "Handbook of scheduling: algorithms, models, and performance analysis." CRC Press, 2004.

Leung, Joseph Y-T., Michael Pinedo, and Guohua Wan. "Competitive two-agent scheduling and its applications." Operations Research 58, no. 2 (2010): 458-469.

Levin, Leonid A. "Универсальные задачи перебора." Problems of Information Transmission 9, no. 3 (1973): 115-116.

Li, Ang, Xiaowei Yang, Srikanth Kandula, and Ming Zhang. "CloudCmp: comparing public cloud providers." In Proceedings of the 10th ACM SIGCOMM conference on Internet measurement, pp. 1-14. ACM, 2010.

Lim, Andrew, Hong Ma, Brian Rodrigues, Sun Teck Tan, and Fei Xiao. "New meta-heuristics for the resource-constrained project scheduling problem." Flexible Services and Manufacturing Journal 25, no. 1-2 (2013): 48-73.

Limoncelli, Tom, Strata R. Chalup, and Christina J. Hogan. The Practice of Cloud System Administration: Designing and Operating Large Distributed Systems. Vol. 2. Pearson Education, 2014.

Litzkow, Michael J., Miron Livny, and Matt W. Mutka. "Condor-a hunter of idle workstations." In Distributed Computing Systems, 1988., 8th International Conference on, pp. 104-111. IEEE, 1988.

Liu, Haikun, Hai Jin, Cheng-Zhong Xu, and Xiaofei Liao. "Performance and energy modeling for live migration of virtual machines." Cluster computing 16, no. 2 (2013): 249-264.

Liu, Howard T., and John A. Silvester. "Dynamic resource allocation scheme for distributed heterogeneous computer systems." U.S. Patent 5,031,089, issued July 9, 1991.

Liu, Peng, and Lixin Tang. "Two-agent scheduling with linear deteriorating jobs on a single machine." In International Computing and Combinatorics Conference, pp. 642-650. Springer Berlin Heidelberg, 2008.

Liu, Xunyun, and Rajkumar Buyya. "D-Storm: Dynamic Resource-Efficient Scheduling of Stream Processing Applications." In Parallel and Distributed Systems (ICPADS), 2017 IEEE 23rd International Conference on, pp. 485-492. IEEE, 2017.

Long, Qingqi, Jie Lin, and Zhixun Sun. "Agent scheduling model for adaptive dynamic load balancing in agent-based distributed simulations." Simulation Modelling Practice and Theory 19, no. 4 (2011): 1021-1034.

Lozi, Jean-Pierre, Baptiste Lepers, Justin Funston, Fabien Gaud, Vivien Quéma, and Alexandra Fedorova. "The Linux scheduler: a decade of wasted cores." In Proceedings of the Eleventh European Conference on Computer Systems, p. 1. ACM, 2016.

Luo, Jun-Zhou, Jia-Hui Jin, Ai-Bo Song, and Fang Dong. "Cloud computing: architecture and key technologies." Journal of China Institute of Communications 32, no. 7 (2011): 3-21.

Malhotra, Rahul, and Prince Jain. "Study and Comparison of CloudSim simulators in the cloud computing." The SIJ Transactions on Computer Science Engineering & its Applications (2013).

Mao, Ming, and Marty Humphrey. "Auto-scaling to minimize cost and meet application deadlines in cloud workflows." In High Performance Computing, Networking, Storage and Analysis (SC), 2011 International Conference for, pp. 1-12. IEEE, 2011.

Marey, Omar, Jamal Bentahar, Ehsan Khosrowshahi-Asl, Khalid Sultan, and Rachida Dssouli. "Decision making under subjective uncertainty in argumentation-based agent negotiation." Journal of Ambient Intelligence and Humanized Computing 6, no. 3 (2015): 307-323.

Marshall, Nick. "Mastering VMware VSphere 6." John Wiley & Sons, 2015.

Marz, Nathan. "A Storm is coming: more details and plans for release." Engineering Blog. Twitter, Inc. August 4, 2011. Available from: https://blog.twitter.com/engineering/en_us/a/2011/a-storm-is-coming-more-details-and-plans-for-release.html Retrieved July 16, 2018.

Mashtizadeh, Ali, Emré Celebi, Tal Garfinkel, and Min Cai. "The design and evolution of live storage migration in VMware ESX." In USENIX ATC, vol. 11, pp. 1-14. 2011.

Mateescu, Gabriel, Wolfgang Gentzsch, and Calvin J. Ribbens. "Hybrid computing—where HPC meets grid and cloud computing." Future Generation Computer Systems 27, no. 5 (2011): 440-453.

McCullough, John C., Yuvraj Agarwal, Jaideep Chandrashekar, Sathyanarayan Kuppuswamy, Alex C. Snoeren, and Rajesh K. Gupta. "Evaluating the effectiveness of model-based power characterization." In USENIX Annual Technical Conf, vol. 20. 2011.

Mechalas, John. "Performance Impact of Intel® Secure Key on OpenSSL." Intel Corporation. July 24, 2012. Available from: https://software.intel.com/en-us/articles/performance-impact-of-intel-secure-key-on-openssl Retrieved December 22, 2017.

Meisner, David, Christopher M. Sadler, Luiz André Barroso, Wolf-Dietrich Weber, and Thomas F. Wenisch. "Power management of online data-intensive services." In Computer Architecture (ISCA), 2011 38th Annual International Symposium on, pp. 319-330. IEEE, 2011.

Mell, Peter, and Tim Grance. "The NIST definition of cloud computing." National Institute of Standards and Technology. September 2011. NIST Special Publication 800-145.

Melnik, Sergey, Andrey Gubarev, Jing Jing Long, Geoffrey Romer, Shiva Shivakumar, Matt Tolton, and Theo Vassilakis. "Dremel: interactive analysis of web-scale datasets." Proceedings of the VLDB Endowment 3, no. 1-2 (2010): 330-339.

Merkel, Dirk. "Docker: lightweight linux containers for consistent development and deployment." Linux Journal 2014, no. 239 (2014): 2.

Milano, Michela, and Andrea Roli. "MAGMA: a multiagent architecture for metaheuristics." IEEE Transactions on Systems, Man, and Cybernetics, Part B (Cybernetics) 34, no. 2 (2004): 925-941.

Mishra, Asit K., Joseph L. Hellerstein, Walfredo Cirne, and Chita R. Das. "Towards characterizing cloud backend workloads: insights from Google compute clusters." ACM SIGMETRICS Performance Evaluation Review 37, no. 4 (2010): 34-41.

Mobini, MD Mahdi, Masoud Rabbani, M. S. Amalnik, Jafar Razmi, and A. R. Rahimi-Vahed. "Using an enhanced scatter search algorithm for a resource-constrained project scheduling problem." Soft Computing 13, no. 6 (2009): 597-610.

Monteserin, Ariel, J. Andrés Díaz-Pace, Ignacio Gatti, and Silvia Schiaffino. "Agent Negotiation Techniques for Improving Quality-Attribute Architectural Tradeoffs." Advances in Practical Applications of Cyber-Physical Multi-Agent Systems: The PAAMS Collection, vol. 10349 (2017): 183-195.

Montresor, Alberto, Hein Meling, and Ozalp Babaoglu. "Messor: Load-balancing through a swarm of autonomous agents." In AP2PC, vol. 2, pp. 125-137. 2002.

Moreno, Ismael Solis, Peter Garraghan, Paul Townend, and Jie Xu. "An approach for characterizing workloads in google cloud to derive realistic resource utilization models." In Service Oriented System Engineering (SOSE), 2013 IEEE 7th International Symposium on, pp. 49-60. IEEE, 2013.

Murray, Derek G., Malte Schwarzkopf, Christopher Smowton, Steven Smith, Anil Madhavapeddy, and Steven Hand. "CIEL: a universal execution engine for distributed data-flow computing." In Proc. 8th ACM/USENIX Symposium on Networked Systems Design and Implementation, pp. 113-126. 2011.

Naik, Nitin. "Building a virtual system of systems using Docker Swarm in multiple clouds." In Systems Engineering (ISSE), 2016 IEEE International Symposium on, pp. 1-3. IEEE, 2016.

Namjoshi, Jyoti, and Archana Gupte. "Service oriented architecture for cloud based travel reservation software as a service." In Cloud Computing, 2009. CLOUD'09. IEEE International Conference on, pp. 147-150. IEEE, 2009.

Nguyen, Ngoc Thanh, Maria Ganzha, and Marcin Paprzycki. "A consensus-based multi-agent approach for information retrieval in internet." In International Conference on Computational Science, pp. 208-215. Springer, Berlin, Heidelberg, 2006.

Noel, Karen, and Michael Tsirkin. "Memory duplication by destination host in virtual machine live migration." U.S. Patent 9,459,902, issued October 4, 2016.

Nong, Q. Q., T. C. E. Cheng, and C. T. Ng. "Two-agent scheduling to minimize the total cost." European Journal of Operational Research 215, no. 1 (2011): 39-44.

Nwana, Hyacinth S. "Software agents: An overview." The knowledge engineering review 11, no. 3 (1996): 205-244.

Obe, Regina, and Leo Hsu. "PostgreSQL: Up and Running: a Practical Guide to the Advanced Open Source Database." O'Reilly Media, Inc. 2017.

Odersky, Martin, Lex Spoon, and Bill Venners. "Programming in Scala: Updated for Scala 2.12." Artima, Inc. 2016.

Oi, Hitoshi. "A preliminary workload analysis of specjvm2008." In Computer Engineering and Technology, 2009. ICCET'09. International Conference on, vol. 2, pp. 13-19. IEEE, 2009.

Othman, Sarah Ben, Hayfa Zgaya, Mariagrazia Dotoli, and Slim Hammadi. "An agent-based Decision Support System for resources' scheduling in Emergency Supply Chains." Control Engineering Practice 59 (2017): 27-43.

Pabla, Chandandeep Singh. "Completely fair scheduler." Linux Journal 2009, no. 184 (2009): 4.

Padoin, Edson L., Márcio Castro, Laércio L. Pilla, Philippe OA Navaux, and Jean-François Méhaut. "Saving energy by exploiting residual imbalances on iterative applications." In High Performance Computing (HiPC), 2014 21st International Conference on, pp. 1-10. IEEE, 2014.

Pascual, Jose, Javier Navaridas, and Jose Miguel-Alonso. "Effects of topology-aware allocation policies on scheduling performance." In Job Scheduling Strategies for Parallel Processing, pp. 138-156. Springer Berlin/Heidelberg, 2009.

Perez-Gonzalez, Paz, and Jose M. Framinan. "A common framework and taxonomy for multicriteria scheduling problems with interfering and competing jobs: Multi-agent scheduling problems." European Journal of Operational Research 235, no. 1 (2014): 1-16.

Pinel, Frédéric, Johnatan E. Pecero, Pascal Bouvry, and Samee U. Khan. "A review on task performance prediction in multi-core based systems." In Computer and Information Technology (CIT), 2011 IEEE 11th International Conference on, pp. 615-620. IEEE, 2011.

Pinheiro, Eduardo, Ricardo Bianchini, Enrique V. Carrera, and Taliver Heath. "Load balancing and unbalancing for power and performance in cluster-based systems." In Workshop on compilers and operating systems for low power, vol. 180, pp. 182-195. 2001.

Pop, Florin, Ciprian Dobre, Gavril Godza, and Valentin Cristea. "A simulation model for grid scheduling analysis and optimization." In Parallel Computing in Electrical Engineering, 2006. PAR ELEC 2006. International Symposium on, pp. 133-138. IEEE, 2006.

Pooranian, Zahra, Mohammad Shojafar, Jemal H. Abawajy, and Ajith Abraham. "An efficient meta-heuristic algorithm for grid computing." Journal of Combinatorial Optimization 30, no. 3 (2015): 413-434.

Prokopec, Aleksandar, Nathan Grasso Bronson, Phil Bagwell, and Martin Odersky. "Concurrent tries with efficient non-blocking snapshots." In Acm Sigplan Notices, vol. 47, no. 8, pp. 151-160. ACM, 2012.

Qiao, Lin, Kapil Surlaker, Shirshanka Das, Tom Quiggle, Bob Schulman, Bhaskar Ghosh, Antony Curtis et al. "On brewing fresh espresso: Linkedin's distributed data serving platform." In Proceedings of the 2013 ACM SIGMOD International Conference on Management of Data, pp. 1135-1146. ACM, 2013.

Ramasubramanian, Manikandan, and Mukheem Ahmed. "Remote-direct-memory-access-based virtual machine live migration." U.S. Patent 9,619,270, issued April 11, 2017.

Ranjbar, Mohammad. "Solving the resource-constrained project scheduling problem using filter-and-fan approach." Applied mathematics and computation 201, no. 1 (2008): 313-318.

Ray, Biplob R., Morshed Chowdhury, and Usman Atif. "Is High Performance Computing (HPC) Ready to Handle Big Data?" In International Conference on Future Network Systems and Security, pp. 97-112. Springer, Cham, 2017.

Reddy, Reddivari Himadeep, Sri Krishna Kumar, Kiran Jude Fernandes, and Manoj Kumar Tiwari. "A Multi-Agent System based simulation approach for planning procurement operations and scheduling with multiple cross-docks." Computers & Industrial Engineering 107 (2017): 289-300.

Reiss, Charles, John Wilkes, and Joseph L. Hellerstein. "Obfuscatory obscanturism: making workload traces of commercially-sensitive systems safe to release." In Network Operations and Management Symposium (NOMS), 2012 IEEE, pp. 1279-1286. IEEE, 2012.

Reiss, Charles, John Wilkes, and Joseph L. Hellerstein. "Google cluster-usage traces: format+ schema." Google, Inc. Version of 2013.05.06, for trace version 2, 2013.

Ren, Gang, Eric Tune, Tipp Moseley, Yixin Shi, Silvius Rus, and Robert Hundt. "Google-wide profiling: A continuous profiling infrastructure for data centers." IEEE micro 30, no. 4 (2010): 65-79.

Richardson, Matthew, and Pedro Domingos. "The Intelligent surfer: Probabilistic Combination of Link and Content Information in PageRank." In NIPS, pp. 1441-1448. 2001.

Rodriguez, Maria Alejandra, and Rajkumar Buyya. "A taxonomy and survey on scheduling algorithms for scientific workflows in IaaS cloud computing environments." Concurrency and Computation: Practice and Experience 29, no. 8 (2017).

Roestenburg, Raymond, Rob Bakker, and Rob Williams. "Akka in action." Manning Publications Co., 2015.

Rybina, Kateryna, Waltenegus Dargie, Subramanya Umashankar, and Alexander Schill. "Modelling the live migration time of virtual machines." In OTM Confederated International Conferences" On the Move to Meaningful Internet Systems", pp. 575-593. Springer, Cham, 2015.

Salfner, Felix, Peter Tröger, and Andreas Polze. "Downtime analysis of virtual machine live migration." In The Fourth International Conference on Dependability (DEPEND 2011). IARIA, pp. 100-105. 2011.

Sapuntzakis, Constantine P., Ramesh Chandra, Ben Pfaff, Jim Chow, Monica S. Lam, and Mendel Rosenblum. "Optimizing the migration of virtual computers." ACM SIGOPS Operating Systems Review 36, no. SI (2002): 377-390.

Sargent, P. "Back to school for a brand new ABC." The Guardian (March, 12), no. 3 (1992): 12-28.

Sarood, Osman, Phil Miller, Ehsan Totoni, and Laxmikant V. Kale. "“Cool” Load Balancing for High Performance Computing Data Centers." IEEE Transactions on Computers 61, no. 12 (2012): 1752-1764.

Savill, John. "Mastering Windows Server 2016 Hyper-V." (2016).

Schaerf, Andrea, Yoav Shoham, and Moshe Tennenholtz. "Adaptive load balancing: A study in multi-agent learning." Journal of artificial intelligence research 2 (1995): 475-500.

Schirmer, Andreas. "A guide to complexity theory in operations research." No. 381. Manuskripte aus den Instituten für Betriebswirtschaftslehre der Universität Kiel, 1995.

Schwarzkopf, Malte, Andy Konwinski, Michael Abd-El-Malek, and John Wilkes. "Omega: flexible, scalable schedulers for large compute clusters." In Proceedings of the 8th ACM European Conference on Computer Systems, pp. 351-364. ACM, 2013.

Shankland, Stephen. "Google uncloaks once-secret server." CNET News, December 11, 2009. Available from: https://www.cnet.com/news/google-uncloaks-once-secret-server-10209580/ Retrieved December 17, 2016.

Sharma, Bikash, Victor Chudnovsky, Joseph L. Hellerstein, Rasekh Rifaat, and Chita R. Das. "Modeling and synthesizing task placement constraints in Google compute clusters." In Proceedings of the 2nd ACM Symposium on Cloud Computing, p. 3. ACM, 2011.

Shen, Siqi, Vincent van Beek, and Alexandru Iosup. "Statistical characterization of business-critical workloads hosted in cloud datacenters." In Cluster, Cloud and Grid Computing (CCGrid), 2015 15th IEEE/ACM International Symposium on, pp. 465-474. IEEE, 2015.

Shi, Dongcai, Jianwei Yin, Wenyu Zhang, Jinxiang Dong, and Dandan Xiong. "A distributed collaborative design framework for multidisciplinary design optimization." In International Conference on Computer Supported Cooperative Work in Design, pp. 294-303. Springer, Berlin, Heidelberg, 2005.

Shirinbab, Sogand, Lars Lundberg, and Dragos Ilie. "Performance comparison of KVM, VMware and XenServer using a large telecommunication application." In Cloud Computing. IARIA XPS Press, 2014.

Shiv, Kumar, Kingsum Chow, Yanping Wang, and Dmitry Petrochenko. "SPECjvm2008 performance characterization." Computer Performance Evaluation and Benchmarking (2009): 17-35.

Shreedhar, Madhavapeddi, and George Varghese. "Efficient fair queueing using deficit round robin." In ACM SIGCOMM Computer Communication Review, vol. 25, no. 4, pp. 231-242. ACM, 1995.

Singh, Ajit. "New York Stock Exchange Oracle Exadata – Our Journey." Oracle, Inc. November 17, 2017. Available from: http://www.oracle.com/technetwork/database/availability/con8821-nyse-2773005.pdf Retrieved June 28, 2018.

Smanchat, Sucha, and Kanchana Viriyapant. "Taxonomies of workflow scheduling problem and techniques in the cloud." Future Generation Computer Systems 52 (2015): 1-12.

Smarr, Larry, and Charles E. Catlett. "Metacomputing." Grid Computing: Making the Global Infrastructure a Reality (2003): 825-835.

Smith, Randall. "Docker Orchestration." Packt Publishing Ltd, 2017.

Sun, Gang, Dan Liao, Vishal Anand, Dongcheng Zhao, and Hongfang Yu. "A new technique for efficient live migration of multiple virtual machines." Future Generation Computer Systems 55 (2016): 74-86.

Sun, Shizhao, and Xiaoguang Liu. "EC-DNN: A new method for parallel training of deep neural networks." Neurocomputing 287 (2018): 118-127.

Tafa, Igli, Elinda Kajo, Ariana Bejleri, Olimpjon Shurdi, and Aleksandër Xhuvani. "The Performance between XEN-HVM, XEN-PV and OPEN-VZ During Live Migration." International Journal of Advanced Computer Science and Applications (2011): 126-132.

Tang, Xuehai, Zhang Zhang, Min Wang, Yifang Wang, Qingqing Feng, and Jizhong Han. "Performance evaluation of light-weighted virtualization for paas in clouds." In International Conference on Algorithms and Architectures for Parallel Processing, pp. 415-428. Springer, Cham, 2014.

Thain, Douglas, Todd Tannenbaum, and Miron Livny. "Distributed computing in practice: the Condor experience." Concurrency and computation: practice and experience 17, no. 2-4 (2005): 323-356.

Torvalds, Linus "Re: Just a second …" The Linux Kernel Mailing List. December 15, 2001. Available from http://tech-insider.org/linux/research/2001/1215.html Retrieved September 27, 2017.

Toshniwal, Ankit, Siddarth Taneja, Amit Shukla, Karthik Ramasamy, Jignesh M. Patel, Sanjeev Kulkarni, Jason Jackson et al. "Storm @Twitter." In Proceedings of the 2014 ACM SIGMOD international conference on Management of data, pp. 147-156. ACM, 2014.

Tsirkin, Michael, and Karen Noel. "Ordered memory pages transmission in virtual machine live migration." U.S. Patent 9,483,414, issued November 1, 2016.

Tuong, N. Huynh, Ameur Soukhal, and J-C. Billaut. "Single-machine multi-agent scheduling problems with a global objective function." Journal of Scheduling 15, no. 3 (2012): 311-321.

Tyagi, Rinki, and Santosh Kumar Gupta. "A Survey on Scheduling Algorithms for Parallel and Distributed Systems." In Silicon Photonics & High Performance Computing, pp. 51-64. Springer, Singapore, 2018.

Urma, Raoul-Gabriel, Mario Fusco, and Alan Mycroft. "Java 8 in Action: Lambdas, Streams, and functional-style programming." Manning Publications, 2014.

Varda, Kenton. "Protocol buffers: Google's data interchange format." Google Open Source Blog, July 7, 2008.

Vavilapalli, Vinod Kumar, Arun C. Murthy, Chris Douglas, Sharad Agarwal, Mahadev Konar, Robert Evans, Thomas Graves et al. "Apache hadoop yarn: Yet another resource negotiator." In Proceedings of the 4th annual Symposium on Cloud Computing, p. 5. ACM, 2013.

Verma, Akshat, Gautam Kumar, Ricardo Koller, and Aritra Sen. "Cosmig: Modeling the impact of reconfiguration in a cloud." In Modeling, Analysis & Simulation of Computer and Telecommunication Systems (MASCOTS), 2011 IEEE 19th International Symposium on, pp. 3-11. IEEE, 2011.

Verma, Abhishek, Luis Pedrosa, Madhukar Korupolu, David Oppenheimer, Eric Tune, and John Wilkes. "Large-scale cluster management at Google with Borg." In Proceedings of the Tenth European Conference on Computer Systems, p. 18. ACM, 2015.

Wang, Cheng, Qianlin Liang, and Bhuvan Urgaonkar. "An empirical analysis of amazon ec2 spot instance features affecting cost-effective resource procurement." ACM Transactions on Modeling and Performance Evaluation of Computing Systems (TOMPECS) 3, no. 2 (2018): 6.

Wang, Guanying, Ali R. Butt, Henry Monti, and Karan Gupta. "Towards synthesizing realistic workload traces for studying the hadoop ecosystem." In Modeling, Analysis & Simulation of Computer and Telecommunication Systems (MASCOTS), 2011 IEEE 19th International Symposium on, pp. 400-408. IEEE, 2011.

Wang, Gong, T. N. Wong, and Xiaohuan Wang. "A hybrid multi-agent negotiation protocol supporting agent mobility in virtual enterprises." Information Sciences 282 (2014): 1-14.

Weinberger, Edward. "Correlated and uncorrelated fitness landscapes and how to tell the difference." Biological cybernetics63, no. 5 (1990): 325-336.

Weiss, Gerhard. "Multiagent Systems. 2nd edition." MIT Press. 2013

White, Tom. "Hadoop: The definitive guide." O'Reilly Media, Inc. 2012.

Wickremasinghe, Bhathiya, and Rajkumar Buyya. "CloudAnalyst: A CloudSim-based tool for modelling and analysis of large scale cloud computing environments." MEDC project report 22, no. 6 (2009): 433-659.

Wilkes, John. "Cluster Management at Google with Borg." GOTO Berlin 2016. November 15, 2016.

Wong, C. S., I. K. T. Tan, R. D. Kumari, J. W. Lam, and W. Fun. "Fairness and interactive performance of O(1) and cfs linux kernel schedulers." In Information Technology, 2008. International Symposium on, vol. 4, pp. 1-8. IEEE, 2008.

Wooldridge, Michael, and Nicholas R. Jennings. "Intelligent agents: Theory and practice." The knowledge engineering review 10, no. 2 (1995): 115-152.

Wu, Yangyang, and Ming Zhao. "Performance modelling of virtual machine live migration." In Cloud Computing (CLOUD), 2011 IEEE International Conference on, pp. 492-499. IEEE, 2011

Wyai, Loh Chee, Cheah WaiShiang, and Marlene Valerie AiSiok Lu. "Agent Negotiation Patterns for Multi Agent Negotiation System." Advanced Science Letters 24, no. 2 (2018): 1464-1469.

Van der Meulen, Rob and Christy Pettey "Gartner Forecasts Worldwide Public Cloud Services Revenue to Reach $260 Billion in 2017." October 12, 2017. Available from: https://www.gartner.com/newsroom/id/3815165 Retrieved February 28, 2018.

Vohra, Deepak. "Scheduling pods on nodes. " In Kubernetes Management Design Patterns, pp. 199-236. Apress, Berkeley, CA. 2017.

Xu, Cheng-Zhong, and Brian Wims. "A mobile agent based push methodology for global parallel computing." Concurrency - Practice and Experience 12, no. 8 (2000): 705-726.

Xydas, Erotokritos, Charalampos Marmaras, and Liana M. Cipcigan. "A multi-agent based scheduling algorithm for adaptive electric vehicles charging." Applied energy 177 (2016): 354-365.

Vagata, Pamela, and Kevin Wilfong. "Scaling the Facebook data warehouse to 300 PB." Facebook, Inc. April 10, 2014. Available from: https://code.fb.com/core-data/scaling-the-facebook-data-warehouse-to-300-pb/ Retrieved June 28, 2018.

Yang, Yongjian, Yajun Chen, Xiaodong Cao, and Jiubin Ju. "Load balancing using mobile agent and a novel algorithm for updating load information partially." In Lecture Notes in Computer Science 3619, pp. 1243-1252. 2005.

Yin, Yunqiang, Shuenn-Ren Cheng, T. C. E. Cheng, Wen-Hung Wu, and Chin-Chia Wu. "Two-agent single-machine scheduling with release times and deadlines." International Journal of Shipping and Transport Logistics 5, no. 1 (2013): 75-94.

Yoo, Andy B., Morris A. Jette, and Mark Grondona. "Slurm: Simple linux utility for resource management." In Workshop on Job Scheduling Strategies for Parallel Processing, pp. 44-60. Springer, Berlin, Heidelberg, 2003.

Younge, Andrew J., Gregor Von Laszewski, Lizhe Wang, Sonia Lopez-Alarcon, and Warren Carithers. "Efficient resource management for cloud computing environments." In Green Computing Conference, 2010 International, pp. 357-364. IEEE, 2010.

Yu, Chenying, and Fei Huan. "Live migration of docker containers through logging and replay." In Advances in Computer Science Research, 3rd International Conference on Mechatronics and Industrial Informatics, pp. 623-626. Atlantis Press, 2015.

Yu, Jia, and Rajkumar Buyya. "A taxonomy of scientific workflow systems for grid computing." ACM Sigmod Record 34, no. 3 (2005): 44-49.

Zaharia, Matei, Dhruba Borthakur, J. Sen Sarma, Khaled Elmeleegy, Scott Shenker, and Ion Stoica. Job scheduling for multi-user mapreduce clusters. Vol. 47. Technical Report UCB/EECS-2009-55, EECS Department, University of California, Berkeley, 2009.

Zaharia, Matei, Mosharaf Chowdhury, Michael J. Franklin, Scott Shenker, and Ion Stoica. "Spark: Cluster computing with working sets." HotCloud 10, no. 10-10 (2010): 95.

Zaharia, Matei, Mosharaf Chowdhury, Tathagata Das, Ankur Dave, Justin Ma, Murphy McCauley, Michael J. Franklin, Scott Shenker, and Ion Stoica. "Resilient distributed datasets: A fault-tolerant abstraction for in-memory cluster computing." In Proceedings of the 9th USENIX conference on Networked Systems Design and Implementation, pp. 2-2. USENIX Association, 2012.

Zakarya, Muhammad, and Lee Gillam. "Energy efficient computing, clusters, grids and clouds: A taxonomy and survey." Sustainable Computing: Informatics and Systems 14 (2017): 13-33.

Zecevic, Petar, and Marko Bonaci. "Spark in Action." (2016).

Zhang, Dongli, Moussa Ehsan, Michael Ferdman, and Radu Sion. "DIMMer: A case for turning off DIMMs in clouds." In Proceedings of the ACM Symposium on Cloud Computing, pp. 1-8. ACM, 2014.

Zhang, Jiao, Fengyuan Ren, Ran Shu, Tao Huang, and Yunjie Liu. "Guaranteeing delay of live virtual machine migration by determining and provisioning appropriate bandwidth." IEEE Transactions on Computers 65, no. 9 (2016): 2910-2917.

Zhang, Qi, Joseph L. Hellerstein, and Raouf Boutaba. "Characterizing task usage shapes in Google's compute clusters." In Proceedings of the 5th international workshop on large scale distributed systems and middleware, pp. 1-6. sn, 2011.

Zhang, Zhuo, Chao Li, Yangyu Tao, Renyu Yang, Hong Tang, and Jie Xu. "Fuxi: a fault-tolerant resource management and job scheduling system at internet scale." Proceedings of the VLDB Endowment 7, no. 13 (2014): 1393-1404.

Zhao, Ming, and Renato J. Figueiredo. "Experimental study of virtual machine migration in support of reservation of cluster resources." In Proceedings of the 2nd international workshop on Virtualization technology in distributed computing, p. 5. ACM, 2007.

Zhou, Jingren. "Big data analytics and intelligence at alibaba cloud." In Proceedings of the Twenty-Second International Conference on Architectural Support for Programming Languages and Operating Systems, pp. 1-1. ACM, 2017.

Zhu, Xiaomin, Chao Chen, Laurence T. Yang, and Yang Xiang. "ANGEL: Agent-based scheduling for real-time tasks in virtualized clouds." IEEE Transactions on Computers 64, no. 12 (2015): 3389-3403.